\documentclass{aa}
\usepackage{graphicx}
\usepackage{txfonts}
\usepackage{natbib}
\usepackage{booktabs}
\usepackage[bookmarks=true,bookmarksopen=true,pdfstartview = FitV,pdfpagelayout = SinglePage,colorlinks = true,linkcolor=blue,citecolor=blue,urlcolor = blue]{hyperref} 
\bibpunct{(}{)}{;}{a}{}{,}
\let\oldbibliography\thebibliography
\renewcommand{\thebibliography}[1]{
  \oldbibliography{#1}
  \setlength{\itemsep}{0pt}
} 
\newcommand{\sgra}{Sgr~A*}
\newcommand{\magn}{SGR~J1745-29}
\DeclareMathOperator\erf{erf}

\begin{document}

\title{Sixteen years of X-ray monitoring of Sagittarius~A*: \\
Evidence for a decay of the faint flaring rate from 2013 August, \\13 months before a rise in the bright flaring rate}
\author{Enmanuelle~Mossoux \inst{\ref{inst1},\ref{inst2}}
\and Nicolas~Grosso \inst{\ref{inst1}}} 
\institute{Observatoire astronomique de Strasbourg, Universit\'e de Strasbourg, CNRS, UMR 7550, 11 rue de l'Universit\'e, F-67000 Strasbourg, France \label{inst1}
\and Groupe d'Astrophysique des Hautes Energies, Institut d'Astrophysique et de G\'eophysique, Universit\'e de Li\`ege, All\'ee du 6 Août, 19c, B\^at B5c, 4000 Li\`ege, Belgium \label{inst2}}
\date{Received  / Accepted  }

\abstract
{ X-ray flaring activity from the closest supermassive black hole \sgra{} located at the center of our Galaxy has been observed since 2000 October\ 26 thanks to the current generation of X-ray facilities.
Recently, in a study of X-ray flaring activity from \sgra{} using Chandra and XMM-Newton public observations from 1999 to 2014 and Swift monitoring in 2014, researchers have argued that the ``bright and very bright'' flaring rate has increased from 2014 August\ 31.}
{As a result of additional observations performed in 2015 with Chandra, XMM-Newton, and Swift (total exposure of 482$\,$ks), we seek to test the significance and persistence of this increase of flaring rate and to determine the threshold of unabsorbed flare flux or fluence leading to any change of flaring rate.}
{We reprocessed the Chandra, XMM-Newton, and Swift data from 1999 to 2015 November\ 2.
From these data, we detected the X-ray flares via the two-step Bayesian blocks algorithm with a prior on the number of change points properly calibrated for each observation.
We improved the Swift data analysis by correcting the effects of the target variable position on the detector and we detected the X-ray flares with a $3\sigma$ threshold on the binned light curves.
The mean unabsorbed fluxes of the 107 detected flares were consistently computed from the extracted spectra and the corresponding calibration files, assuming the same spectral parameters.
We constructed the observed distribution of flare fluxes and durations from the XMM-Newton and Chandra detections.
We corrected this observed distribution from the detection biases to estimate the intrinsic distribution of flare fluxes and durations.
From this intrinsic distribution, we determined the average flare detection efficiency for each XMM-Newton, Chandra, and Swift observation.
We finally applied the Bayesian blocks algorithm on the arrival times of the flares corrected from the corresponding efficiency.}
{We confirm a constant overall flaring rate from 1999 to 2015 and a rise in the flaring rate by a factor of three for the most luminous and most energetic flares from 2014 August\ 31, i.e., about four months after the pericenter passage of the Dusty S-cluster Object (DSO)/G2 close to \sgra{}.
In addition, we identify a decay of the flaring rate for the less luminous and less energetic flares from 2013 August and November, respectively, i.e., about 10 and 7 months before the pericenter passage of the DSO/G2 and 13 and 10 months before the rise in the bright flaring rate.}
{The decay of the faint flaring rate is difficult to explain in terms of the tidal disruption of a dusty cloud since it occurred well before the pericenter passage of the DSO/G2, whose stellar nature is now well established.
Moreover, a mass transfer from the DSO/G2 to \sgra{} is not required to produce the rise in the bright flaring rate since the energy saved by the decay of the number of faint flares during a long period of time may be later released by several bright flares during a shorter period of time.
}


\titlerunning{Sixteen years of X-ray monitoring of Sagittarius~A*}
\maketitle

\section{Introduction} 
The center of the Milky Way hosts the closest supermassive black hole (SMBH) named \sgra{} at a distance of 8$\,$kpc \citep{genzel10,falcke13}.
The bolometric luminosity of \sgra{} is $10^{-9}$ times smaller than the Eddington luminosity $\mathrm{\textit{L}_{Edd}}=3\times 10^{44}\ \mathrm{erg}\,\mathrm{s^{-1}}$ \citep{yuan03} for a SMBH mass of $M=4 \times 10^6\ \mathrm{\textit{M}_{\sun}}$ \citep{schodel02,ghez08,gillessen09}.
Above this steady emission, \sgra{} experiences some temporal increases of flux in X-rays \citep[e.g.,][]{baganoff01,porquet03,porquet08,neilsen13}, near-infrared \citep[e.g.,][]{genzel03,yusef-zadeh06b,dodds-eden09,witzel12} and sub-millimeter/radio \citep[e.g.,][]{zhao03,yusef-zadeh06c,yusef-zadeh08,marrone08,yusef-zadeh09}.
The near-infrared flare spectra are well reproduced by the synchrotron process \citep{eisenhauer05,eckart06} and the sub-millimeter/radio may be explained by the adiabatically expanding plasmon model \citep{vanderlaan66,yusef-zadeh06c}, but the radiative processes for the creation of the X-ray activity are still debated.
Moreover, several mechanisms can explain the origin of eruptions in X-rays and infrared: a shock produced by the interaction between orbiting stars and hot accretion flow \citep{nayakshin03,nayakshin04}, 
a hotspot model \citep{broderick05,eckart06b,meyer06,trippe07}, a Rossby instability producing magnetized plasma bubbles in the hot accretion flow \citep{tagger06,liu06}, 
an additional heating of electrons near the black hole due to processes such as accretion instability or magnetic reconnection \citep{baganoff01,markoff01,yuan03,yuan09}, 
an increase of accretion rate when some fresh material reaches the close environment of the black hole \citep{yuan03,czerny13}, and tidal disruption of asteroids \citep{cadez06,cadez08,kostic09,zubovas12}.

The study of a large number of flares is valuable to constrain the radiative processes and emission mechanisms at the origin of the flaring activity from \sgra{}.
Moreover, the survey of the Dusty S-cluster Object (DSO)/G2 on its way toward \sgra{} has increased the number of observations of the SMBH.
\citet{valencias15} showed that DSO/G2 is an 1--2$\,M_{\odot}$ pre-main sequence star with an accretion disk producing the Br$\gamma$ emission line by magnetospheric accretion onto the stellar photosphere and has survived to its pericenter passage at about 2000$\,$R$_\mathrm{s}$ from \sgra{} on 2014 April\ 20 (2014 March\ 1--2014 June\ 10).

The first statistical study on the X-ray flares of \sgra{} was made by \citet{neilsen13} thanks to the 2012 Chandra X-ray Visionary Project (\textit{XVP}).
During this campaign 39 flares with a 2--10$\,$keV observed luminosity larger than $10^{34}\,\mathrm{erg}\,\mathrm{s^{-1}}$ were detected using a Gaussian flare fitting on the light curves binned on 300$\,$s, resulting in an observed X-ray flaring rate of $1.1^{+0.2}_{-0.1}$ flare per day.
The flares detected with the Gaussian fitting method are limited to a minimum duration of 400$\,$s and a minimum peak count rate of $0.015\,$ACIS-S3$\mathrm{\,count\,s^{-1}}$ (corresponding to a mean flux in 2--8$\,$keV of $0.6\times 10^{-12}\,\mathrm{erg\,s^{-1}\,cm^{-2}}$ with their spectral parameters)  due to the Poissonian noise of the non-flaring light curve and the limitations that the authors put on their Gaussian shape to avoid any spurious detection.
\citet{neilsen13} also tested the Bayesian blocks algorithm used by \citet{nowak12} to analyze the individual photon arrival times (Scargle 2002, priv.\ comm.) and detected 45 flares, 34 of which were also found with their Gaussian fitting method.

\citet{ponti15} studied the flaring rate with the Python implementation of the Bayesian blocks algorithm\footnote{This Python program can be found at \href{https://jakevdp.github.io/blog/2012/09/12/dynamic-programming-in-python/}{https://jakevdp.github.io/blog/2012/09/12/dynamic-programming-in-python/}} \citep{scargle13} by merging the XMM-Newton and Chandra observations where \sgra{} was observed with an off-axis angle lower than $2\arcmin$ from September\ 1999 to October\ 2014 and the 2014 Swift observations. 
These authors reported an increase of the bright and very bright flaring rate (corresponding arbitrarily to flares with an absorbed fluence larger than $50\times 10^{-10}\ \mathrm{erg\ cm^{-2}}$) from 2014 August\ 31 until the end of the 2014 X-rays observations on November\ 2 with a level of $2.52$ flare per day, i.e., $9.3$ times larger than the bright flaring rate observed from 1999 to 2014 August\ 31.
However, they only used the 2014 Swift monitoring to determine the change of flaring rate; but from 2006 to 2013, six X-ray flares were detected during the Swift monitoring of 985$\,$ks \citep{degenaar15}, which should be included to investigate the significance of the detection of the flaring rate change.

Finally, \citet{yuan16} also carried out statistical studies on the X-ray flares observed by Chandra from 1999 to 2012.
They detected the X-ray flares using a Gaussian fitting on the individual photon arrival times.
The detection efficiency of this method was presented in their Fig.~3 as a function of the flare duration and fluence.
This method becomes more efficient as the flare fluence increases but is less efficient for the detection of flares longer than 10$\,$ks.

However, these different studies present several issues in their data analyses, especially for  flare detection.
Firstly, the authors never correct of the bias of their detection methods\footnote{In \citet{yuan16}, the efficiency of their detection method is computed but never used.}, which would lead to an intrinsic flaring rate that is higher than those observed.
Indeed, all of the detection methods presented above are less efficient in the detection of the faintest and shortest flares.
This issue is very important for the simultaneous study of data from different instruments (for example, in \citealt{ponti15} and \citealt{yuan16}) that have different sensitivities and angular resolutions leading to different efficiencies for  flare detection and an incoherence in the overall flaring rate.

Secondly, the Bayesian blocks method uses a prior on the number of changes of the flaring rate (named change point by \citealt{scargle98}) to control the rate of false positive detections.
As stated by \citet{scargle13}, this prior ``depends on only the number of data points and the adopted value of [false positive rate]''.
It thus needs to be calibrated using simulations of event lists containing the same number of counts as those studied.
The Python implementation of the Bayesian blocks algorithm used by \citet{ponti15} works with the geometric prior given in Eq.~21 of \citet{scargle13}.
However, as told by \citet{scargle13}, this geometric prior was obtained for a given range of number of events (which was unfortunately unspecified).
This may explain the inconsistency between the false positive rate adopted by \citet{ponti15} and their resulting false detection probability appearing in their Sect.~5.5.
Indeed, they tested how many times a spurious change of flaring rate is detected by simulating event lists containing the same number of flare arrival times drawn in a uniform distribution and applying the Bayesian blocks algorithm to determine how many times a change of flaring rate is detected.
Using a false positive rate of 0.3\% and the geometric prior, they reported a probability of false detection of 0.1\%, which points out an unreliable calibration between the false positive rate and the prior.

Thirdly, \citet{ponti15} used WebPIMMS for the computation of the flare flux.
But WebPIMMS considers the effective area and the redistribution matrix computed for an on-axis source and for the full detector field of view.
However, the flare spectra were extracted from circular regions of 1.25 or 10$\arcsec$ radius centered on \sgra{}.
Since the point spread function (PSF) extraction fraction is not corrected by WebPIMMS, the inferred unabsorbed flux is systematically underestimated by these authors.
Finally, none of these previous works studied the impact on the flare detection efficiency of the overlap between the flare duration and observing time, i.e., the edge effects when a flare begins before the observation start or ends after the observation stop.

In this work, we use the two-step Bayesian blocks method \citep{mossoux14,mossoux15} with a proper prior calibration since we believe this method to be most efficient for flare detection.
Indeed, contrary to the Gaussian fitting method used by \citet{neilsen13}, the Bayesian blocks method is applied directly on the event lists and is able to detect flares that are shorter than $400\,$s (see Fig.~A.2 of \citealt{mossoux16}).
Moreover, comparing the efficiency of the method of \citet{yuan16} with those of the Bayesian blocks method, we stress that the Bayesian blocks method is more efficient for the detection of long flares. 
For the shortest and faintest flares, the method of \citet{yuan16} detects more features than the Bayesian blocks method but these authors did not control their false positive rate.
We also determined the flare detection efficiency by taking the edge effects into account in our simulations.
We also use the spectral fitting program ISIS \citep{isis} and the effective area and redistribution matrix files associated with the spectrum extraction region to consistently compute the mean unabsorbed flux of the X-ray flares.

Owing to the 2015 Swift monitoring and 2015 Chandra and XMM-Newton observations, there are about 459$\,$ks of additional observations of \sgra{} allowing us to investigate the persistence and significance of the bright flaring rate argued by \citet{ponti15} based on only 200$\,$ks of observations from 2014 August\ 31.
After reducing the 1999--2015 data of XMM-Newton, Chandra, and Swift (Sect.~\ref{observation}), we search for flares using the two-step Bayesian blocks algorithm \citep{mossoux14,mossoux15} for XMM-Newton and Chandra and the method proposed by \citet{degenaar13} that is optimized for the Swift observations (Sect.~\ref{flare}).
We then compute their mean unabsorbed fluxes with the spectral parameters computed by \citet{nowak12} for the brightest X-ray flares (Sect.~\ref{flux}).
This method of taking the effects of the off-axis angle into account allows us to study a large number of observations without a drastic limitation on the off-axis angle of \sgra{}.
To correct the flare detection bias for each observation, we compute the flux and duration distribution of the flares observed with XMM-Newton and Chandra and correct it from the merged detection efficiency of the Bayesian block algorithm to determine the intrinsic flux-duration distribution (Sect.~\ref{distr:distriiiiiiiiiii}).
From this intrinsic distribution, we compute the average flare detection efficiency associated with each XMM-Newton, Chandra, and Swift observation and investigate the existence of a flux or fluence threshold leading to a change in the unbiased X-ray flaring rate observed from 1999 to 2015 using the Bayesian blocks algorithm and the relevant prior calibration (Sect.~\ref{flare_rate}).
We discuss the physical origin of a change of flaring rate in Sect.~\ref{discussion} and summarize our results in Sect.~\ref{conclusion}.

\section{Observations and data reduction}
\label{observation}
In this work, we extend the flaring analysis to the 1999--2015 XMM-Newton and Chandra observations, where \sgra{} was observed with an off-axis angle lower than 8$\arcmin,$ and to the overall 2006--2015 Swift observations since \sgra{} was mainly observed with an off-axis angle lower than 8$\arcmin$.
Theoretically, our data reduction and analysis methods do not have any limitations on the off-axis angle but considering larger off-axis angles might lead to more confusion with the diffuse emission of the Galactic center.
We retrieved the public observations of \sgra{} made with XMM-Newton, Chandra, and Swift from the XMM-Newton Science Archive (XSA)\footnote{\href{http://www.cosmos.esa.int/web/xmm-newton/xsa}{http://www.cosmos.esa.int/web/xmm-newton/xsa}}, the Chandra Search and Retrieval interface (ChaSeR)\footnote{\href{http://cda.harvard.edu/chaser}{http://cda.harvard.edu/chaser}} and the Swift Archive Download Portal\footnote{\href{http://www.swift.ac.uk/swift\_portal}{http://www.swift.ac.uk/swift\_portal}}, respectively.
Our XMM-Newton, Chandra, and Swift data sample has a total exposure time that is about $2.1\,$Ms longer than the $6.9\,$Ms considered previously.

\subsection{XMM-Newton observations}
XMM-Newton \citep{jansen01} has observed the Galactic center since 2000 September with the EPIC/pn \citep{strueder01} and EPIC/MOS1 and MOS2 \citep{turner01} cameras.
The 54 observations of \sgra{} from 2000 September\ to 2015 April\ have a total effective exposure of about $2.2\ \mathrm{Ms}$. 
The observation start and end times corresponding to the earliest good time intervals (GTI) start and the latest GTI stop of the three cameras are reported in Table \ref{table:xmm} in Universal Time (UT).
The conversion from the Terrestrial Time (TT) registered aboard XMM-Newton to UT is computed using NASA's HEASARC Tool xTime\footnote{The website of xTime is: \href{http://heasarc.gsfc.nasa.gov/cgi-bin/Tools/xTime/xTime.pl}{http://heasarc.gsfc.nasa.gov/cgi-bin/Tools/xTime/xTime.pl}}.
The duration of the observations reported in Table \ref{table:xmm} is the sum of the GTI.
Most of the observations were made in frame window mode with the medium filter\footnote{Exceptions are the 2000 September\ 21, 2001 September\ 4, 2004 March\ 28 and 30 observations, where EPIC/pn was in frame window extended mode leading to a lower time resolution (199.1$\,$ms instead of 73.4$\,$ms);
the 2014 April\ 3 observations, where EPIC/MOS1 and MOS2 observed in small window mode leading to a better time resolution (0.3$\,$s instead of 2.6$\,$s) but a smaller part of the central CCD observing ($100 \times 100$~pixels);
the 2002 February\ 26 and October\ 3 observations, where EPIC/pn observed with the thick filter and the 2008 March\ 3 and September\ 23, where the three cameras observed with the thin filter.}.

The XMM-Newton data reduction is the same as presented in, for example, \citet{mossoux14}.
We created the event lists for the MOS and pn cameras using the \texttt{emchain} and \texttt{epchain} tasks from the Science Analysis Software (SAS) package (version 14.0; Current Calibration files of 2015 June 13).
We suppressed the time ranges when the soft-proton flare count rate in the full detector light curve in the $2-10\,$keV energy range is larger than $0.009$ and $0.004\ \mathrm{count\ s^{-1}\ arcmin^{-2}}$ for pn and MOS, respectively.
For the MOS cameras, we selected the single, double, triple, and quadruple events (\texttt{PATTERN$\leq 12$}) and used the bit mask \texttt{\#XMMEA\_SM} to reject the dead columns and bad pixels.
For the pn camera, we selected the single and double events (\texttt{PATTERN$\leq 4$}) and used the more drastic bit mask \texttt{FLAG==0} to reject the dead columns and bad pixels.

The source+background (src+bkg) extraction region is a $10\arcsec$-radius disk centered on the Very-Long-Baseline Interferometry (VLBI) radio position of \sgra: RA(J2000)=$17^{\mathrm{h}}45^{\mathrm{m}}40\fs{}0409$, Dec(J2000)=$-29^{\circ} 00\arcmin 28\farcs{}118$ \citep{reid99}.
This region allows us to extract 50\% of the energy at $1.5\,$keV on-axis.
We did not register the EPIC coordinates again since the absolute astrometry for the EPIC cameras ($1.2\arcsec$; \citealt{EPIC_calibration_status_document}) is very small compared to the size of this extraction region and the PSF half power diameter (HPD).

For observations in frame window (extended) mode, the bkg extraction region is a $\approx 3\arcmin \times 3\arcmin$ region at $\approx 4\arcmin$-north of \sgra{}.
For observations in small window mode, the background extraction region is a $\approx 3\arcmin \times 3\arcmin$ area at $\approx 7\arcmin$-east of \sgra{} (i.e., on the adjacent CCD).
The X-ray sources in the background region were detected using the SAS task \texttt{edetect\_chain} and filtered out.

\subsection{Chandra observations}
Chandra has observed the Galactic center since 1999 September\ with the ACIS-I and ACIS-S cameras \citep{garmire03}.
The 121 observations of \sgra{} from 1999 September\ to 2015 October\ have a total effective exposure of about $5.8\ \mathrm{Ms}$.
The effective observation start and end times reported in Table~\ref{table:chandra1} in UT correspond to the earliest GTI start and the latest GTI stop.
The ACIS-S observations of the 2012 \textit{XVP} campaign, i.e.,  2013 May 25, and June 6 and 9 and  2015 August\ 11, were made with the High Energy Transmission Grating (HETG), which disperses the source events on the detector.
The ACIS-S observations on 2013 May 12, June 4 and after 2013 July 2 were made with an 1/8 subarray of 128 rows to increase the time resolution in order to reduce the pile-up during the bright flare.
The other observations were made with ACIS-I.

The data reduction was carried out with the Chandra Interactive Analysis of Observations (CIAO) package (version 4.7) and the calibration database (CALDB; version 4.6.9).
The level 1 data were reprocessed via the CIAO script \texttt{chandra\_repro,} which creates a bad pixel file, flags afterglow events, and filters the event patterns, afterglow events, and bad-pixel events.
For observations without HETG, the src+bkg events were extracted from a $1\farcs25$-radius disk centered on the VLBI radio position of \sgra{}.
For the HETG observations, the diffraction order was determined with the CIAO task \texttt{tg\_resolve\_events}.
We then extracted the zero-order events from the $1\farcs25$-radius disk centered on \sgra{} and the $\pm1$-order events from wide rectangle of $2\farcs5$ width centered on the \sgra{} position \citep{nowak12,neilsen13}.
The position angles of the dispersed spectra are given in the region extension of the level 1 data event list.
The bkg region is a $8\farcs2$ disk at $0\farcm54$ south of \sgra{}.

\subsection{Swift observations}
Swift \citep{gherels04} has regularly observed the Galactic center since 2006 with the X-ray telescope (XRT) (PI: N.~Degenaar).
This camera observes between $0.2$ and 10$\,$keV in windowed timing mode or photon counting mode depending on the brightness of the source.
The former observing mode uses only 1D imaging to increase the timing resolution of the data, whereas the latter observing mode delivers the 2D photon positions for the entire XRT field of view ($23\farcm6 \times23\farcm6$) with a time resolution of $2.5\,$s.
The XRT has an effective area of 110$\ \mathrm{cm^2}$ at $1.5\,$keV, an absolute-astrometry uncertainty of 3$\arcsec$, and a spatial resolution of $18\arcsec$ HPD on-axis at 1.5$\,$keV \citep{burrows05}.
The log of each yearly campaign is given in Table~\ref{table:swift}.

\subsubsection{Swift monitoring of \sgra{}}
The results of the Swift monitoring of \sgra{} until 2011 October 25 were reported in \citet{degenaar13}.
The authors computed the mean of the light curve of \sgra{} between $0.3$ and 10$\,$keV of $0.011$~count~s$^{-1}$ with a standard deviation of $\sigma=6.7\times 10^{-3}\ \mathrm{counts}~\mathrm{s^{-1}}$ .
Six X-ray flares with an unabsorbed luminosity larger than $7 \times 10^{34}\ \mathrm{erg}~\mathrm{s^{-1}}$ were observed during these 821~ks of observation using a GTI-binned detection method with a $3\sigma$ threshold leading to a flaring rate of 0.63 flares per day.
The results of the 2012, 2013, and 2014 Swift monitoring were reported in \citet{degenaar15}.
One flare was observed on 2014 September\ 9 with an unabsorbed luminosity of $(1.4\pm0.4) \times 10^{35}\ \mathrm{erg}~\mathrm{s^{-1}}$ during the 510~ks of these three years of observations.

On 2016 February\ 6, a new X-ray transient SWIFT~J174540.7-290015 was detected at 16$''$ north of \sgra{} with a 2--10$\,$keV flux of $1.0\times 10^{-10}\,\mathrm{erg\, s^{-1}}$ \citep{reynolds16}.
This source was identified as a low-mass X-ray binary located near or beyond the Galactic center \citep{ponti16}.
On 2016 May 28, a new X-ray transient SWIFT~J174540.2-290037 was detected in the Swift observations at 10$''$ south of \sgra{} with an unabsorbed 2--10$\,$keV flux of about $(7\pm2)\times10^{-11}\,\mathrm{erg\,s^{-1}\,cm^{-2}}$ \citep{degenaar16}.
Since these two new transient sources have a large X-ray flux showing long-term variations, they contaminate the \sgra{} light curves observed by Swift.
The large flux variations observed in the short-exposure light curve from \sgra{} may thus not be identified as a \sgra{} flare or an accretion burst from the transient sources.
We thus only use the Swift observations from 2006 to 2015 to study the \sgra{} flares.

\begin{figure}
\centering
\includegraphics[width=4.4cm,angle=90]{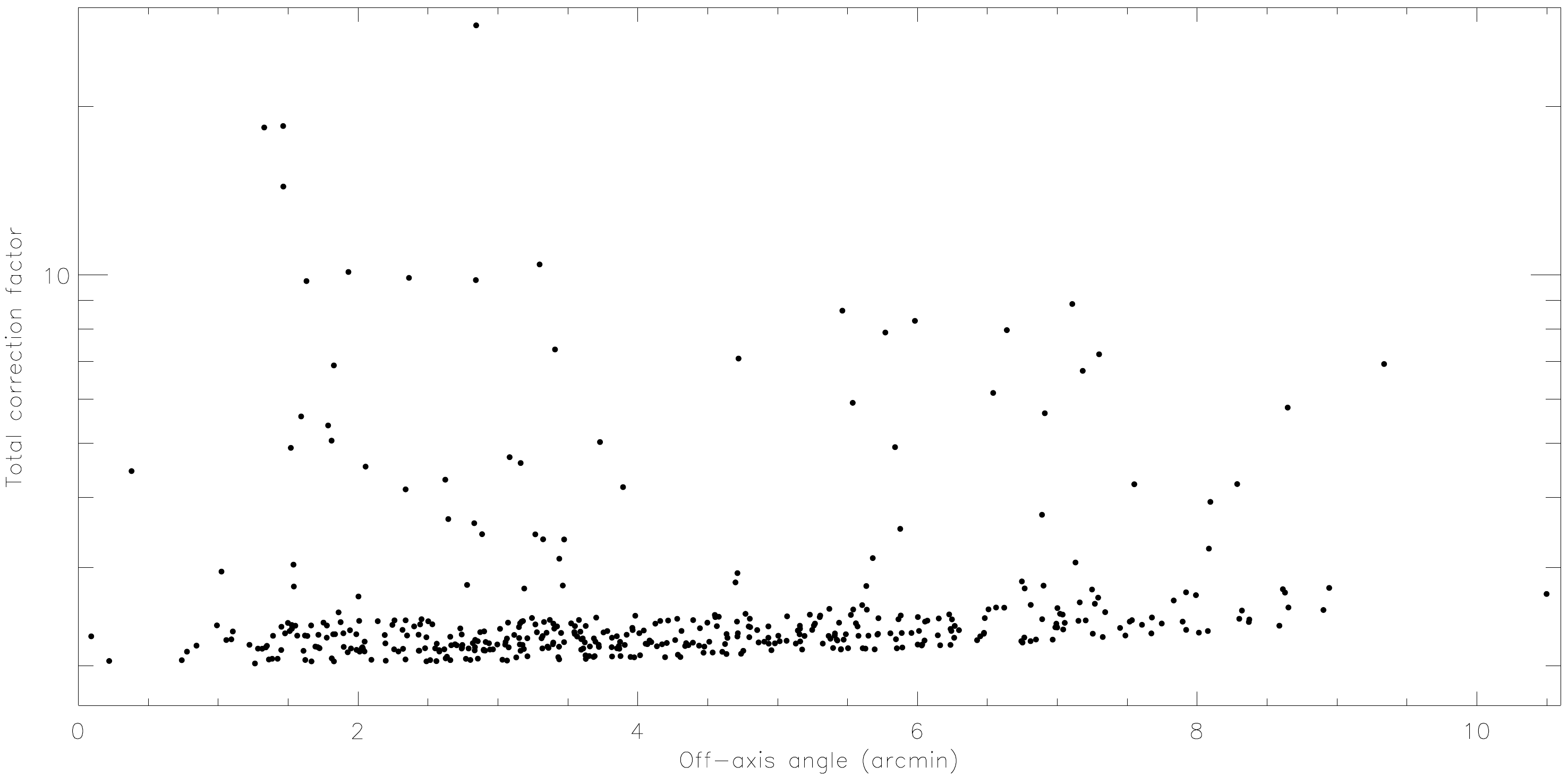}
\caption[Total correction factor of the Swift count rate]{Total correction factor (including bad pixels and dead columns, PSF extraction fraction, and vignetting) for the Swift count rate of \sgra{} vs. off-axis angle for all Swift observations of the Galactic center from 2006 to 2015.}
\label{corr_fact_sw}
\medskip
\centering
\includegraphics[width=4.6cm,angle=90]{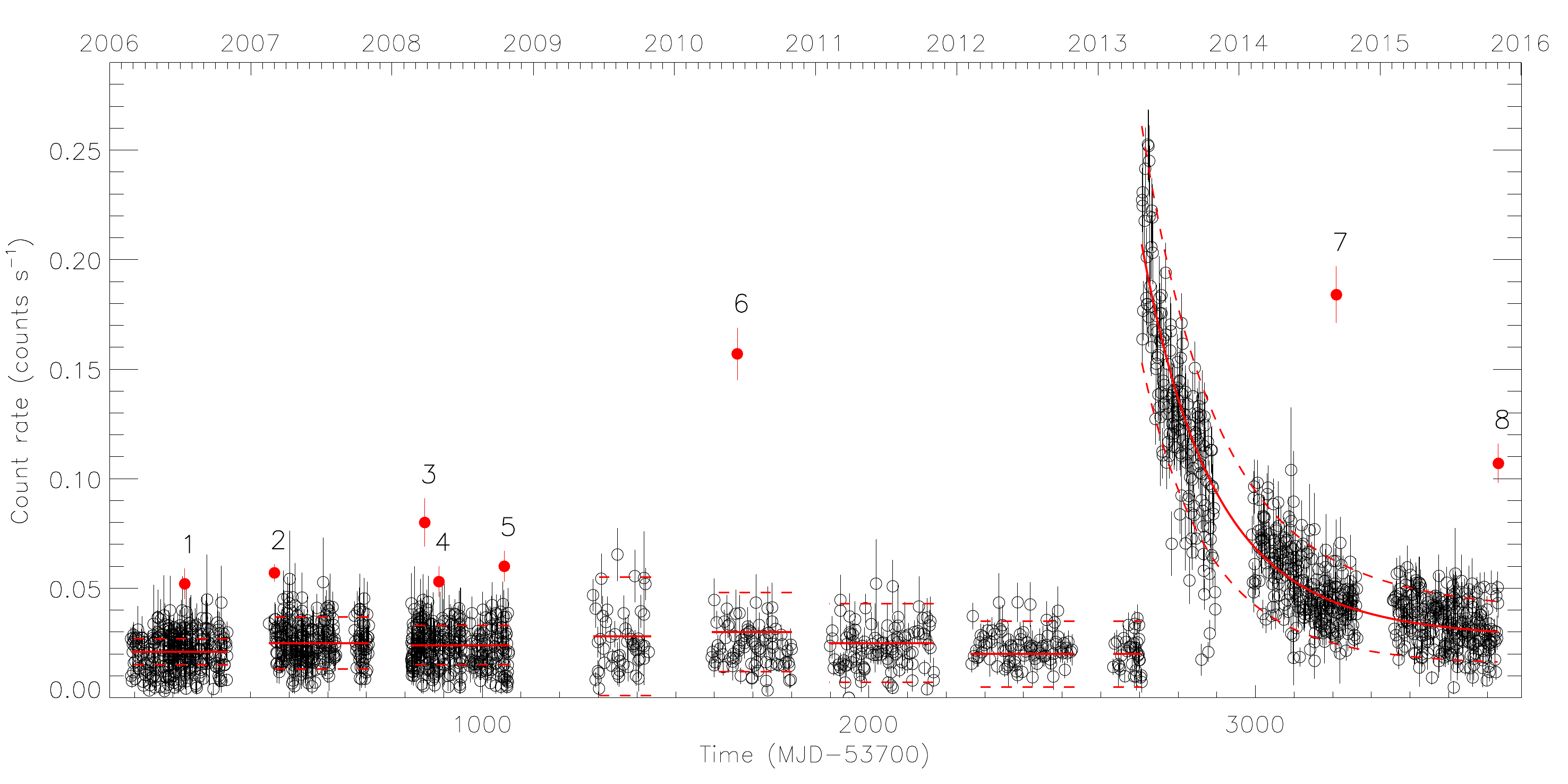}
\caption[Swift light curve of \sgra{} from 2006 to 2015]{Swift/XRT light curve of \sgra{} from 2006 to 2015.
The red points are the X-ray flares with the label corresponding to the flare number in Table~\ref{table:swift}.
The red lines are the non-flaring level of each yearly campaign with their $3\sigma$ threshold for the flare detection in dashed lines.}
\label{fig:swift}
\end{figure}

\subsubsection{Improving the data reduction method}
We reprocessed the level 1 data of the Swift observations made in photon counting mode with the data reduction method of \citet{degenaar13}.
We used the HEASOFT task \texttt{XRTPIPELINE} (v$0.13.1$) and the calibration files released on 2014 June 12 to reject the hot and bad pixels and select the grades between 0 and 12.
From the resulting level 2 data, we used the HEASOFT task \texttt{XSelect} (v$2.4$c) to extract events recorded in a disk of 10$\arcsec$ radius centered on the VLBI radio position of \sgra{}.
Since Swift is on a low-Earth orbit located below the radiation belts, the instrumental background caused by the soft-proton flares is negligible and we thus do not need a background extraction region. 

The target position on the Swift detector is not fixed. Indeed, the off-axis angle of \sgra{} can be as large as 10$\farcm$5 (corresponding to the edge of the field of view), leading to an increase of the PSF width and vignetting; moreover, \sgra{} may be located close to a bad column or bad pixel causing event losses.
To improve the \citet{degenaar13} data reduction, we thus correct the event losses from the variable PSF and vignetting at $2.77\,$keV (the median energy emitted in the 10$\arcsec$ extraction region) running the HEASOFT task \texttt{XRTLCCORR} (v0.3.8).
This task computes the correction factors that have to be applied to the light curve count rates for each 10$\,$s interval.
Figure~\ref{corr_fact_sw} shows the mean correction factor computed for each Swift observation as a function of the off-axis angle of \sgra{}.
This correction factor is different from one observation to an other, varying from 2 to 24.
The correction factor is minimum on-axis with a slightly increasing trend with the off-axis angle because of the increase of the PSF width and the vignetting.
The mean value of the correction factor is $2.8$ but the correction factor increases when \sgra{} is located close to a bad column or pixel leading to a large standard deviation of the correction factor ($2.1$).
Applying the correction factors on the count rates from \sgra{} leads to a higher non-flaring level compared to those computed in \citet{degenaar13}: for the observations between 2006 and 2011, when there is no contamination by transient sources, we find an average count rate level of about $0.027\pm0.004\ \mathrm{counts\,s^{-1}}$ in the 2--10$\,$keV energy band (see Fig.~\ref{fig:swift}) instead of $0.011\pm0.007\ \mathrm{counts\,s^{-1}}$ in the 0.3--10$\,$keV energy band.
This increase of the corrected non-flaring level would lead to a decay of the flare detection efficiency by the Bayesian blocks algorithm but the count rate standard deviation is $1.6$ times lower than computed before since we corrected the count rate bias owing to the bad pixels and dead columns, the PSF extraction fraction, and the vignetting.

\begin{figure}
\centering
\includegraphics[height=8cm,angle=90]{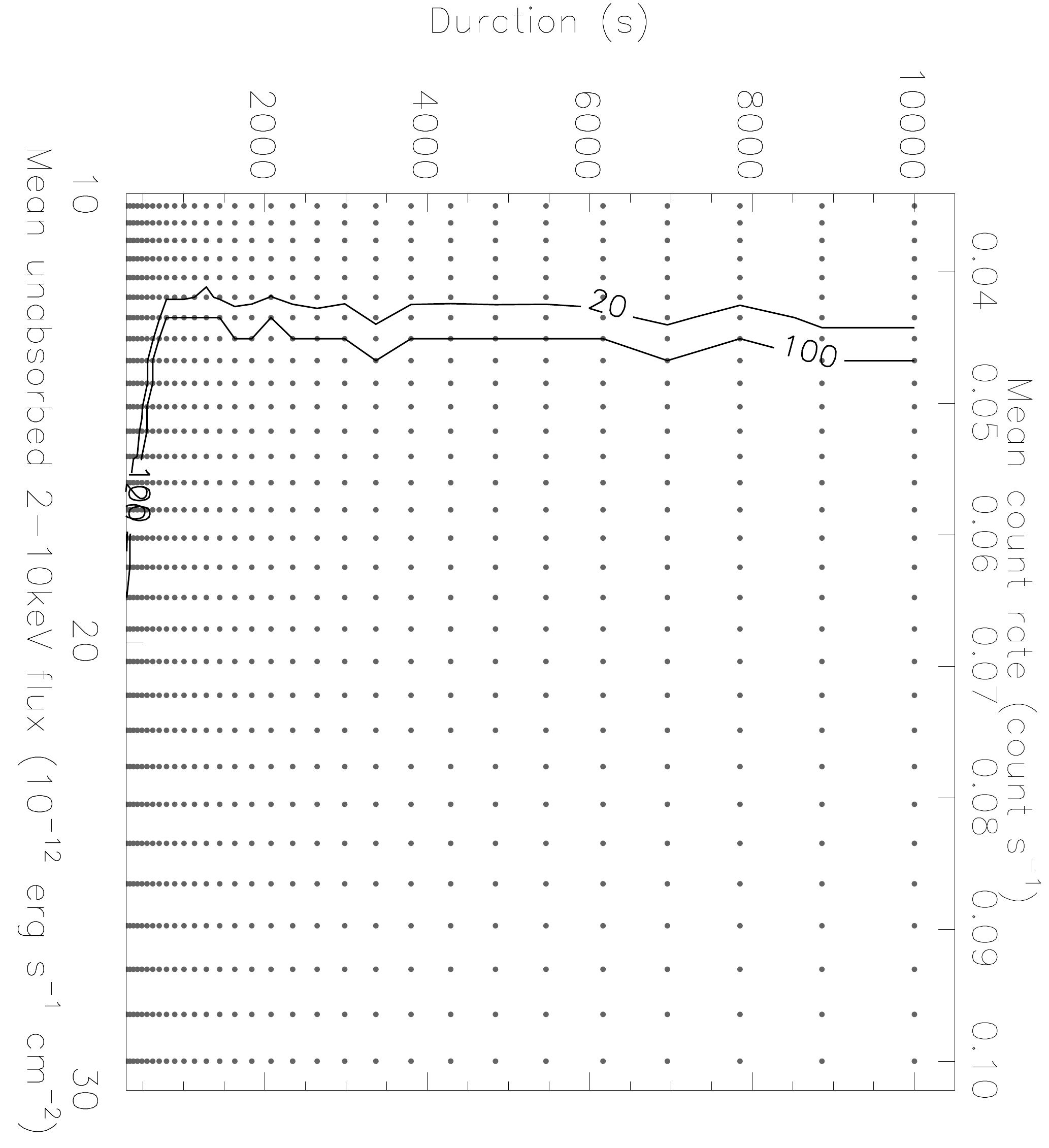}
\includegraphics[trim=0.cm 0.cm 0.cm 0.cm,clip,height=8cm,angle=90]{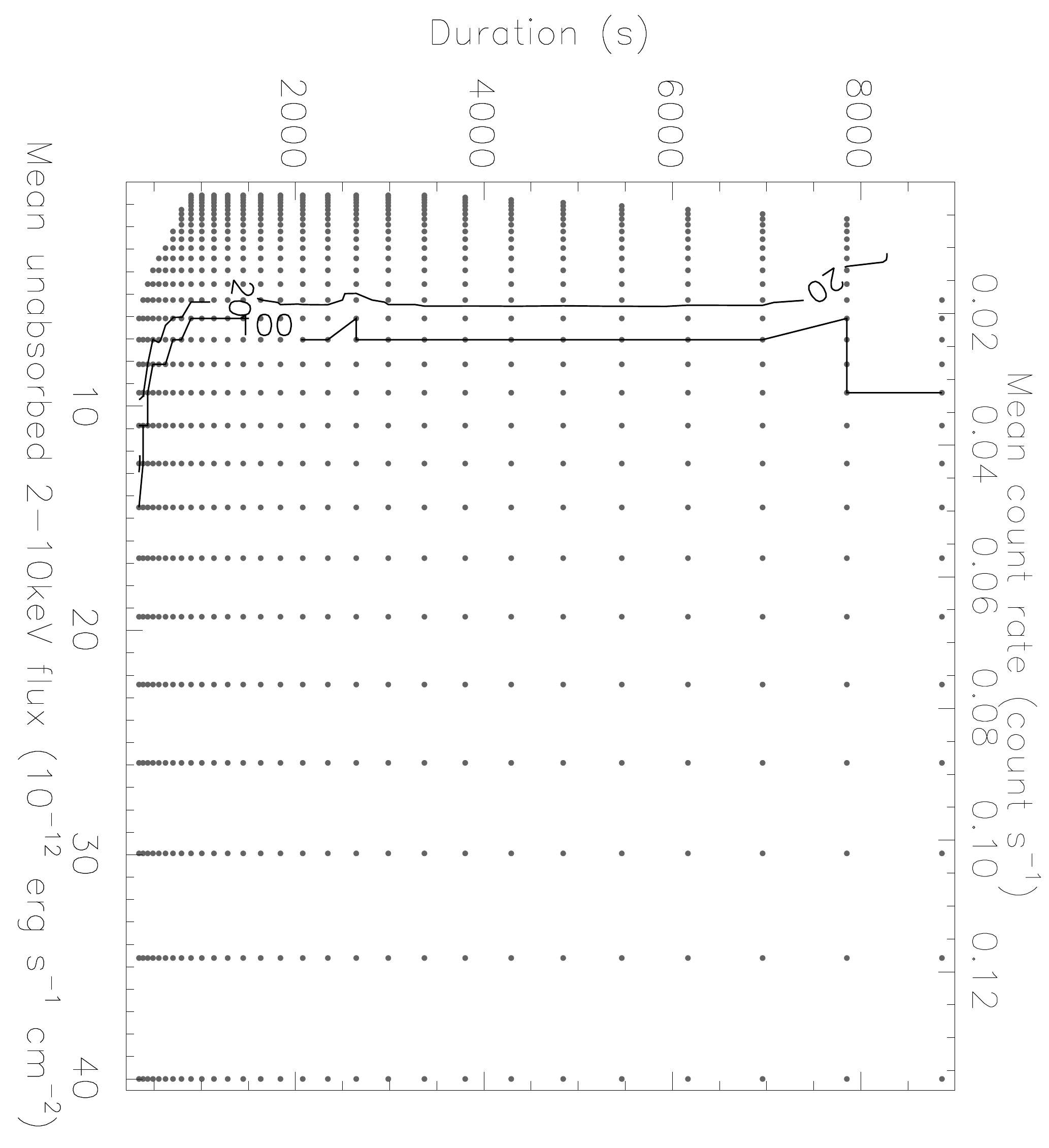}
\caption[Flare detection efficiency of the Bayesian blocks algorithm in the Swift observations]{Flare detection efficiency of the Bayesian blocks algorithm (top panel) and the \citet{degenaar13} detection method (bottom panel) in the Swift observations.
The points are the simulation grid for the Gaussian flare light curve above a non-flaring level of $0.027\,\mathrm{counts\,s^{-1}}$ in the 2--10$\,$keV energy range.
The contour levels are the detection probabilities in percent.}
\label{fig:BB_swift}
\end{figure}

\section{Systematic flare detection}
\label{flare}
\begin{figure*}
\centering
\includegraphics[width=9.3cm,angle=90]{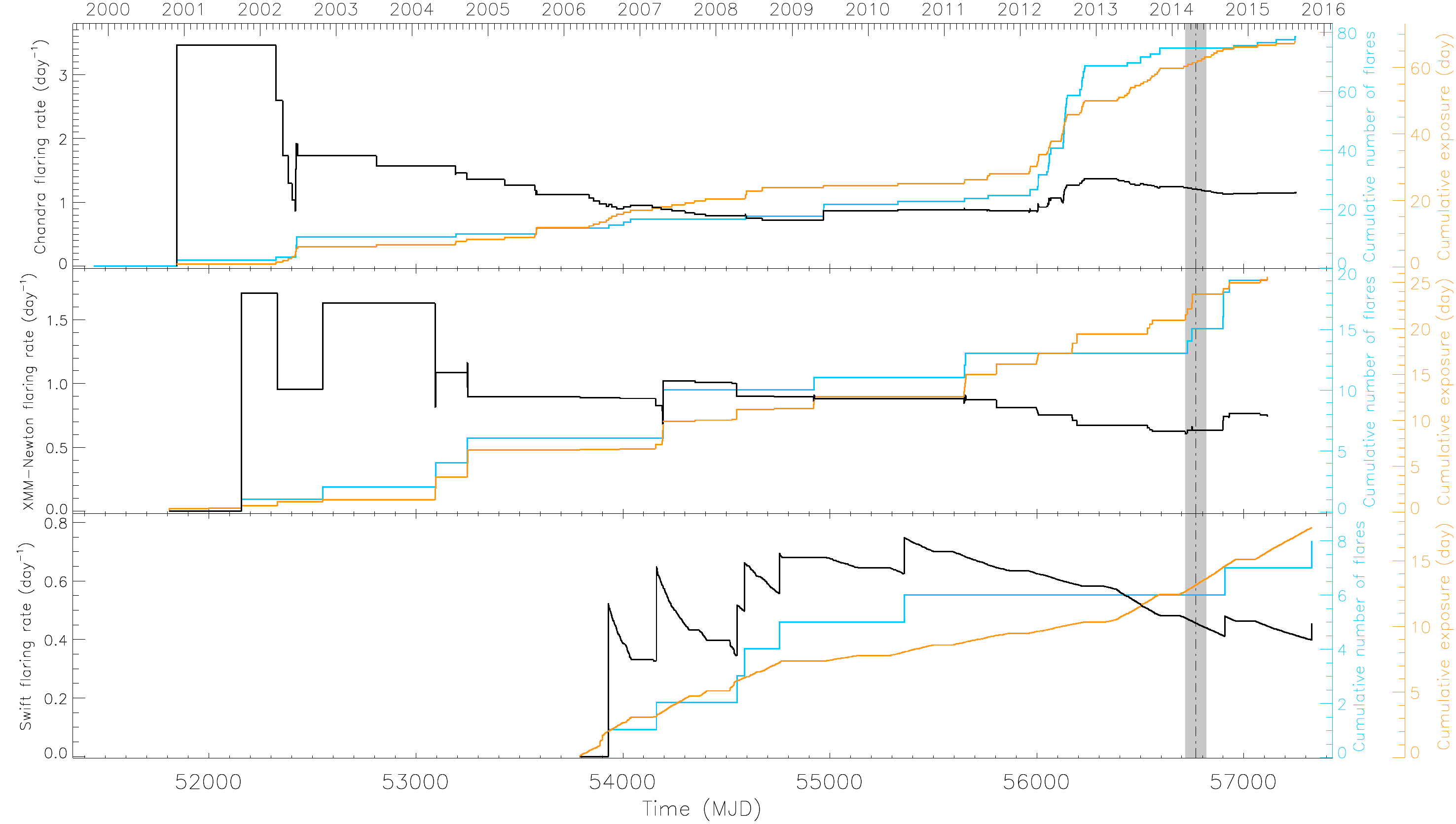}
\caption[Evolution of the mean X-ray flaring rate from 1999 to 2015]{Evolution of the mean X-ray flaring rate from 1999 to 2015 by Chandra (top panel), XMM-Newton (middle panel), and Swift (bottom panel).
The vertical gray stripe with the dot-dashed line is the time range of the DSO/G2 pericenter passage \citep{valencias15}.
The blue and orange lines are the cumulative number of flares and the cumulative observing time, respectively; the black line is the ratio of these values corresponding to the mean flaring rate.}
\label{fig:arrival_time_indiv}
\end{figure*}
\subsection{XMM-Newton and Chandra observations}
To detect X-ray flares observed with XMM-Newton and Chandra, we applied the Bayesian blocks method developed by \citet{scargle98} and refined by \citet{scargle13} on the individual photon arrival times of the src+bkg and bkg event lists with a false positive rate for the flare detection of $0.1\%$.
The result of the Bayesian blocks algorithm is an optimal segmentation of the source event list with blocks of constant count rate separated by the change points.
The event list preparation for the application of the Bayesian block algorithm is the same as explained in \citet{mossoux14}.
For the XMM-Newton observations, we first took care to associate the arrival time of each event with the center of the observational frame during which it was recorded since the randomization of the event arrival time in the frame duration by the data reduction tasks is arbitrary and not reproducible.
The events are thus separated by an integer number of frame durations. 
If several events were recorded during the same frame, we considered that these events are characterized by the same arrival time. 
We then filtered out the frames affected by ionizing particles (i.e., the bad time intervals) by merging the GTIs to obtain a continuous event flux as observed by XMM-Newton or Chandra. 
We divided the continuous event list into Voronoi cells whose start and end times are half of the interval between two adjacent events.
We defined the beginning and end of the first and last cell as the observation start and stop, respectively.
The count rate in each Voronoi cell is thus the total number of events in the cell divided by the cell duration.
We then corrected  the CCD livetime (i.e., the ratio between the integration time and CCD readout time) by applying a weight on the duration of the Voronoi cells.

To apply the Bayesian blocks algorithm with a consistent false positive rate, we calibrated the prior number of change points ($ncp\_prior$) for the number of events in the src+bkg and bkg event lists and the desired false positive rate.
Following the method proposed by \citet{scargle13} and used in \citet{mossoux14}, we simulated 100 Poisson fluxes with a mean count rate corresponding to the non-flaring level of each observation and containing a number of uniformly distributed events that is the same as in the considered event list. 
For each set of 100 simulations, we increased  $ncp\_prior$ from 3 to 8 by a step of 0.1 and we computed the number of false positives detected.
The value of $ncp\_prior$ that corresponds to the considered event list is thus the value that retrieves the desired false positive rate (here, $p_\mathrm{1} = 0.03,$ leading to a false positive rate for the flare detection of $1-p_\mathrm{1}^2=0.1\%$).

We applied the two-step Bayesian blocks algorithm of \citet{mossoux14,mossoux15} on the resulting event lists to correct for any detector flaring background as proposed by \citet{scargle13b}. We first applied the algorithm on the src+bkg event list, whereas the bkg contribution at each event arrival time is estimated by applying the Bayesian blocks algorithm on the background event list.
We then applied the algorithm on the src+bkg event list where the Voronoi intervals are weighted by the ratio of the src+bkg and background-subtracted src+bkg contributions. 

The non-flaring level of \sgra{} is defined by the count rate of the longest Bayesian block (leading to the lower error on the count rate) while the flares are associated with the higher Bayesian block count rates.
The mean count rate of a flare is the mean count rate of the flaring blocks subtracted from the non-flaring level.
The flares observed by Chandra and XMM-Newton and detected by the Bayesian blocks algorithm are represented in \ref{fig:xmm_flare} and Fig.~\ref{fig:chandra_flare1}.
A comparison with the flare characteristics observed by \citet{ponti15} is given in Appendix~\ref{app_comp}.

In 2004, the XMM-Newton observations revealed an artificial increase of the non-flaring level due to the transient X-ray emission of the low-mass X-ray eclipsing binary located at $2\farcs9$ south from \sgra{} \citep{porquet05}.
Moreover, the \sgra{} light curve also showed dips due to the eclipses of the X-ray binary that were retrieved by the Bayesian block algorithm (see the fourth panel of Fig.~\ref{fig:xmm_flare}).
During the observations made from the 2013 April\ 25, the non-flaring level of \sgra{} was also artificially increased because of the burst phase of the Galactic center magnetar \magn{} located at only 2$\farcs$4 southeast from \sgra{} \citep{degenaar13,dwelly13,kennea13a}.
The most prominent effect of the increase of the non-flaring level is a decay of the sensitivity to the detection of the faintest flares \citep{mossoux16}.

\subsection{Swift observations}
\label{eff_swift}
Owing to the low Earth orbit, the duration of Swift observations are about 1$\,$ks, which is short compared to the flare observed durations (from some hundred of seconds to more than 10$\,$ks).
We tested the effect of this short exposure on the detection probability of the flares with the Bayesian blocks algorithm.
We first simulated two non-flaring event lists with a typical exposure of 1$\,$ks and a Poisson flux with a non-flaring level of $0.027\,\mathrm{counts\,s^{-1}}$ in the 2--10$\,$keV energy range.
We then simulated a third event list with a Gaussian flare above this non-flaring level using for the sampling 30 mean count rates from 0.035 to $0.1\,\mathrm{counts\,s^{-1}}$ and 30 durations from 300$\,$s to 10$\,$ks in logarithmic scale.
We finally extracted a time range of 1$\,$ks from different part of the simulated flare to create a typical Swift event list of a flare; the center of the time range is defined to divide the flare duration into 10 time ranges.
We applied the Bayesian blocks algorithm on the three (non-flaring, flaring, and non-flaring) concatenated event lists and computed how many times the algorithm found two change points.
The mean count rates of the flares are converted to the mean unabsorbed fluxes using the averaged conversion factor between the mean count rates and mean unabsorbed fluxes in the 2--10$\,$keV energy band of $293.5\times 10^{-12}\,\mathrm{erg\,s^{-1}\,cm^{-2}/XRT\,count\,s^{-1}}$ , which is computed for $N_\mathrm{H}=14.3\times 10^{22}\,\mathrm{cm^{-2}}$ and $\Gamma=2$ via ISIS, and the effective area, which is computed for the 10$\arcsec$ extraction region and the redistribution matrix file that corresponds to the 2006 September\ 15 Swift observation where \sgra{} was on-axis.
The resulting detection probability, shown in the top panel of Fig.~\ref{fig:BB_swift}, has two different regimes with a small range of mean unabsorbed flux where the detection probability jumps from 20\% to 100\%.
For flare durations longer than 800$\,$s, the X-ray flares are either nearly undetected (detection probability lower than 20\%) or always detected with a mean unabsorbed flux limit of about $0.044\,\mathrm{counts\,s^{-1}}$ , which corresponds to $13.2\times 10^{-12}\,\mathrm{erg\,s^{-1}\,cm^{-2}}$.
The flare detection efficiency decreases with the decay of the flare duration with a 100\% detection probability at $0.044\,\mathrm{counts\,s^{-1}}$ for a flare duration of 800$\,$s and $0.065\,\mathrm{counts\,s^{-1}}$ for a flare duration of 300$\,$s.
The Bayesian blocks algorithm thus detects flares with a duration longer than the observing time less efficiently and detects only flares with a mean unabsorbed flux larger than $13.2\times 10^{-12}\,\mathrm{erg\,s^{-1}\,cm^{-2}}$ when the flare duration is larger than the observation exposure.

To assess the detection efficiency of the \citet{degenaar13} method for the Swift observations, we simulated several $1\,$ks event lists as done previously, but we now work on a logarithmic mean unabsorbed flux grid of 30 points between $0.6$ and $40.0\times 10^{-12}\,\mathrm{erg\,s^{-1}\,cm^{-2}}$ and a logarithmic duration grid of 30 points between $300\,$s and $10.1\,$ks to cover the duration and flux ranges of the overall observed flares (see next sections).
These simulations are carried out for each Swift non-flaring level observed from 2006 to 2015.
We then applied the \citet{degenaar13} detection method to compute how many times the flare is detected.
The resulting detection efficiencies $p_{obs}$ for the 2006--2012 observations (i.e., without transient sources) are shown in the bottom panel of Fig.~\ref{fig:BB_swift}.
As for the flare detection with the Bayesian blocks method, the detection efficiency jumps from 20 to 100\% in a small range of mean unabsorbed flux.
But the flux limits for 100\% detection (about $7\times 10^{-12}\,\mathrm{erg\,s^{-1}\,cm^{-2}}$) in the \citet{degenaar13} detection method are well below those of the Bayesian blocks method, making the former more efficient for flare detection with Swift.

Therefore, we used the GTI-binned method of \citet{degenaar13}, which is optimized to detect the X-ray flares for the Swift observing setup.
We first selected the src events in the 2$-10\,$keV energy band to build the \sgra{} light curves binned on each GTI.
We rejected the GTIs whose exposure is lower than 100$\,$s since the error bar on the count rate during this short exposure is large.
For the observations between 2006 and 2012, the non-flaring level from the src event list in each yearly campaign is computed as the ratio between the number of events recorded during each campaign and the corresponding yearly exposures.
A light curve bin is associated with a flare if the lower limit on the count rate in this observation is larger than the non-flaring level of the corresponding yearly campaign plus three times the standard deviation of the yearly campaign light curve.
During the 2013, 2014, and 2015 Swift campaigns, the non-flaring level observed in the \sgra{} light curves displays large variations due to the presence of the Galactic center magnetar (see Fig.~2 of \citealt{lynch14}).
The non-flaring level during these campaigns is fitted using two exponential power laws following \citet{lynch14}$\,$\footnote{We cannot directly use their fit since they did not correct from the losses caused by the bad pixels and dead columns, the PSF extraction fraction, and the vignetting.}, i.e., 
\begin{equation}
   \begin{array}{ll}
    CR= & (0.246\pm0.009)\,e^{-\frac{t-t_0}{(66.2\pm3.5)\,\mathrm{d}}} \\
    & +(0.012\pm0.05)\,e^{-\frac{t-t_0}{(79.0\pm9.7)\,\mathrm{d}}}\\
    & +(0.027\pm0.004)\ \mathrm{counts\ s^{-1}},
    \end{array}
\end{equation}
 with $t_0=56406$~MJD.
During these three campaigns, a flare is detected if the mean count rate during the observation is larger than this count rate fit plus three times the $1\sigma$ error.
The mean count rate of a flare detected with Swift is the mean count rate of the observation subtracted from the non-flaring level.
The flares detected with Swift are represented in Fig.~\ref{fig:swift}.

\subsection{X-ray flares detected from 1999 to 2015}
The time of the start and end of the flares observed by XMM-Newton, Chandra, and Swift,  as well as the non-flaring levels, are given in Tables~\ref{table:xmm}, \ref{table:chandra1} and \ref{table:swift} of Appendix~\ref{app_log}, respectively.
In total, 107 X-ray flares were detected between 1999 and 2015:  19 flares with XMM-Newton, 80 flares with Chandra, and 8 flares with Swift.
The mean flare duration is 2739$\,$s, the standard deviation is 2210$\,$s, and the median is 2018$\,$s, which  implies that the flare durations have a nearly homogeneous distribution without preferred value.
The cumulative number of flares is given in Fig.~\ref{fig:arrival_time_indiv} (blue line) as a function of time (with observing gaps) for Chandra (top panel), XMM-Newton (middle panel), and Swift (bottom panel).
The flare times are computed as $(t_\mathrm{start}+t_\mathrm{end})/2$ with $t_\mathrm{start}$ and $t_\mathrm{end}$ indicating the start and end times of the flare.
We also represent the cumulative exposure (orange line) for each instrument in this figure.
The mean flaring rate is then computed as the ratio between these two curves (black line).
The mean flaring rates observed by each instrument on 2015 November\ are different; these are $1.15\pm0.13$, $0.78\pm0.17,$ and $0.45\pm0.16$ flare per day for Chandra, XMM-Newton, and Swift, respectively.
This is because of the different sensitivity of the cameras and the different non-flaring levels observed by the instruments, which depend on both the instrument sensitivity and angular resolution.
It is thus necessary to correct the detection bias due to these heterogeneous sensitivities to study consistently the flaring rate obtained by the combination of three instruments. 
To assess the detection efficiency for the three instruments, we used two characteristics of flares that are independent of the instruments: the flare duration (already computed in this section) and mean unabsorbed flux (see Sect.~\ref{flux}).

\section{X-ray flare fluxes}
\label{flux}
To correctly compute the mean unabsorbed fluxes of the X-ray flares observed with XMM-Newton and Chandra, we extracted their spectra, ancillary files (arf), and response matrix files (rmf) with the SAS script \texttt{especget} for XMM-Newton and the CIAO script \texttt{specextract} for Chandra.
For the Swift observations, because of the short exposure time, we extracted the flare spectra during the entire observation via the HEASOFT task \texttt{XSelect}, and we created the corresponding arf via \texttt{xrtmkarf} (version 0.6.3).
The rmf were taken in the calibration database\footnote{\href{http://heasarc.gsfc.nasa.gov/FTP/caldb/data/swift/xrt/cpf/rmf/}{http://heasarc.gsfc.nasa.gov/FTP/caldb/data/swift/xrt/cpf/rmf/}}.
The non-flaring spectrum was extracted from the closest in-time observation.

We grouped the flare spectra with a minimum of one count with \texttt{grppha} to fit them with an absorbed power law created with \textit{TBnew} \citep{wilms00} and \textit{pegpwrlw} with a dust scattering modeled thanks to \textit{dustscat} \citep{predehl95} using the Cash statistic \citep{cash79} in \texttt{ISIS}.
For the XMM-Newton and Swift observations, we fit the spectra with the values of the hydrogen column density ($N_\mathrm{H}$) and the power-law index ($\Gamma$) fixed to those computed for the two brightest X-ray flares observed with XMM-Newton and the 2012 February\ 9 bright Chandra flare: $N_\mathrm{H}=14.3\times 10^{22}\ \mathrm{cm^{-2}}$ and $\Gamma=2$ \citep{porquet03,porquet08,nowak12}.
Only the mean unabsorbed flux between 2 and 10$\,$keV is a free parameter.
The resulting mean unabsorbed fluxes of each X-ray flare observed by XMM-Newton and Swift are given in Tables~\ref{table:xmm} and \ref{table:swift} of Appendix~\ref{app_log}.

For the Chandra observations, the pile-up must be taken into account.
The pile-up is due to the arrival of more than one photon per pixel island during the same readout frame.
The multiple photons are either recorded as a unique photon of merged (higher) energy or they produce the pattern (or grade) migration of the event leading to these photons not classified as an X-ray event anymore.
In the latter case, a dip appears in the center of the PSF image of a bright source. 
We use the pile-up model of \citet{davis01b} that is available in \texttt{ISIS} with the photon migration parameter $\alpha=1$ \citep{nowak12,neilsen13} for a PSF fraction of 95\% corresponding to the $1\farcs25$ extraction region.
We fit the spectra with this pile-up model applied on the absorbed power-law model with the fixed $N_\mathrm{H}$ and $\Gamma$ reported above and a free mean unabsorbed flux between 2 and 10~keV.
Table~\ref{table:chandra1} of Appendix~\ref{app_log} reports the resulting mean unabsorbed fluxes observed by Chandra between 2 and 10$\,$keV.

\begin{figure*}
\centering
\includegraphics[width=13.5cm,angle=90]{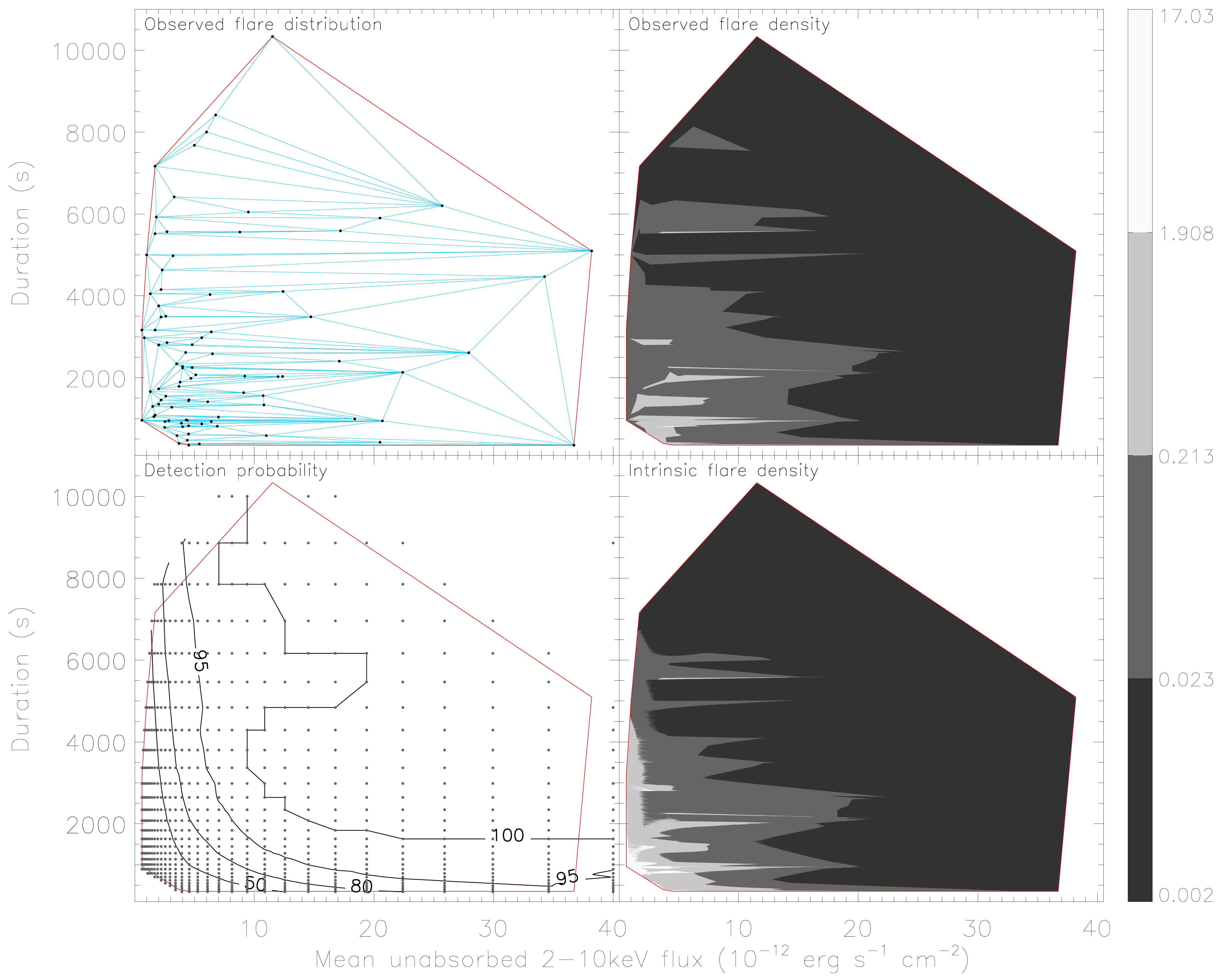}
\caption[Flux--duration distribution of the X-ray flares from \sgra{}]{Flux--duration distribution of the X-ray flares from \sgra{}.
\textit{Top left panel:} The observed flare flux--duration distribution observed with XMM-Newton and Chandra from 1999 to 2015 (black dots) using a false positive rate for the flare detection of $0.1\%$.
The blue lines are the corresponding Delaunay triangles.
The red lines define the convex hull.
\textit{Top right panel:} The observed flare density distribution is shown.
The filled contours are indicated in logarithmic scale and the corresponding color bar is represented in the right-hand side of the figure in unit of $\mathrm{10^{10}\ s^{-1}\ erg^{-1}\,s\,cm^2}$.
\textit{Bottom left panel:} The merged detection efficiency with a false positive rate for the flare detection of $0.1\%$ for XMM-Newton and Chandra from 1999 to 2015 in percent.
The dots represent the simulation grid.
\textit{Bottom right panel:} The intrinsic flare density distribution corrected from the observing bias is shown.
The filled contours use the same logarithmic scale as in the top right panel.
}
\label{fig:detect}
\end{figure*}

Three flares observed with XMM-Newton and Chandra begin before the start of the observation and three other flares end after the end of the observation.
According to the phase of the flare that is not observed, this leads to a lower or upper limit on the mean unabsorbed flux.
Indeed, assuming a Gaussian flare, if we only observe the end of the decay phase or the beginning of the rise phase, the resulting mean unabsorbed flux is a lower limit on its actual value; if we observe the end of the rise phase and decay phase or the rise phase and the beginning of the decay phase, the resulting mean unabsorbed flux is an upper limit on its actual value.
For the eight flares observed with Swift, the duration of the flares are associated with the observing time, thereby leading to a lower limit on the mean unabsorbed flux if the flare duration is lower than the exposure.
If the flare duration is larger than the exposure, the orientation of the limit depends on which  part of the flare is observed.
Hereafter, we consider these lower or upper limits on the mean unabsorbed flux as the actual value of the flare flux.
The averaged mean unabsorbed flux for the X-ray flares observed by XMM-Newton, Chandra, and Swift is $8.4\times 10^{-12}\,\mathrm{erg\,s^{-1}\,cm^{-2}}$ with a standard deviation of $10.0\times 10^{-12}\,\mathrm{erg\,s^{-1}\,cm^{-2}}$ while the median is $4.5\times 10^{-12}\,\mathrm{erg\,s^{-1}\,cm^{-2}}$.
The observed distribution of the mean unabsorbed flux is thus skewed toward the faintest flares.
However, the different detection sensitivities of the instruments according to the flare mean unabsorbed flux and duration biases the observed distribution toward the highest and longest flares.
We thus need to correct of the detection sensitivities to study correctly the merged duration and mean unabsorbed flux distribution.

\section{Intrinsic flare distribution}
\label{distr:distriiiiiiiiiii}
To determine the intrinsic flare distribution, we computed the density distribution of the flares observed only with XMM-Newton and Chandra since the characteristics of the flares observed with Swift are not sufficiently constrained.
We then corrected this observed flare density of the merged detection bias of XMM-Newton and Chandra.

\subsection{Observed flare distribution}
\label{distr:obs}
The observed flare density is computed from the mean unabsorbed fluxes and durations of the X-ray flares observed with XMM-Newton and Chandra from 1999 to 2015 using the Delaunay tessellation field estimator (DTFE; \citealt{schaap00,weygaert09}).
We constructed the minimum triangulation of the Delaunay tessellation (blue lines in the top left panel of Fig.~\ref{fig:detect}).
The density associated with a given flare position is then computed via the Delaunay triangles connected to this flare and conserving the total flare number in the reconstructed density field.
We computed for each flare, $i$, the area $W_i=\sum A_k$ with, $A_k$, which is the area of the triangle $k$ whose the vertex is the flare $i$ at the location $\pmb{x}_i$.
The flare density per surface unit in the mean unabsorbed flux duration plane that is associated with the flare $i$ is $d_i=3/W_i$ .
The discretized map of the flare density is linearly interpolated inside the convex hull of the observed flare set at a point $\pmb{x}$ in the Delaunay triangle $m$: $d_\mathrm{obs}=d_i+ \bigtriangledown d|_m(\pmb{x}-\pmb{x}_i)$ with $d|_m$ the estimated constant density gradient within $m$.
The resulting filled contour map of the observed flare density distribution is shown in the top right panel of Fig.~\ref{fig:detect} with the density levels of the observed flares in logarithmic scale.

\subsection{X-ray flare detection efficiency}
\label{distr:eff}
\begin{figure}
\centering
\includegraphics[trim=2.6cm 1.cm 0.cm 0.cm,clip,width=10cm]{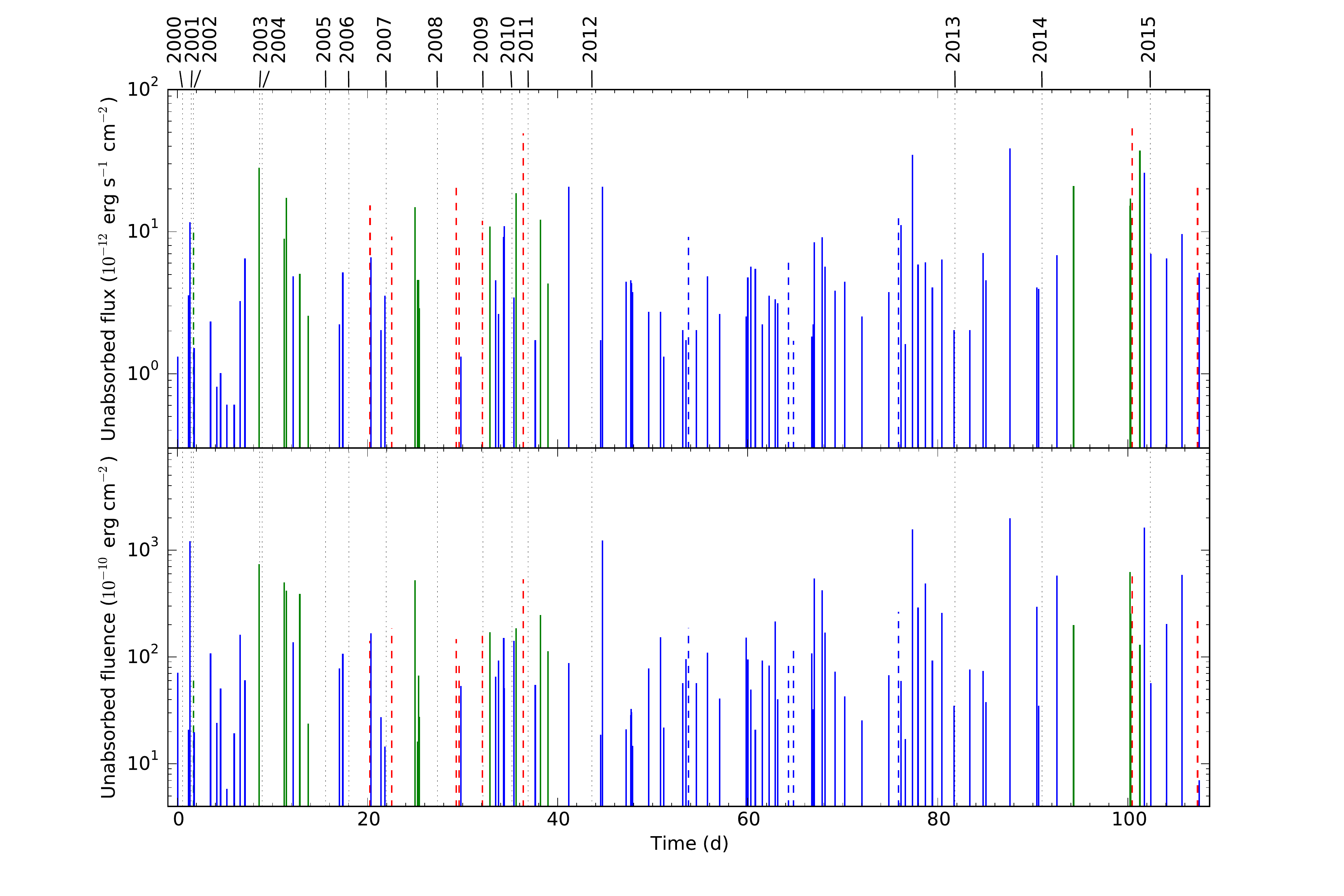}
\caption[Temporal distribution of the flare fluxes and fluences]{Temporal distribution of the flare fluxes and fluences.
The mean arrival times of the flares without observing gaps and with the correction of the average flare detection efficiency are represented by vertical lines. 
The dotted lines are the time of the beginning of the first observation of the year.
The blue, green, and red lines are the Chandra, XMM-Newton, and Swift flares, respectively.
The dashed lines are only lower or upper limits on the flare flux and fluence due to the truncated flare duration when it begins before the start of the observation or ends after the stop of the observation.
\textit{Top panel:} The mean unabsorbed flux distribution is shown.
\textit{Bottom panel:} The mean unabsorbed fluence distribution is shown.}
\label{fig:arrival_time_orig}
\centering
\includegraphics[trim=2.6cm 1.cm 0.cm 0.cm,clip,width=10cm]{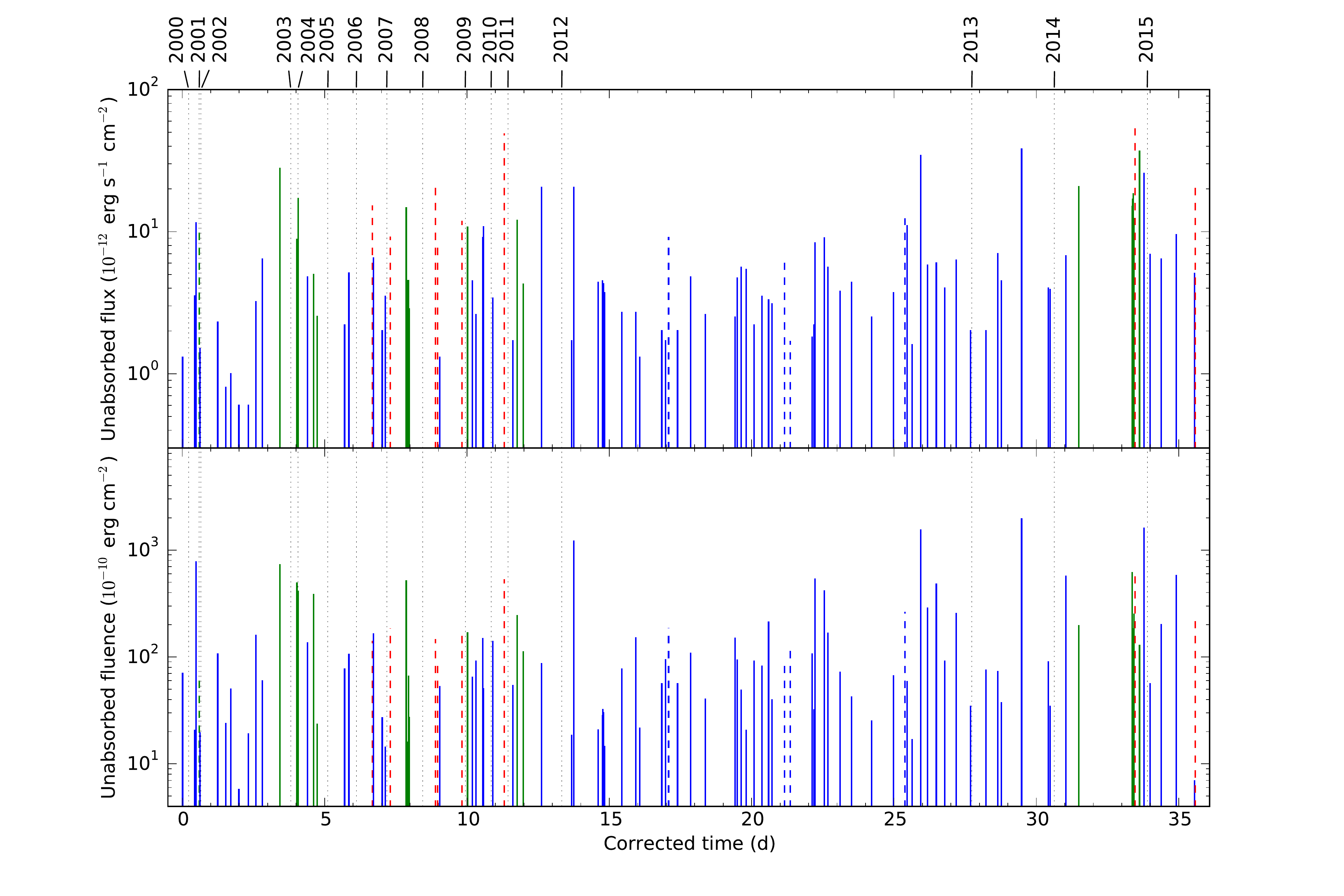}
\caption[Temporal distribution of the flare fluxes and fluences corrected from the sensitivity bias]{Temporal distribution of the flare fluxes and fluences corrected from the sensitivity bias.
See caption of Fig.~\ref{fig:arrival_time_orig} for details.}
\label{fig:arrival_time}
\end{figure}
The detection efficiency of the X-ray flares depends on the instrument sensitivity, non-flaring level, and observing time.
We used the flare durations and mean unabsorbed fluxes to compute the detection efficiency of the Bayesian blocks algorithm at each point in this 2D parameter space for each XMM-Newton and Chandra observation ($p_{obs} \le 1$).
We defined a logarithmic mean unabsorbed flux sampling of 30 points between $0.6$ and $40.0\times 10^{-12}\,\mathrm{erg\,s^{-1}\,cm^{-2}}$ and a logarithmic duration sampling of 30 points between $300\,$s and $10.5\,$ks to cover the duration and flux ranges of the overall observed flares.
We only analyzed the 2D grid points located inside and close to the convex hull (see bottom left panel of Fig.~\ref{fig:detect}).
The mean unabsorbed fluxes were converted into mean count rates using the average ratio of the count rate to unabsorbed flux computed for the flares detected with each instrument, i.e., 111.3, 248.2, and $148.1\times 10^{-12}\,\mathrm{erg\,s^{-1}\,cm^{-2}/count\,s^{-1}}$ for XMM-Newton/EPIC pn, Chandra/ACIS-S3 subarray, and Chandra/ACIS-I, respectively.
For each grid point, we simulated 200 event lists with Poisson flux reproducing a Gaussian-flare light curve with different mean count rates and durations (defined from $-2\sigma$ to $+2\sigma$ as in \citealt{neilsen13} with $\sigma$ the Gaussian standard deviation) above each of these non-flaring levels.
The detection efficiency depends strongly on the overlap between the flare duration and observing time, i.e., the edge effects.
We thus first define the time range of the simulated event list as $T_\mathrm{exp}+2T_\mathrm{flare}$, where $T_\mathrm{exp}$ is the observing time and $T_\mathrm{flare}$ is the flare duration.
We then drew, for each simulation, the time of the flare maximum as uniformly distributed between $T_\mathrm{flare}/2$ and $T_\mathrm{exp}+3T_\mathrm{flare}/2$ and simulated the event list (see Appendix~\ref{app:poisson} for details about the event list simulations).
We finally selected the events whose arrival times are between $T_\mathrm{flare}$ and $T_\mathrm{exp}-T_\mathrm{flare}$ to create our final event list.
We then applied the Bayesian blocks algorithm on each of these final event lists to compute how many times the algorithm detects the flare for a false positive rate for the flare detection of $0.1\%$.

Since the flares are described with parameters that are independent from the telescope instruments, we were able to combine the $p_{obs}$ of each instrument and each observation computed on the same grid.
We firstly weighted the local detection efficiencies according to the exposure time of the corresponding observation since the impact of the detection efficiency on the number of observed flare depends on the exposure.
We finally summed the weighted local detection efficiencies to determine the merged (weighted mean) local detection efficiency of XMM-Newton and Chandra shown in the bottom left panel of Fig.~\ref{fig:detect} with the grid points.
The merged local detection efficiency along the border of the convex hull was computed by a linear interpolation between the merged local detection efficiency on either side of the convex hull.

\subsection{Correction of the observed flare distribution}
\label{distr:corr}
The observed flare distribution was finally corrected from the merged local detection efficiency to compute the intrinsic flares distribution.
The observed flare distribution at each point grid $\pmb{x}$ was then corrected by the merged local detection efficiency $p_\mathrm{merged}(\pmb{x}) \le 1$ as $d_\mathrm{intr}(\pmb{x})=d_\mathrm{obs}(\pmb{x})/p_\mathrm{merged}(\pmb{x})$ (see Eq.~17 of \citealt{weygaert09}).
The intrinsic flare distribution is shown with filled contour in logarithmic scale in the bottom right panel of Fig.~\ref{fig:detect}.
The intrinsic flare distribution is now highest for the faintest and shortest flares.

\section{Temporal distribution of the X-ray flares from 1999 to 2015}
\label{flare_rate}
We then combined the overall XMM-Newton, Chandra, and Swift observations and removed the observational gaps to create a continuous exposure containing the times of the 107 flares detected. 
The observational overlays were also removed to keep only the most sensitive instrument.
Figure~\ref{fig:arrival_time_orig} shows the flare arrival times without observing gaps over the total exposure time of 107.6$\,$days (corresponding to 9.3$\,$Ms).
The height of each vertical line representing a flare corresponds to the mean unabsorbed flux (top panel) and fluence (mean unabsorbed flux times duration; bottom panel) between 2 and 10$\,$keV.
We thus observe a flaring rate of $0.98\pm 0.09$ flare per day, which is lower but statistically consistent with the flaring rate deduced by \citet{neilsen13}, which was limited to the 2012 Chandra \textit{XVP} campaign, since XMM-Newton and Swift are less sensitive to fainter and shorter flares.
We thus needed to correct the flare count rate from the flare detection bias due to the heterogeneous instrumental sensitivities.

\subsection{Correction of the sensitivity bias}
To correct the temporal flare distribution of the sensitivity bias, for each observation we determined the average flare detection efficiency $\eta_{obs}$ by applying the detection efficiencies $p_{obs}$ computed in the Sect.~\ref{eff_swift} for Swift and Sect.~\ref{distr:eff} for XMM-Newton and Chandra.
The intrinsic flare distribution $d_\mathrm{intr}(\pmb{x})$ at each grid point $\pmb{x}$ is affected by $p_{obs}(\pmb{x}) \le 1 $, thus leading to the observation of only a percentage of this flare density.
By computing the ratio between the 2D integral on the convex hull of the intrinsic flare distribution affected by the local detection efficiency for a given non-flaring level and the intrinsic flare distribution, we assessed the average flare detection efficiency $\eta_{obs}<1$ corresponding to this observation,
\begin{equation}
   \eta_{obs}=\frac{\int\!\int d_\mathrm{intr}(\pmb{x})\, p_{obs}(\pmb{x})\,d\pmb{x}}{\int\!\int d_\mathrm{intr}(\pmb{x})\,d\pmb{x}}\, .
\end{equation}
The values of $\eta_{obs}$ are reported in Tables~\ref{table:xmm}, \ref{table:chandra1} and \ref{table:swift}.

We thus obtained a set of merged observations from XMM-Newton, Chandra, and Swift, each containing $N \ge 0$ flares, with their corresponding exposure and average flare detection efficiency $\eta_{obs}$.
To correct the flaring rate from the instrumental sensitivity, for each observing time $T$ we computed the corrected observing time as $T_\mathrm{corr}=T\,\eta_{obs}$ , thus leading to a higher and unbiased flaring count rate in the corresponding observation. 
Figure~\ref{fig:arrival_time} shows the flares times without observing gaps over the total corrected exposure time of 35.6$\,$days.

\subsection{Study of the unbiased X-ray flaring rate}
We divided the corrected exposure time in Voronoi cells each containing one flare and whose the separation times are the mean time between two consecutive flares.
We applied the Bayesian blocks algorithm on the Voronoi cells with a false positive rate for the change point detection of $p_\mathrm{1} = 0.05$ and the corresponding $ncp\_prior=$4.17 calibrated for 107 flares uniformly distributed during 35.6~days, i.e., corresponding to a Poisson flux.
The overall flaring activity is constant, with a flaring rate of 3.0$\pm$0.3 flares per day.
This is significantly higher than those computed by \citet{neilsen13} since we corrected the sensitivity bias.

We now investigate the existence of a flux or fluence threshold that leads to a change in the flaring rate.

\subsubsection{Flux threshold for a change of flaring rate}
\label{flux_thres}
Two methods can be used to look for a flux threshold: first, the top-to-bottom search where, at each step, we remove the flare with the highest unabsorbed flux, but we keep the corresponding exposure time and update the Voronoi cells; and, second, the bottom-to-top search where, at each step, we remove the flare with the lowest unabsorbed flux or fluence.
At each step, we apply the Bayesian blocks algorithm with a false probability rate of $p_\mathrm{1}=0.05$ on the resulting flare list and we repeat this operation until the algorithm found a flaring rate change.
The $ncp\_prior$ is calibrated at each step according to the number of remaining flares to ensure a significance of at least 95\% of any detected change point.
Since we cannot argue that one of these two methods is better than the other, we tested both.

\begin{figure}
\centering
\includegraphics[trim=2.6cm 1.cm 0.cm 0.cm,clip,width=9.7cm]{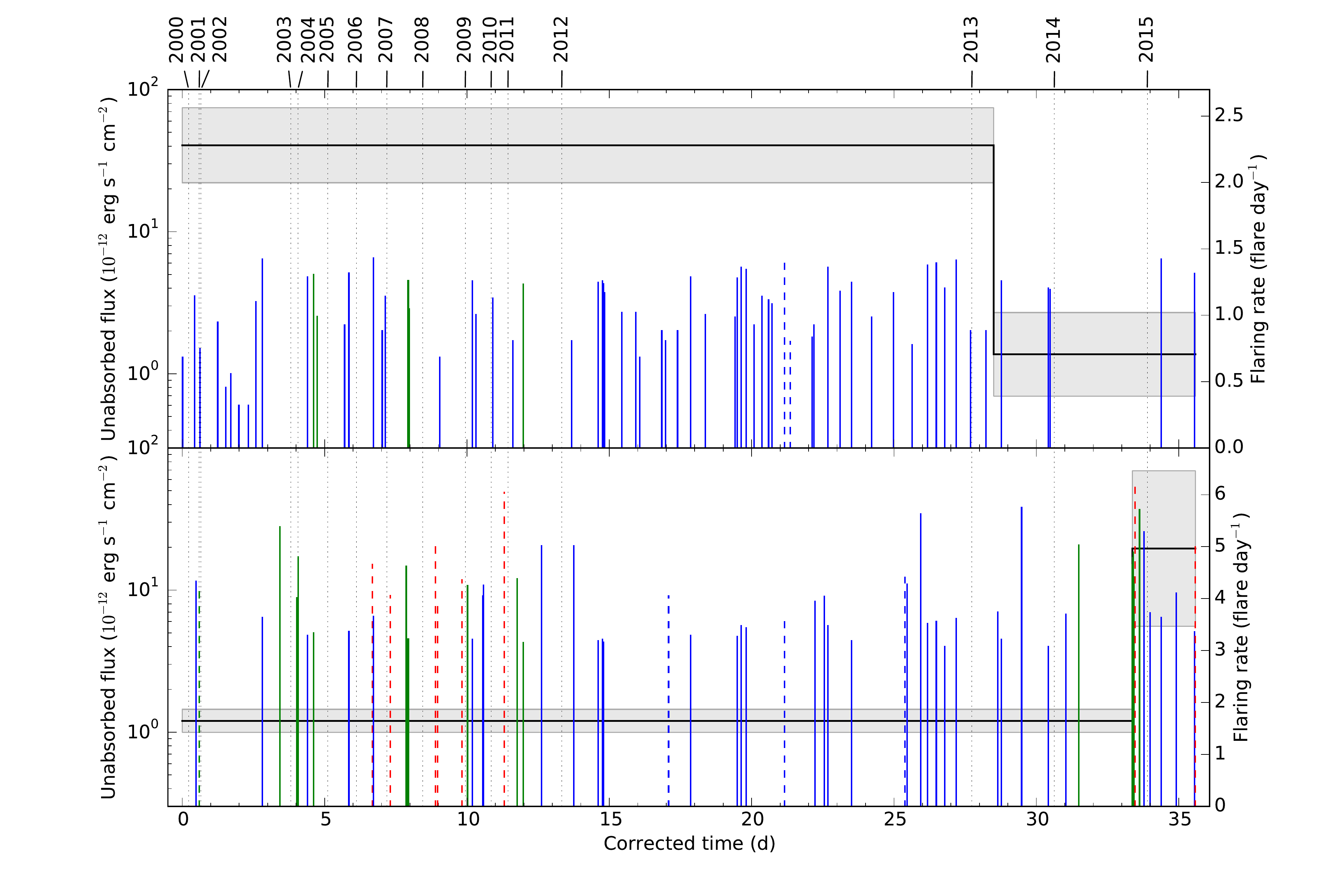}
\caption[X-ray flaring rate from 1999 to 2015 computed by the Bayesian blocks algorithm for the most luminous and less luminous flares]{X-ray flaring rate from 1999 to 2015 computed by the Bayesian blocks algorithm for the most luminous and less luminous flares.
See caption of Fig.~\ref{fig:arrival_time_orig} for the description of the flares and Table~\ref{table:summary} for the values of the thresholds.
The Bayesian blocks are indicated with thick black lines.
\textit{Top panel:} The results for the top-to-bottom search are shown.
\textit{Bottom panel:} The results for the bottom-to-top search are shown.
See text for details.}
\label{fig:BB_flux}
\centering
\includegraphics[trim=2.6cm 1.cm 0.cm 0.cm,clip,width=9.7cm]{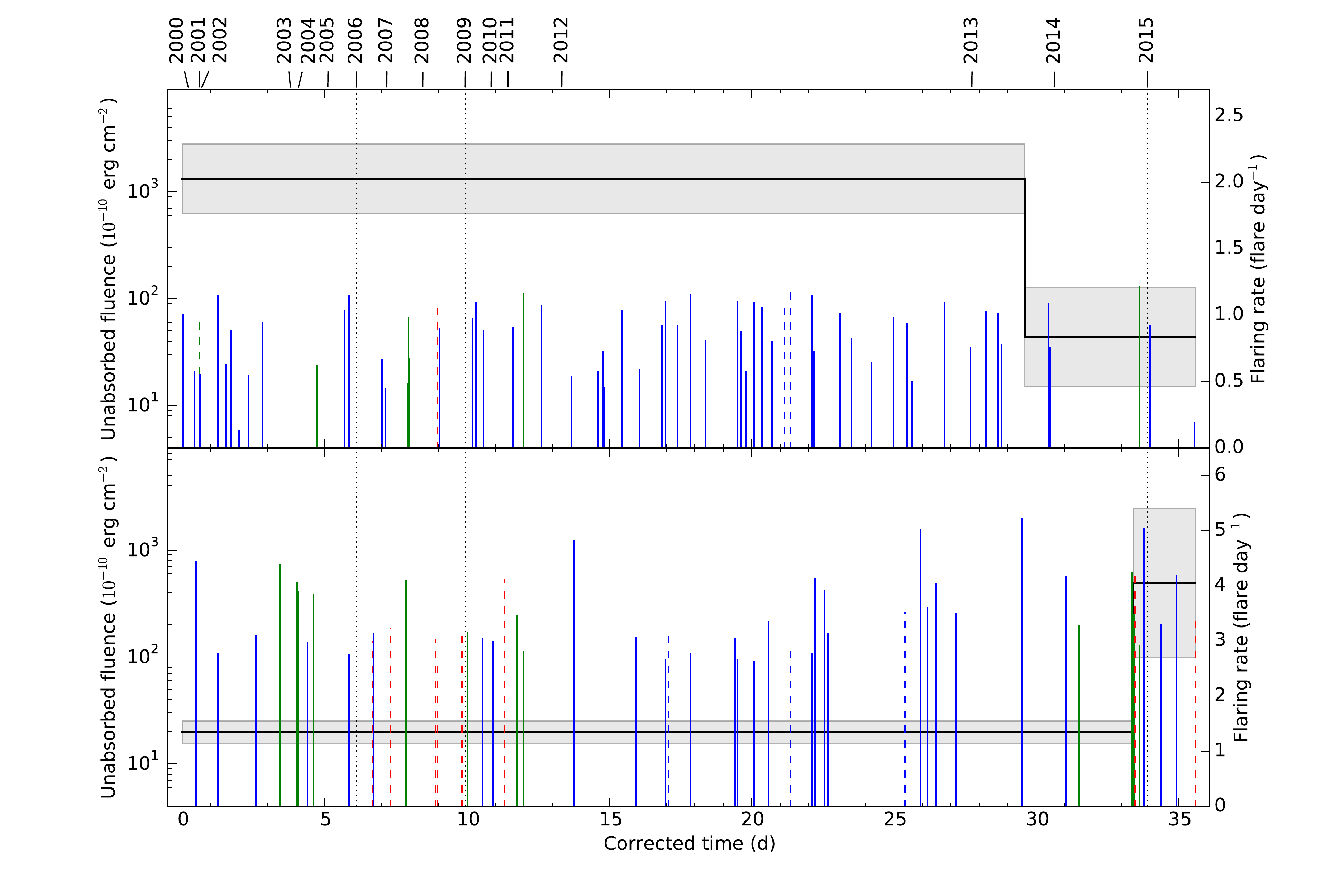}
\caption[X-ray flaring rate from 1999 to 2015 computed by the Bayesian blocks algorithm for the most energetic and less energetic flares]{X-ray flaring rate from 1999 to 2015 computed by the Bayesian blocks algorithm for the most energetic and less energetic flares.
See caption of Fig.~\ref{fig:BB_flux} for details.}
\label{fig:BB_fluence}
\end{figure}

We first performed a top-to-bottom search.
A change of flaring rate is detected at 28.5 days, i.e., between the Chandra flare on 2013 May 25 and the second Chandra flare on 2013 July 27 (flares \#70 and \#72 in Table~\ref{table:chandra1} and Fig.~\ref{fig:chandra_flare1}) considering only 70 flares with a mean unabsorbed flux lower than or equal to $6.5\times 10^{-12}\ \mathrm{erg\ s^{-1}\ cm^{-2}}$ (the less luminous flares) with $p_\mathrm{1} = 0.05$ and the corresponding $ncp\_prior=4.18$.
The resulting Bayesian blocks are shown in the top panel of Fig.~\ref{fig:BB_flux} where only these 70 flares are shown.
The first block contains 65 flares while the second block contains 5 flares.
The flaring rate decreases from $2.3 \pm 0.3$ to $0.7 \pm 0.3$ flares per day.
By decreasing the false probability rate, this flaring rate change is detected for $p_\mathrm{1}>0.034$, which leads to a significance of  $1-p_\mathrm{1}=96.6$\%.
To compare the two flaring rates, we computed the p-value for the null hypothesis that the flaring rates are the same (i.e., a rate ratio of 1) considering a Poisson process for the flare arrival times \citep{gehrels86,fay10}.
The p-value to compare the 65 flares that occurs in 28.5 days and the 5 flares that occurs in 7.09 days is 0.006.
The ratio between the two flaring rates is 3.2 and the 95\% confidence interval is 1.3-10.3.

We then performed the bottom-to-top search by recursively removing the flare with the lowest unabsorbed flux and applying the Bayesian blocks algorithm.
We found one change of flaring rate by considering only 66 flares with a mean unabsorbed flux larger than or equal to $4.0\times 10^{-12}\ \mathrm{erg\ s^{-1}\ cm^{-2}}$ (the most luminous flares) with $p_\mathrm{1} = 0.05$ and the corresponding $ncp\_prior=4.31$.
The resulting Bayesian blocks are shown in the bottom panel of Fig.~\ref{fig:BB_flux} where only these 66 flares are shown.
The change of flaring rate happened on 2014 August 31 (33.36 days) between the two XMM-Newton flares \#16 and \#17 in Table~\ref{table:xmm} and Fig.~\ref{fig:xmm_flare}.
The blocks contain 55 and 11 flares that correspond to flaring rates of $1.6\pm0.2$ and $5.0\pm1.5$ flares per day.
This flaring rate change is still detected until a false positive rate of $p_\mathrm{1} = 0.048$ ($ncp\_prior=4.35$), thus leading to a significance for this change point of $1-p_1=95.2$\%.
The p-value comparing the 55 flares that occurs in 33.36 days and the 11 flares that occurs in 2.22 days is 0.005.
The ratio between the two flaring rates is 3.0 and the 95\% confidence interval is $1.4-5.8$.

A summary of these results is given in Table~\ref{table:summary}.

To assess the probability of a false positive, we simulated a homogeneous Poisson flux of flare arrival times using the method described in Appendix~\ref{app:poisson}.
We created 500 sets of 107 arrival times uniformly distributed in 0--35.6$\,$days.
We also considered a constant flux distribution, i.e., we created 500 sets of 107 fluxes uniformly distributed in $0.6-59.4\times 10^{-12}\ \mathrm{erg\ s^{-1}\ cm^{-2}}$ that we associated with the flare arrival times.
We then performed the top-to-bottom and bottom-to-top searches and recorded the flaring rate changes.
We detected a change of flaring rate in both the top-to-bottom and bottom-to-top search for only 42 trials, corresponding to 8.4\% of the simulations.
In these 42 subsets, none of the change points were detected after 37 and 41 cuts as is the case for our observations; but for 71\% of these sets (i.e., 32 sets), the change points were detected after more than 77 trials, which is a greater number than in our observations.
This implies that the joint probability to observe a change point in subsamples containing 70 and 65 flares is lower than $(1/500)^2=4\times 10^{-6}$.
Moreover, for 38 of the 42 subsets (90\%), the time intervals between the two change points are between $-2$ and $0\,$days.
Only one of the 42 subsets (2\%) has a time interval comprised between 4.5 and 5$\,$days as observed in our observations.
In the light of these results, the probability that the change points found in our observations by the two search methods are due the detection of a false positive is lower than $1.2\times 10^{-5}$.
We can thus state that the change of flaring rate that we observe is likely due to a change in the flux distribution.

\subsubsection{Fluence threshold for a change of flaring rate}
\label{fluence_thres}

\begin{table}
\centering
\caption[Summary of the change of X-ray flaring rates detected between 1999 and 2015]{Summary of the change of X-ray flaring rates detected between 1999 and 2015.}
\label{table:summary}
\resizebox{0.49\textwidth}{!}{
\begin{tabular}{@{}lcc@{}}
\hline
\hline
  & Top-to-bottom & Bottom-to-top \\
\hline
 Flux threshold ($10^{-12}\,\mathrm{erg\,s^{-1}\,cm^{-2}}$) & $\le6.5$ & $\ge4.0$ \\
 Number of less and most luminous flares  & 70 & 66 \\
 Corrected time of the change point & 28.5 & 33.4 \\
 Date of the change point & 2013 May 25--July 27 & 2014 August 31 \\
 First block (Flare per day) & $2.3\pm0.3$ & $1.6\pm0.2$ \\
 Second block (Flare per day) & $0.7\pm0.3$ & $5.0\pm1.5$ \\
 Significance (\%) & 96.6 & 95.2 \\
\hline
 Fluence threshold ($10^{-10}\,\mathrm{erg\,cm^{-2}}$) & $\le128.1$ & $\ge91.3$ \\
 Number of less and most energetic flares  & 65 & 54 \\
 Corrected time of the change point & 29.6 & 33.4 \\
 Date of the change point & 2013 July 27--October 28 & 2014 August 31 \\
 First block (Flare per day) & $2.0\pm0.3$ & $1.2\pm0.2$ \\
 Second block (Flare per day) & $0.8\pm0.4$ & $4.1\pm1.3$ \\
 Significance (\%) & 95.1 & 95.1 \\
\hline
\end{tabular}
}
\end{table}

We carried out the same study with the unabsorbed fluence.
We first performed the top-to-bottom search: a change of flaring rate was found considering only 65 flares with an unabsorbed fluence lower than or equal to $128.1\times 10^{-10}\,\mathrm{erg\ cm^{-2}}$ (the less energetic flares) with $p_\mathrm{1} = 0.05$ and the corresponding $ncp\_prior=4.12$.
The resulting Bayesian blocks are shown in the top panel of Fig.~\ref{fig:BB_fluence} where only these 65 flares are shown.
The first block contains 60 flares while the second one contains 5 flares.
The change of flaring rate happens between the second Chandra flare on 2013 July 27 (flare \#72 in Table~\ref{table:chandra1} and Fig.~\ref{fig:chandra_flare1}) and the first Chandra flare on 2013 October\ 28 (flare \#74 in Table~\ref{table:chandra1} and Fig.~\ref{fig:chandra_flare1}) (29.59 days).
The corresponding flaring rates are $2.0\pm0.3$ and $0.8\pm0.4$ flares per day.
We detected this flaring rate change for a decreasing false positive rate until $p_\mathrm{1} = 0.049$ ($ncp\_prior=4.14$), which leads to a probability that this change of flaring rate is a real rate of   $1-p_1=95.1$\%.
The p-value comparing the 60 flares that occur in 29.59~days and the 5 flares that occur in 5.99~days is 0.05.
The ratio between the two flaring rates is 2.4 and the 95\% confidence interval is $1.0-7.7$.

For the bottom-to-top search, a change of flaring rate was detected considering only 54 flares with a mean unabsorbed fluence larger than or equal to $91.3\times 10^{-10}\,\mathrm{erg\ cm^{-2}}$ (the most energetic flares) with $p_\mathrm{1} = 0.05$ and the corresponding $ncp\_prior=4.01$.
The resulting Bayesian blocks are shown in the bottom panel of Fig.~\ref{fig:BB_fluence} where only these 54 flares are shown.
The two blocks are described by flaring rates of $1.2 \pm 0.2$ and $4.1 \pm 1.3$ flares per day.
The change of flaring rate happens on 2014 August\ 31 (33.36~days) between the two XMM-Newton flares \#16 and \#17 in Table~\ref{table:xmm} and Fig.~\ref{fig:xmm_flare}.
This flaring rate change was detected for a decreasing false positive rate until $p_\mathrm{1} = 0.049$ ($ncp\_prior=4.03$), which leads to a probability that this change of flaring rate is real of $1-p_1=95.1\%$.
This increase of flaring rate for the most energetic flares occurs at the same date as the increase of the flaring rate for the most luminous flares.
The p-value comparing the 45 flares that occur in 33.36~days and the 9 flares that occur in 2.22~days is 0.011.
The ratio between the two flaring rates is 3.0 and the 95\% confidence interval is $1.3-6.2$.

A summary of these results is given in Table~\ref{table:summary}.

By performing the same simulations as described previously, the probability that the change points found for observations by the two search methods are due to the detection of a false positive is lower than $6.5\times 10^{-5}$.
We can thus state that the change of flaring rate that we observe is likely due to a change in the fluence distribution.

\section{Discussion}
\label{discussion}
Our high flaring rate Bayesian block for the most energetic flares identifies the same five flares that created the increase of flaring rate in \citet{ponti15} plus three additional flares observed in 2015 with Chandra (flares \#78 and \#79) and Swift (flare \#8).
The start of the high flaring rate happened 131~days (80--181~days) after the DSO/G2 pericenter passage near \sgra{} (computed with the DSO/G2 pericenter passage determined by \citealt{valencias15}).
As argued in \citet{mossoux16}, if some material from DSO/G2 was accreted toward \sgra{}, the increase of flux should not be observed before the end of 2017 considering a pericenter distance of 2000$\,R_\mathrm{s}$ and an efficiency of the mechanism of angular momentum transport of $\alpha=0.1$.
Two interpretations can thus be proposed to explain this increase of flaring rate. Firstly, the increase of flaring rate could be due to the accretion of matter from the DSO/G2 onto \sgra{} considering an efficiency of the mechanism of angular momentum transport of at least $0.6$.
Secondly, the increase of flaring rate could be explained by the increase of the efficiency of the mechanisms producing the X-ray flares, such as a Rossby instability producing magnetized plasma bubbles in the hot accretion flow \citep{tagger06,liu06}, additional heating of electrons due to accretion instability or magnetic reconnection \citep{baganoff01,markoff01,yuan03,yuan09}, or 
the tidal disruption of an asteroid \citep{cadez06,cadez08,kostic09,zubovas12}.

Interestingly, the decay of the less luminous and less energetic flares occurs about 300 and 220~days before the DSO/G2 pericenter passage near \sgra{}, therefore, about 13 and 10 months before the increase of the most luminous and most energetic flaring rate, respectively.
For comparison, we compute the energy saved during the decay of the flaring rate of less energetic flares occurring between 2013 July 27 and October\ 28 and the energy lost during the increase of the flaring rate of most energetic flares after 2014 August\ 31 as
\begin{equation}
 E_\mathrm{saved}<F_1\,\int^{T}_{t_1}\Delta CR_1\, dt=F_1\,\Delta CR_1\, (T-t_1)
\end{equation}
and
\begin{equation}
 E_\mathrm{lost}>F_2\,\int^{T}_{t_2}\Delta CR_2\, dt=F_2\,\Delta CR_2\, (T-t_2)\, ,
\end{equation}
where $T=35.6$ are the corrected days, $F_1$ and $F_2$ the fluence thresholds, $t_1$ and $t_2$ the corrected days of the change points, and $\Delta CR_1$ and $\Delta CR_1$ the absolute values of the difference on the flaring rate between the first and second block (see Table~\ref{table:summary}).
For the less energetic flares, $F_1=128.1\times 10^{-10}\,\mathrm{erg\ cm^{-2}}$, $\Delta CR_1=1.2\pm0.5$ flare per day, and $T-t_1=6.0\pm0.8$ corrected days, which leads to a saved energy of $E_\mathrm{saved}<(9.2\pm4.8)\times 10^{-8}\,\mathrm{erg\ cm^{-2}}$.
For the most energetic flares, $F_2=91.3\times 10^{-10}\,\mathrm{erg\ cm^{-2}}$, $\Delta CR_2=2.8\pm1.4$ flare per day and $T-t_2=2.214\pm0.005$ corrected days, which leads to a lost energy of $E_\mathrm{lost}>(5.6\pm2.7)\times 10^{-8}\,\mathrm{erg\ cm^{-2}}$.
Therefore, the energy saved by the decrease of the number of less energetic flares during several corrected days could be released by a few bright flares during a shorter period.
This energy could be stored in the distortions of the magnetic field lines and then released during a magnetic reconnection event.
This is reminiscent of the behavior of earthquakes, in which stresses produce several small events during a long period of time or may accumulate before releasing in a large event.
The input of fresh accreting material from the DSO/G2 is thus not needed to explain this large increase of the most luminous and most energetic flares.

\section{Conclusions}
\label{conclusion}
The Swift campaigns and Chandra and XMM-Newton observations of \sgra{} from 1999 to 2015 have allowed us to compute the intrinsic distribution of the mean unabsorbed flare flux and duration and to study the significance of a change flaring rate.
The 96 X-ray flares observed with Chandra and XMM-Newton were detected via the two-step Bayesian blocks algorithm \citep{mossoux14,mossoux15} and 8 X-ray flares observed with Swift were detected via an improvement of the \citet{degenaar13} detection method.
By correcting the observed flare flux and duration distribution from the merged local detection efficiency of XMM-Newton and Chandra, we have been able to estimate the intrinsic flare flux and duration distributions, which are maximum for the smallest and shortest flares.
The flaring rate observed by Chandra, XMM-Newton and Swift together has then been corrected from the average flare detection efficiency of the corresponding instruments.

No significant change of flaring rate is found with the Bayesian blocks algorithm considering the overall flares, which lead to an intrinsic flaring rate of $3.0\pm0.3$ flares per day.
However, we identify, for the first time, a significant decay of the flaring rate (with probability larger than 95.1\%) for the less luminous (fainter than $6.5\times 10^{-12}\ \mathrm{erg\ s^{-1}\ cm^{-2}}$) and less energetic (lower than $128.1\times 10^{-10}\,\mathrm{erg\ cm^{-2}}$) flares by a factor of 3.2 (1.3--10.3, 95\% confidence limits) and 2.4 (1.0--7.7, 95\% confidence limits) after 2013 May 25 and 2013 July 27, respectively (see Table~\ref{table:summary}).
These decays occur about 300 and 220~days before the pericenter passage of the DSO/G2, which implies that this change of flaring rate is difficult to explain by the passage of the DSO/G2 near \sgra{}.

We confirm a significant increase of the flaring rate (with probability larger than 95.1\%) for the most luminous (brighter than $4.0\times 10^{-12}\ \mathrm{erg\ s^{-1}\ cm^{-2}}$) and most energetic (larger than $91.3\times 10^{-10}\,\mathrm{erg\ cm^{-2}}$) flares by a factor of 3.0 (95\% confidence limits of 1.4--5.8 and 1.3--6.2), respectively, from 2014 August\ 31 until 2015 November\ 2 (i.e., the last observation).

The energy released during this increase of bright flaring rate could come from the energy saved during the decay of the faintest flares.
The input of fresh accreting material from the DSO/G2 is thus not needed to explain this large increase of the most luminous and most energetic flares.

\begin{acknowledgements}
We thank the PIs that obtained since 1999 the X-ray observations of \sgra{} used in this work.
E.M. acknowledge Universit\'{e} de Strasbourg for her IdEx PhD grant.
This work made use of public data from the Swift data archive, and data supplied by the UK Swift Science Data Center at the University of Leicester. 
Swift is supported at Penn State University by NASA Contract NAS5-00136. 
This research has made use of the XRT Data Analysis Software (XRTDAS) developed under the responsibility of the ASI Science Data Center (ASDC), Italy.
This work is also based on public data from the XMM-Newton project, which is an ESA Science Mission with instruments and contributions directly funded by ESA member states and the USA (NASA).
This work also uses public data obtained from the Chandra Data Archive.
\end{acknowledgements}

\bibliographystyle{aa}
\bibliography{biblio_centre_galactique.bib}

\appendix
\section{The observation log and the X-ray flares detected from 1999 to 2015}
\label{app_log}

\begin{table*}[ht]
\caption[Observation log of the XMM-Newton observations and the detected X-ray flares in 1999--2015]{Observation log of public XMM-Newton observations and the X-ray flares detected in this work.}
\centering
\scalebox{0.6}{
\label{table:xmm}
\begin{tabular}{@{}lllcccccrrcrc@{}}
\hline
\hline
\multicolumn{7}{c}{Observations} & \multicolumn{6}{c}{Flares} \\
\multicolumn{7}{l}{\rule[0.5ex]{45.6em}{0.55pt}} & \multicolumn{6}{r}{\rule[0.5ex]{34.5em}{0.55pt}} \\
\multicolumn{11}{c}{} & \multicolumn{2}{c}{Mean} \\
\multicolumn{11}{c}{} & \multicolumn{2}{r}{\rule[0.5ex]{18em}{0.55pt}} \\
\multicolumn{1}{c}{ObsID} & \multicolumn{1}{c}{PI} & \multicolumn{1}{c}{Start} & \multicolumn{1}{c}{End} & \multicolumn{1}{c}{Duration}  & \multicolumn{1}{c}{Non-flaring level} & \multicolumn{1}{c}{$ \eta_{obs}\,$\tablefootmark{a}} & \multicolumn{1}{c}{\#} & \multicolumn{1}{c}{Start\tablefootmark{b}} & \multicolumn{1}{c}{Stop\tablefootmark{b}} & \multicolumn{1}{c}{Duration} & \multicolumn{1}{c}{Count rate\tablefootmark{c}} & \multicolumn{1}{c}{Flux\tablefootmark{d}} \\
 & & \multicolumn{1}{c}{(UT)} & \multicolumn{1}{c}{(UT)} & \multicolumn{1}{c}{(ks)} & \multicolumn{1}{c}{($\mathrm{count\ s^{-1}}$)} & (\%) & & \multicolumn{1}{c}{(UT)} & \multicolumn{1}{c}{(UT)} & \multicolumn{1}{c}{(s)} & \multicolumn{1}{c}{($\mathrm{count\ s^{-1}}$)} & \multicolumn{1}{c}{($10^{-12}\ \mathrm{erg\ s^{-1}\ cm^{-2}}$)}\\
\hline  
112970601\tablefootmark{e} &  M. Turner  & 2000-09-17 18:41:04 & 2000-09-17 19:13:58 &2.0&  $0.102 \pm 0.005$  & 2.94 & & \dots\dots\dots & \dots\dots\dots & \dots\dots & \dots\dots\dots\dots\dots & \dots\dots\\
112970501\tablefootmark{f,h} &  M. Turner  & 2000-09-21 09:21:08 & 2000-09-21 15:16:37 &24.9&  $0.029 \pm 0.001$ & 24.2 & & \dots\dots\dots & \dots\dots\dots & \dots\dots & \dots\dots\dots\dots\dots & \dots\dots \\
112971601\tablefootmark{f} &  M. Turner  & 2001-03-31 11:31:01 & 2001-03-31 12:40:31 &4.0&  $0.038 \pm 0.002$  & 13.0 & & \dots\dots\dots & \dots\dots\dots & \dots\dots & \dots\dots\dots\dots\dots &\dots\dots\\
112972101\tablefootmark{h} &  M. Turner  & 2001-09-04 02:34:33 & 2001-09-04 08:41:10 &21.7&  $0.099 \pm 0.003$  & 21.6 & 1& 08:29:45 & >08:41:10 & >685 & $0.105 \pm 0.021$ & $29.6$\\
111350101\tablefootmark{k} &  B. Aschenbach  & 2002-02-26 06:40:39 & 2002-02-26 17:55:35 &40.0&  $0.105 \pm 0.001$ & 22.1 & & \dots\dots\dots & \dots\dots\dots & \dots\dots & \dots\dots\dots\dots\dots & \dots\dots\\
111350301\tablefootmark{k} &  B. Aschenbach  & 2002-10-03 07:15:59 & 2002-10-03 11:34:19 &15.4&  $0.100 \pm 0.002$ & 21.6 & 2& 10:08:32 & 10:52:01 & 2609 & $0.289 \pm 0.018$ & $27.9$\\
202670501\tablefootmark{h} &  A. Goldwurm  & 2004-03-28 16:44:29 & 2004-03-30 03:25:39 &105.6&  $0.200 \pm 0.001$ & 20.6 & & \dots\dots\dots & \dots\dots\dots & \dots\dots & \dots\dots\dots\dots\dots & \dots\dots\\
202670601\tablefootmark{h} &  A. Goldwurm  & 2004-03-30 17:16:48 & 2004-04-01 03:35:08 &107.0&  $0.187 \pm 0.001$ & 20.1 & 3& 40:38:22 & 42:10:58 & 5556 & $0.079 \pm 0.007$ & $8.79$\\
 \rule[0.5ex]{4.5em}{0.55pt} & \rule[0.5ex]{6em}{0.55pt} & \rule[0.5ex]{8.5em}{0.55pt} & \rule[0.5ex]{8.5em}{0.55pt} & \rule[0.5ex]{2.5em}{0.55pt} & \rule[0.5ex]{5.5em}{0.55pt} & \rule[0.5ex]{1.5em}{0.55pt} & 4& 46:51:25 & 47:31:27 & 2402 & $0.154 \pm 0.012$ & $17.1$\\
202670701 &  A. Goldwurm  & 2004-08-31 03:33:48 & 2004-09-01 16:05:43 &127.5&  $0.190 \pm 0.002$ & 14.6 & 5& 08:48:44 & 10:56:42 & 7678 & $0.075 \pm 0.008$ & $4.99$\\
 \rule[0.5ex]{4.5em}{0.55pt} & \rule[0.5ex]{6em}{0.55pt} & \rule[0.5ex]{8.5em}{0.55pt} & \rule[0.5ex]{8.5em}{0.55pt} & \rule[0.5ex]{2.5em}{0.55pt} & \rule[0.5ex]{5.5em}{0.55pt}& \rule[0.5ex]{1.5em}{0.55pt} & 6& 30:36:54 & 30:54:29 & 927 & $0.023 \pm 0.010$ & $2.53$\\
202670801 &  A. Goldwurm  & 2004-09-02 03:23:36 & 2004-09-03 16:03:37 &130.8&  $0.157 \pm 0.001$ &20.1 & & \dots\dots\dots & \dots\dots\dots & \dots\dots & \dots\dots\dots\dots\dots & \dots\dots\\
302882601 &  R. Wijnands  & 2006-02-27 04:26:53 & 2006-02-27 05:49:46 &4.9&  $0.109 \pm 0.004$ &10.6 & & \dots\dots\dots & \dots\dots\dots & \dots\dots & \dots\dots\dots\dots\dots & \dots\dots\\
302884001 &  R. Wijnands  & 2006-09-08 17:18:55 & 2006-09-08 18:42:09 &5.0&  $0.092 \pm 0.003$ &11.3 & & \dots\dots\dots & \dots\dots\dots & \dots\dots & \dots\dots\dots\dots\dots & \dots\dots\\
506291201\tablefootmark{e} &  R. Wijnands  & 2007-02-27 06:07:31 & 2007-02-27 16:51:07 &38.6&  $0.048 \pm 0.001$ &23.7 & & \dots\dots\dots & \dots\dots\dots & \dots\dots & \dots\dots\dots\dots\dots & \dots\dots\\
402430701 &  D. Porquet  & 2007-03-30 21:27:07 & 2007-03-31 06:28:47 &32.3&  $0.121 \pm 0.002$ &23.4 & & \dots\dots\dots & \dots\dots\dots & \dots\dots & \dots\dots\dots\dots\dots & \dots\dots\\
402430301 &  D. Porquet  & 2007-04-01 15:06:44 & 2007-04-02 17:05:07 &101.3&  $0.111 \pm 0.001$ &21.8 & & \dots\dots\dots & \dots\dots\dots & \dots\dots & \dots\dots\dots\dots\dots & \dots\dots\\
402430401 &  D. Porquet  & 2007-04-03 16:43:24 & 2007-04-04 19:48:15 &86.4&  $0.105 \pm 0.001$&21.9  & 7& 29:11:21 & 30:09:27 & 3486 & $0.216 \pm 0.014$ & $14.7$\\
 \rule[0.5ex]{4.5em}{0.55pt} & \rule[0.5ex]{6em}{0.55pt} & \rule[0.5ex]{8.5em}{0.55pt} & \rule[0.5ex]{8.5em}{0.55pt} & \rule[0.5ex]{2.5em}{0.55pt} & \rule[0.5ex]{5.5em}{0.55pt}& \rule[0.5ex]{1.5em}{0.55pt} & 8& 35:32:10 & 35:38:01 & 351 & $0.060 \pm 0.009$ & $4.52$\\
 \rule[0.5ex]{4.5em}{0.55pt} & \rule[0.5ex]{6em}{0.55pt} & \rule[0.5ex]{8.5em}{0.55pt} & \rule[0.5ex]{8.5em}{0.55pt} & \rule[0.5ex]{2.5em}{0.55pt} & \rule[0.5ex]{5.5em}{0.55pt}& \rule[0.5ex]{1.5em}{0.55pt} & 9& 38:27:12 & 38:51:31 & 1458 & $0.061 \pm 0.009$ & $4.53$\\
 \rule[0.5ex]{4.5em}{0.55pt} & \rule[0.5ex]{6em}{0.55pt} & \rule[0.5ex]{8.5em}{0.55pt} & \rule[0.5ex]{8.5em}{0.55pt} & \rule[0.5ex]{2.5em}{0.55pt} & \rule[0.5ex]{5.5em}{0.55pt}& \rule[0.5ex]{1.5em}{0.55pt} & 10& 40:44:52 & 40:59:51 & 959 & $0.042 \pm 0.005$ & $2.85$\\
504940201 &  R. Wijnands  & 2007-09-06 10:27:56 & 2007-09-06 13:39:06 &11.1&  $0.109 \pm 0.002$ &20.5 & & \dots\dots\dots & \dots\dots\dots & \dots\dots & \dots\dots\dots\dots\dots & \dots\dots\\
511000301\tablefootmark{j} &  R. Wijnands  & 2008-03-03 23:47:45 & 2008-03-04 01:19:25 &5.1&  $0.111 \pm 0.004$ &12.3 & & \dots\dots\dots & \dots\dots\dots & \dots\dots & \dots\dots\dots\dots\dots & \dots\dots\\
505670101 &  A. Goldwurm  & 2008-03-23 17:21:01 & 2008-03-24 20:17:25 &96.6&  $0.110 \pm 0.001$  &21.7& & \dots\dots\dots & \dots\dots\dots & \dots\dots & \dots\dots\dots\dots\dots & \dots\dots\\
511000401\tablefootmark{j} &  R. Wijnands  & 2008-09-23 15:53:29 & 2008-09-23 17:08:17 &5.1&  $0.096 \pm 0.004$ &27.2 & & \dots\dots\dots & \dots\dots\dots & \dots\dots & \dots\dots\dots\dots\dots & \dots\dots\\
554750401 &  A. Goldwurm  & 2009-04-01 01:17:45 & 2009-04-01 11:58:54 &38.0&  $0.105 \pm 0.001$  &21.9& & \dots\dots\dots & \dots\dots\dots & \dots\dots & \dots\dots\dots\dots\dots & \dots\dots\\
554750501 &  A. Goldwurm  & 2009-04-03 01:55:00 & 2009-04-03 13:43:39 &42.4&  $0.101 \pm 0.001$ &27.2 & 11& 08:47:39 & 09:13:39 & 1560 & $0.059 \pm 0.010$ & $10.7$\\
554750601 &  A. Goldwurm  & 2009-04-05 03:52:26 & 2009-04-05 13:02:10 &32.8&  $0.104 \pm 0.002$ &27.2 & & \dots\dots\dots & \dots\dots\dots & \dots\dots & \dots\dots\dots\dots\dots & \dots\dots\\
604300601 &  D. Porquet  & 2011-03-28 08:11:53 & 2011-03-28 21:15:04 &45.2&  $0.092 \pm 0.001$  &22.1& & \dots\dots\dots & \dots\dots\dots & \dots\dots & \dots\dots\dots\dots\dots & \dots\dots\\
604300701 &  D. Porquet  & 2011-03-30 09:25:00 & 2011-03-30 21:12:57 &42.3&  $0.099 \pm 0.001$ &22.2 & 12& 17:42:01 & 18:15:46 & 2025 & $0.119 \pm 0.010$ & $11.9$\\
604300801 &  D. Porquet  & 2011-04-01 09:01:06 & 2011-04-01 19:09:54 &37.3&  $0.090 \pm 0.002$ &22.5 & & \dots\dots\dots & \dots\dots\dots & \dots\dots & \dots\dots\dots\dots\dots & \dots\dots\\
604300901 &  D. Porquet  & 2011-04-03 08:14:00 & 2011-04-03 18:55:51 &36.5&  $0.098 \pm 0.002$ &22.4 & 13& 07:51:24 & 08:34:59 & 2615 & $0.104 \pm 0.008$ & $4.26$\\
604301001 &  D. Porquet  & 2011-04-05 07:31:48 & 2011-04-05 20:26:33 &48.1&  $0.089 \pm 0.002$ &22.3 & & \dots\dots\dots & \dots\dots\dots & \dots\dots & \dots\dots\dots\dots\dots & \dots\dots\\
658600101 &  C. Darren Dowell  & 2011-08-31 23:36:30 & 2011-09-01 13:04:17 &47.6&  $0.098 \pm 0.001$ &21.6 & & \dots\dots\dots & \dots\dots\dots & \dots\dots & \dots\dots\dots\dots\dots & \dots\dots\\
658600201 &  C. Darren Dowell  & 2011-09-01 20:25:57 & 2011-09-02 10:44:22 &51.3&  $0.095 \pm 0.001$ &23.5 & & \dots\dots\dots & \dots\dots\dots & \dots\dots & \dots\dots\dots\dots\dots & \dots\dots\\
674600601 &  A. Goldwurm  & 2012-03-13 04:14:14 & 2012-03-13 09:47:24 &19.6&  $0.096 \pm 0.002$&21.7  & & \dots\dots\dots & \dots\dots\dots & \dots\dots & \dots\dots\dots\dots\dots & \dots\dots\\
674600701 &  A. Goldwurm  & 2012-03-15 05:09:04 & 2012-03-15 09:10:51 &14.0&  $0.094 \pm 0.002$ &23.0 & & \dots\dots\dots & \dots\dots\dots & \dots\dots & \dots\dots\dots\dots\dots & \dots\dots\\
674601101 &  A. Goldwurm  & 2012-03-17 03:21:30 & 2012-03-17 10:09:54 &25.7&  $0.101 \pm 0.003$ &22.8 & & \dots\dots\dots & \dots\dots\dots & \dots\dots & \dots\dots\dots\dots\dots & \dots\dots\\
674600801 &  A. Goldwurm  & 2012-03-19 04:14:14 & 2012-03-19 10:12:43 &21.0&  $0.096 \pm 0.002$ &25.7 & & \dots\dots\dots & \dots\dots\dots & \dots\dots & \dots\dots\dots\dots\dots & \dots\dots\\
674601001 &  A. Goldwurm  & 2012-03-21 03:52:26 & 2012-03-21 10:07:06 &22.0&  $0.094 \pm 0.002$ &23.3 & & \dots\dots\dots & \dots\dots\dots & \dots\dots & \dots\dots\dots\dots\dots & \dots\dots\\
694640301 &  R. Terrier  & 2012-08-31 11:42:07 & 2012-08-31 22:57:43 &40.0&  $0.078 \pm 0.001$ &23.0 & & \dots\dots\dots & \dots\dots\dots & \dots\dots & \dots\dots\dots\dots\dots & \dots\dots\\
694640401\tablefootmark{f} &  R. Terrier  & 2012-09-02 19:09:49 & 2012-09-03 09:34:03 &53.0&  $0.010 \pm 0.0002$ &22.3 & & \dots\dots\dots & \dots\dots\dots & \dots\dots & \dots\dots\dots\dots\dots & \dots\dots\\
694641001\tablefootmark{f} &  R. Terrier  & 2012-09-23 20:42:07 & 2012-09-24 09:36:52 &46.0&  $0.015 \pm 0.0002$&23.2  & & \dots\dots\dots & \dots\dots\dots & \dots\dots & \dots\dots\dots\dots\dots & \dots\dots\\
694641101\tablefootmark{f} &  R. Terrier  & 2012-09-24 10:38:50 & 2012-09-24 21:53:44 &40.0&  $0.068 \pm 0.001$ &15.1 & & \dots\dots\dots & \dots\dots\dots & \dots\dots & \dots\dots\dots\dots\dots & \dots\dots\\
724210201 &  G. Ponti  & 2013-08-30 20:52:40 & 2013-08-31 12:26:18 &55.6&  $0.534 \pm 0.003$ &15.7 & & \dots\dots\dots & \dots\dots\dots & \dots\dots & \dots\dots\dots\dots\dots & \dots\dots\\
700980101 &  D. Haggard  & 2013-09-10 04:12:07 & 2013-09-10 14:11:46 &35.7&  $0.538 \pm 0.003$&15.9  & & \dots\dots\dots & \dots\dots\dots & \dots\dots & \dots\dots\dots\dots\dots & \dots\dots\\
724210501 &  G. Ponti  & 2013-09-22 21:54:32 & 2013-09-23 09:17:52 &39.4&  $0.506 \pm 0.003$ &16.1 & & \dots\dots\dots & \dots\dots\dots & \dots\dots & \dots\dots\dots\dots\dots & \dots\dots\\
723410301 &  N. Grosso  & 2014-02-28 18:18:41 & 2014-03-01 08:53:15 &51.9&  $0.320 \pm 0.002$ &23.3 & & \dots\dots\dots & \dots\dots\dots & \dots\dots & \dots\dots\dots\dots\dots & \dots\dots\\
723410401 &  N. Grosso  & 2014-03-10 14:49:09 & 2014-03-11 05:57:28 &54.0&  $0.312 \pm 0.002$ &18.6 & 14& 16:44:48 & 19:03:51 & 8468 & $0.119 \pm 0.008$ & $6.76$\\
723410501 &  N. Grosso  & 2014-04-02 03:42:35 & 2014-04-02 20:22:19 &54.9&  $0.287 \pm 0.002$ &18.2 & 15& 16:53:00 & 17:08:44 & 944 & $0.200 \pm 0.013$ & $20.7$\\
690441801\tablefootmark{i} & G.L. Isra$\ddot{\mathrm{e}}$l & 2014-04-03 05:48:45 & 2014-04-04 05:01:14 &83.5& $0.294 \pm 0.002$ &18.2& & \dots\dots\dots & \dots\dots\dots & \dots\dots & \dots\dots\dots\dots\dots & \dots\dots\\
743630201\tablefootmark{i} & G. Ponti & 2014-08-30 20:00:24 & 2014-08-31 04:54:26 & 28.5 & $0.170 \pm 0.003$ &18.2& 16& 23:46:11 & 25:19:18 & 5587 & $0.219 \pm 0.008$ & $17.2$ \\
 \rule[0.5ex]{4.5em}{0.55pt} & \rule[0.5ex]{6em}{0.55pt} & \rule[0.5ex]{8.5em}{0.55pt} & \rule[0.5ex]{8.5em}{0.55pt} & \rule[0.5ex]{2.5em}{0.55pt} & \rule[0.5ex]{5.5em}{0.55pt}& \rule[0.5ex]{1.5em}{0.55pt} & 17& 28:36:49 & 28:53:19 & 990 & $0.234 \pm 0.023$ & $18.4$ \\
743630301 & G. Ponti & 2014-08-31 21:03:54 & 2014-09-01 04:01:15 & 22.3 & $0.169 \pm 0.003$&18.5 & 18& 25:21:16 & 25:55:05 & 2029 & $0.135 \pm 0.012$ & $12.4$ \\
743630401 & G. Ponti & 2014-09-27 19:47:57 & 2014-09-28 02:57:18 & 22.9 & $0.177 \pm 0.002$&18.6 & & \dots\dots\dots & \dots\dots\dots & \dots\dots & \dots\dots\dots\dots\dots & \dots\dots\\
743630501 & G. Ponti & 2014-09-28 21:42:09 & 2014-09-29 08:12:51 & 33.7 & $0.167 \pm 0.002$ &18.1& 19& 30:06:58 & 30:12:47 & 349 & $0.160 \pm 0.031$ & $14.7$ \\
743630601 & G. Ponti & 2015-02-26 06:58:40 & 2015-02-26 15:26:25 & 27.1 & $0.152 \pm 0.002$&18.9 & & \dots\dots\dots & \dots\dots\dots & \dots\dots & \dots\dots\dots\dots\dots & \dots\dots\\
743630701\tablefootmark{g}~~ & G. Ponti & 2015-03-31 10:25:12 & 2015-03-31 10:26:38 & 0.1 & $0.253 \pm 0.058$&5.71 & & \dots\dots\dots & \dots\dots\dots & \dots\dots & \dots\dots\dots\dots\dots & \dots\dots\\
743630801 & G. Ponti & 2015-04-01 09:14:43 & 2015-04-01 15:55:24 & 21.5 & $0.164 \pm 0.002$ &19.2& & \dots\dots\dots & \dots\dots\dots & \dots\dots & \dots\dots\dots\dots\dots & \dots\dots\\
743630901 & G. Ponti & 2015-04-02 09:39:43 & 2015-04-02 11:35:50 & 6.2 & $0.182 \pm 0.004$ &10.6& & \dots\dots\dots & \dots\dots\dots & \dots\dots & \dots\dots\dots\dots\dots & \dots\dots\\
\hline
\end{tabular}
}
\tablefoot{
\tablefoottext{a} {The average flare detection efficiency above the corresponding non-flaring level.}
\tablefoottext{b} {The flare start and end times are given in hh:mm:ss since the day of the observation start.
Flares beginning or ending at the start or stop of the observation lead to a lower limit on the flare duration and a lower or upper limit on the flare mean count rate and mean flux.
The flux value of these flares were taken equal to this limit in the flaring rate study.}
\tablefoottext{c} {The flare mean count rates are computed after subtraction of the non-flaring level.}
\tablefoottext{d} {Mean unabsorbed flux between 2 and 10$\,$keV determined for $N_\mathrm{H}=14.3\times 10^{22}\ \mathrm{cm^{-2}}$ and $\Gamma=2$.}
\tablefoottext{e} {For this observation, the Galactic center was observed only with EPIC/pn.}
\tablefoottext{f} {For these observations, the Galactic center was observed only with EPIC/MOS1 and 2.}
\tablefoottext{g} {The data transfer from XMM-Newton to the Earth during this observation was affected by the GALILEO launch and Early Orbit Phase.}
\tablefoottext{h} {Frame window extended mode.}
\tablefoottext{i} {Small window.}
\tablefoottext{j} {Thin filter.}
\tablefoottext{k} {Thick filter.}
}
\normalsize
\end{table*}

\begin{table*}[ht]
\caption[Observation log of the Chandra observations and the detected X-ray flares in 1999--2015]{Observation log of public Chandra observations and the X-ray flares detected in this work.}
\centering
\scalebox{0.55}{
\label{table:chandra1}
\begin{tabular}{@{}rlllcrcccrrcrc@{}}
\hline
\hline
\multicolumn{8}{c}{Observations} & \multicolumn{6}{c}{Flares} \\
\multicolumn{8}{l}{\rule[0.5ex]{50.5em}{0.55pt}} & \multicolumn{6}{r}{\rule[0.5ex]{34.5em}{0.55pt}} \\
\multicolumn{12}{c}{} & \multicolumn{2}{c}{Mean} \\
\multicolumn{12}{c}{} & \multicolumn{2}{r}{\rule[0.5ex]{18em}{0.55pt}} \\
\multicolumn{1}{c}{ObsID} & \multicolumn{1}{c}{PI} & \multicolumn{1}{c}{Start} & \multicolumn{1}{c}{End} & \multicolumn{1}{c}{Duration}  & \multicolumn{1}{c}{Instrument} & \multicolumn{1}{c}{Non-flaring level} & \multicolumn{1}{c}{$\eta_{obs}$\tablefootmark{a}} & \multicolumn{1}{c}{\#} & \multicolumn{1}{c}{Start\tablefootmark{b}} & \multicolumn{1}{c}{Stop\tablefootmark{b}} & \multicolumn{1}{c}{Duration} & \multicolumn{1}{c}{Count rate\tablefootmark{c}} & \multicolumn{1}{c}{Flux\tablefootmark{d}} \\
 & & \multicolumn{1}{c}{(UT)} & \multicolumn{1}{c}{(UT)} & \multicolumn{1}{c}{(ks)} & & \multicolumn{1}{c}{($\mathrm{count\ s^{-1}}$)} & (\%) & & \multicolumn{1}{c}{(UT)} & \multicolumn{1}{c}{(UT)} & \multicolumn{1}{c}{(s)} & \multicolumn{1}{c}{($\mathrm{count\ s^{-1}}$)} &\multicolumn{1}{c}{($10^{-12}\ \mathrm{erg\ s^{-1}\ cm^{-2}}$)}\\
\hline  
242&  G. Garmire  & 1999-09-21 02:40:49 & 1999-09-21 17:03:17 &46.5&  ACIS-I3  &  $0.0048 \pm 0.0001$ &42.1 & 1& <02:40:49 & 04:10:23 & >5374 & $0.004 \pm 0.001$ & $1.30$\\
1561& F. Baganoff    & 2000-10-26 19:05:19 & 2001-07-14 05:56:28 &49.9&  ACIS-I3  &  $0.0059 \pm 0.0008$ &42.5 & 2& 26:36:54 & 26:46:39 & 585 & $0.030 \pm 0.014$ & $3.52$\\
\rule[0.5ex]{2.5em}{0.55pt} & \rule[0.5ex]{6em}{0.55pt} & \rule[0.5ex]{8.5em}{0.55pt} & \rule[0.5ex]{8.5em}{0.55pt} & \rule[0.5ex]{2.5em}{0.55pt} & \rule[0.5ex]{7em}{0.55pt} & \rule[0.5ex]{6.5em}{0.55pt} & \rule[0.5ex]{1.5em}{0.55pt} & 3& 27:55:35 & 30:46:46 & 10335 & $0.110 \pm 0.004$ & $11.5$\\
2951&  G. Garmire  & 2002-02-19 14:26:32 & 2002-02-19 18:32:35 &12.5&  ACIS-I3  &  $0.0039 \pm 0.0009$ &36.8 & 4& 15:48:32 & 16:10:09 & 1297 & $0.012 \pm 0.001$ & $1.50$ \\
2952&  G. Garmire  & 2002-03-23 12:23:04 & 2002-03-23 16:10:07 &12.0&  ACIS-I3  &  $0.0053 \pm 0.0007$ &40.1 & & \dots\dots\dots & \dots\dots\dots & \dots\dots & \dots\dots\dots\dots\dots & \dots\dots \\
2953&  G. Garmire  & 2002-04-19 10:57:39 & 2002-04-19 14:13:34 &11.7&  ACIS-I3  &  $0.0042 \pm 0.0006$ &37.2 & & \dots\dots\dots & \dots\dots\dots & \dots\dots & \dots\dots\dots\dots\dots & \dots\dots \\
2954&  G. Garmire  & 2002-05-07 09:23:04 & 2002-05-07 13:18:12 &12.6&  ACIS-I3  &  $0.0047 \pm 0.0006$ &39.3 & & \dots\dots\dots & \dots\dots\dots & \dots\dots & \dots\dots\dots\dots\dots & \dots\dots \\
2943&  F. Baganoff  & 2002-05-22 23:17:41 & 2002-05-23 09:55:42 &38.2&  ACIS-I3  &  $0.0054 \pm 0.0003$ &42.3 & & \dots\dots\dots & \dots\dots\dots & \dots\dots & \dots\dots\dots\dots\dots & \dots\dots \\
3663&  F. Baganoff  & 2002-05-24 11:49:02 & 2002-05-24 22:56:10 &38.5&  ACIS-I3  &  $0.0056 \pm 0.0003$ &42.2 & 5& 19:06:04 & 20:23:14 & 4630 & $0.015 \pm 0.002$ & $2.31$\\
3392&  F. Baganoff  & 2002-05-25 15:13:52 & 2002-05-27 14:32:44 &168.9&  ACIS-I3  &  $0.0052 \pm 0.0005$&43.3  & 6& 28:04:29 & 28:54:04 & 2975 & $0.018 \pm 0.009$ & $0.87$\\
\rule[0.5ex]{2.5em}{0.55pt} & \rule[0.5ex]{6em}{0.55pt} & \rule[0.5ex]{8.5em}{0.55pt} & \rule[0.5ex]{8.5em}{0.55pt} & \rule[0.5ex]{2.5em}{0.55pt} & \rule[0.5ex]{7em}{0.55pt} & \rule[0.5ex]{6.5em}{0.55pt} & \rule[0.5ex]{1.5em}{0.55pt} & 7& 37:37:32 & 39:02:56 & 5000 & $0.015 \pm 0.009$ & $1.00$\\
\rule[0.5ex]{2.5em}{0.55pt} & \rule[0.5ex]{6em}{0.55pt} & \rule[0.5ex]{8.5em}{0.55pt} & \rule[0.5ex]{8.5em}{0.55pt} & \rule[0.5ex]{2.5em}{0.55pt} & \rule[0.5ex]{7em}{0.55pt} & \rule[0.5ex]{6.5em}{0.55pt} & \rule[0.5ex]{1.5em}{0.55pt} & 8& 53:33:16 & 53:49:15 & 959 & $0.024 \pm 0.008$ & $0.69$\\
3393&  F. Baganoff  & 2002-05-28 05:33:33 & 2002-05-30 02:33:05 &160.1&  ACIS-I3  &  $0.0048 \pm 0.0003$&43.3  & 9& 15:10:11 & 16:02:55 & 3164 & $0.081 \pm 0.040$ & $0.61$\\
\rule[0.5ex]{2.5em}{0.55pt} & \rule[0.5ex]{6em}{0.55pt} & \rule[0.5ex]{8.5em}{0.55pt} & \rule[0.5ex]{8.5em}{0.55pt} & \rule[0.5ex]{2.5em}{0.55pt} & \rule[0.5ex]{7em}{0.55pt} & \rule[0.5ex]{6.5em}{0.55pt} & \rule[0.5ex]{1.5em}{0.55pt} & 10& 29:40:49 & 31:03:47 & 4978 & $0.027 \pm 0.012$ & $3.28$\\
\rule[0.5ex]{2.5em}{0.55pt} & \rule[0.5ex]{6em}{0.55pt} & \rule[0.5ex]{8.5em}{0.55pt} & \rule[0.5ex]{8.5em}{0.55pt} & \rule[0.5ex]{2.5em}{0.55pt} & \rule[0.5ex]{7em}{0.55pt} & \rule[0.5ex]{6.5em}{0.55pt} & \rule[0.5ex]{1.5em}{0.55pt} & 11& 2:37:15 & 42:52:43 & 928 & $0.058 \pm 0.016$ & $6.40$\\
3665&  F. Baganoff  & 2002-06-03 01:22:29 & 2002-06-04 03:23:00 &91.1&  ACIS-I3  &  $0.0050 \pm 0.0002$ &43.1 & & \dots\dots\dots & \dots\dots\dots & \dots\dots & \dots\dots\dots\dots\dots & \dots\dots \\
3549&  G. Garmire  & 2003-06-19 18:26:46 & 2003-06-20 01:52:50 &25.1&  ACIS-I3  &  $0.0055 \pm 0.0004$& 42.3 & & \dots\dots\dots & \dots\dots\dots & \dots\dots & \dots\dots\dots\dots\dots & \dots\dots \\
4683&  G. Garmire  & 2004-07-05 22:32:02 & 2004-07-06 12:54:49 &50.2&  ACIS-I3  &  $0.0049 \pm 0.0003$  &42.4& & \dots\dots\dots & \dots\dots\dots & \dots\dots & \dots\dots\dots\dots\dots & \dots\dots \\
4684&  G. Garmire  & 2004-07-06 22:27:16 & 2004-07-07 12:50:57 &50.2&  ACIS-I3  &  $0.0056 \pm 0.0004$ &42.7 & 12& 27:17:55 & 28:04:40 & 2805 & $0.039 \pm 0.022$ & $4.87$\\
5360&  F. Baganoff  & 2004-08-28 12:02:59 & 2004-08-28 13:59:10 &5.2&  ACIS-I3  &  $0.0036 \pm 0.0008$ &33.7 & & \dots\dots\dots & \dots\dots\dots & \dots\dots & \dots\dots\dots\dots\dots & \dots\dots \\
6113&  F. Baganoff  & 2005-02-27 06:23:57 & 2005-02-27 08:27:17 &4.9&  ACIS-I3  &  $0.0054 \pm 0.0011$ &34.1 & & \dots\dots\dots & \dots\dots\dots & \dots\dots & \dots\dots\dots\dots\dots & \dots\dots \\
5950&  F. Baganoff  & 2005-07-24 19:56:25 & 2005-07-25 10:05:43 &49.2&  ACIS-I3  &  $0.0052 \pm 0.0003$ &42.4 & & \dots\dots\dots & \dots\dots\dots & \dots\dots & \dots\dots\dots\dots\dots  & \dots\dots\\
5951&  F. Baganoff  & 2005-07-27 19:06:08 & 2005-07-28 08:25:32 &45.2&  ACIS-I3  &  $0.0048 \pm 0.0003$ &42.2 & & \dots\dots\dots & \dots\dots\dots & \dots\dots & \dots\dots\dots\dots\dots & \dots\dots \\
5952&  F. Baganoff  & 2005-07-29 19:48:58 & 2005-07-30 09:05:36 &45.9&  ACIS-I3  &  $0.0055 \pm 0.0003$ &42.5 & 13& 26:31:09 & 27:29:10 & 3481 & $0.016 \pm 0.002$ & $2.16$\\
5953&  F. Baganoff  & 2005-07-30 19:37:18 & 2005-07-31 09:10:32 &46.0&  ACIS-I3  &  $0.0052 \pm 0.0002$ &42.4 & 14& 22:13:27 & 22:47:55 & 2068 & $0.043 \pm 0.004$ & $5.13$\\
5954&  F. Baganoff  & 2005-08-01 20:13:00 & 2005-08-02 01:16:15 &18.1&  ACIS-I3  &  $0.0042 \pm 0.0005$&40.2  & & \dots\dots\dots & \dots\dots\dots & \dots\dots & \dots\dots\dots\dots\dots & \dots\dots \\
6639&  F. Baganoff  & 2006-04-11 05:31:13 & 2006-04-11 07:06:03 &4.5&  ACIS-I3  &  $0.0044 \pm 0.0011$ &34.5 & & \dots\dots\dots & \dots\dots\dots & \dots\dots & \dots\dots\dots\dots\dots & \dots\dots \\
6640&  F. Baganoff  & 2006-05-03 22:24:25 & 2006-05-04 00:22:07 &5.2&  ACIS-I3  &  $0.0076 \pm 0.0013$ &36.4 & & \dots\dots\dots & \dots\dots\dots & \dots\dots & \dots\dots\dots\dots\dots & \dots\dots \\
6641&  F. Baganoff  & 2006-06-01 16:05:47 & 2006-06-01 17:55:45 &5.1&  ACIS-I3  &  $0.0097 \pm 0.0014$ &36.2 & & \dots\dots\dots & \dots\dots\dots & \dots\dots & \dots\dots\dots\dots\dots & \dots\dots \\
6642&  F. Baganoff  & 2006-07-04 10:59:35 & 2006-07-04 12:51:17 &5.2&  ACIS-I3  &  $0.0070 \pm 0.0012$ &35.7 & & \dots\dots\dots & \dots\dots\dots & \dots\dots & \dots\dots\dots\dots\dots & \dots\dots \\
6363&  F. Baganoff  & 2006-07-17 03:56:11 & 2006-07-17 12:41:06 &30.2&  ACIS-I3  &  $0.0042 \pm 0.0004$&41.8  & 15& 05:52:05 & 06:35:07 & 2516 & $0.055 \pm 0.007$ & $6.49$\\
6643&  F. Baganoff  & 2006-07-30 14:28:24 & 2006-07-30 16:21:53 &5.0&  ACIS-I3  &  $0.0042 \pm 0.0009$ &33.8 & & \dots\dots\dots & \dots\dots\dots & \dots\dots & \dots\dots\dots\dots\dots & \dots\dots \\
6644&  F. Baganoff  & 2006-08-22 05:52:40 & 2006-08-22 07:46:29 &5.0&  ACIS-I3  &  $0.0054 \pm 0.0011$ &35.1 & & \dots\dots\dots & \dots\dots\dots & \dots\dots & \dots\dots\dots\dots\dots & \dots\dots \\
6645&  F. Baganoff  & 2006-09-25 13:48:17 & 2006-09-25 15:41:27 &5.2&  ACIS-I3  &  $0.0061 \pm 0.0009$ &20.7 & 16& 14:00:33 & 14:23:03 & 1350 & $0.008 \pm 0.003$ & $2.07$\\
6646&  F. Baganoff  & 2006-10-29 03:43:05 & 2006-10-29 05:12:30 &5.2&  ACIS-I3  &  $0.0071 \pm 0.0012$ &36.3 & 17& <03:43:05 & 03:49:51 & $>407$ & $0.012 \pm 0.006$ & $3.45$ \\
7554&  F. Baganoff  & 2007-02-11 06:15:10 & 2007-02-11 08:14:16 &5.1&  ACIS-I3  &  $0.0044 \pm 0.0009$ &36.7 & & \dots\dots\dots & \dots\dots\dots & \dots\dots & \dots\dots\dots\dots\dots & \dots\dots \\
7555&  F. Baganoff  & 2007-03-25 22:53:57 & 2007-03-26 00:50:14 &5.2&  ACIS-I3  &  $0.0055 \pm 0.0011$ &34.3 & & \dots\dots\dots & \dots\dots\dots & \dots\dots & \dots\dots\dots\dots\dots & \dots\dots \\
7556&  F. Baganoff  & 2007-05-17 01:02:59 & 2007-05-17 03:11:34 &5.0&  ACIS-I3  &  $0.0060 \pm 0.0011$ &35.7 & & \dots\dots\dots & \dots\dots\dots & \dots\dots & \dots\dots\dots\dots\dots & \dots\dots \\
7557&  F. Baganoff  & 2007-07-20 02:25:15 & 2007-07-20 04:27:51 &5.0&  ACIS-I3  &  $0.0047 \pm 0.0006$ &35.7 & & \dots\dots\dots & \dots\dots\dots & \dots\dots & \dots\dots\dots\dots\dots & \dots\dots \\
7558&  F. Baganoff  & 2007-09-02 20:17:30 & 2007-09-02 22:01:29 &5.0&  ACIS-I3  &  $0.0072 \pm 0.0012$ &34.5 & & \dots\dots\dots & \dots\dots\dots & \dots\dots & \dots\dots\dots\dots\dots & \dots\dots \\
7559&  F. Baganoff  & 2007-10-26 10:02:16 & 2007-10-26 11:50:28 &5.1&  ACIS-I3  &  $0.0050 \pm 0.0010$ &35.8 & & \dots\dots\dots & \dots\dots\dots & \dots\dots & \dots\dots\dots\dots\dots & \dots\dots \\
9169&  F. Yusef-zadeh  & 2008-05-05 03:50:56 & 2008-05-05 12:05:56 &28.0&  ACIS-I3  &  $0.0055 \pm 0.0005$ &34.8 & 18& 10:35:14 & 11:42:44 & 4050 & $0.006 \pm 0.002$ & $1.27$\\
9170&  F. Yusef-zadeh  & 2008-05-06 02:58:17 & 2008-05-06 10:58:05 &27.1&  ACIS-I3  &  $0.0050 \pm 0.0004$ &42.6 & & \dots\dots\dots & \dots\dots\dots & \dots\dots & \dots\dots\dots\dots\dots & \dots\dots \\
9171&  F. Yusef-zadeh  & 2008-05-10 03:15:52 & 2008-05-10 11:24:06 &28.0&  ACIS-I3  &  $0.0048 \pm 0.0004$ &42.3 & & \dots\dots\dots & \dots\dots\dots & \dots\dots & \dots\dots\dots\dots\dots & \dots\dots \\
9172&  F. Yusef-zadeh  & 2008-05-11 03:34:30 & 2008-05-11 11:42:23 &27.8&  ACIS-I3  &  $0.0053 \pm 0.0005$ &42.0 & & \dots\dots\dots & \dots\dots\dots & \dots\dots & \dots\dots\dots\dots\dots & \dots\dots \\
9174&  F. Yusef-zadeh  & 2008-07-25 21:48:55 & 2008-07-26 06:25:59 &29.2&  ACIS-I3  &  $0.0044 \pm 0.0003$ &56.0 & & \dots\dots\dots & \dots\dots\dots & \dots\dots & \dots\dots\dots\dots\dots & \dots\dots \\
9173&  F. Yusef-zadeh  & 2008-07-26 21:18:02 & 2008-07-27 05:27:58 &28.1&  ACIS-I3  &  $0.0038 \pm 0.0004$ &42.5 & & \dots\dots\dots & \dots\dots\dots & \dots\dots & \dots\dots\dots\dots\dots & \dots\dots \\
10556&  F. Baganoff  & 2009-05-18 02:18:24 & 2009-05-19 10:22:34 &114.0&  ACIS-I3  &  $0.0053 \pm 0.0002$ &43.1 & 19 & 02:34:59 & 02:58:54 & 1435 & $0.031 \pm 0.007$ & $4.03$\\
\rule[0.5ex]{2.5em}{0.55pt} & \rule[0.5ex]{6em}{0.55pt} & \rule[0.5ex]{8.5em}{0.55pt} & \rule[0.5ex]{8.5em}{0.55pt} & \rule[0.5ex]{2.5em}{0.55pt} & \rule[0.5ex]{7em}{0.55pt} & \rule[0.5ex]{6.5em}{0.55pt} & \rule[0.5ex]{1.5em}{0.55pt} & 20& 09:41:05 & 10:39:26 & 3501 & $0.019 \pm 0.003$ & $2.57$\\
\rule[0.5ex]{2.5em}{0.55pt} & \rule[0.5ex]{6em}{0.55pt} & \rule[0.5ex]{8.5em}{0.55pt} & \rule[0.5ex]{8.5em}{0.55pt} & \rule[0.5ex]{2.5em}{0.55pt} & \rule[0.5ex]{7em}{0.55pt} & \rule[0.5ex]{6.5em}{0.55pt} & \rule[0.5ex]{1.5em}{0.55pt} & 21& 23:03:43 & 23:30:56 & 1633 & $0.091 \pm 0.007$ & $9.07$\\
\rule[0.5ex]{2.5em}{0.55pt} & \rule[0.5ex]{6em}{0.55pt} & \rule[0.5ex]{8.5em}{0.55pt} & \rule[0.5ex]{8.5em}{0.55pt} & \rule[0.5ex]{2.5em}{0.55pt} & \rule[0.5ex]{7em}{0.55pt} & \rule[0.5ex]{6.5em}{0.55pt} & \rule[0.5ex]{1.5em}{0.55pt} & 22& 24:29:22 & 24:51:37 & 465 & $0.091 \pm 0.015$ & $10.8$\\
11843&  G. Garmire  & 2010-05-13 02:11:23 & 2010-05-14 00:41:47 &80.0&  ACIS-I3  &  $0.0059 \pm 0.0003$ &42.9 & 23& 03:30:04 & 04:38:30 & 4106 & $0.036 \pm 0.026$ & $3.36$\\
13016&  F. Baganoff  & 2011-03-29 10:29:11 & 2011-03-29 15:56:33 &18.1&  ACIS-I3  &  $0.0035 \pm 0.0005$ &37.5 & 24& 10:40:51 & 11:33:35 & 3164 & $0.009 \pm 0.003$ & $1.70$\\
13017&  F. Baganoff  & 2011-03-31 10:28:17 & 2011-03-31 15:58:39 &18.1&  ACIS-I3  &  $0.0047 \pm 0.0005$&40.5  & & \dots\dots\dots & \dots\dots\dots & \dots\dots & \dots\dots\dots\dots\dots & \dots\dots \\
13508&  R. Terrier  & 2011-07-19 01:21:58 & 2011-07-19 10:38:44 &31.9&  ACIS-I0  &  $0.0028 \pm 0.0003$&37.8  & & \dots\dots\dots & \dots\dots\dots & \dots\dots & \dots\dots\dots\dots\dots & \dots\dots \\
12949&  R. Terrier  & 2011-07-21 07:14:23 & 2011-07-22 00:19:59 &59.2&  ACIS-I0  &  $0.0030 \pm 0.0001$&41.2  & 25& 18:04:31 & 18:11:33 & 422 & $0.045 \pm 0.011$ & $20.5$\\
13438&  R. Terrier  & 2011-07-29 05:32:16 & 2011-07-30 00:31:56 &67.1&  ACIS-I0  &  $0.0019 \pm 0.0001$ &36.9 & & \dots\dots\dots & \dots\dots\dots & \dots\dots & \dots\dots\dots\dots\dots & \dots\dots \\
13850&  F. Baganoff  & 2012-02-06 00:36:15 & 2012-02-06 17:53:58 &60.1&  ACIS-S3/HETG  &  $0.0061 \pm 0.0003$ &39.3 & & \dots\dots\dots & \dots\dots\dots & \dots\dots & \dots\dots\dots\dots\dots & \dots\dots \\
14392&  F. Baganoff  & 2012-02-09 06:15:50 & 2012-02-09 23:18:07 &59.2&  ACIS-S3/HETG  &  $0.0054 \pm 0.0003$ &39.4 & 26& 10:38:58 & 10:57:02 & 1084 & $0.016 \pm 0.002$ & $1.74$\\
\rule[0.5ex]{2.5em}{0.55pt} & \rule[0.5ex]{6em}{0.55pt} & \rule[0.5ex]{8.5em}{0.55pt} & \rule[0.5ex]{8.5em}{0.55pt} & \rule[0.5ex]{2.5em}{0.55pt} & \rule[0.5ex]{7em}{0.55pt} & \rule[0.5ex]{6.5em}{0.55pt} & \rule[0.5ex]{1.5em}{0.55pt} & 27& 14:25:32 & 16:03:51 & 5899 & $0.109 \pm 0.004$ & $20.5$\\
14394&  F. Baganoff  & 2012-02-10 03:15:10 & 2012-02-10 08:50:27 &18.1&  ACIS-S3/HETG  &  $0.0065 \pm 0.0005$ &39.7 & & \dots\dots\dots & \dots\dots\dots & \dots\dots & \dots\dots\dots\dots\dots & \dots\dots \\
14393&  F. Baganoff  & 2012-02-11 10:12:07 & 2012-02-11 22:19:03 &41.5&  ACIS-S3/HETG  &  $0.0077 \pm 0.0004$ &40.3 & & \dots\dots\dots & \dots\dots\dots & \dots\dots & \dots\dots\dots\dots\dots & \dots\dots \\
13856&  F. Baganoff  & 2012-03-15 08:44:14 & 2012-03-15 20:24:26 &40.1&  ACIS-S3/HETG  &  $0.0055 \pm 0.0004$&40.1  & & \dots\dots\dots & \dots\dots\dots & \dots\dots & \dots\dots\dots\dots\dots & \dots\dots \\
13857&  F. Baganoff  & 2012-03-17 08:56:51 & 2012-03-17 20:27:57 &39.6&  ACIS-S3/HETG  &  $0.0066 \pm 0.0005$ &39.9 & 28& 16:04:36 & 16:12:26 & 471 & $0.031 \pm 0.006$ & $4.41$\\
13854&  F. Baganoff  & 2012-03-20 10:12:19 & 2012-03-20 17:06:09 &23.1&  ACIS-S3/HETG  &  $0.0081 \pm 0.0011$ & 38.3& 29& 11:40:52 & 11:51:17 & 625 & $0.046 \pm 0.006$ & $4.47$\\
\rule[0.5ex]{2.5em}{0.55pt} & \rule[0.5ex]{6em}{0.55pt} & \rule[0.5ex]{8.5em}{0.55pt} & \rule[0.5ex]{8.5em}{0.55pt} & \rule[0.5ex]{2.5em}{0.55pt} & \rule[0.5ex]{7em}{0.55pt} & \rule[0.5ex]{6.5em}{0.55pt} & \rule[0.5ex]{1.5em}{0.55pt} & 30& 12:40:17 & 12:53:38 & 801 & $0.047 \pm 0.006$ & $4.00$\\
\rule[0.5ex]{2.5em}{0.55pt} & \rule[0.5ex]{6em}{0.55pt} & \rule[0.5ex]{8.5em}{0.55pt} & \rule[0.5ex]{8.5em}{0.55pt} & \rule[0.5ex]{2.5em}{0.55pt} & \rule[0.5ex]{7em}{0.55pt} & \rule[0.5ex]{6.5em}{0.55pt} & \rule[0.5ex]{1.5em}{0.55pt} & 31& 14:02:04 & 14:18:11 & 697 & $0.042 \pm 0.006$ & $4.34$\\
\rule[0.5ex]{2.5em}{0.55pt} & \rule[0.5ex]{6em}{0.55pt} & \rule[0.5ex]{8.5em}{0.55pt} & \rule[0.5ex]{8.5em}{0.55pt} & \rule[0.5ex]{2.5em}{0.55pt} & \rule[0.5ex]{7em}{0.55pt} & \rule[0.5ex]{6.5em}{0.55pt} & \rule[0.5ex]{1.5em}{0.55pt} & 32& 16:21:56 & 16:28:29 & 393 & $0.095 \pm 0.008$ & $3.72$\\
14413&  F. Baganoff  & 2012-03-21 06:43:00 & 2012-03-21 11:08:58 &14.7&  ACIS-S3/HETG  &  $0.0064 \pm 0.0006$ &39.1 & & \dots\dots\dots & \dots\dots\dots & \dots\dots & \dots\dots\dots\dots\dots & \dots\dots \\
13855&  F. Baganoff  & 2012-03-22 11:23:50 & 2012-03-22 17:29:22 &20.1&  ACIS-S3/HETG  &  $0.0067 \pm 0.0006$ &39.3 & & \dots\dots\dots & \dots\dots\dots & \dots\dots & \dots\dots\dots\dots\dots & \dots\dots \\
14414&  F. Baganoff  & 2012-03-23 17:47:45 & 2012-03-24 00:00:18 &20.1&  ACIS-S3/HETG  &  $0.0060 \pm 0.0005$ &39.3 & & \dots\dots\dots & \dots\dots\dots & \dots\dots & \dots\dots\dots\dots\dots & \dots\dots \\
13847&  F. Baganoff  & 2012-04-30 16:17:14 & 2012-05-02 11:37:48 &154.1&  ACIS-S3/HETG  &  $0.0067 \pm 0.0002$&38.5  & 33& 36:21:51 & 37:09:26 & 2855 & $0.016 \pm 0.001$ & $2.71$\\
14427&  F. Baganoff  & 2012-05-06 19:59:28 & 2012-05-07 18:51:38 &80.1&  ACIS-S3/HETG  &  $0.0059 \pm 0.0005$&39.6  & 34& 26:18:42 & 27:51:28 & 5566 & $0.015 \pm 0.002$ & $2.66$\\
\rule[0.5ex]{2.5em}{0.55pt} & \rule[0.5ex]{6em}{0.55pt} & \rule[0.5ex]{8.5em}{0.55pt} & \rule[0.5ex]{8.5em}{0.55pt} & \rule[0.5ex]{2.5em}{0.55pt} & \rule[0.5ex]{7em}{0.55pt} & \rule[0.5ex]{6.5em}{0.55pt} & \rule[0.5ex]{1.5em}{0.55pt} & 35& 35:16:52 & 35:44:31 & 1659 & $0.013 \pm 0.003$ & $1.31$\\
13848&  F. Baganoff  & 2012-05-09 12:01:48 & 2012-05-10 15:41:05 &98.2&  ACIS-S3/HETG  &  $0.0066 \pm 0.0002$ &39.8 & & \dots\dots\dots & \dots\dots\dots & \dots\dots & \dots\dots\dots\dots\dots & \dots\dots \\
13849&  F. Baganoff  & 2012-05-11 03:17:40 & 2012-05-13 05:39:54 &178.7&  ACIS-S3/HETG  &  $0.0071 \pm 0.0003$&38.5  & 36& 16:37:37 & 17:24:19 & 2802 & $0.019 \pm 0.003$ & $2.01$\\
\rule[0.5ex]{2.5em}{0.55pt} & \rule[0.5ex]{6em}{0.55pt} & \rule[0.5ex]{8.5em}{0.55pt} & \rule[0.5ex]{8.5em}{0.55pt} & \rule[0.5ex]{2.5em}{0.55pt} & \rule[0.5ex]{7em}{0.55pt} & \rule[0.5ex]{6.5em}{0.55pt} & \rule[0.5ex]{1.5em}{0.55pt} & 37& 24:20:31 & 25:52:29 & 5518 & $0.009 \pm 0.002$ & $1.72$\\
\rule[0.5ex]{2.5em}{0.55pt} & \rule[0.5ex]{6em}{0.55pt} & \rule[0.5ex]{8.5em}{0.55pt} & \rule[0.5ex]{8.5em}{0.55pt} & \rule[0.5ex]{2.5em}{0.55pt} & \rule[0.5ex]{7em}{0.55pt} & \rule[0.5ex]{6.5em}{0.55pt} & \rule[0.5ex]{1.5em}{0.55pt} & 38& 31:41:37 & 32:15:35 & 2038 & $0.021 \pm 0.004$ & $9.20$\\
\rule[0.5ex]{2.5em}{0.55pt} & \rule[0.5ex]{6em}{0.55pt} & \rule[0.5ex]{8.5em}{0.55pt} & \rule[0.5ex]{8.5em}{0.55pt} & \rule[0.5ex]{2.5em}{0.55pt} & \rule[0.5ex]{7em}{0.55pt} & \rule[0.5ex]{6.5em}{0.55pt} & \rule[0.5ex]{1.5em}{0.55pt} & 39& 51:10:55 & 51:57:34 & 2799 & $0.041 \pm 0.005$ & $2.08$\\
13846&  F. Baganoff  & 2012-05-16 10:40:15 & 2012-05-17 02:18:07 &56.2&  ACIS-S3/HETG  &  $0.0064 \pm 0.0003$ &39.8 & & \dots\dots\dots & \dots\dots\dots & \dots\dots & \dots\dots\dots\dots\dots & \dots\dots \\
14438&  F. Baganoff  & 2012-05-18 04:27:35 & 2012-05-18 12:10:09 &25.8&  ACIS-S3/HETG  &  $0.0060 \pm 0.0005$ &39.5 & & \dots\dots\dots & \dots\dots\dots & \dots\dots & \dots\dots\dots\dots\dots & \dots\dots \\
13845&  F. Baganoff  & 2012-05-19 10:41:18 & 2012-05-21 00:48:07 &135.3&  ACIS-S3/HETG  &  $0.0062 \pm 0.0002$ &39.6 & 40& 13:50:55 & 14:28:21 & 2246 & $0.013 \pm 0.003$ & $4.79$\\
\rule[0.5ex]{2.5em}{0.55pt} & \rule[0.5ex]{6em}{0.55pt} & \rule[0.5ex]{8.5em}{0.55pt} & \rule[0.5ex]{8.5em}{0.55pt} & \rule[0.5ex]{2.5em}{0.55pt} & \rule[0.5ex]{7em}{0.55pt} & \rule[0.5ex]{6.5em}{0.55pt} & \rule[0.5ex]{1.5em}{0.55pt} & 41& 44:48:27 & 45:14:16 & 1548 & $0.056 \pm 0.004$ & $2.62$\\
14460&  F. Baganoff  & 2012-07-09 22:33:10 & 2012-07-10 05:47:47 &24.1&  ACIS-S3/HETG  &  $0.0050 \pm 0.0006$&38.2  & & \dots\dots\dots & \dots\dots\dots & \dots\dots & \dots\dots\dots\dots\dots & \dots\dots \\
13844&  F. Baganoff  & 2012-07-10 23:10:04 & 2012-07-11 05:21:09 &20.1&  ACIS-S3/HETG  &  $0.0051 \pm 0.0005$ &39.5 & & \dots\dots\dots & \dots\dots\dots & \dots\dots & \dots\dots\dots\dots\dots & \dots\dots \\
14461&  F. Baganoff  & 2012-07-12 05:47:45 & 2012-07-12 19:58:25 &51.0&  ACIS-S3/HETG  &  $0.0073 \pm 0.0004$ &40.1 & & \dots\dots\dots & \dots\dots\dots & \dots\dots & \dots\dots\dots\dots\dots & \dots\dots \\
13853&  F. Baganoff  & 2012-07-14 00:36:15 & 2012-07-14 21:05:13 &73.7&  ACIS-S3/HETG  &  $0.0057 \pm 0.0003$ &38.1 & & \dots\dots\dots & \dots\dots\dots & \dots\dots & \dots\dots\dots\dots\dots & \dots\dots \\
13841&  F. Baganoff  & 2012-07-17 21:05:19 & 2012-07-18 10:04:59 &45.1&  ACIS-S3/HETG  &  $0.0064 \pm 0.0004$ &38.9 & & \dots\dots\dots & \dots\dots\dots & \dots\dots & \dots\dots\dots\dots\dots & \dots\dots \\
14465&  F. Baganoff  & 2012-07-18 23:23:20 & 2012-07-19 11:43:25 &44.3&  ACIS-S3/HETG  &  $0.0057 \pm 0.0004$ & 38.7& 42& <23:23:20 & 25:01:37 & >5957 & $0.012 \pm 0.002$ & $2.51$\\
\rule[0.5ex]{2.5em}{0.55pt} & \rule[0.5ex]{6em}{0.55pt} & \rule[0.5ex]{8.5em}{0.55pt} & \rule[0.5ex]{8.5em}{0.55pt} & \rule[0.5ex]{2.5em}{0.55pt} & \rule[0.5ex]{7em}{0.55pt} & \rule[0.5ex]{6.5em}{0.55pt} & \rule[0.5ex]{1.5em}{0.55pt} & 43& 28:18:15 & 28:51:17 & 1982 & $0.012 \pm 0.003$ & $4.75$\\
14466&  F. Baganoff  & 2012-07-20 12:37:09 & 2012-07-21 01:32:24 &45.1&  ACIS-S3/HETG  &  $0.0066 \pm 0.0004$ &40.0 & 44& 13:12:19 & 13:26:49 & 870 & $0.067 \pm 0.009$ & $5.61$\\
\rule[0.5ex]{2.5em}{0.55pt} & \rule[0.5ex]{6em}{0.55pt} & \rule[0.5ex]{8.5em}{0.55pt} & \rule[0.5ex]{8.5em}{0.55pt} & \rule[0.5ex]{2.5em}{0.55pt} & \rule[0.5ex]{7em}{0.55pt} & \rule[0.5ex]{6.5em}{0.55pt} & \rule[0.5ex]{1.5em}{0.55pt} & 45& 24:27:22 & 24:33:42 & 380 & $0.028 \pm 0.008$ & $5.39$\\
13842&  F. Baganoff  & 2012-07-21 11:52:09 & 2012-07-23 17:42:01 &191.8&  ACIS-S3/HETG  &  $0.0059 \pm 0.0003$ &38.1 & 46& 28:31:33 & 29:40:45 & 4152 & $0.031 \pm 0.003$ & $2.21$\\
\rule[0.5ex]{2.5em}{0.55pt} & \rule[0.5ex]{6em}{0.55pt} & \rule[0.5ex]{8.5em}{0.55pt} & \rule[0.5ex]{8.5em}{0.55pt} & \rule[0.5ex]{2.5em}{0.55pt} & \rule[0.5ex]{7em}{0.55pt} & \rule[0.5ex]{6.5em}{0.55pt} & \rule[0.5ex]{1.5em}{0.55pt} & 47& 45:52:48 & 46:31:48 & 2340 & $0.049 \pm 0.005$ & $3.52$\\
\rule[0.5ex]{2.5em}{0.55pt} & \rule[0.5ex]{6em}{0.55pt} & \rule[0.5ex]{8.5em}{0.55pt} & \rule[0.5ex]{8.5em}{0.55pt} & \rule[0.5ex]{2.5em}{0.55pt} & \rule[0.5ex]{7em}{0.55pt} & \rule[0.5ex]{6.5em}{0.55pt} & \rule[0.5ex]{1.5em}{0.55pt} & 48& 60:14:54 & 62:01:49 & 6415 & $0.015 \pm 0.001$ & $3.34$\\
\hline
\end{tabular}
}
\normalsize
\end{table*}

\setcounter{table}{1}
\begin{table*}[!ht]
\caption[Continued]{Continued.}
\centering
\scalebox{0.55}{
\begin{tabular}{@{}rlllcrcccrrcrc@{}}
\hline
\hline
\multicolumn{8}{c}{Observations} & \multicolumn{6}{c}{Flares} \\
\multicolumn{8}{l}{\rule[0.5ex]{50.5em}{0.55pt}} & \multicolumn{6}{r}{\rule[0.5ex]{34.5em}{0.55pt}} \\
\multicolumn{12}{c}{} & \multicolumn{2}{c}{Mean} \\
\multicolumn{12}{c}{} & \multicolumn{2}{r}{\rule[0.5ex]{18em}{0.55pt}} \\
\multicolumn{1}{c}{ObsID} & \multicolumn{1}{c}{PI} & \multicolumn{1}{c}{Start} & \multicolumn{1}{c}{End} & \multicolumn{1}{c}{Duration}  & \multicolumn{1}{c}{Instrument} & \multicolumn{1}{c}{Non-flaring level} & \multicolumn{1}{c}{$\eta_{obs}$\tablefootmark{a}} & \multicolumn{1}{c}{\#} & \multicolumn{1}{c}{Start\tablefootmark{b}} & \multicolumn{1}{c}{Stop\tablefootmark{b}} & \multicolumn{1}{c}{Duration} & \multicolumn{1}{c}{Count rate\tablefootmark{c}} & \multicolumn{1}{c}{Flux\tablefootmark{d}} \\
 & & \multicolumn{1}{c}{(UT)} & \multicolumn{1}{c}{(UT)} & \multicolumn{1}{c}{(ks)} & & \multicolumn{1}{c}{($\mathrm{count\ s^{-1}}$)} & (\%) & & \multicolumn{1}{c}{(UT)} & \multicolumn{1}{c}{(UT)} & \multicolumn{1}{c}{(s)} & \multicolumn{1}{c}{($\mathrm{count\ s^{-1}}$)} &\multicolumn{1}{c}{($10^{-12}\ \mathrm{erg\ s^{-1}\ cm^{-2}}$)}\\
\hline
13839&  F. Baganoff  & 2012-07-24 07:02:13 & 2012-07-26 08:21:38 &176.3&  ACIS-S3/HETG  &  $0.0067 \pm 0.0003$ &38.9 & 49& 09:19:45 & 09:41:03 & 1278 & $0.040 \pm 0.006$ & $3.11$\\
\rule[0.5ex]{2.5em}{0.55pt} & \rule[0.5ex]{6em}{0.55pt} & \rule[0.5ex]{8.5em}{0.55pt} & \rule[0.5ex]{8.5em}{0.55pt} & \rule[0.5ex]{2.5em}{0.55pt} & \rule[0.5ex]{7em}{0.55pt} & \rule[0.5ex]{6.5em}{0.55pt} & \rule[0.5ex]{1.5em}{0.55pt} & 50& 36:33:24 & 36:56:55 & 1411 & $0.066 \pm 0.007$ & $6.11$\\
\rule[0.5ex]{2.5em}{0.55pt} & \rule[0.5ex]{6em}{0.55pt} & \rule[0.5ex]{8.5em}{0.55pt} & \rule[0.5ex]{8.5em}{0.55pt} & \rule[0.5ex]{2.5em}{0.55pt} & \rule[0.5ex]{7em}{0.55pt} & \rule[0.5ex]{6.5em}{0.55pt} & \rule[0.5ex]{1.5em}{0.55pt} & 51& 48:07:41 & 50:07:09 & 7168 & $0.078 \pm 0.005$ & $1.69$\\
13840&  F. Baganoff  & 2012-07-26 20:02:14 & 2012-07-28 17:39:12 &162.5&  ACIS-S3/HETG  &  $0.0069 \pm 0.0002$ &39.3 & 52& 59:06:44 & 60:45:28 & 5924 & $0.007 \pm 0.001$ & $1.80$\\
\rule[0.5ex]{2.5em}{0.55pt} & \rule[0.5ex]{6em}{0.55pt} & \rule[0.5ex]{8.5em}{0.55pt} & \rule[0.5ex]{8.5em}{0.55pt} & \rule[0.5ex]{2.5em}{0.55pt} & \rule[0.5ex]{7em}{0.55pt} & \rule[0.5ex]{6.5em}{0.55pt} & \rule[0.5ex]{1.5em}{0.55pt} & 53& 63:17:33 & 63:41:47 & 1454 & $0.014 \pm 0.002$ & $2.18$\\
14432&  F. Baganoff  & 2012-07-30 12:56:09 & 2012-07-31 10:12:43 &74.3&  ACIS-S3/HETG  &  $0.0059 \pm 0.0003$ &39.9 & 54& <12:56:09 & 14:42:31 & >6442 & $0.003 \pm 0.001$ & $8.29$\\
\rule[0.5ex]{2.5em}{0.55pt} & \rule[0.5ex]{6em}{0.55pt} & \rule[0.5ex]{8.5em}{0.55pt} & \rule[0.5ex]{8.5em}{0.55pt} & \rule[0.5ex]{2.5em}{0.55pt} & \rule[0.5ex]{7em}{0.55pt} & \rule[0.5ex]{6.5em}{0.55pt} & \rule[0.5ex]{1.5em}{0.55pt} & 55& 32:56:20 & >34:12:43 & >4583 & $0.051 \pm 0.004$ & $9.08$\\
13838&  F. Baganoff  & 2012-08-01 17:28:12 & 2012-08-02 21:55:51 &99.6&  ACIS-S3/HETG  &  $0.0068 \pm 0.0003$ &40.4 & 56& 24:19:15 & 25:08:53 & 2977 & $0.024 \pm 0.006$ & $5.64$\\
13852&  F. Baganoff  & 2012-08-04 02:37:07 & 2012-08-05 22:37:20 &156.6&  ACIS-S3/HETG  &  $0.0072 \pm 0.0003$ &39.6 & 57& 07:37:35 & 08:09:05 & 1890 & $0.042 \pm 0.005$ & $3.75$\\
\rule[0.5ex]{2.5em}{0.55pt} & \rule[0.5ex]{6em}{0.55pt} & \rule[0.5ex]{8.5em}{0.55pt} & \rule[0.5ex]{8.5em}{0.55pt} & \rule[0.5ex]{2.5em}{0.55pt} & \rule[0.5ex]{7em}{0.55pt} & \rule[0.5ex]{6.5em}{0.55pt} & \rule[0.5ex]{1.5em}{0.55pt} & 58& 32:07:01 & 32:22:59 & 958 & $0.016 \pm 0.004$ & $4.44$\\
14439&  F. Baganoff  & 2012-08-06 22:16:11 & 2012-08-08 05:44:50 &111.7&  ACIS-S3/HETG  &  $0.0064 \pm 0.0002$ &39.6 & 59& 51:14:47 & 51:31:32 & 1005 & $0.009 \pm 0.003$ & $2.49$\\
14462&  F. Baganoff  & 2012-10-06 16:32:00 & 2012-10-08 06:19:59 &133.4&  ACIS-S3/HETG  &  $0.0063 \pm 0.0003$&39.5  & 60& 28:20:29 & 28:50:16 & 1787 & $0.024 \pm 0.004$ & $3.67$\\
\rule[0.5ex]{2.5em}{0.55pt} & \rule[0.5ex]{6em}{0.55pt} & \rule[0.5ex]{8.5em}{0.55pt} & \rule[0.5ex]{8.5em}{0.55pt} & \rule[0.5ex]{2.5em}{0.55pt} & \rule[0.5ex]{7em}{0.55pt} & \rule[0.5ex]{6.5em}{0.55pt} & \rule[0.5ex]{1.5em}{0.55pt} & 61& 52:37:38 & 53:13:08 & 2130 & $0.021 \pm 0.002$ & $12.4$\\
14463&  F. Baganoff  & 2012-10-16 00:50:55 & 2012-10-16 09:46:00 &30.8&  ACIS-S3/HETG  &  $0.0066 \pm 0.0006$ &32.6 & 62& 05:46:23 & 05:54:11 & 535 & $0.102 \pm 0.019$ & $11.0$\\
13851&  F. Baganoff  & 2012-10-16 18:48:39 & 2012-10-18 01:03:03 &107.1&  ACIS-S3/HETG  &  $0.0058 \pm 0.0003$&40.2  & 63& 26:17:47 & 26:35:13 & 1046 & $0.047 \pm 0.007$ & $1.61$\\
\rule[0.5ex]{2.5em}{0.55pt} & \rule[0.5ex]{6em}{0.55pt} & \rule[0.5ex]{8.5em}{0.55pt} & \rule[0.5ex]{8.5em}{0.55pt} & \rule[0.5ex]{2.5em}{0.55pt} & \rule[0.5ex]{7em}{0.55pt} & \rule[0.5ex]{6.5em}{0.55pt} & \rule[0.5ex]{1.5em}{0.55pt} & 64& 43:47:49 & 45:02:19 & 4470 & $0.073 \pm 0.005$ & $34.3$\\
15568&  F. Baganoff  & 2012-10-18 08:54:33 & 2012-10-18 19:35:13 &36.1&  ACIS-S3/HETG  &  $0.0062 \pm 0.0004$ &40.1 & 65& 18:13:26 & >19:35:13 & >4907 & $0.006 \pm 0.002$ & $5.82$\\
13843&  F. Baganoff  & 2012-10-22 16:00:07 & 2012-10-24 02:07:34 &120.7&  ACIS-S3/HETG  &  $0.0066 \pm 0.0003$ &40.1 & 66& 33:11:43 & 35:25:03 & 8000 & $0.031 \pm 0.004$ & $6.11$\\
15570&  F. Baganoff  & 2012-10-25 03:29:12 & 2012-10-25 23:11:05 &68.7&  ACIS-S3/HETG  &  $0.0061 \pm 0.0003$ &39.9 & 67& 05:37:50 & 06:15:44 & 2274 & $0.027 \pm 0.005$ & $4.03$\\
14468&  F. Baganoff  & 2012-10-29 23:42:19 & 2012-10-31 17:01:14 &146.1&  ACIS-S3/HETG  &  $0.0058 \pm 0.0002$ &39.8 & 68& 31:09:43 & 32:16:53 & 4030 & $0.019 \pm 0.002$ & $6.33$\\
\rule[0.5ex]{2.5em}{0.55pt} & \rule[0.5ex]{6em}{0.55pt} & \rule[0.5ex]{8.5em}{0.55pt} & \rule[0.5ex]{8.5em}{0.55pt} & \rule[0.5ex]{2.5em}{0.55pt} & \rule[0.5ex]{7em}{0.55pt} & \rule[0.5ex]{6.5em}{0.55pt} & \rule[0.5ex]{1.5em}{0.55pt} & 69& 61:44:54 & 62:13:41 & 1727 & $0.023 \pm 0.001$ & $2.01$\\
14941&  F. Baganoff  & 2013-04-06 01:21:15 & 2013-04-06 07:14:49 &20.1&  ACIS-I3  &  $0.0039 \pm 0.0004$ &39.9 & & \dots\dots\dots & \dots\dots\dots & \dots\dots & \dots\dots\dots\dots\dots & \dots\dots \\
14942&  F. Baganoff  & 2013-04-14 15:41:11 & 2013-04-14 21:49:30 &20.1&  ACIS-I3  &  $0.0051 \pm 0.0005$&41.6  & & \dots\dots\dots & \dots\dots\dots & \dots\dots & \dots\dots\dots\dots\dots & \dots\dots \\
14702&  N. Rea  & 2013-05-12 10:36:44 & 2013-05-12 15:34:02 &15.1&  ACIS-S3/subarray  &  $0.0236 \pm 0.0013$ &37.5 & & \dots\dots\dots & \dots\dots\dots & \dots\dots & \dots\dots\dots\dots\dots & \dots\dots \\
15040&  D. Haggard  & 2013-05-25 11:36:12 & 2013-05-25 18:48:48 &24.4&  ACIS-S3/HETG  &  $0.0033 \pm 0.0003$&36.2  & 70& 17:23:43 & 18:26:18 & 3750 & $0.006 \pm 0.002$ & $2.12$\\
14703&  N. Rea  & 2013-06-04 08:43:31 & 2013-06-04 14:27:14 &18.6&  ACIS-S3/subarray  &  $0.0094 \pm 0.0007$&39.1  & & \dots\dots\dots & \dots\dots\dots & \dots\dots & \dots\dots\dots\dots\dots & \dots\dots \\
15651&  D. Haggard  & 2013-06-05 21:30:38 & 2013-06-06 01:47:52 &14.1&  ACIS-S3/HETG  &  $0.0032 \pm 0.0005$ &34.3 & & \dots\dots\dots & \dots\dots\dots & \dots\dots & \dots\dots\dots\dots\dots & \dots\dots \\
15654&  D. Haggard  & 2013-06-09 04:23:04 & 2013-06-09 07:36:37 &9.3&  ACIS-S3/HETG  &  $0.0027 \pm 0.0005$ &29.4 & & \dots\dots\dots & \dots\dots\dots & \dots\dots & \dots\dots\dots\dots\dots & \dots\dots \\
14946&  F. Baganoff  & 2013-07-02 06:47:30 & 2013-07-02 12:43:53 &20.1&  ACIS-S3/subarray  &  $0.0099 \pm 0.0007$ &39.1 & & \dots\dots\dots & \dots\dots\dots & \dots\dots & \dots\dots\dots\dots\dots & \dots\dots \\
15041&  D. Haggard  & 2013-07-27 01:27:10 & 2013-07-27 15:52:18 &50.1&  ACIS-S3/subarray  &  $0.0141 \pm 0.0006$&38.8  & 71& 03:29:36 & 03:46:53 & 1037 & $0.021 \pm 0.021$ & $7.01$\\
\rule[0.5ex]{2.5em}{0.55pt} & \rule[0.5ex]{6em}{0.55pt} & \rule[0.5ex]{8.5em}{0.55pt} & \rule[0.5ex]{8.5em}{0.55pt} & \rule[0.5ex]{2.5em}{0.55pt} & \rule[0.5ex]{7em}{0.55pt} & \rule[0.5ex]{6.5em}{0.55pt} & \rule[0.5ex]{1.5em}{0.55pt} & 72& 11:03:11 & 11:16:55 & 924 & $0.019 \pm 0.016$ & $4.47$\\
15042&  D. Haggard  & 2013-08-11 22:55:23 & 2013-08-12 13:05:40 &49.4&  ACIS-S3/subarray  &  $0.0138 \pm 0.0005$ &38.9 & & \dots\dots\dots & \dots\dots\dots & \dots\dots & \dots\dots\dots\dots\dots & \dots\dots \\
14945&  F. Baganoff  & 2013-08-31 10:10:43 & 2013-08-31 16:26:04 &20.1&  ACIS-S3/subarray  &  $0.0082 \pm 0.0006$&37.9  & & \dots\dots\dots & \dots\dots\dots & \dots\dots & \dots\dots\dots\dots\dots & \dots\dots \\
15043&  D. Haggard  & 2013-09-14 00:03:23 & 2013-09-14 14:16:41 &50.1&  ACIS-S3/subarray  &  $0.0090 \pm 0.0006$ &39.6 & 73& 02:02:00 & 04:29:39 & 5097 & $0.523 \pm 0.010$ & $38.2$\\
14944&  F. Baganoff  & 2013-09-20 07:00:52 & 2013-09-20 13:16:13 &20.1&  ACIS-S3/subarray  &  $0.0144 \pm 0.0008$ &39.6 & & \dots\dots\dots & \dots\dots\dots & \dots\dots & \dots\dots\dots\dots\dots & \dots\dots \\
15044&  D. Haggard  & 2013-10-04 17:22:26 & 2013-10-05 06:58:38 &47.1&  ACIS-S3/subarray  &  $0.0102 \pm 0.0004$ &39.0 & & \dots\dots\dots & \dots\dots\dots & \dots\dots & \dots\dots\dots\dots\dots & \dots\dots \\
14943&  F. Baganoff  & 2013-10-17 15:38:04 & 2013-10-17 21:41:46 &20.1&  ACIS-S3/subarray  &  $0.0080 \pm 0.0002$&39.4  & & \dots\dots\dots & \dots\dots\dots & \dots\dots & \dots\dots\dots\dots\dots & \dots\dots \\
14704&  N. Rea  & 2013-10-23 08:52:40 & 2013-10-23 20:41:18 &40.1&  ACIS-S3/subarray  &  $0.0093 \pm 0.0009$&39.7  & & \dots\dots\dots & \dots\dots\dots & \dots\dots & \dots\dots\dots\dots\dots & \dots\dots \\
15045&  D. Haggard  & 2013-10-28 14:30:21 & 2013-10-29 04:59:07 &50.1&  ACIS-S3/subarray  &  $0.0087 \pm 0.0005$ &39.5 & 74& 16:12:19 & 16:48:34 & 2238 & $0.023 \pm 0.004$ & $3.95$\\
\rule[0.5ex]{2.5em}{0.55pt} & \rule[0.5ex]{6em}{0.55pt} & \rule[0.5ex]{8.5em}{0.55pt} & \rule[0.5ex]{8.5em}{0.55pt} & \rule[0.5ex]{2.5em}{0.55pt} & \rule[0.5ex]{7em}{0.55pt} & \rule[0.5ex]{6.5em}{0.55pt} & \rule[0.5ex]{1.5em}{0.55pt} & 75& 19:55:36 & 20:10:22 & 886 & $0.021 \pm 0.016$ & $3.90$ \\
16508&  D. Haggard  & 2014-02-21 11:35:47 & 2014-02-22 01:23:57 &47.9&  ACIS-S3/subarray  &  $0.0084 \pm 0.0004$ &39.6 & & \dots\dots\dots & \dots\dots\dots & \dots\dots & \dots\dots\dots\dots\dots & \dots\dots \\
16211&  D. Haggard  & 2014-03-14 10:16:20 & 2014-03-14 23:43:24 &46.1&  ACIS-S3/subarray  &  $0.0052 \pm 0.0003$ &39.5 & & \dots\dots\dots & \dots\dots\dots & \dots\dots & \dots\dots\dots\dots\dots & \dots\dots \\
16212&  D. Haggard  & 2014-04-04 02:24:32 & 2014-04-04 16:47:09 &50.1&  ACIS-S3/subarray  &  $0.0058 \pm 0.0003$ &39.4 & & \dots\dots\dots & \dots\dots\dots & \dots\dots & \dots\dots\dots\dots\dots & \dots\dots \\
16213&  D. Haggard  & 2014-04-28 02:42:49 & 2014-04-28 17:11:46 &49.6&  ACIS-S3/subarray  &  $0.0068 \pm 0.0003$ &39.4 & & \dots\dots\dots & \dots\dots\dots & \dots\dots & \dots\dots\dots\dots\dots & \dots\dots \\
16214&  D. Haggard  & 2014-05-20 00:17:16 & 2014-05-20 14:46:55 &50.1&  ACIS-S3/subarray  &  $0.0062 \pm 0.0003$ &38.8 & & \dots\dots\dots & \dots\dots\dots & \dots\dots & \dots\dots\dots\dots\dots & \dots\dots \\
16210&  D. Haggard  & 2014-06-03 02:56:53 & 2014-06-03 08:38:29 &18.8&  ACIS-S3/subarray  &  $0.0063 \pm 0.0006$ &39.9 & & \dots\dots\dots & \dots\dots\dots & \dots\dots & \dots\dots\dots\dots\dots & \dots\dots \\
16597&  D. Haggard  & 2014-07-04 20:45:38 & 2014-07-05 02:18:48 &18.2&  ACIS-S3/subarray  &  $0.0075 \pm 0.0006$ &39.7 & & \dots\dots\dots & \dots\dots\dots & \dots\dots & \dots\dots\dots\dots\dots & \dots\dots \\
16215&  D. Haggard  & 2014-07-16 22:41:39 & 2014-07-17 11:47:38 &45.7&  ACIS-S3/subarray  &  $0.0069 \pm 0.0004$ &40.2 & & \dots\dots\dots & \dots\dots\dots & \dots\dots & \dots\dots\dots\dots\dots & \dots\dots \\
16216&  D. Haggard  & 2014-08-02 03:29:56 & 2014-08-02 17:07:33 &47.1&  ACIS-S3/subarray  &  $0.0069 \pm 0.0004$ &32.5 & & \dots\dots\dots & \dots\dots\dots & \dots\dots & \dots\dots\dots\dots\dots & \dots\dots \\
16217&  D. Haggard  & 2014-08-30 04:47:01 & 2014-08-30 15:43:10 &38.1&  ACIS-S3/subarray  &  $0.0097 \pm 0.0006$ &32.9 & & \dots\dots\dots & \dots\dots\dots & \dots\dots & \dots\dots\dots\dots\dots & \dots\dots \\
16218&  D. Haggard  & 2014-10-20 08:20:07 & 2014-10-20 19:57:00 &40.1&  ACIS-S3/subarray  &  $0.0061 \pm 0.0006$&39.9  & 76& 13:21:51 & 15:05:12 & 6201 & $0.127 \pm 0.088$ & $25.7$\\
16963&  G. Garmire  & 2015-02-13 01:00:03 & 2015-02-13 07:56:53 &25.1&  ACIS-S3/subarray  &  $0.0058 \pm 0.0005$ &39.3 & 77& 06:03:19 & 06:16:51 & 812 & $0.033 \pm 0.066$ & $6.86$\\
16966&  G. Garmire  & 2015-05-14 08:45:36 & 2015-05-14 16:25:37 &22.7&  ACIS-S3/subarray  &  $0.0034 \pm 0.0004$ &39.7 & 78& 12:01:25 & 12:53:24 & 3119 & $0.052 \pm 0.007$ & $6.44$\\
17857&  G. Ponti  & 2015-08-11 17:18:08 & 2015-08-13 03:14:33 &120.0&  ACIS-S3/HETG  &  $0.0037 \pm 0.0002$ &39.6 & 79& 34:09:00 & 35:49:47 & 6047 & $0.039 \pm 0.003$ & $6.44$\\
16965&  G. Garmire  & 2015-08-17 10:34:39 & 2015-08-17 18:12:03 &25.1&  ACIS-S3/subarray  &  $0.0062 \pm 0.0005$ &39.7 & & \dots\dots\dots & \dots\dots\dots & \dots\dots & \dots\dots\dots\dots\dots & \dots\dots \\
16964&  G. Garmire  & 2015-10-21 06:03:49 & 2015-10-21 13:22:14 &22.6&  ACIS-S3/subarray  &  $0.0050 \pm 0.0004$ &39.3 & 80 & 13:19:58 & >13:22:14 & >136 & $0.039 \pm 0.017$ & $5.07$ \\
\hline
\end{tabular}
}
\tablefoot{
\tablefoottext{a} {The average flare detection efficiency above the corresponding non-flaring level.}
\tablefoottext{b} {The flare start and end times are given in hh:mm:ss since the day of the observation start.
Flares beginning or ending at the start or stop of the observation lead to a lower limit on the flare duration and a lower or upper limit on the flare mean count rate and mean flux.
The flux value of these flares were taken equal to this limit in the flaring rate study.}
\tablefoottext{c} {The flare mean count rates are computed after subtraction of the non-flaring level.}
\tablefoottext{d} {Mean unabsorbed flux between 2 and 10$\,$keV determined for $N_\mathrm{H}=14.3\times 10^{22}\ \mathrm{cm^{-2}}$ and $\Gamma=2$.}\\
\textbf{}
}
\normalsize

\caption[Observation log of the Swift observations and the detected X-ray flares in 2006--2015]{Observation log of public Swift observations and the X-ray flares detected in this work.}
\centering
\scalebox{0.55}{
\label{table:swift}
\begin{tabular}{@{}llcccccccccc@{}}
\hline
\hline
\multicolumn{6}{c}{Observations} & \multicolumn{6}{c}{Flares} \\
\multicolumn{6}{l}{\rule[0.5ex]{37.5em}{0.55pt}} & \multicolumn{6}{r}{\rule[0.5ex]{45.5em}{0.55pt}} \\
\multicolumn{10}{c}{} & \multicolumn{2}{c}{Mean} \\
\multicolumn{10}{c}{} & \multicolumn{2}{r}{\rule[0.5ex]{22em}{0.55pt}} \\
\multicolumn{1}{c}{First} & \multicolumn{1}{c}{Last} & \multicolumn{1}{c}{Number} & \multicolumn{1}{c}{Total exposure} & \multicolumn{1}{c}{Non-flaring level} & \multicolumn{1}{c}{$\eta_{obs}$\tablefootmark{a}} & \multicolumn{1}{c}{\#} & \multicolumn{1}{c}{Start} & \multicolumn{1}{c}{Stop} & \multicolumn{1}{c}{Duration\tablefootmark{b}} & \multicolumn{1}{c}{Count rate\tablefootmark{c}} & \multicolumn{1}{c}{Flux\tablefootmark{d}}\\
\multicolumn{1}{c}{(UT)} & \multicolumn{1}{c}{(UT)} & & \multicolumn{1}{c}{(ks)} &\multicolumn{1}{c}{($\mathrm{count\ s^{-1}}$)} & (\%) & \multicolumn{1}{c}{(UT)} & \multicolumn{1}{c}{(UT)} & \multicolumn{1}{c}{(s)} &\multicolumn{1}{c}{($\mathrm{count\ s^{-1}}$)}  &\multicolumn{1}{c}{($10^{-12}\ \mathrm{erg\ s^{-1}\ cm^{-2}}$)}\\
\hline  
2006-02-24 22:55:12 & 2006-11-02 14:22:34 & 198 & $261.7$ & $0.021\pm0.002$ & 24.6 & 1 & 2006-07-13 21:57:36 & 2006-07-13 23:39:50 & 924 & $0.031\pm0.007$ & $15.3$\\
2007-02-16 21:38:52 & 2007-11-02 13:52:19 & 163 & $174.6$ & $0.025\pm0.004$ & 24.6 & 2 & 2007-03-03 00:38:21 & 2007-03-03 02:34:56 & 2018 & $0.012\pm0.004$ & $9.21$\\
2008-02-19 23:02:24 & 2008-10-30 09:14:24 & 161 & $199.3$ & $0.024\pm0.003$ & 24.6 & 3 & 2008-03-25 20:24:00 & 2008-03-25 23:36:58 & 707 & $0.056\pm0.011$ & $20.8$\\
\rule[0.5ex]{8.5em}{0.55pt} & \rule[0.5ex]{8.5em}{0.55pt} & \rule[0.5ex]{1.5em}{0.55pt} & \rule[0.5ex]{2.5em}{0.55pt} & \rule[0.5ex]{5.5em}{0.55pt} & \rule[0.5ex]{1.5em}{0.55pt} & 4 & 2008-05-01 14:15:22 & 2008-05-01 20:39:50 & 1056 & $0.029\pm0.007$ & $8.66$\\
\rule[0.5ex]{8.5em}{0.55pt} & \rule[0.5ex]{8.5em}{0.55pt} & \rule[0.5ex]{1.5em}{0.55pt} & \rule[0.5ex]{2.5em}{0.55pt} & \rule[0.5ex]{5.5em}{0.55pt} & \rule[0.5ex]{1.5em}{0.55pt} & 5 & 2008-10-17 17:15:22 & 2008-10-17 22:09:07 & 1369 & $0.036\pm0.007$ & $11.9$\\
2009-06-04 07:23:31 & 2009-11-01 21:37:26 & 36 & $34.64$ & $0.028\pm0.009$ & 24.6 &  & \dots\dots\dots\dots\dots\dots\dots & \dots\dots\dots\dots\dots\dots\dots & \dots\dots & \dots\dots\dots\dots\dots & \dots\dots\\
2010-04-07 01:10:34 & 2010-10-31 10:10:34 & 62 & $70.33$ & $0.030\pm0.006$ & 24.6 & 6 & 2010-06-12 10:23:31 & 2010-06-12 12:07:12& 1081 & $0.127\pm0.012$ & $49.3$\\
2011-02-04 16:53:46 & 2011-11-02 15:38:53 & 81 & $76.78$ & $0.025\pm0.006$ & 24.6 &  & \dots\dots\dots\dots\dots\dots\dots & \dots\dots\dots\dots\dots\dots\dots & \dots\dots & \dots\dots\dots\dots\dots & \dots\dots\\
2012-02-05 20:12:29 & 2012-10-31 23:21:07 & 79 & $73.95$ & $0.020\pm0.005$ & 24.6 &  & \dots\dots\dots\dots\dots\dots\dots & \dots\dots\dots\dots\dots\dots\dots & \dots\dots & \dots\dots\dots\dots\dots & \dots\dots\\
2013-02-03 22:26:24 & 2013-10-31 01:17:46 & 191 & $185.4$ & $0.145\pm0.009$ & 20.0\tablefootmark{e} &  & \dots\dots\dots\dots\dots\dots\dots & \dots\dots\dots\dots\dots\dots\dots & \dots\dots & \dots\dots\dots\dots\dots & \dots\dots\\
2014-02-03 18:57:36 & 2014-11-02 12:56:09 & 236 & $231.4$ & $0.056\pm0.007$ & 24.3 & 7 & 2014-09-09 11:41:17 & 2014-09-09 11:58:34 & 975 & $0.128\pm0.013$ & $59.4$\\
2015-02-03 00:18:43 & 2015-11-02 15:14:24 & 231 & $211.1$ & $0.033\pm0.004$ & 24.6 & 8 & 2015-11-02 14:58:34 & 2015-11-02 15:14:24 & 993 & $0.074\pm0.009$& $22.3$\\
\hline
\end{tabular}
}
\tablefoot{
\tablefoottext{a} {The average flare detection efficiency above the corresponding non-flaring level.}
\tablefoottext{b} {The flare duration corresponds to the corresponding observing time.}
\tablefoottext{c} {The flare mean count rates are computed after subtraction of the non-flaring level.}
\tablefoottext{d} {Mean unabsorbed flux between 2 and 10$\,$keV determined for $N_\mathrm{H}=14.3\times 10^{22}\ \mathrm{cm^{-2}}$ and $\Gamma=2$.}
\tablefoottext{e} {Mean flare detection efficiency. 
Owing to the decay phase of the Galactic center magnetar, the average flare detection efficiency increases during this campaign: 16.2, 17.9, 18.5, 21.7, 22.5, and 23.3\% during 2013 April\ 4--May 11, May 12--17, May 18--29, May 30--June 28, June 29--September\ 7 and September\ 8--October\ 31, respectively.}
}
\normalsize
\end{table*}
\newpage
\textbf{}
\newpage
\textbf{}
\newpage
\textbf{}
\newpage
\textbf{}
\newpage
\textbf{}

\section{X-ray flares detected from 1999 to 2015 with XMM-Newton, Chandra, and Swift}
\label{app_bb}
\begin{figure*}[!ht]
\centering
\begin{tabular}{@{}ccccc@{}}
\includegraphics[width=0.20\textwidth,angle=90]{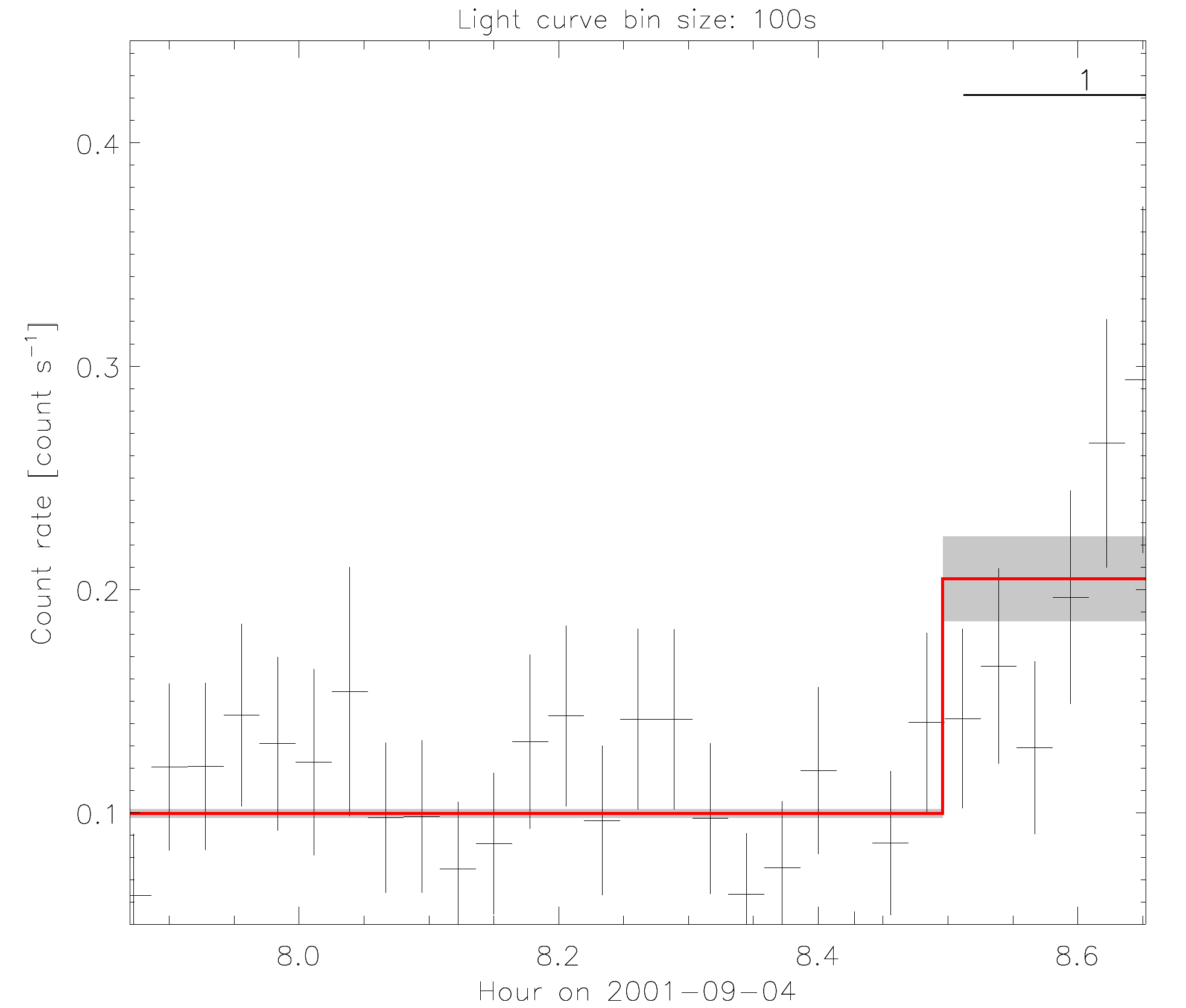}&
\includegraphics[width=0.20\textwidth,angle=90]{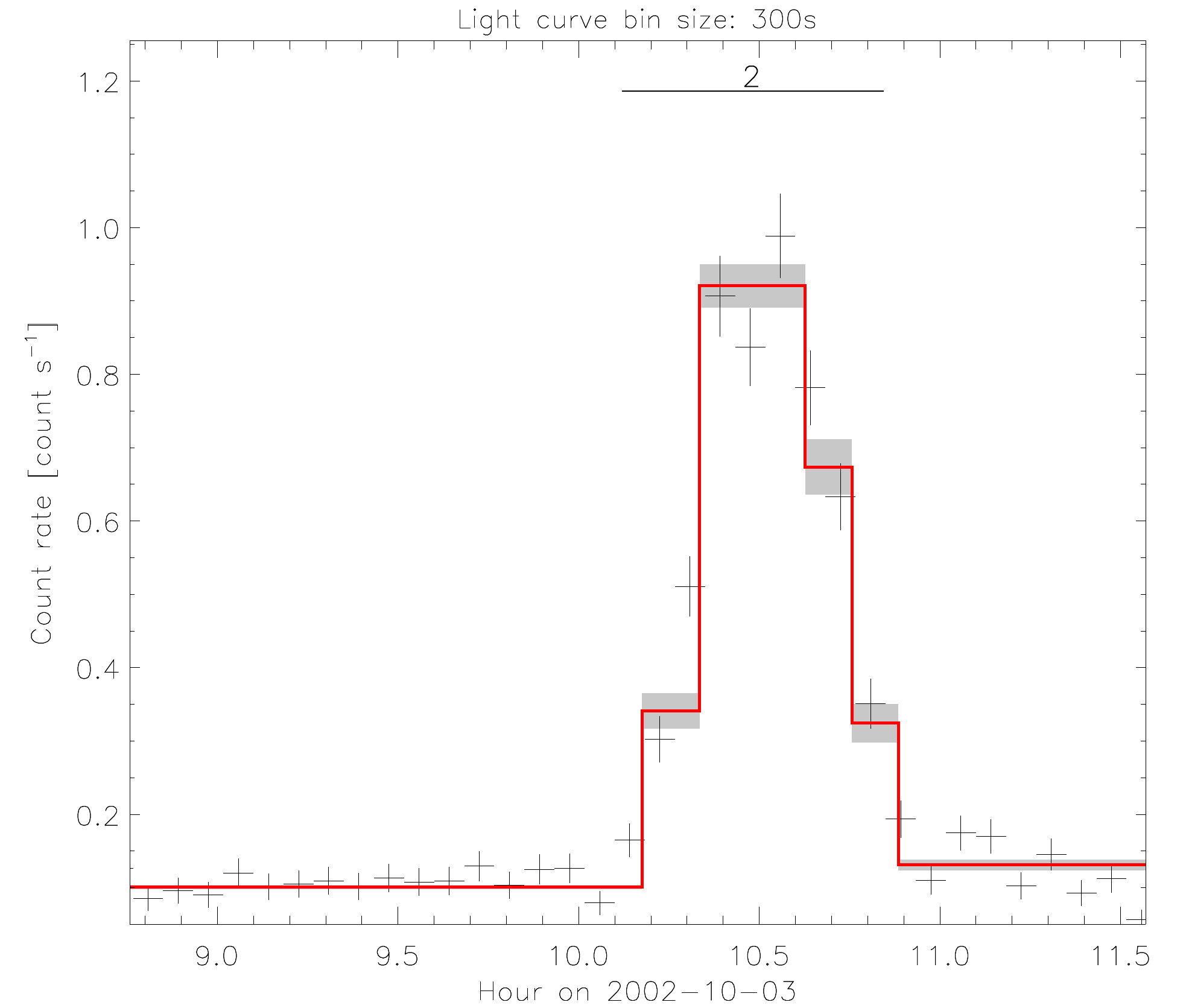}&
\includegraphics[width=0.20\textwidth,angle=90]{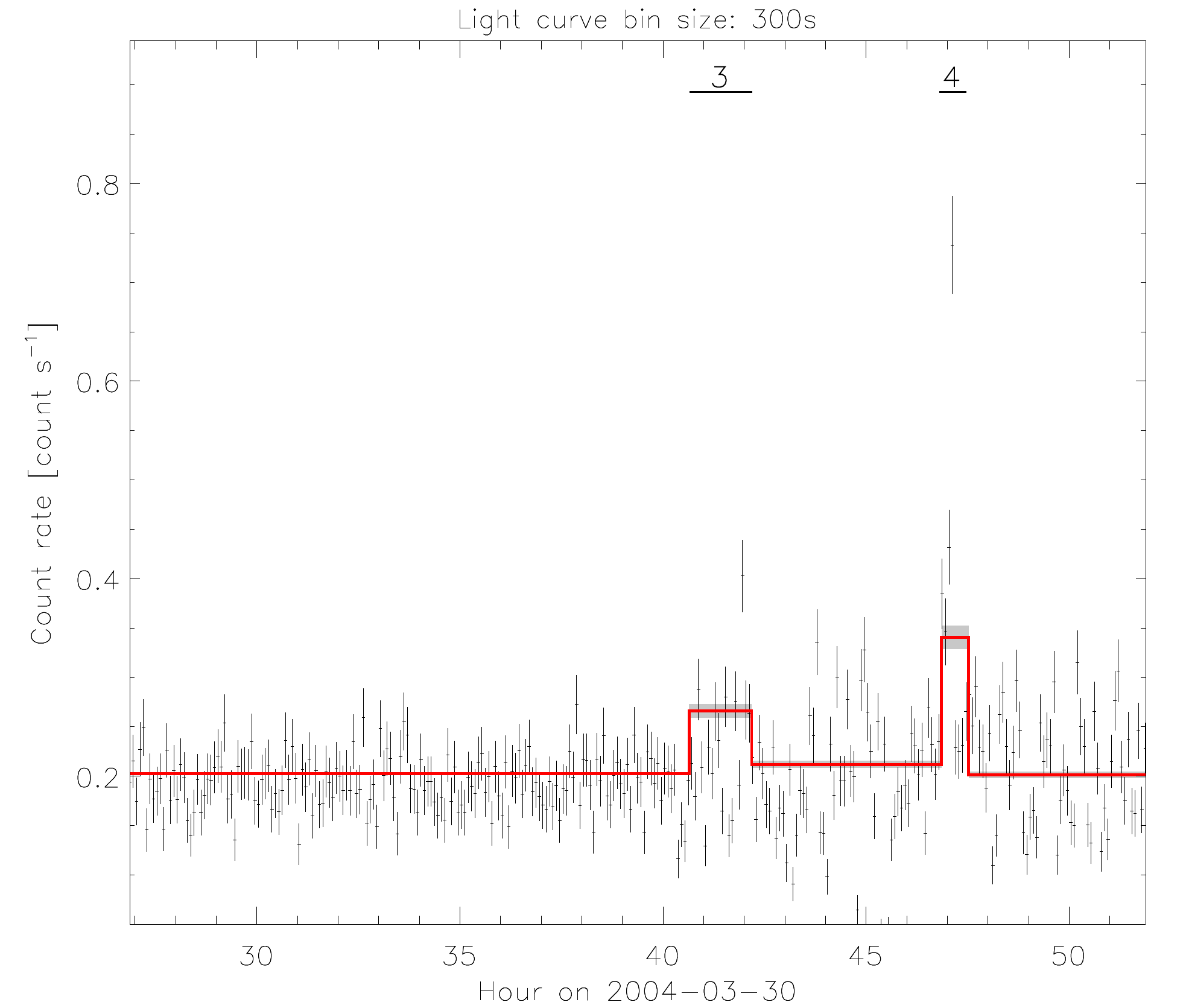}&
\includegraphics[width=0.20\textwidth,angle=90]{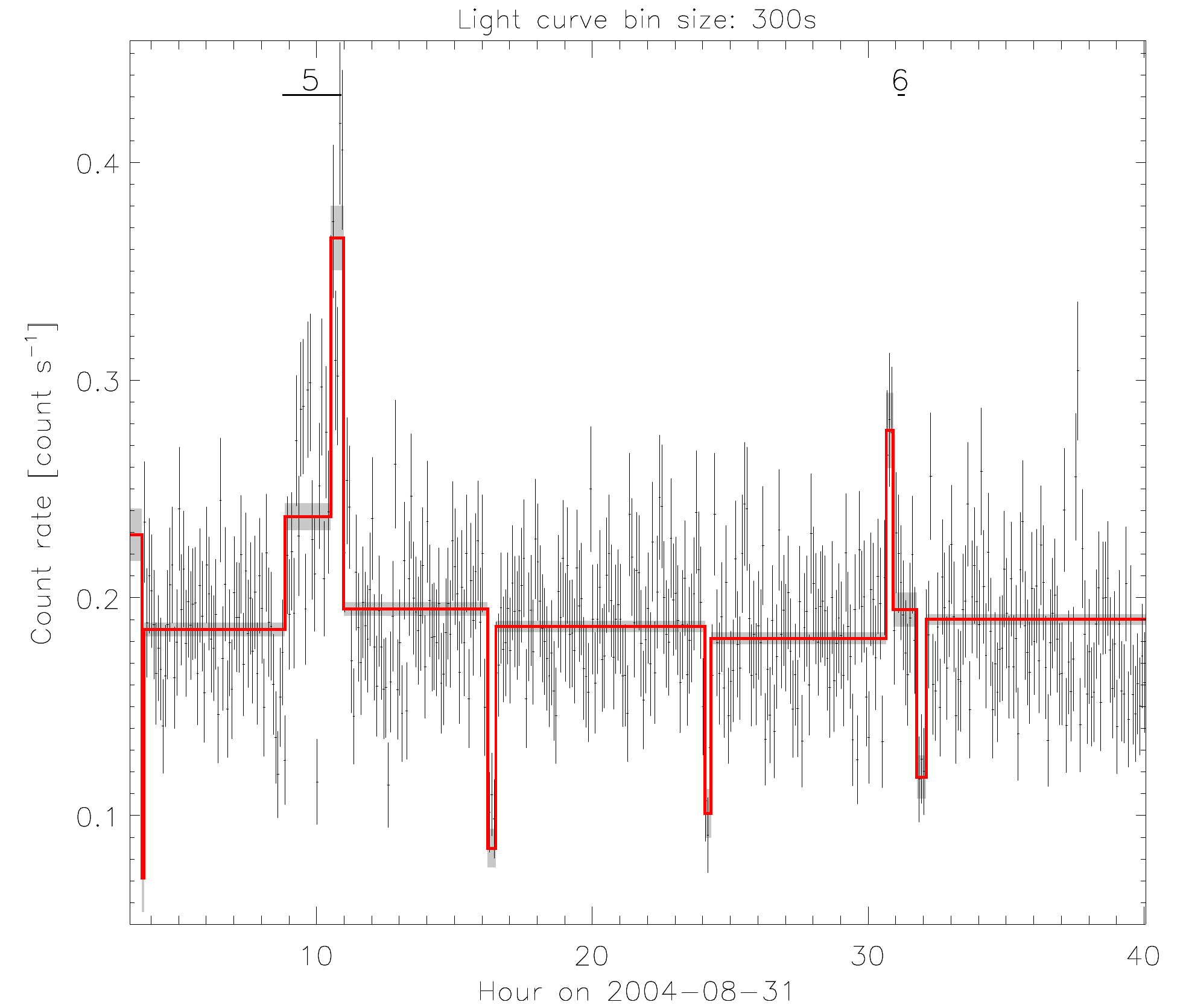}\\
\includegraphics[width=0.20\textwidth,angle=90]{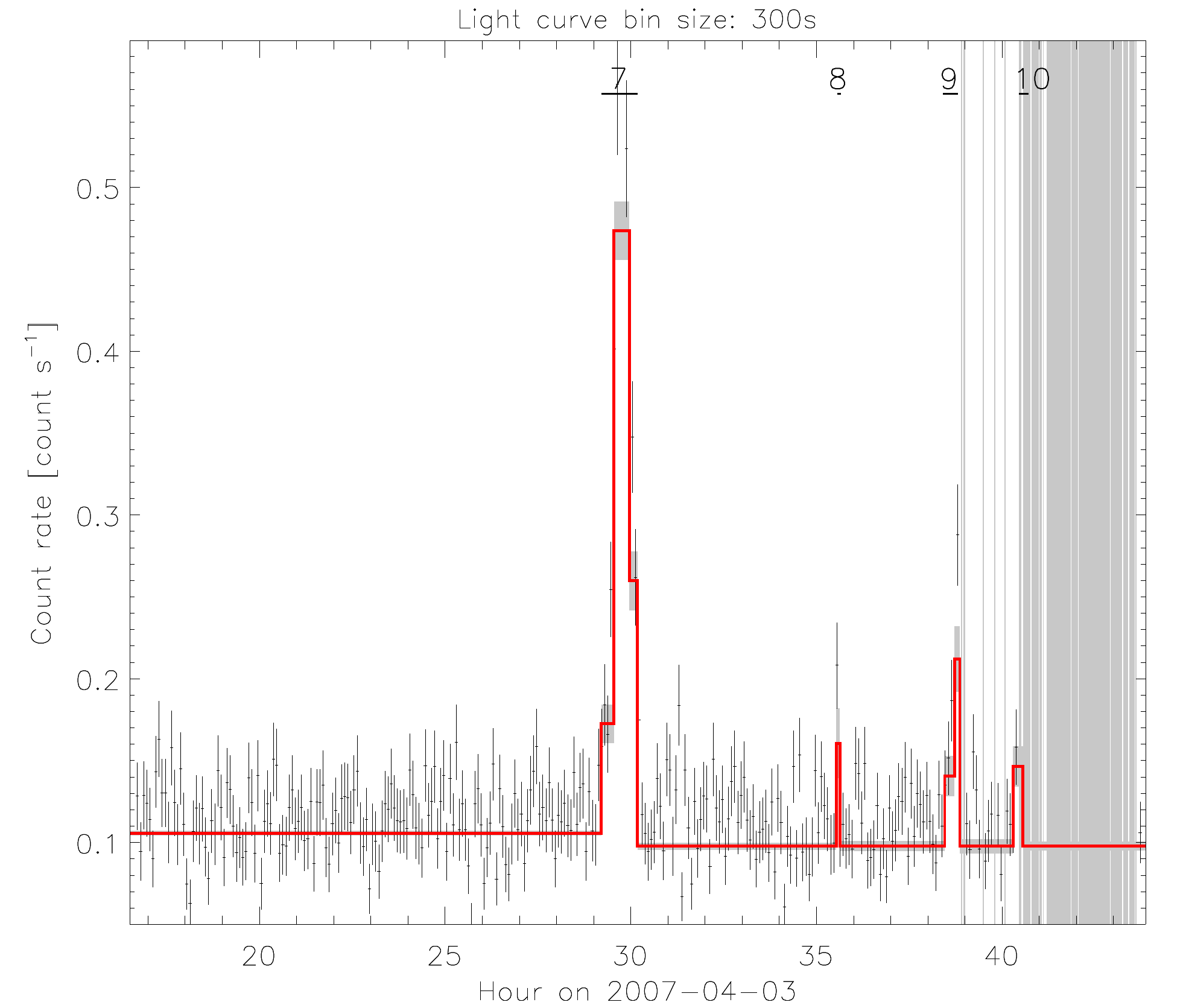}&
\includegraphics[width=0.20\textwidth,angle=90]{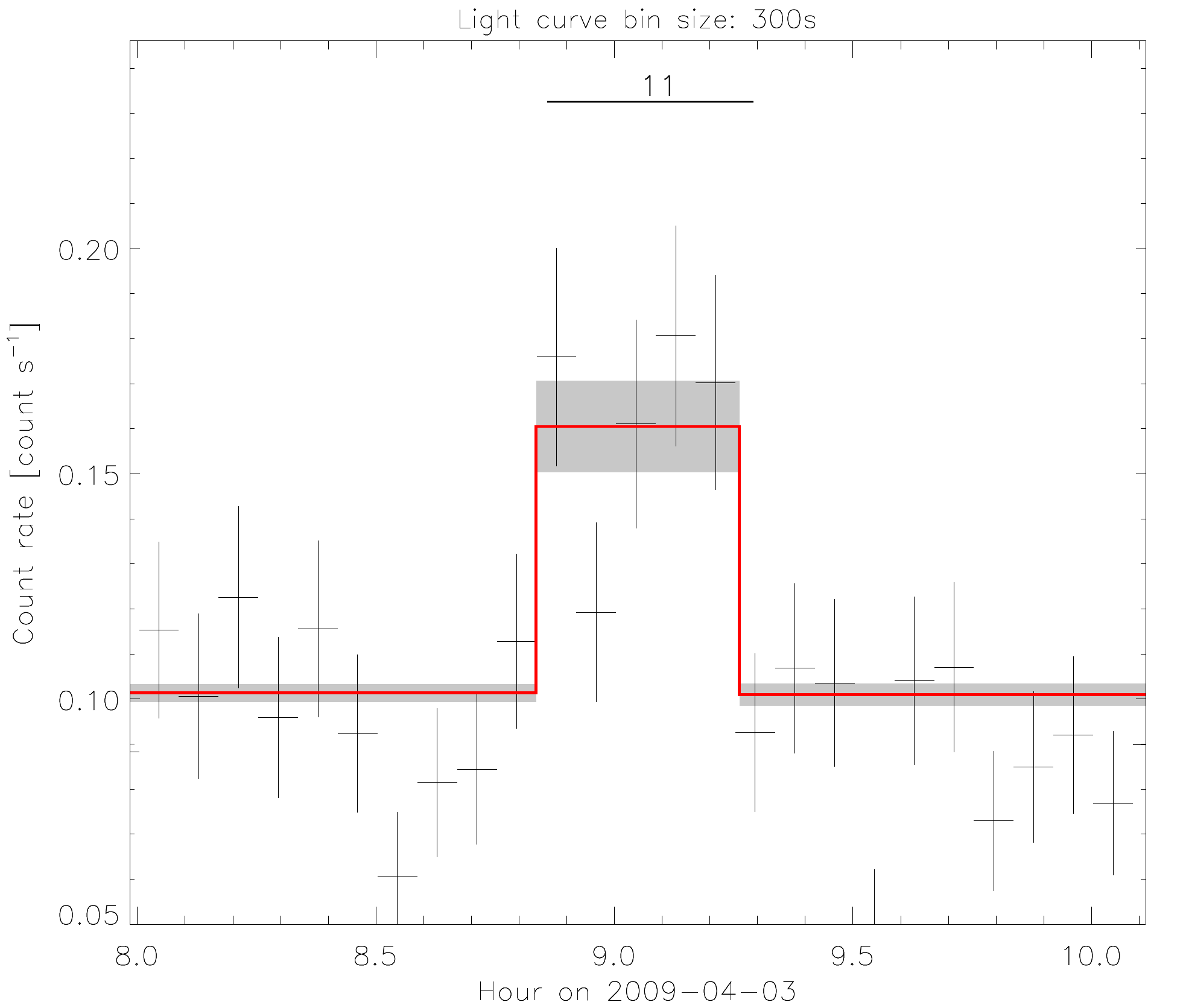}&
\includegraphics[width=0.20\textwidth,angle=90]{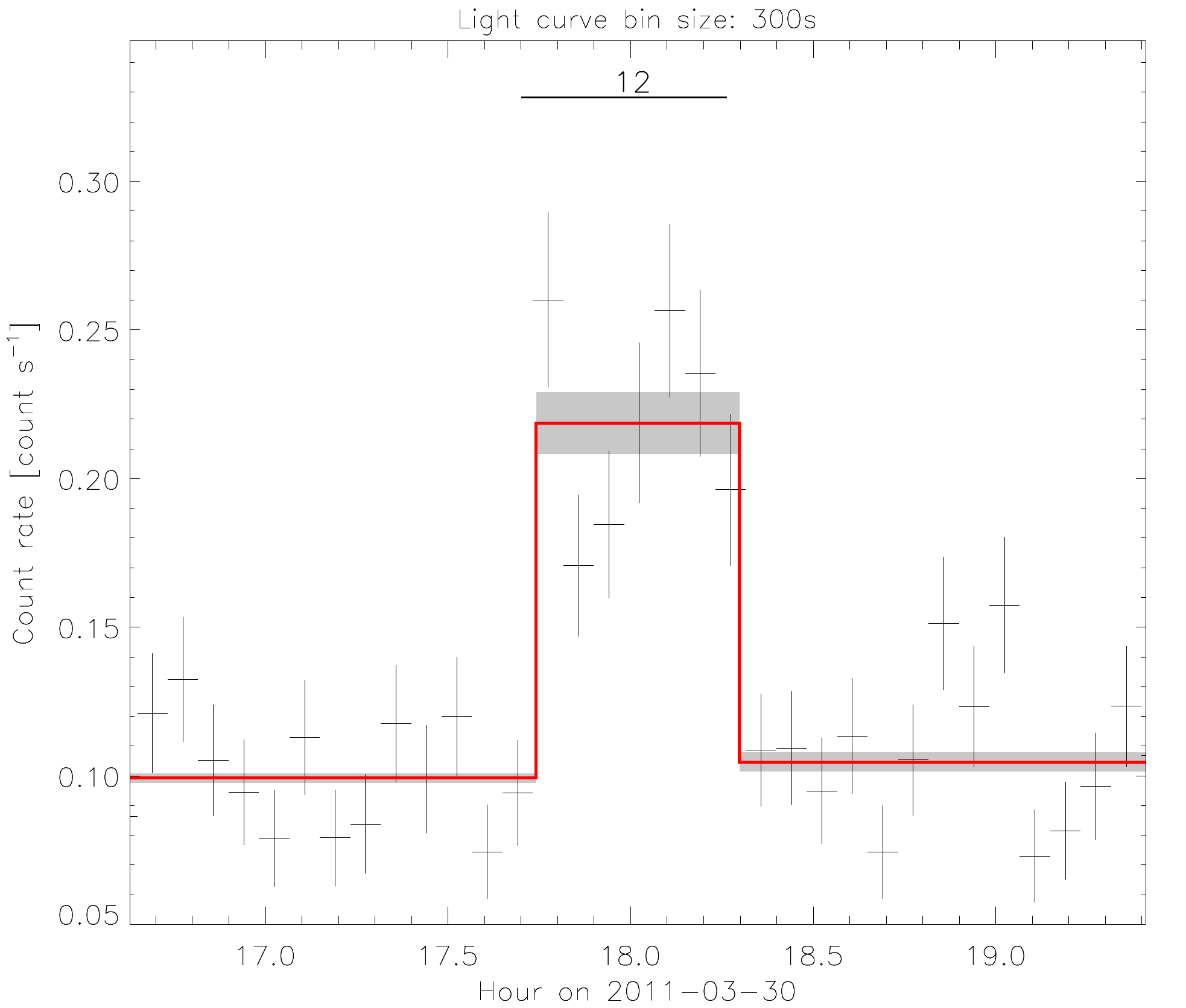}&
\includegraphics[width=0.20\textwidth,angle=90]{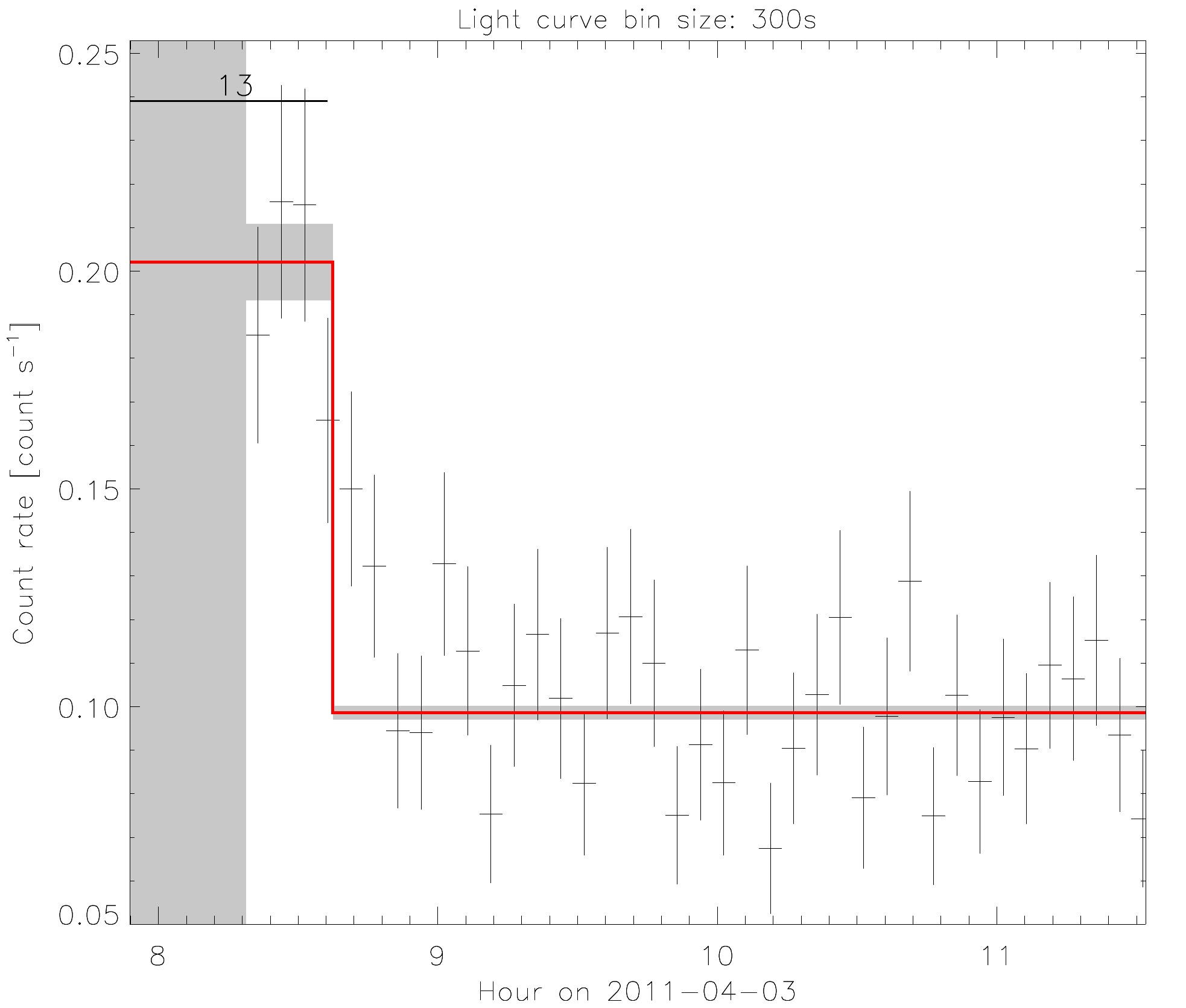}\\
\includegraphics[width=0.20\textwidth,angle=90]{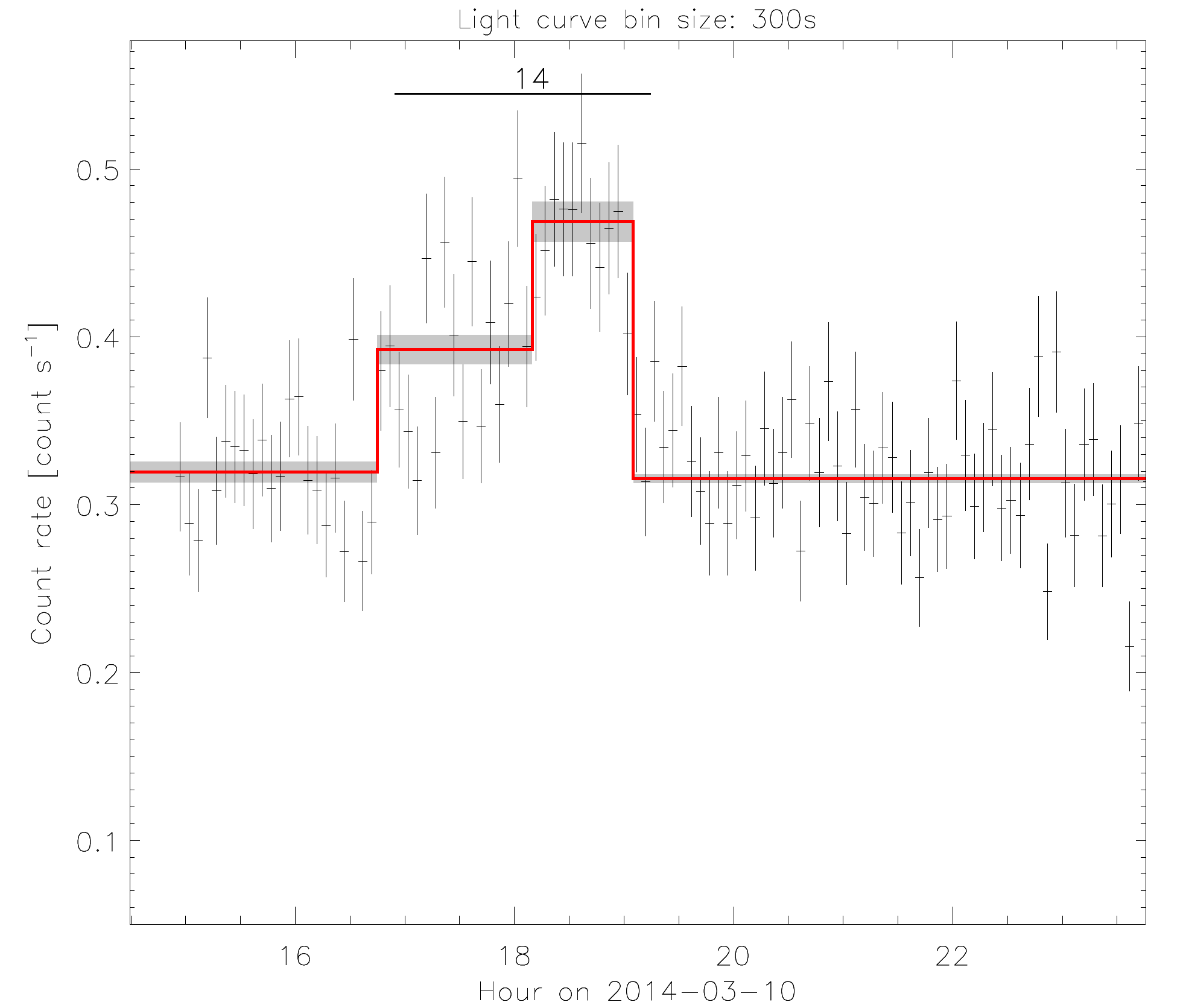}&
\includegraphics[width=0.20\textwidth,angle=90]{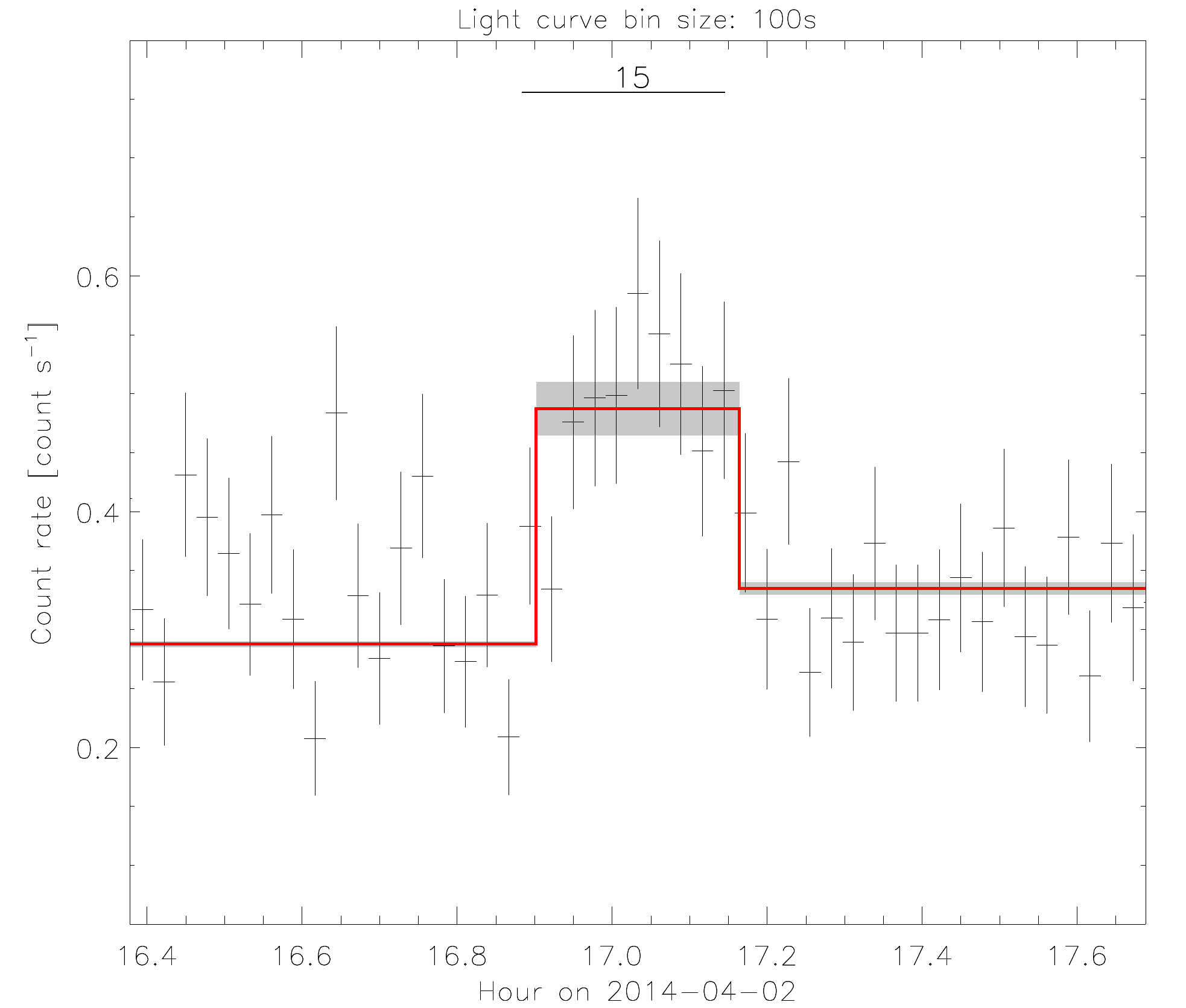}&
\includegraphics[width=0.20\textwidth,angle=90]{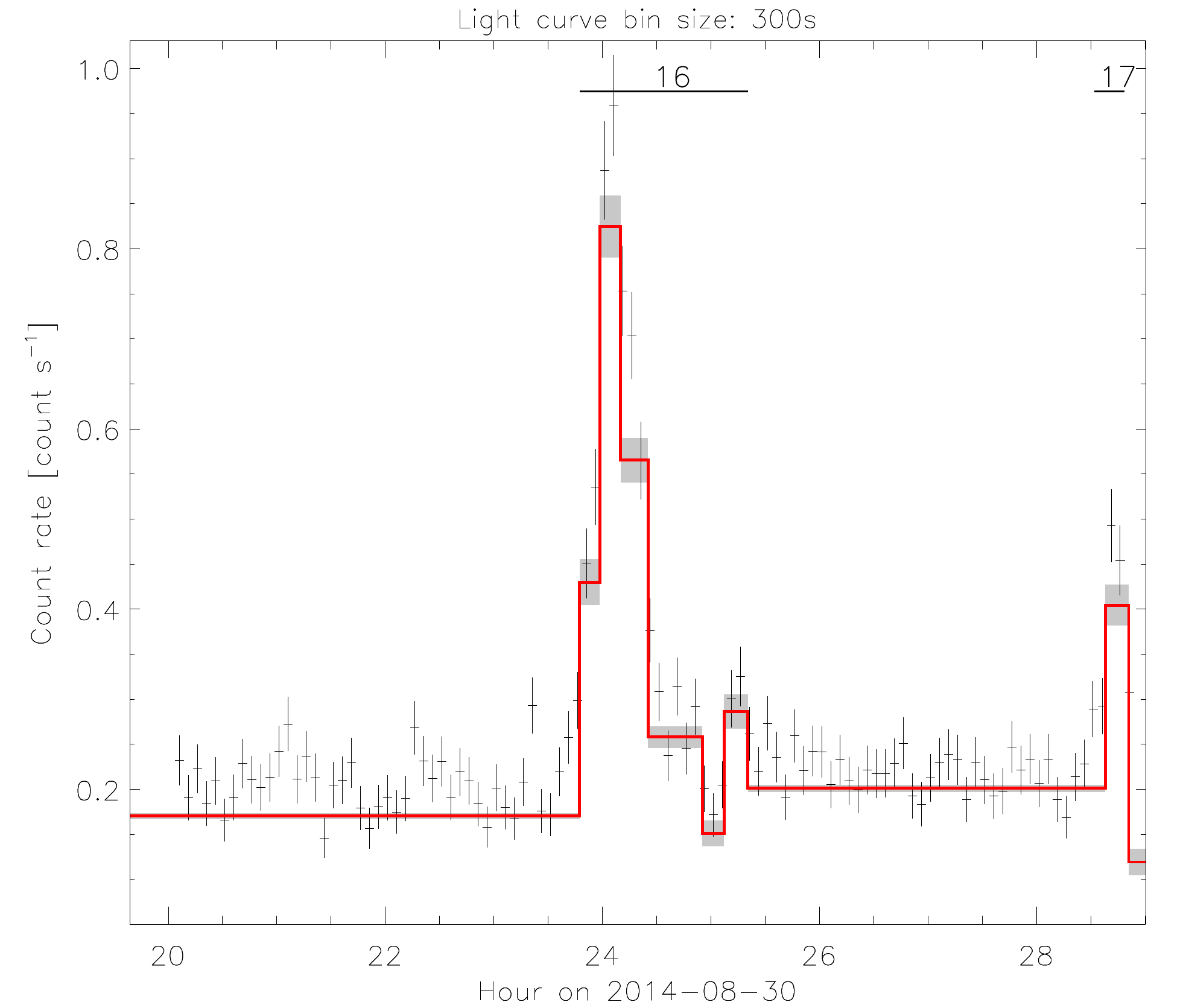}&
\includegraphics[width=0.20\textwidth,angle=90]{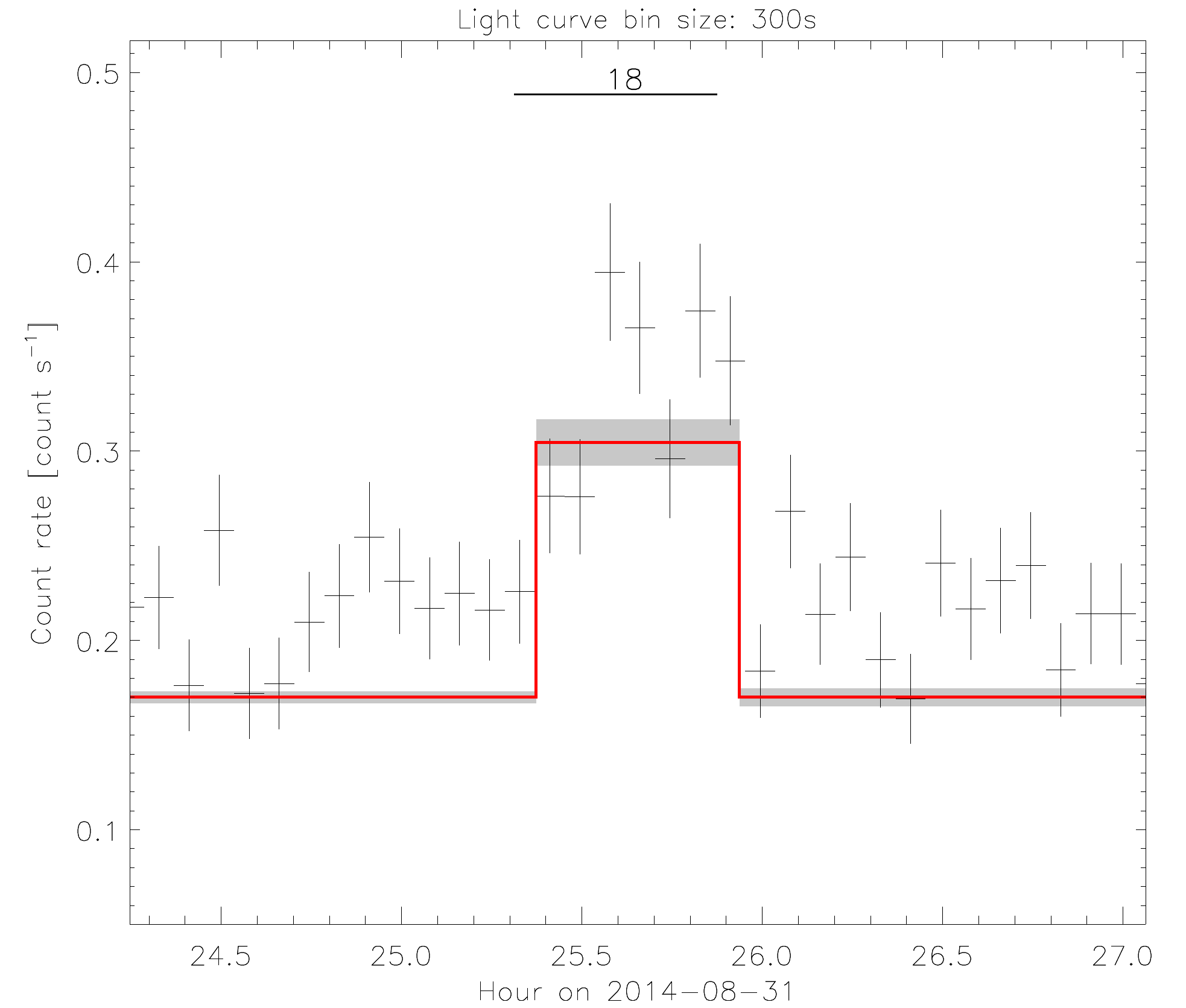}\\
\includegraphics[width=0.20\textwidth,angle=90]{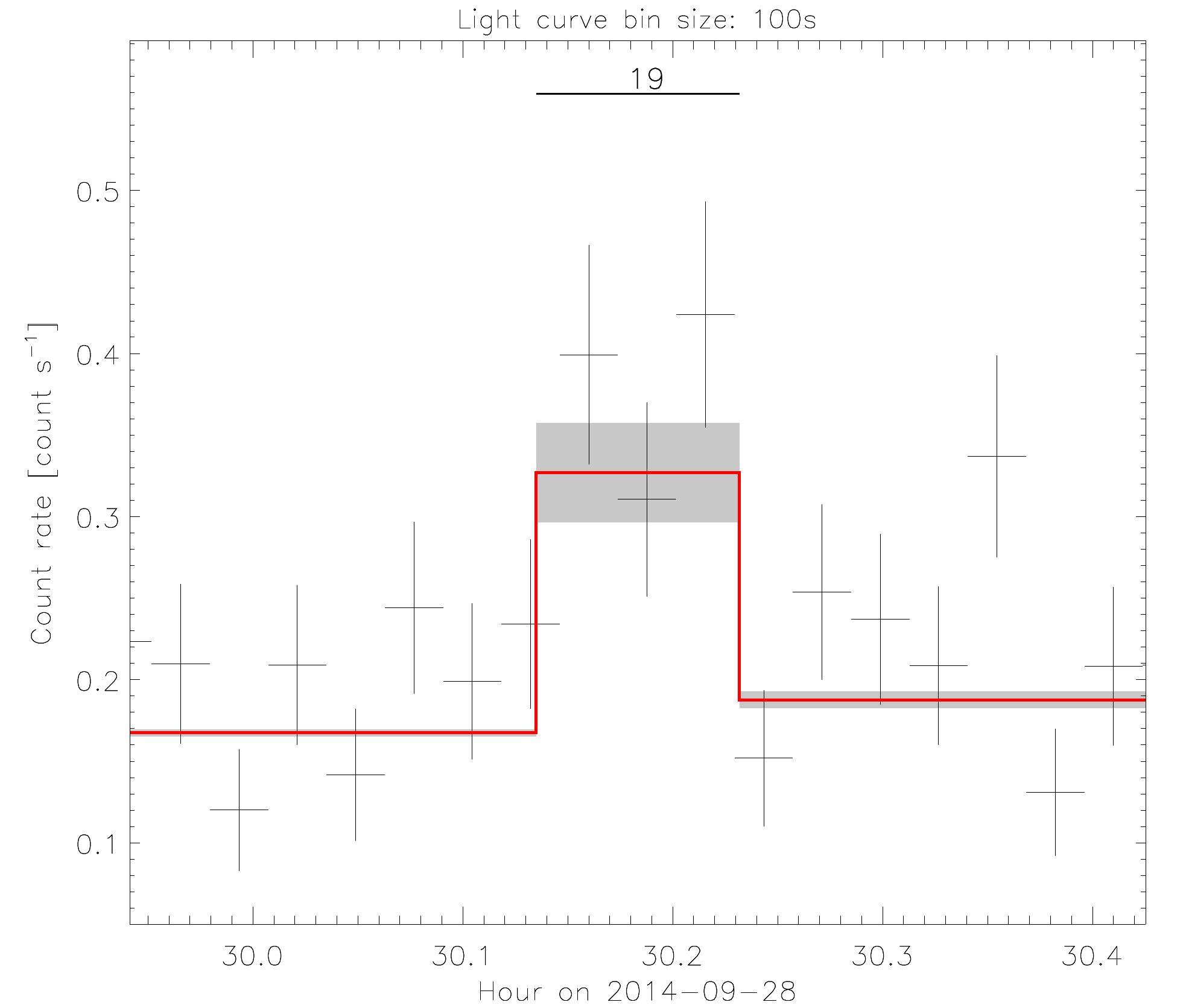}
\end{tabular}
\caption{XMM-Newton flares detected using the Bayesian blocks algorithm with a false positive rate for the flare detection of 0.1\% from 2000 September\ to 2015 April.
The black crosses are the light curves and their error bars.
The bin size of the light curves are reported on the top of each figure.
The red lines are the Bayesian blocks with their errors in horizontal gray rectangles.
The vertical gray stripe is the bad time interval.
Each flare is labeled with the index corresponding to the flare number in Table~\ref{table:xmm}.
The horizontal line is the flare duration.
See caption of Fig.~\ref{fig:chandra_flare1} for details.}
\label{fig:xmm_flare}
\end{figure*}

\newpage
\textbf{}
\begin{figure*}[!ht]
\centering
\begin{tabular}{@{}ccccc@{}}
\includegraphics[width=0.20\textwidth,angle=90]{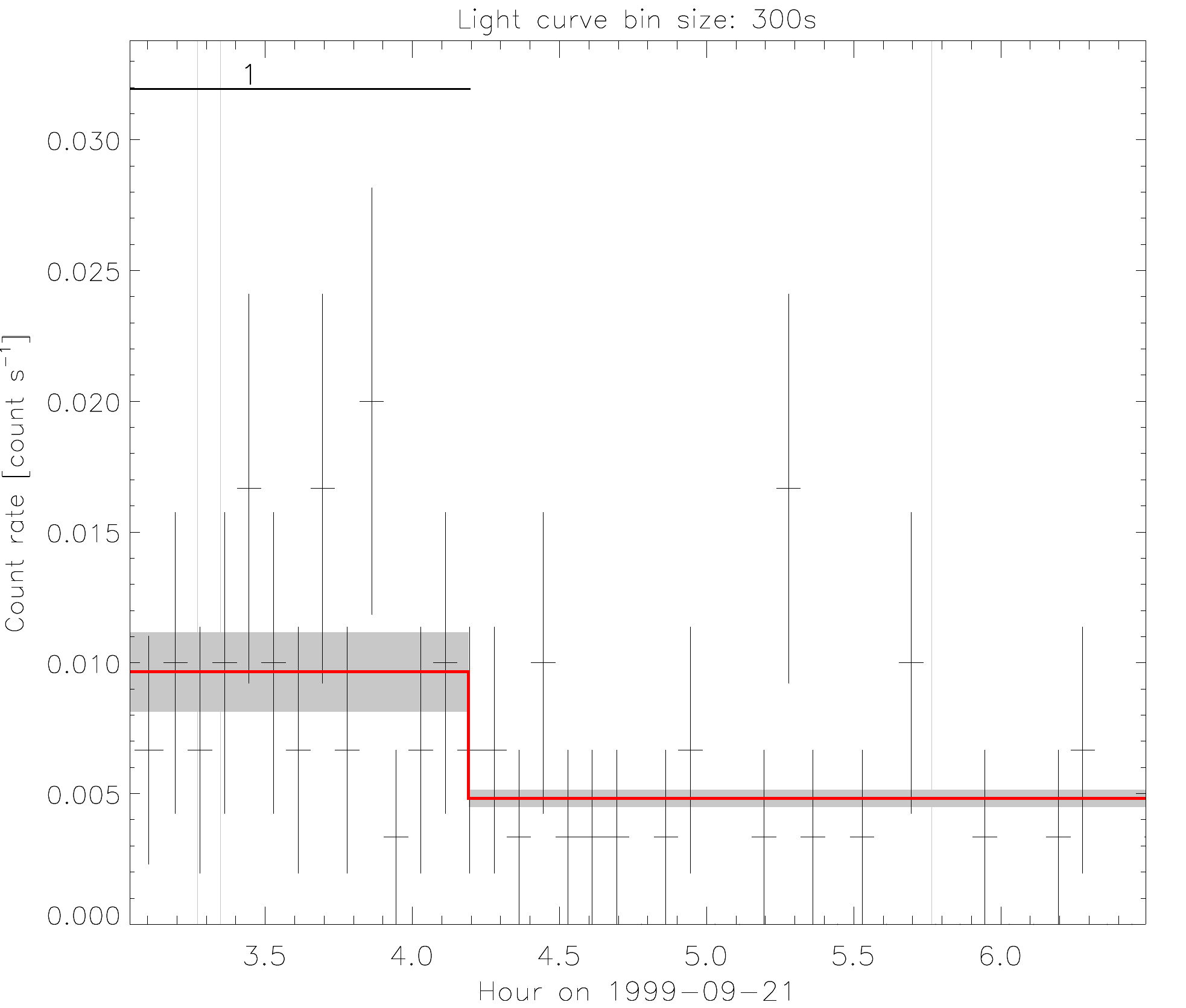}&
\includegraphics[width=0.20\textwidth,angle=90]{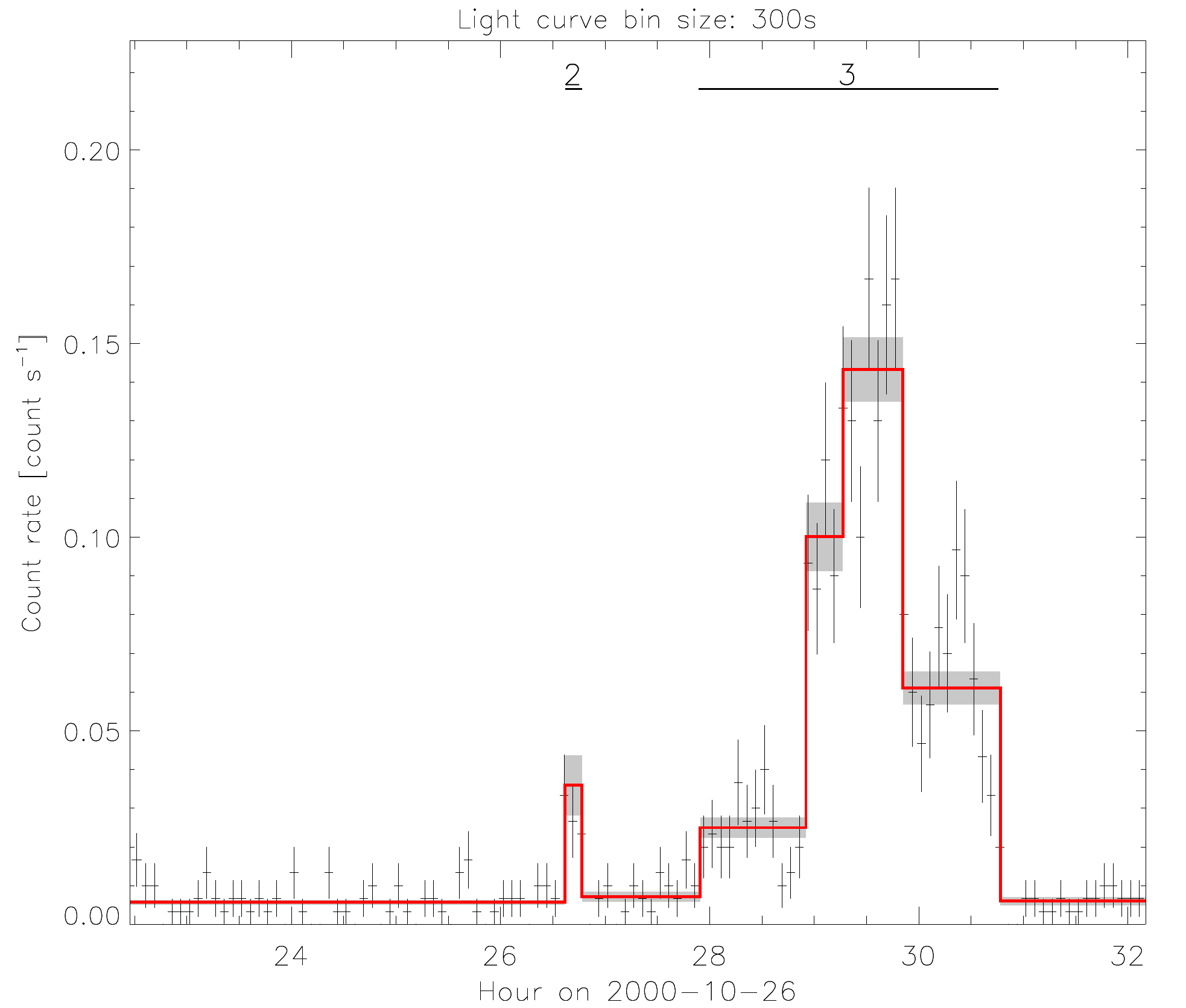}&
\includegraphics[width=0.20\textwidth,angle=90]{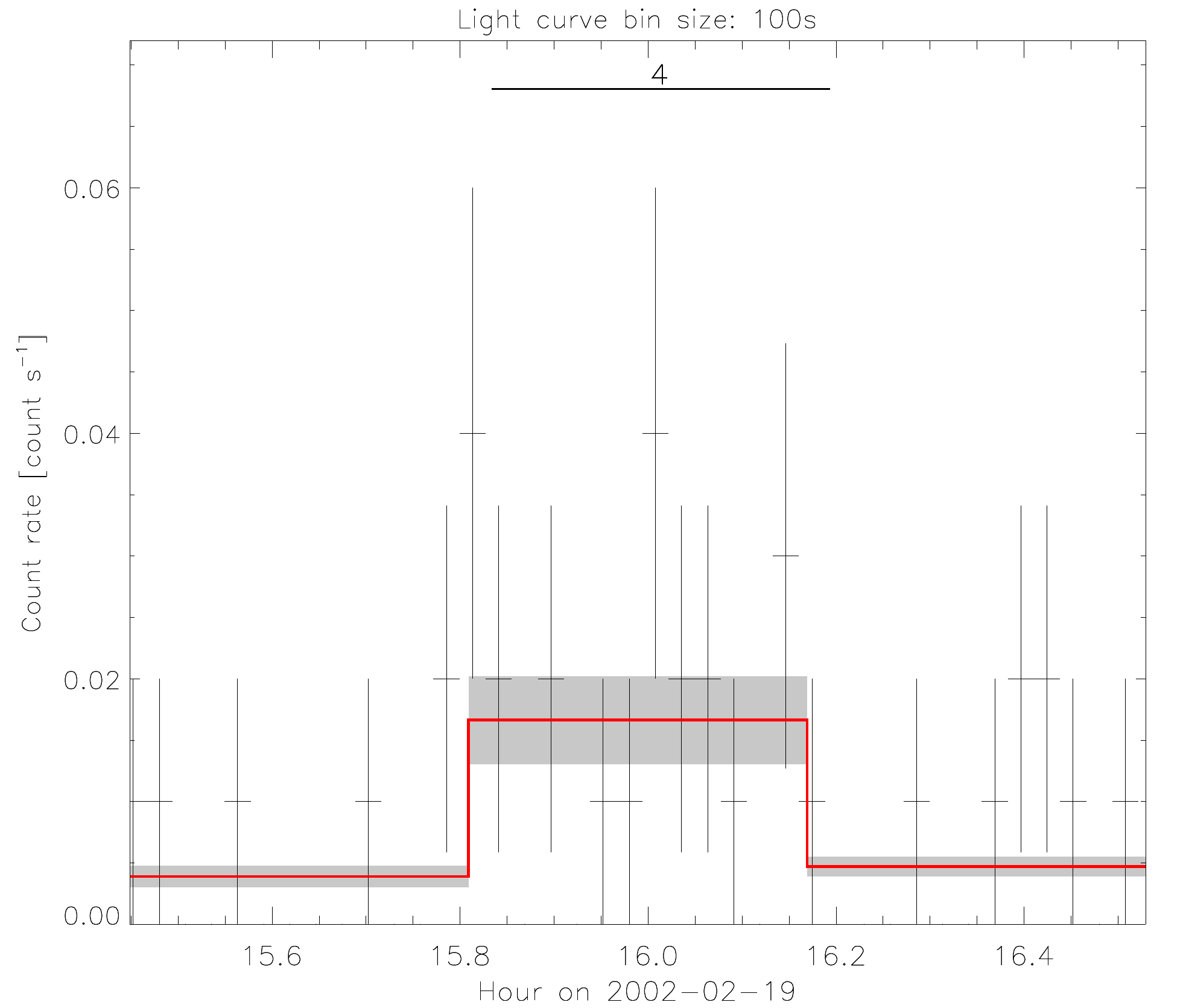}&
\includegraphics[width=0.20\textwidth,angle=90]{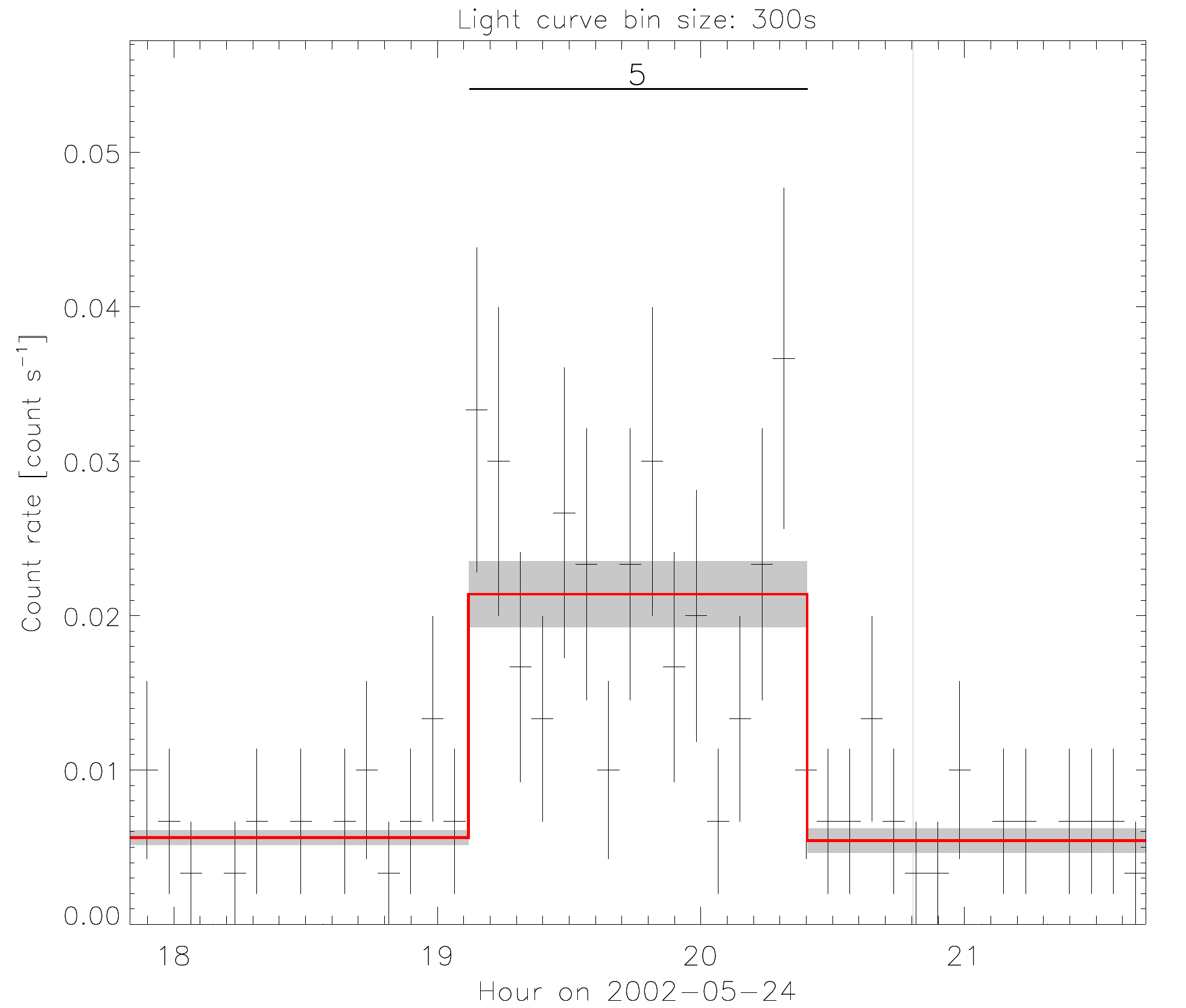}\\
\includegraphics[width=0.20\textwidth,angle=90]{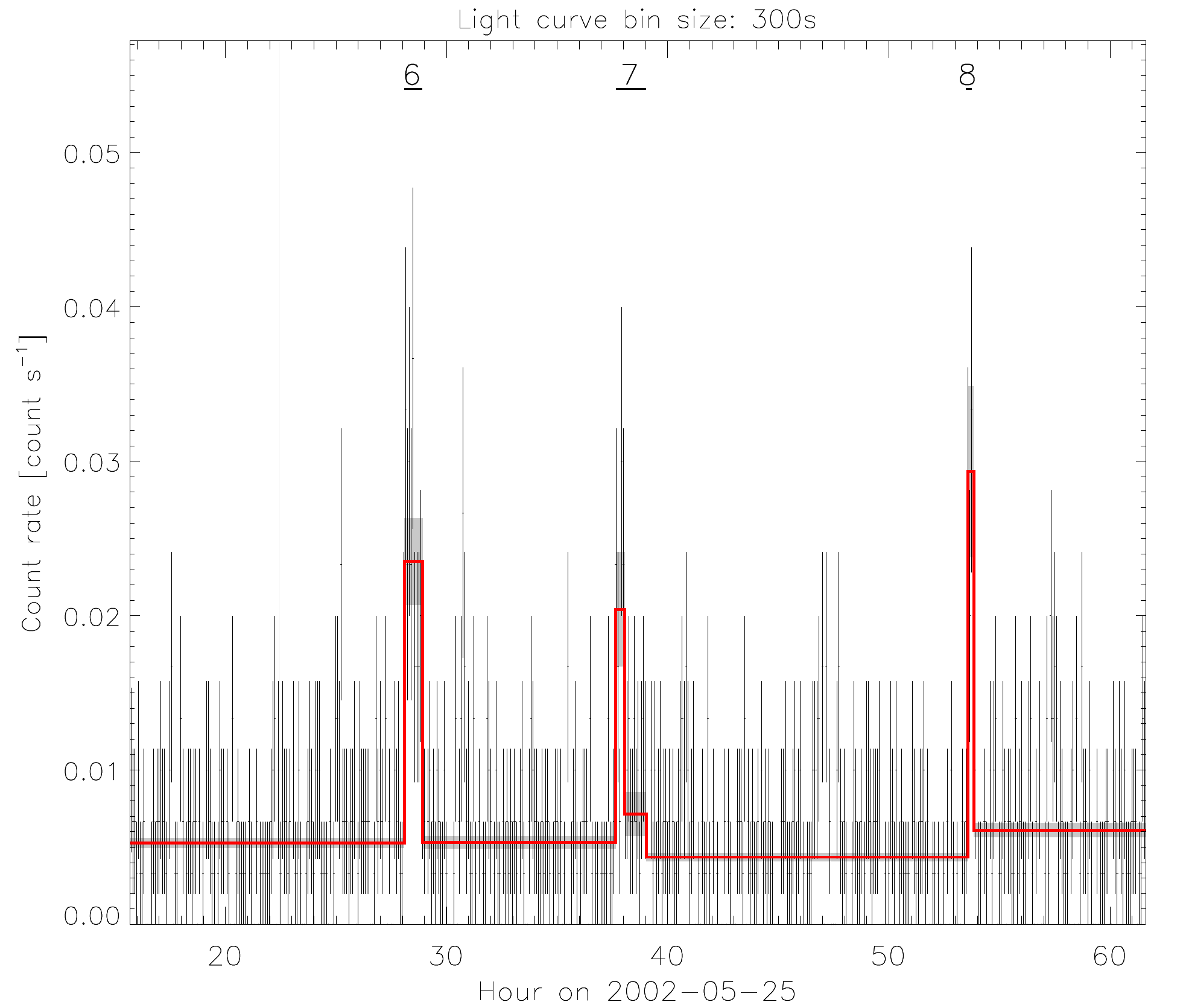}&
\includegraphics[width=0.20\textwidth,angle=90]{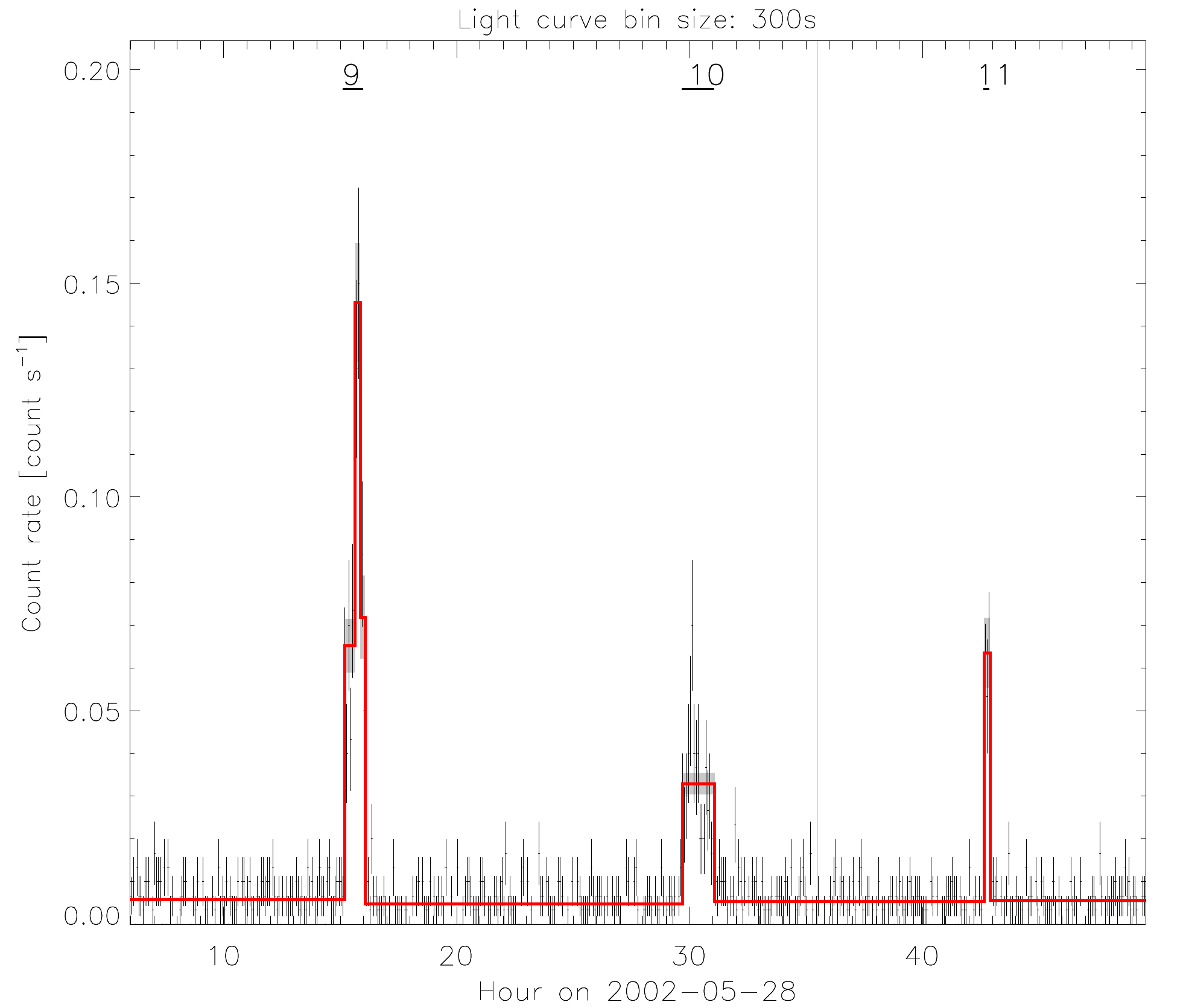}&
\includegraphics[width=0.20\textwidth,angle=90]{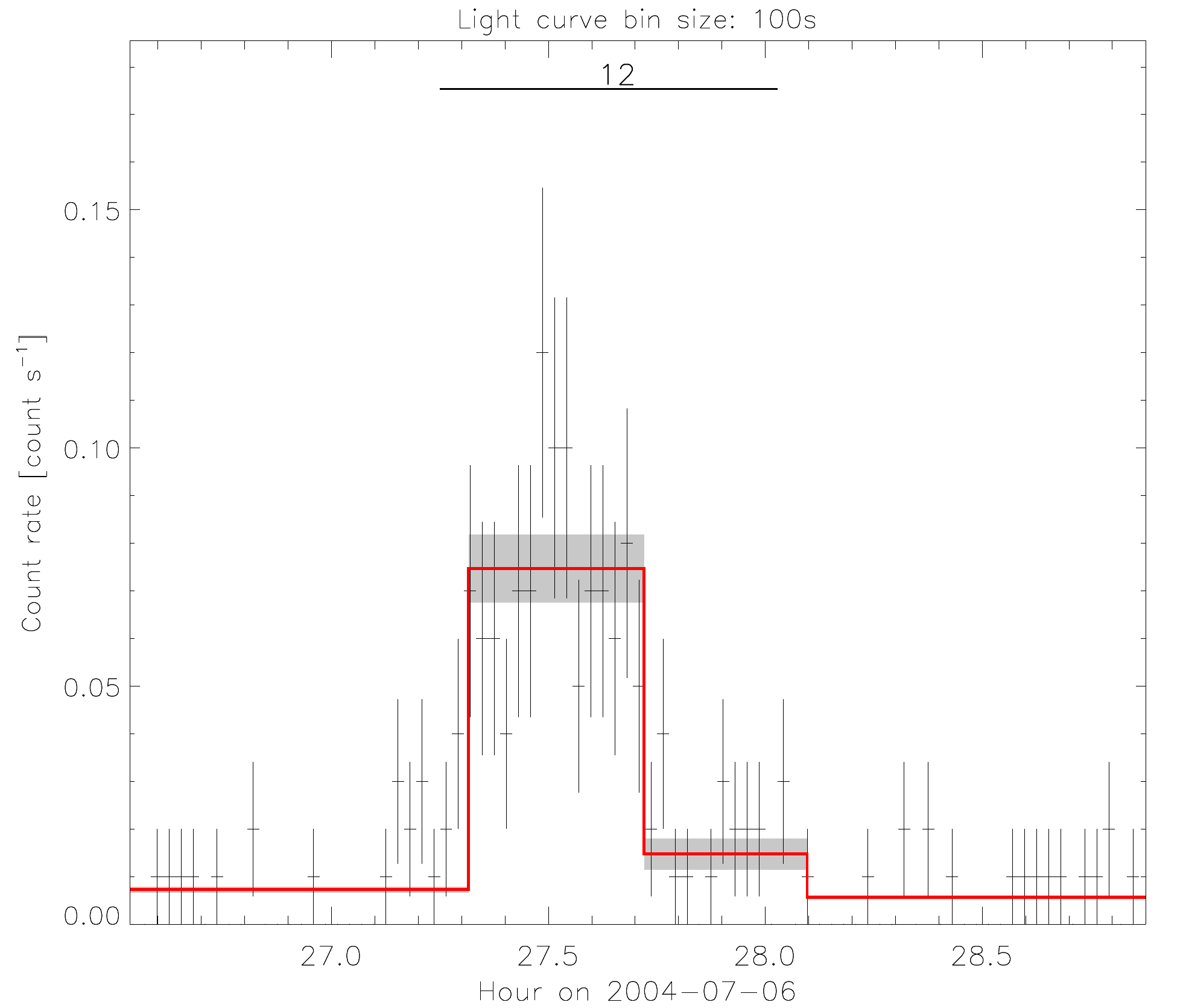}&
\includegraphics[width=0.20\textwidth,angle=90]{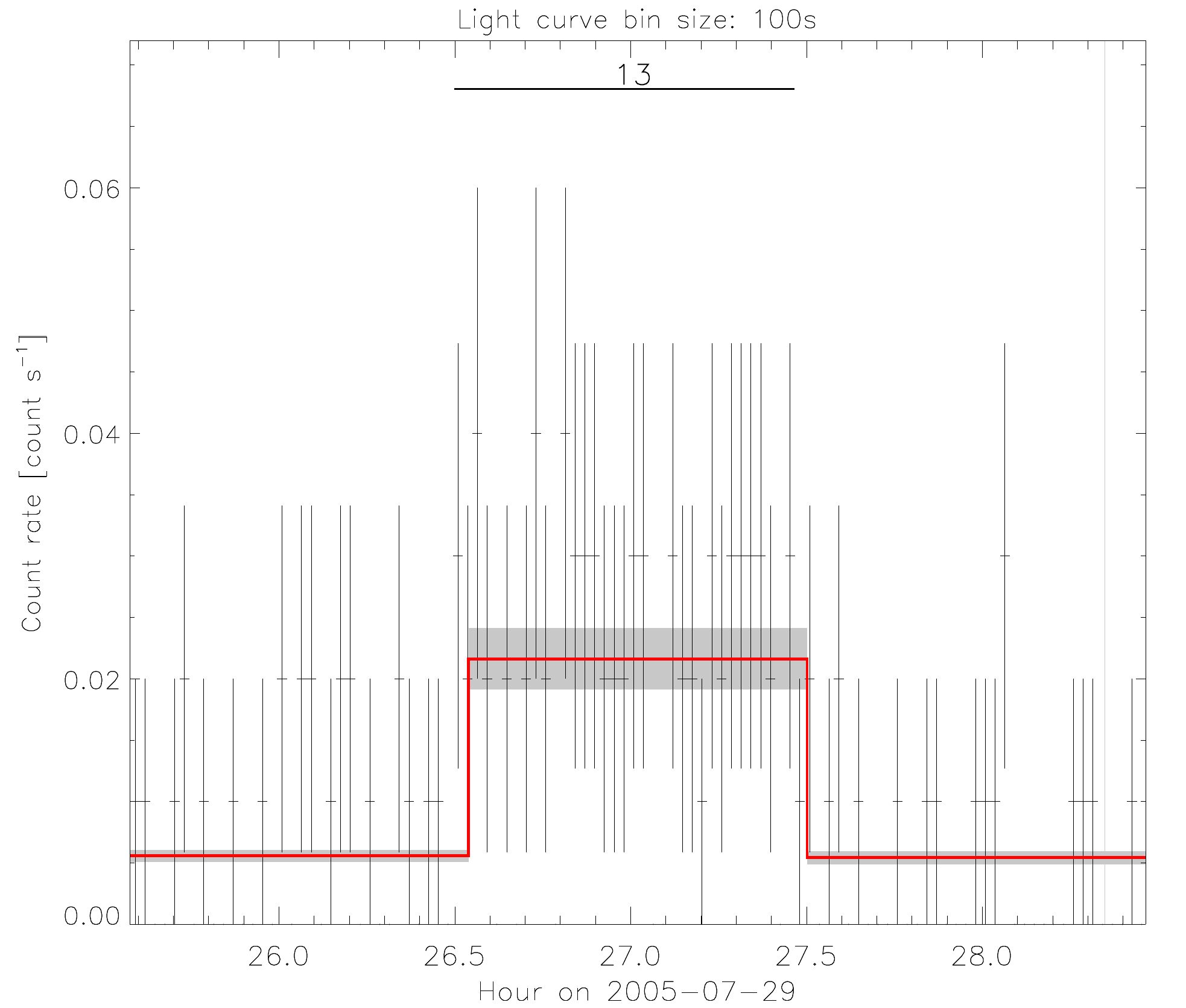}\\
\includegraphics[width=0.20\textwidth,angle=90]{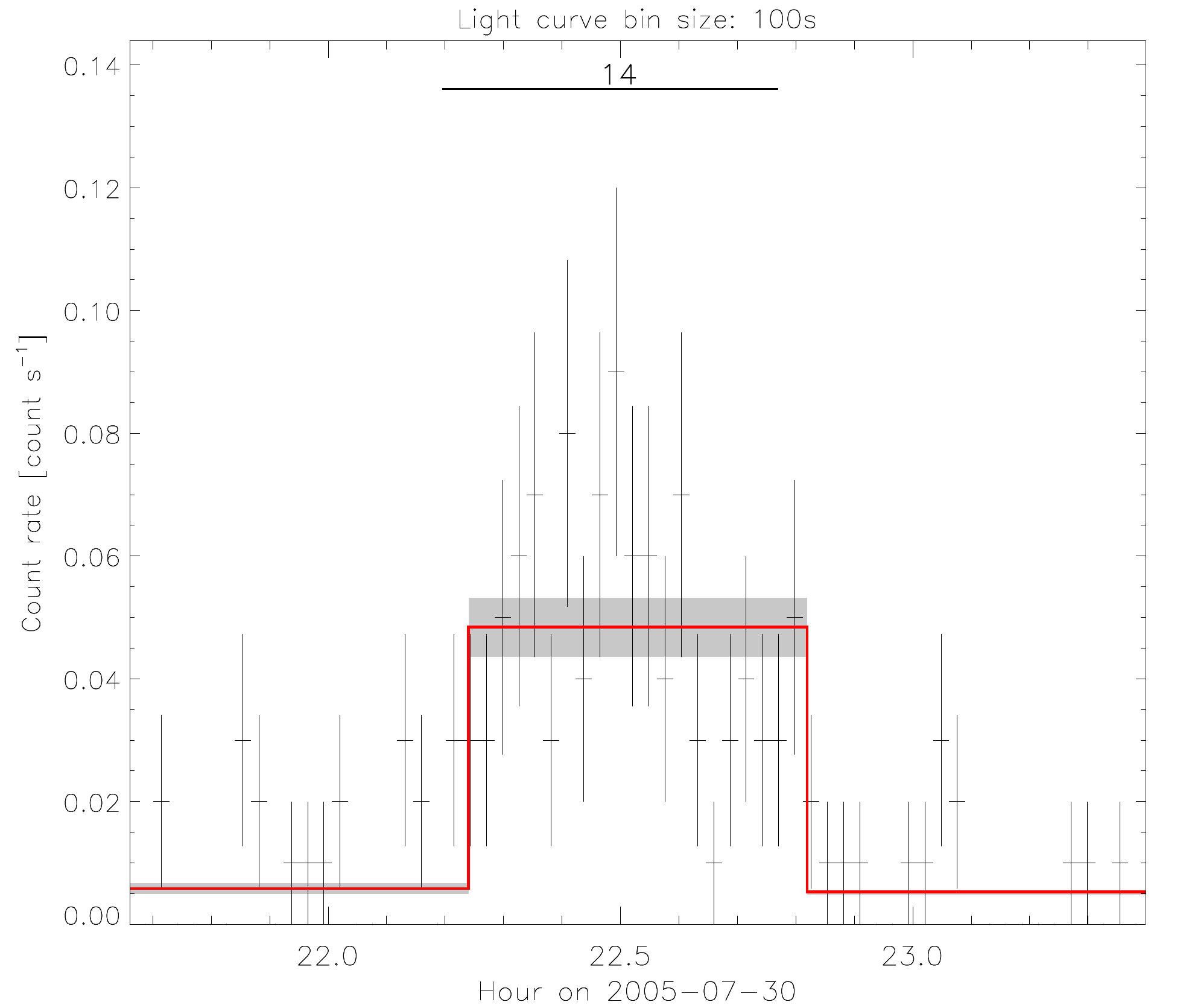}&
\includegraphics[width=0.20\textwidth,angle=90]{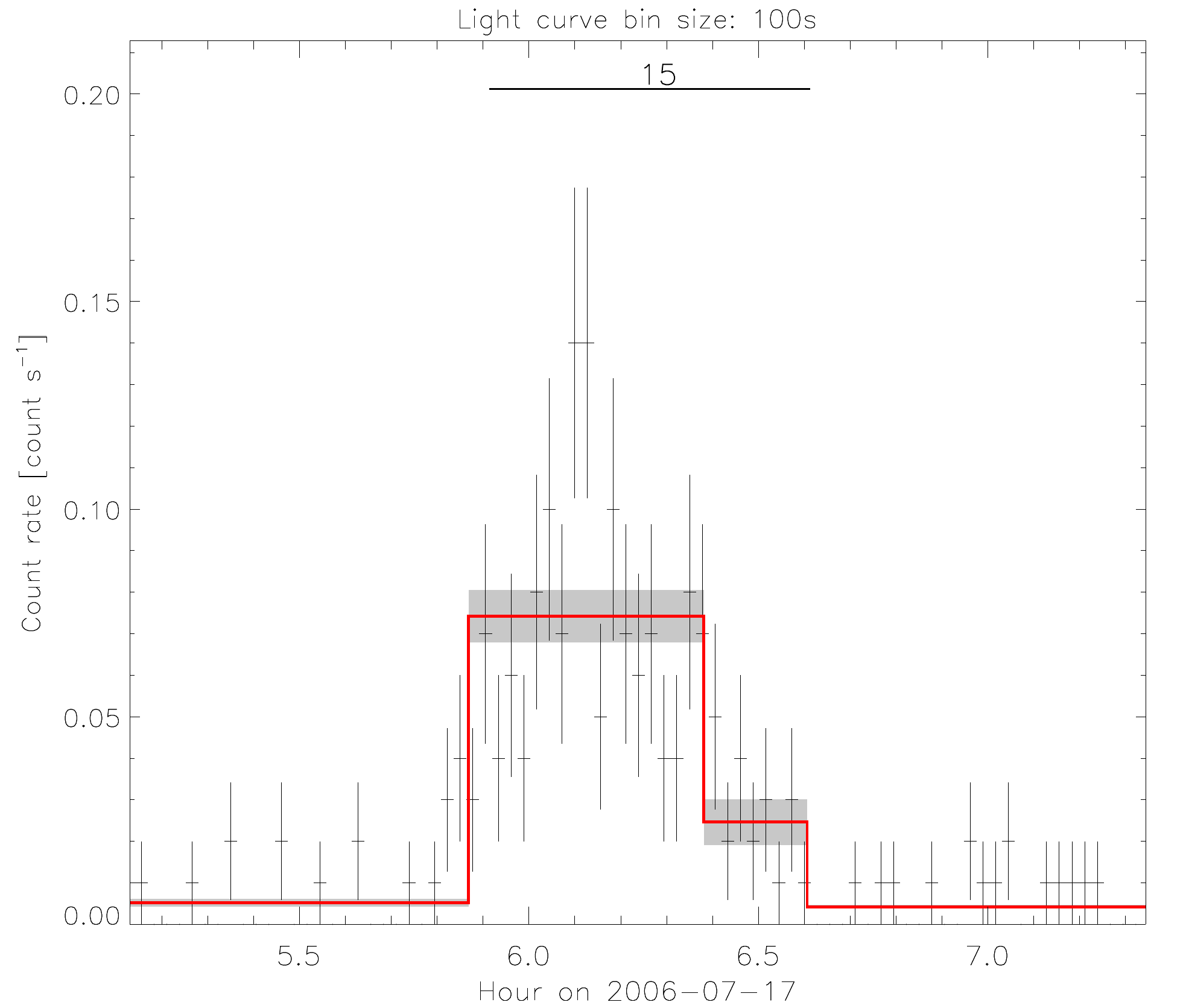}&
\includegraphics[width=0.20\textwidth,angle=90]{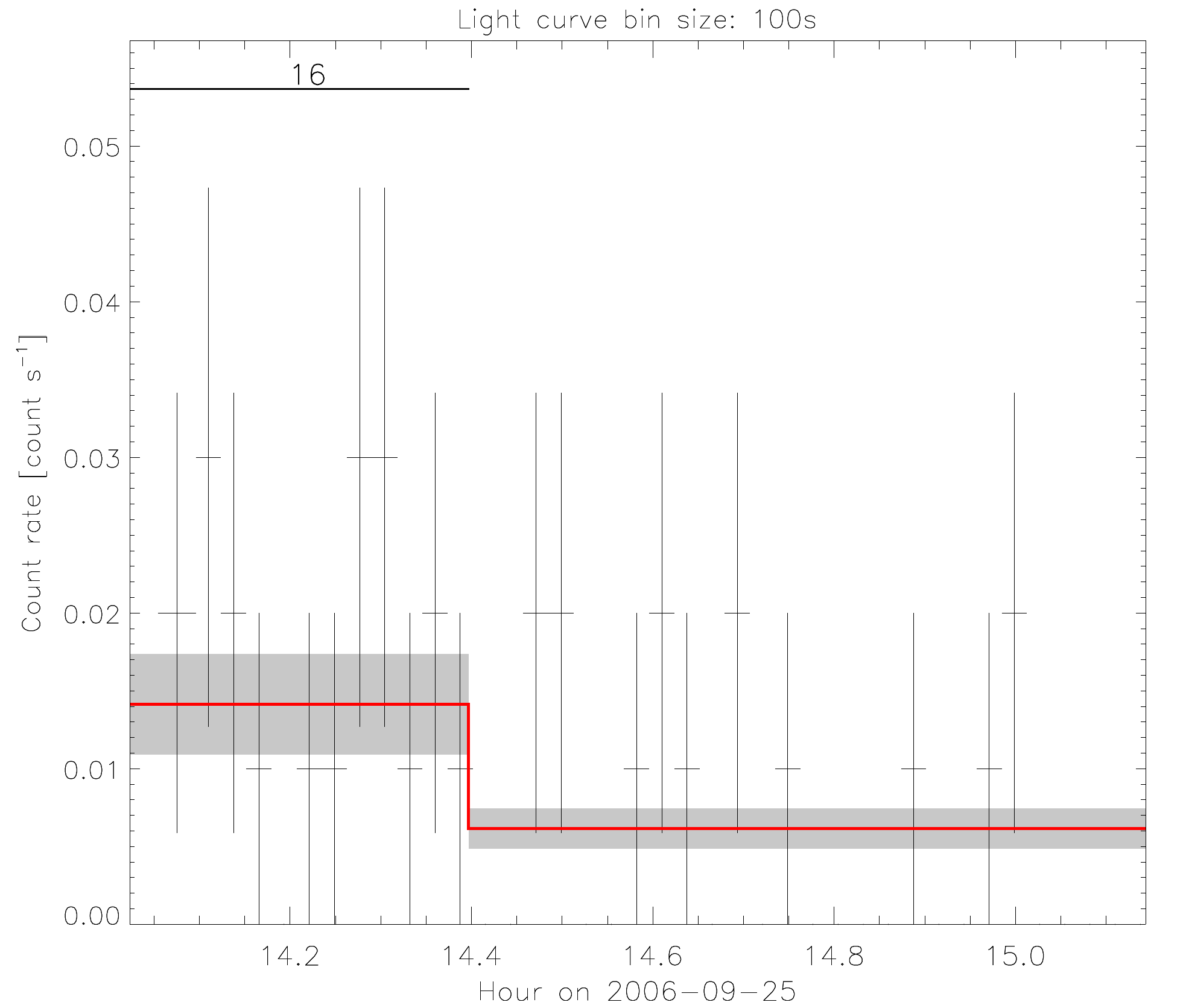}&
\includegraphics[width=0.20\textwidth,angle=90]{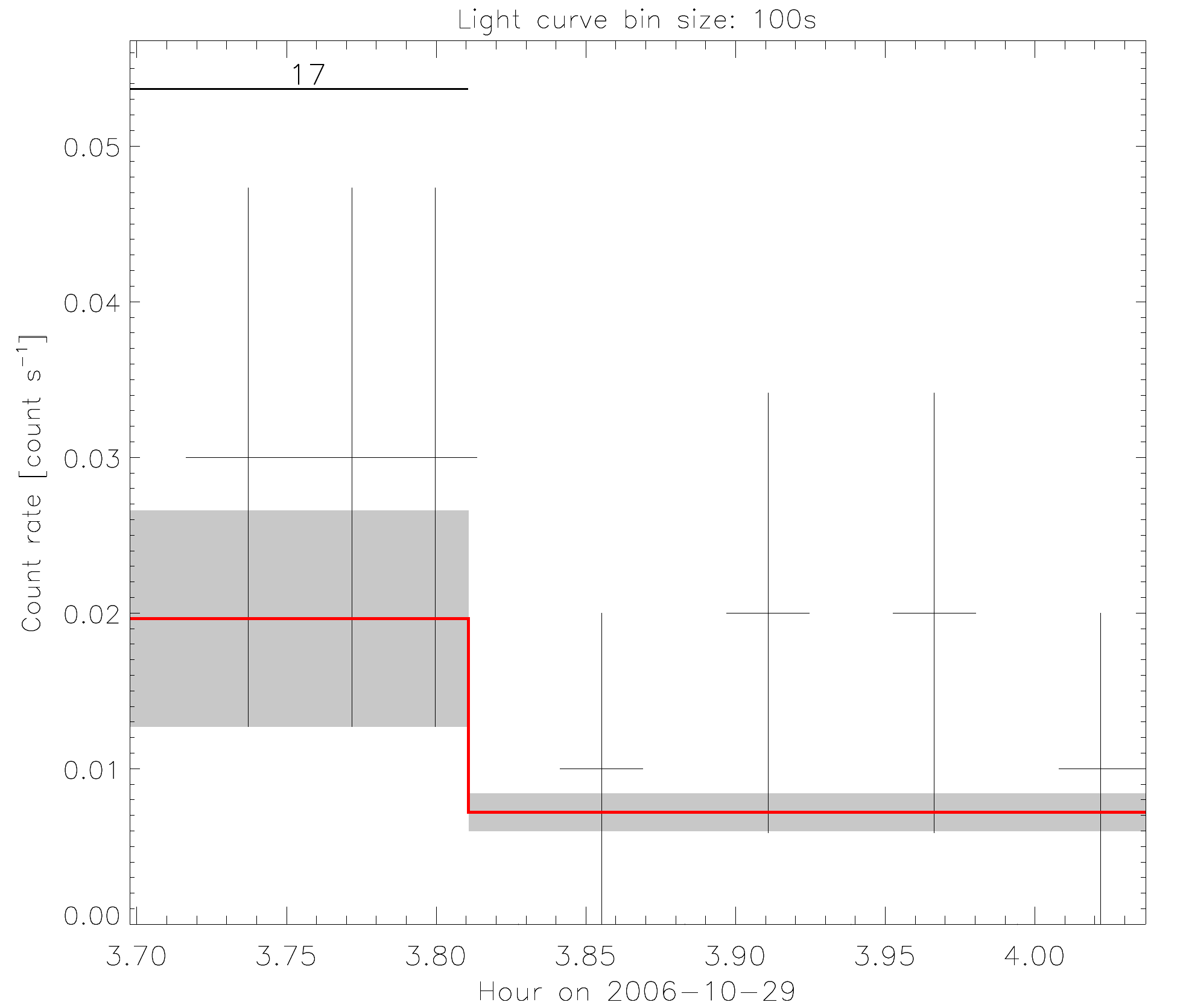}\\
\includegraphics[width=0.20\textwidth,angle=90]{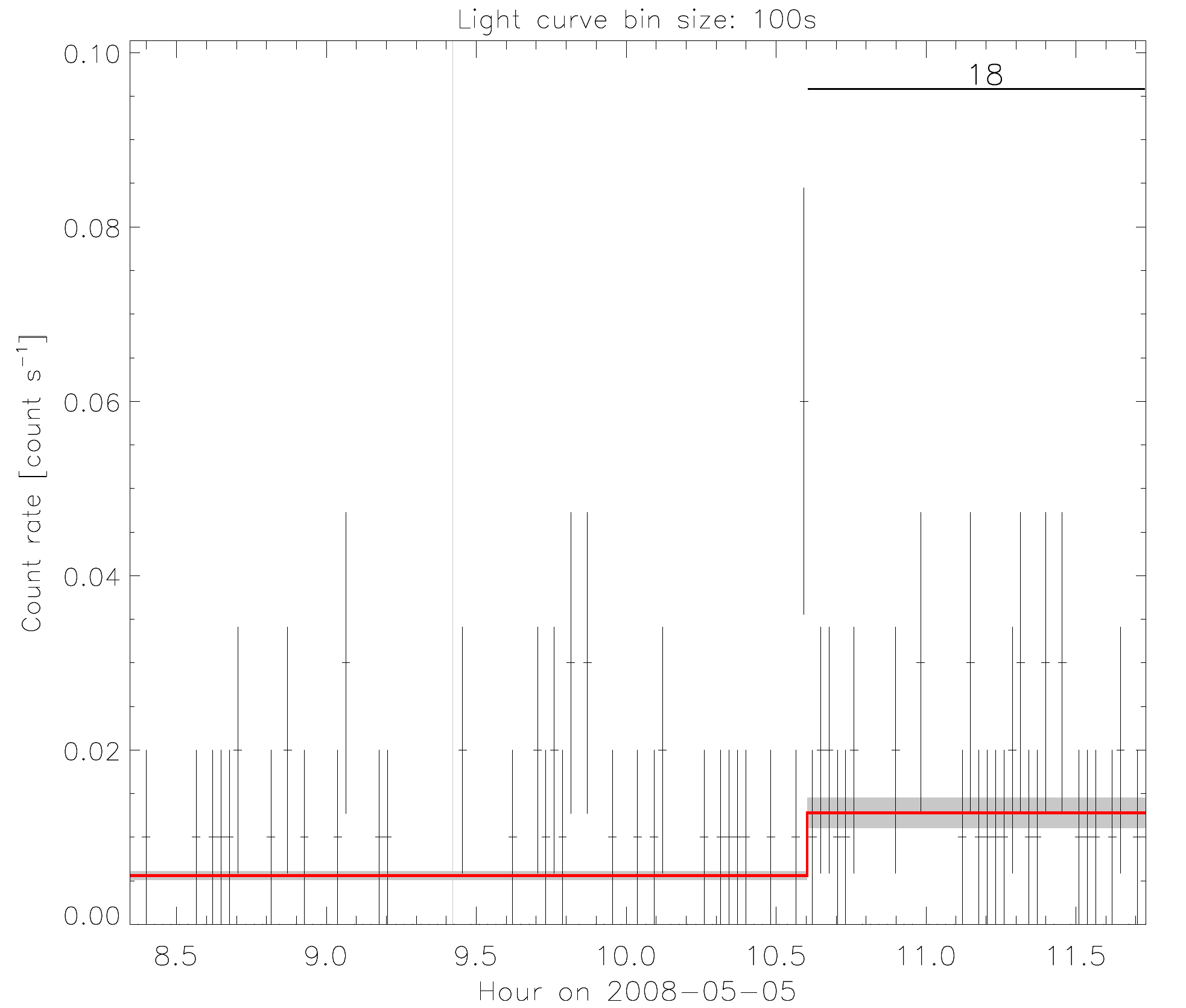}&
\includegraphics[width=0.20\textwidth,angle=90]{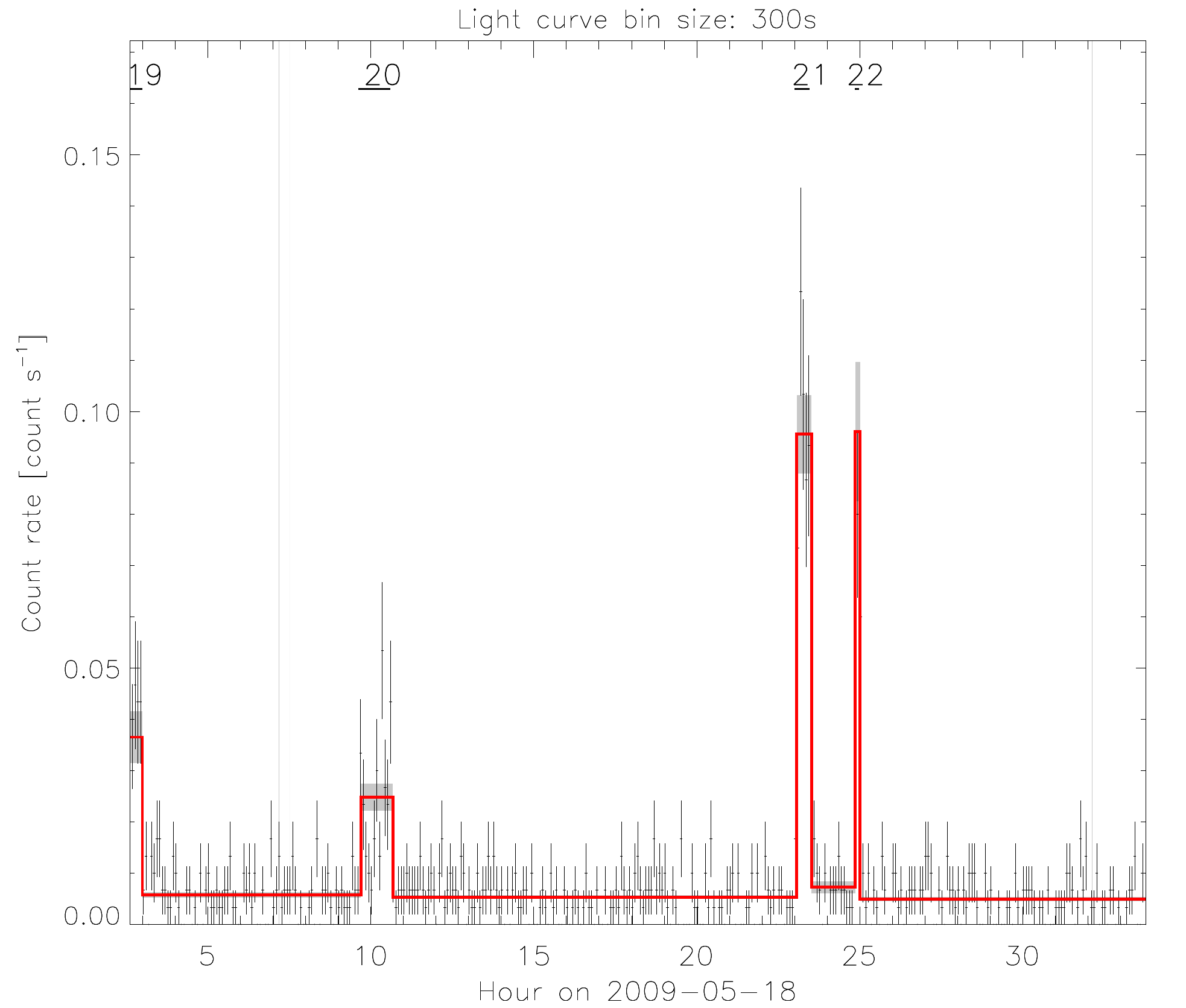}&
\includegraphics[width=0.20\textwidth,angle=90]{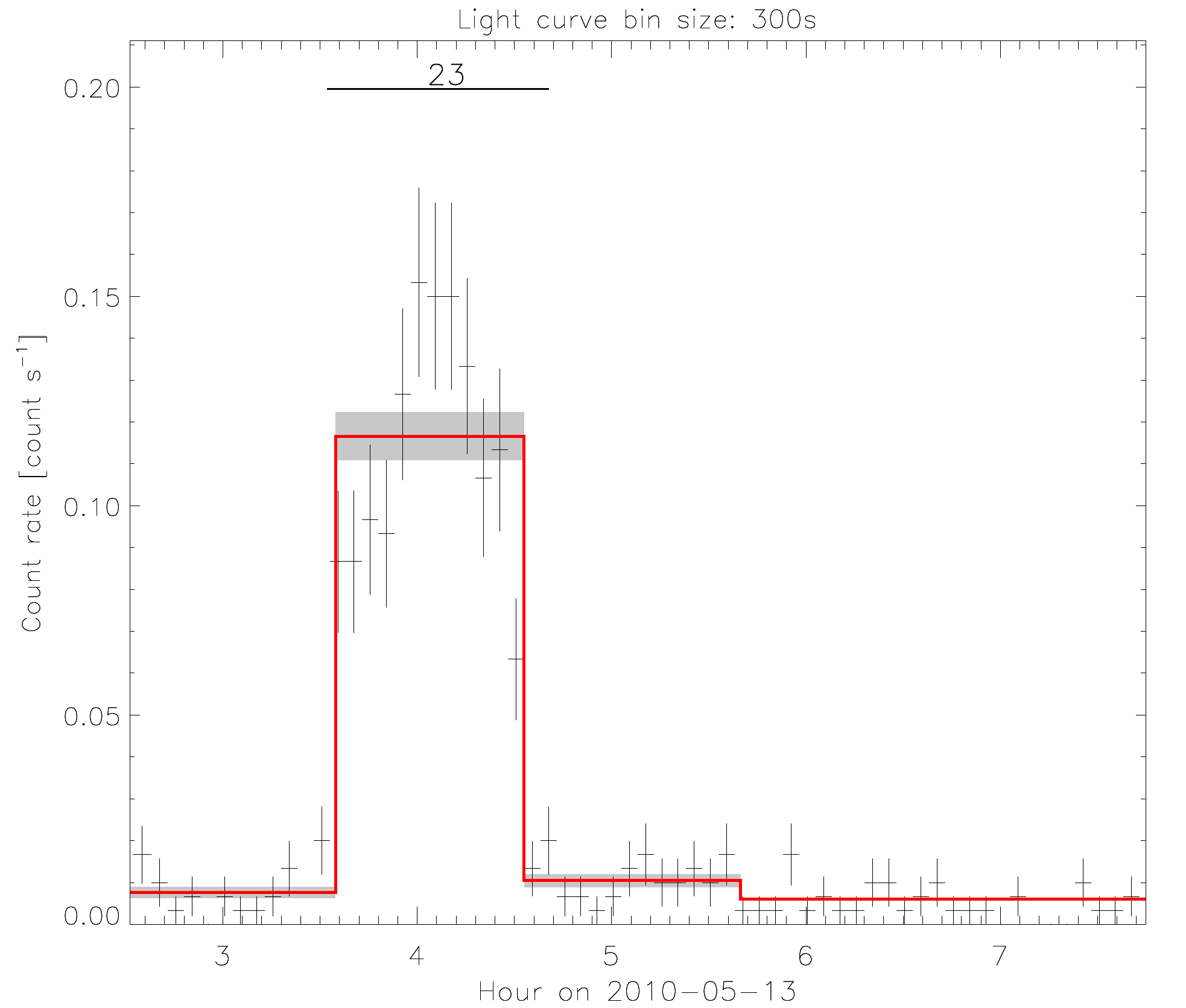}&
\includegraphics[width=0.20\textwidth,angle=90]{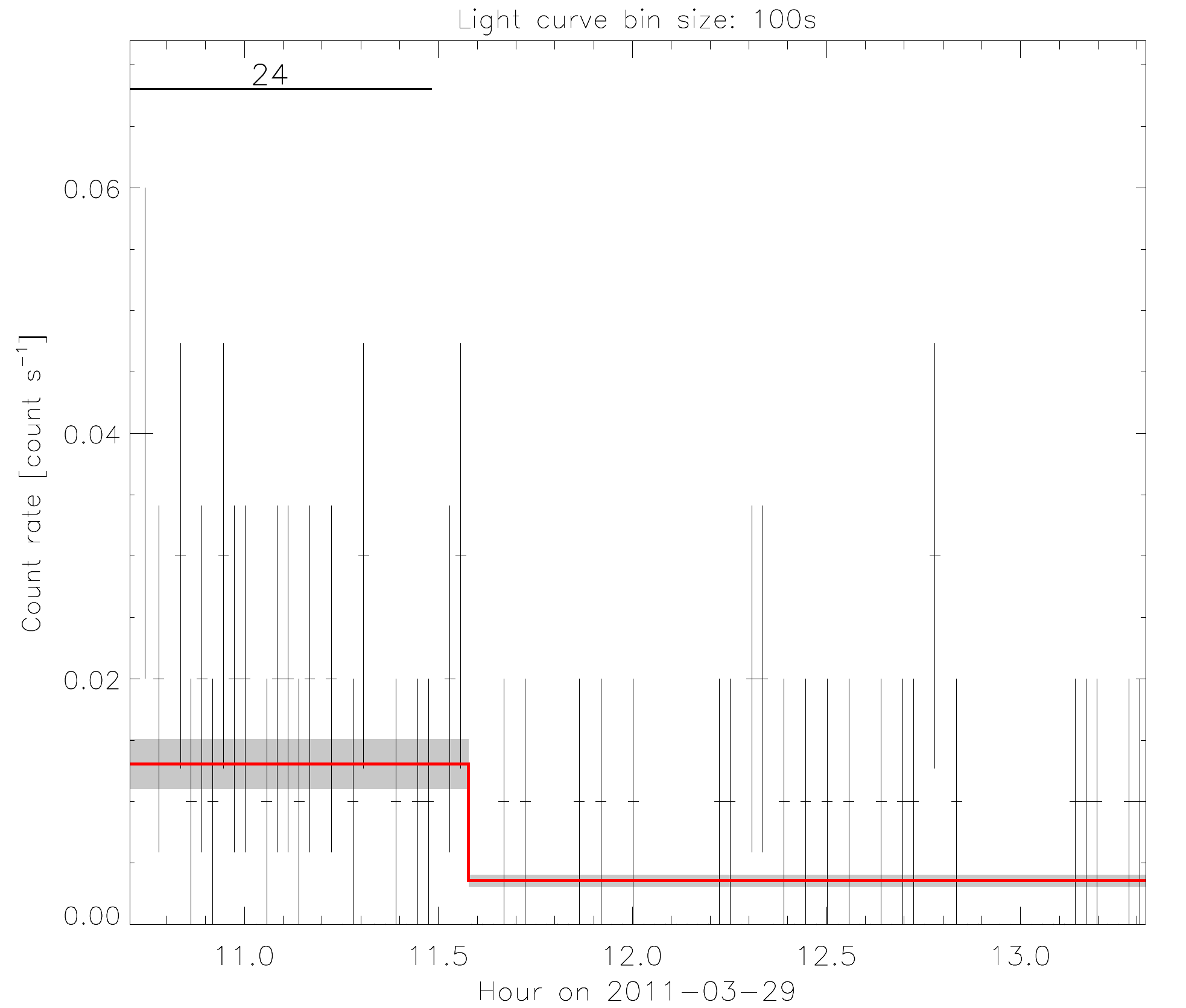}\\
\includegraphics[width=0.20\textwidth,angle=90]{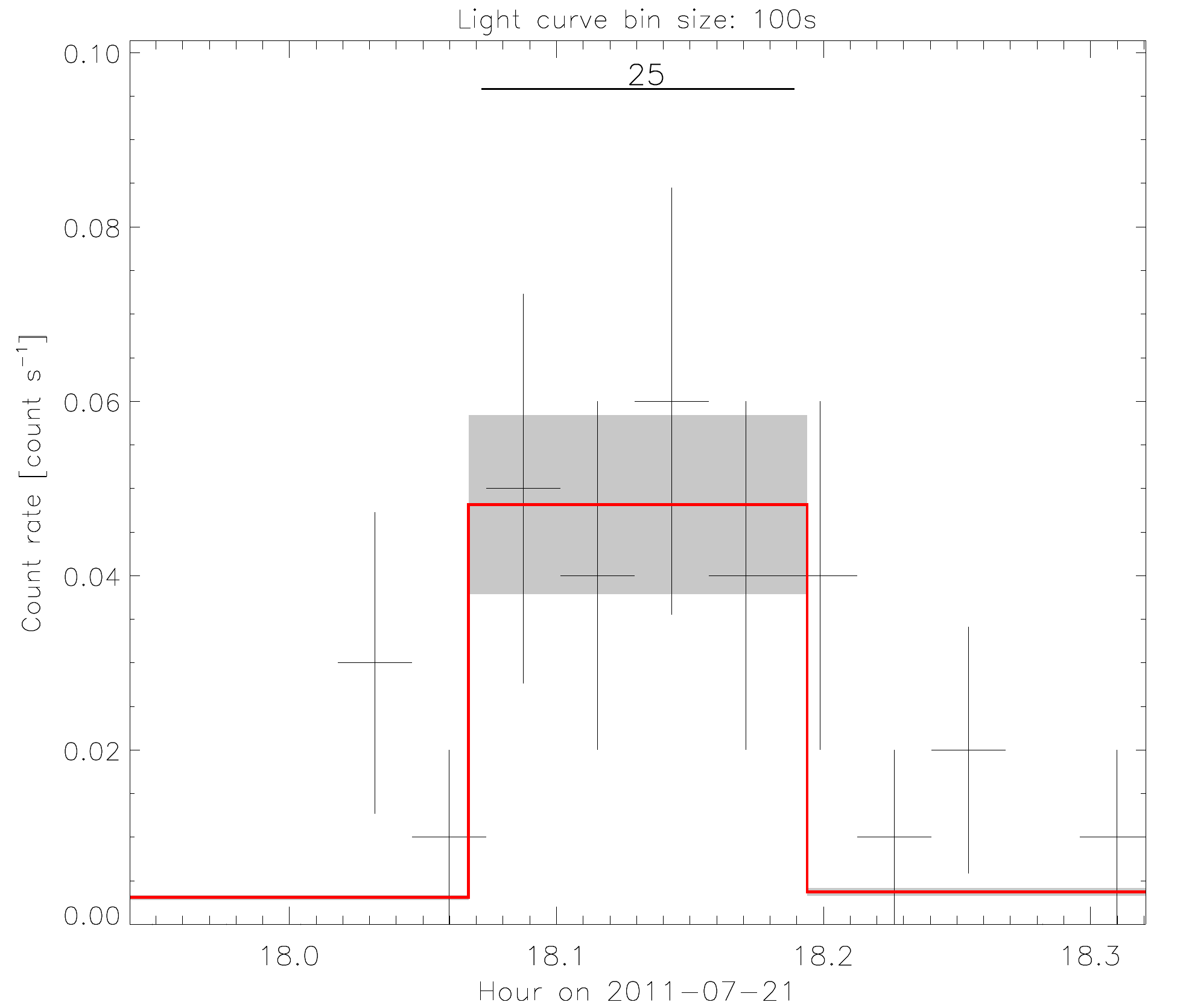}&
\includegraphics[width=0.20\textwidth,angle=90]{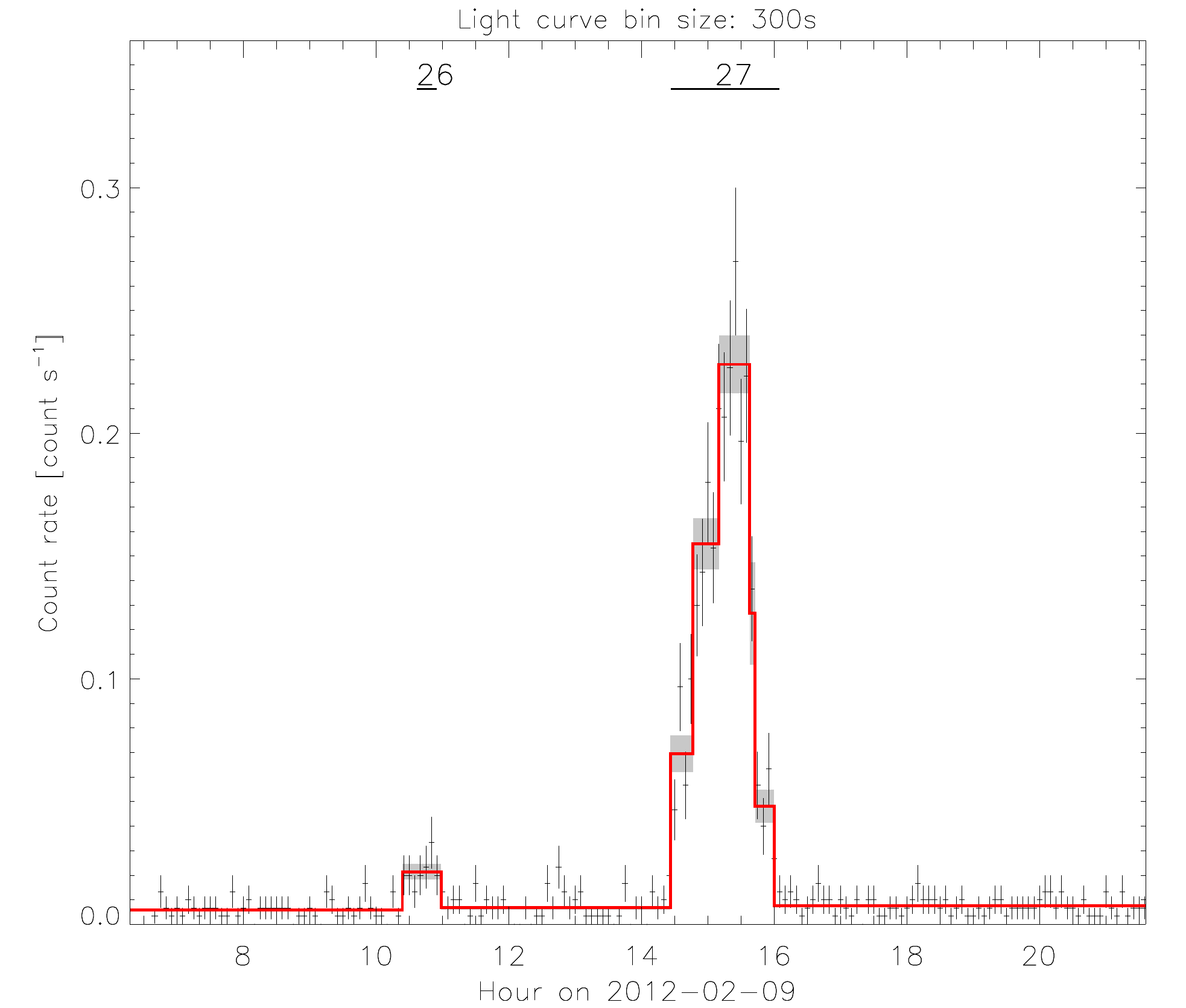}&
\includegraphics[width=0.20\textwidth,angle=90]{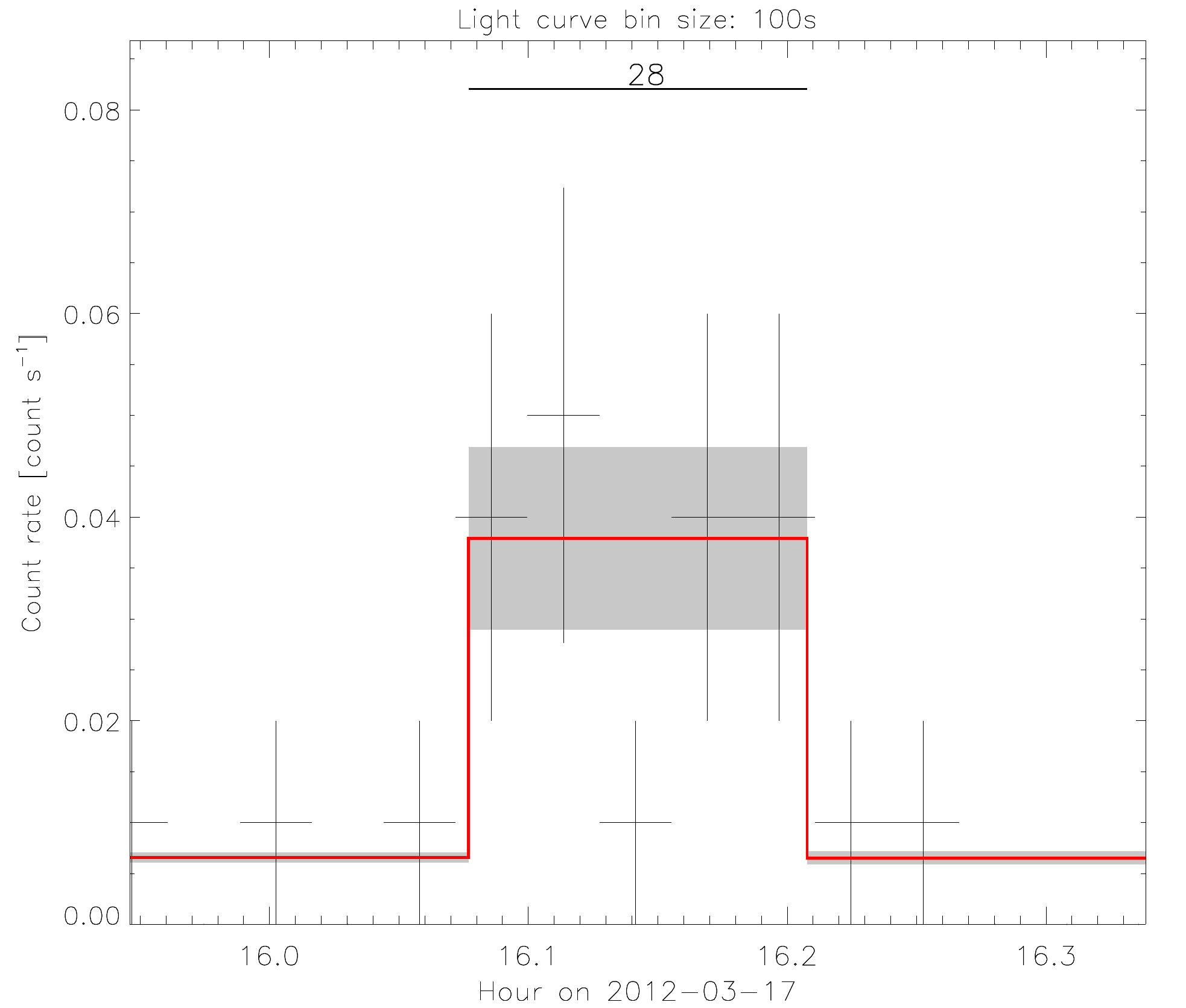}&
\includegraphics[width=0.20\textwidth,angle=90]{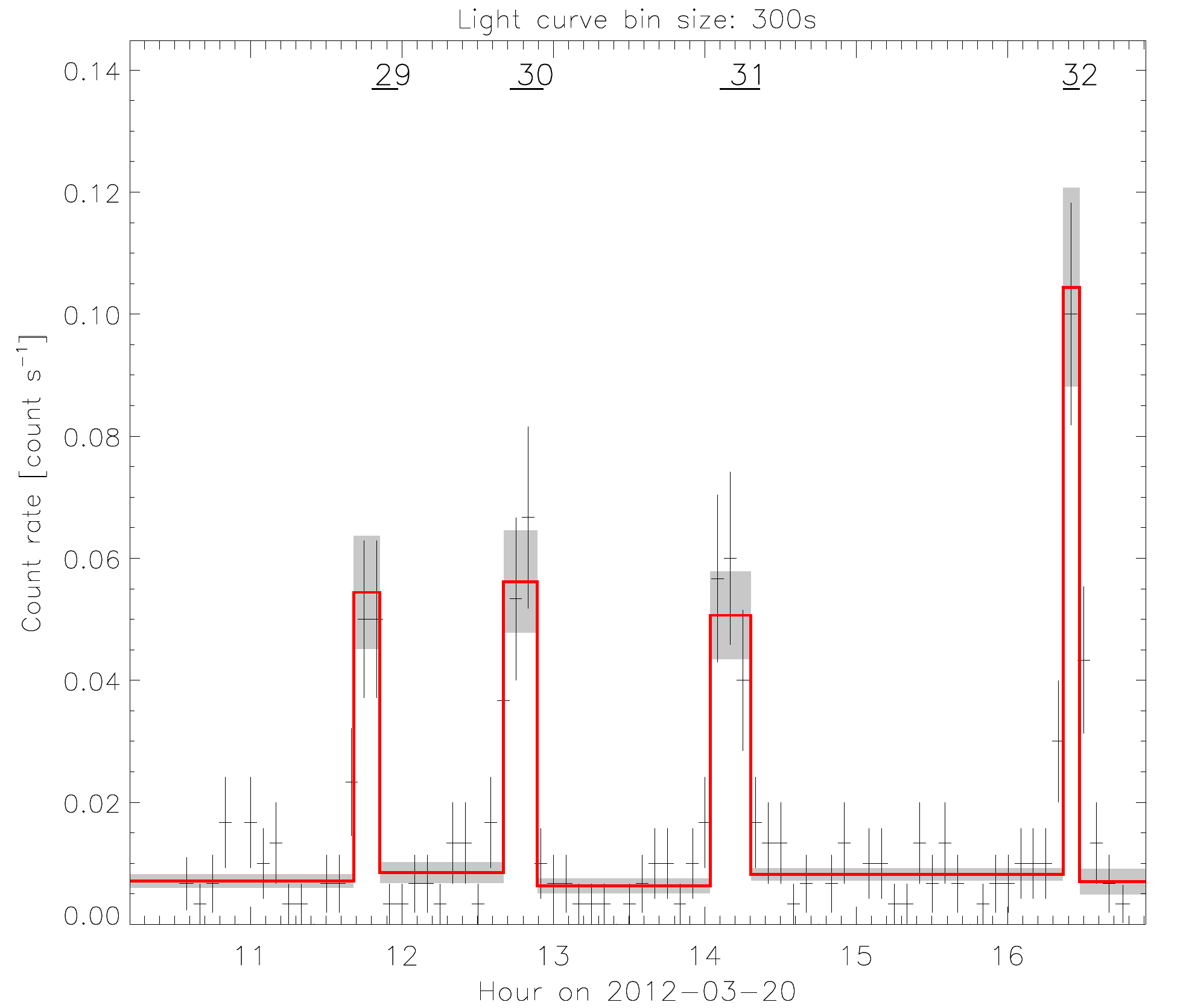}\\
\includegraphics[width=0.20\textwidth,angle=90]{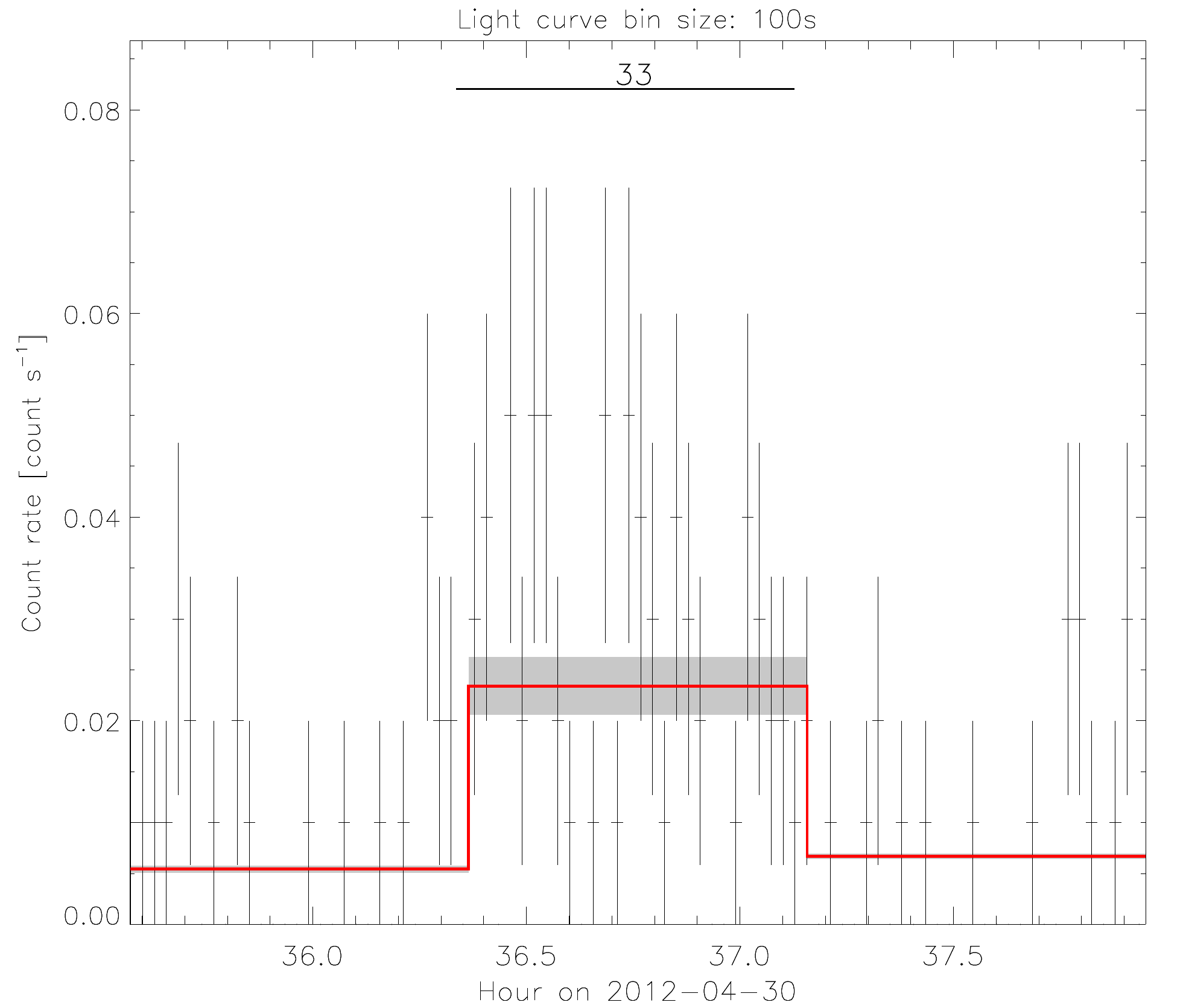}&
\includegraphics[width=0.20\textwidth,angle=90]{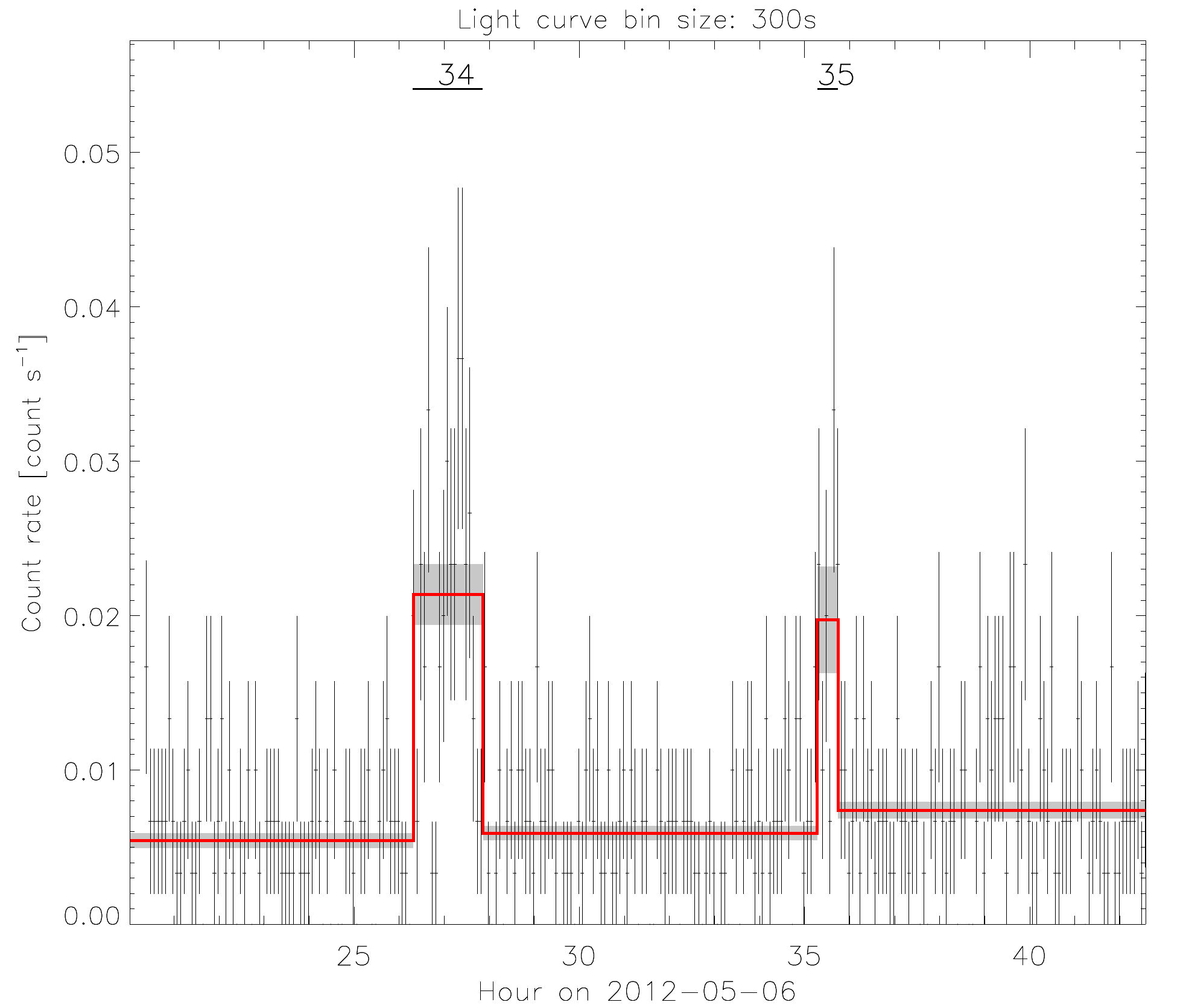}&
\includegraphics[width=0.20\textwidth,angle=90]{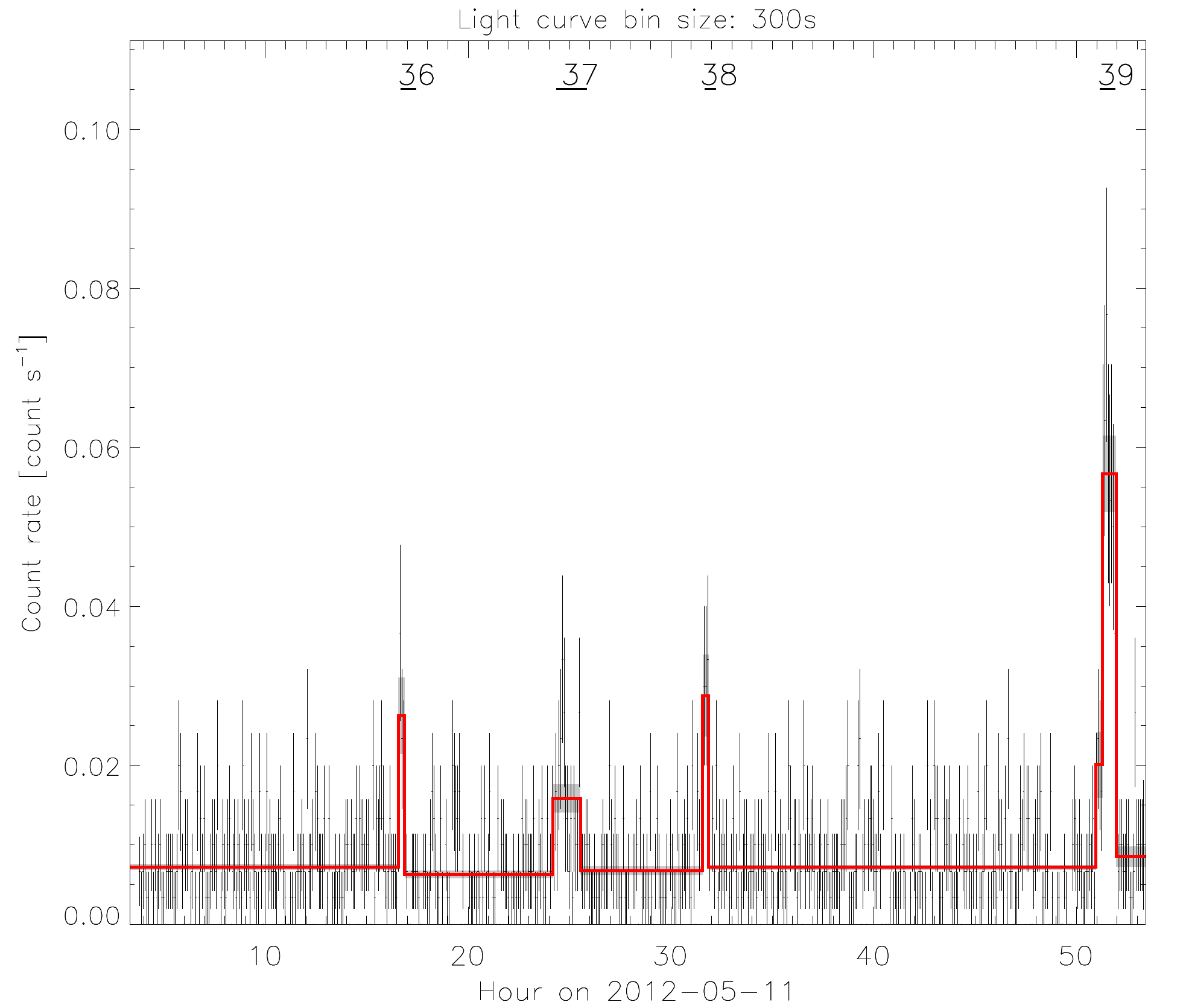}&
\includegraphics[width=0.20\textwidth,angle=90]{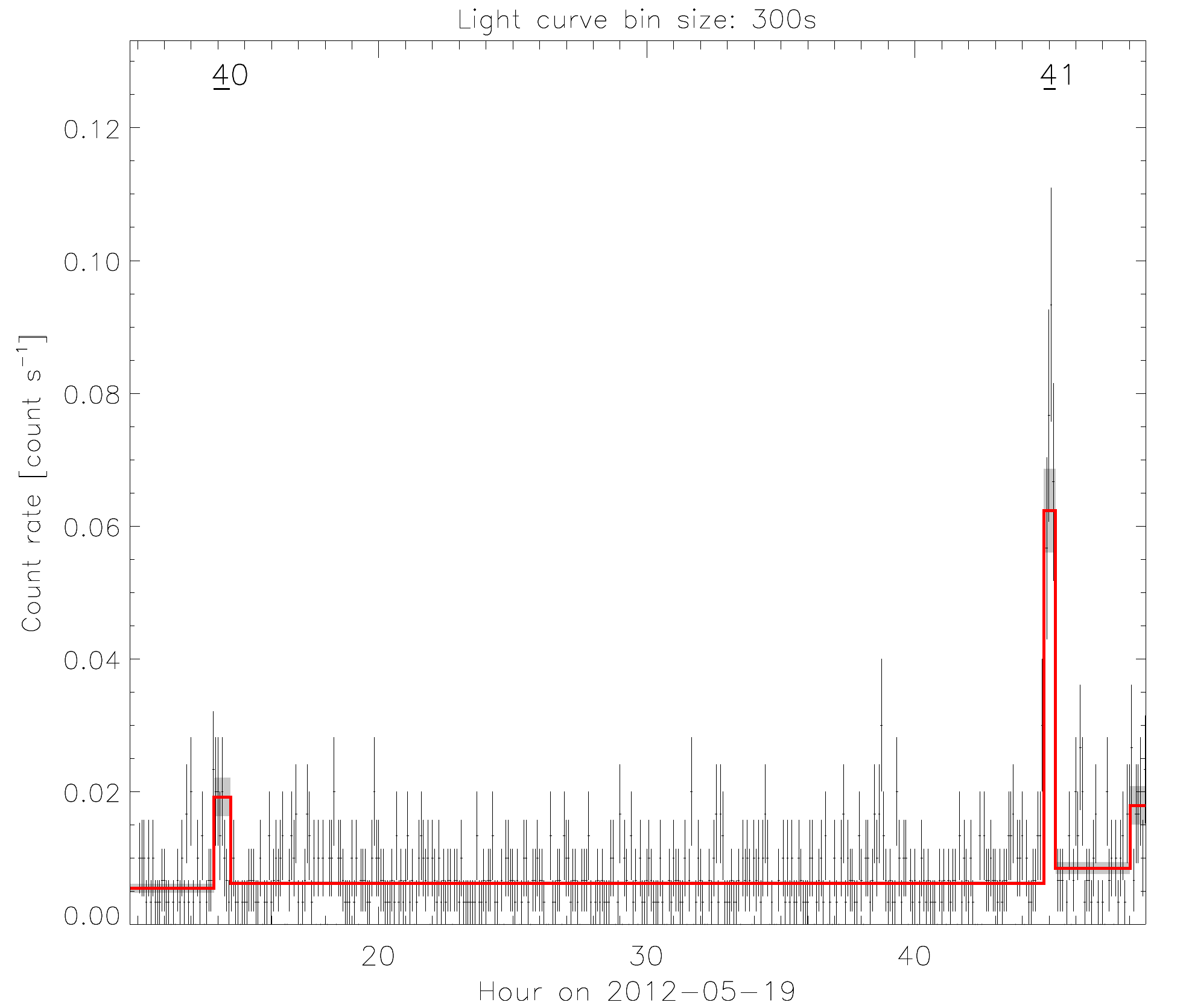}
\end{tabular}
\caption[Chandra flares detected using the Bayesian blocks algorithm with a false positive rate for the flare detection of 0.1\% from 1999 to 2015 October 21.]{Chandra flares detected using the Bayesian blocks algorithm with a false positive rate for the flare detection of 0.1\% from 1999 to 2015 October\ 21.
Each flare is labeled with the index corresponding to the flare number in Tables~\ref{table:chandra1}.
See caption of Fig.~\ref{fig:xmm_flare} for details.}
\label{fig:chandra_flare1}
\end{figure*}

\setcounter{figure}{1}
\begin{figure*}[!ht]
\centering
\begin{tabular}{@{}ccccc@{}}
\includegraphics[width=0.20\textwidth,angle=90]{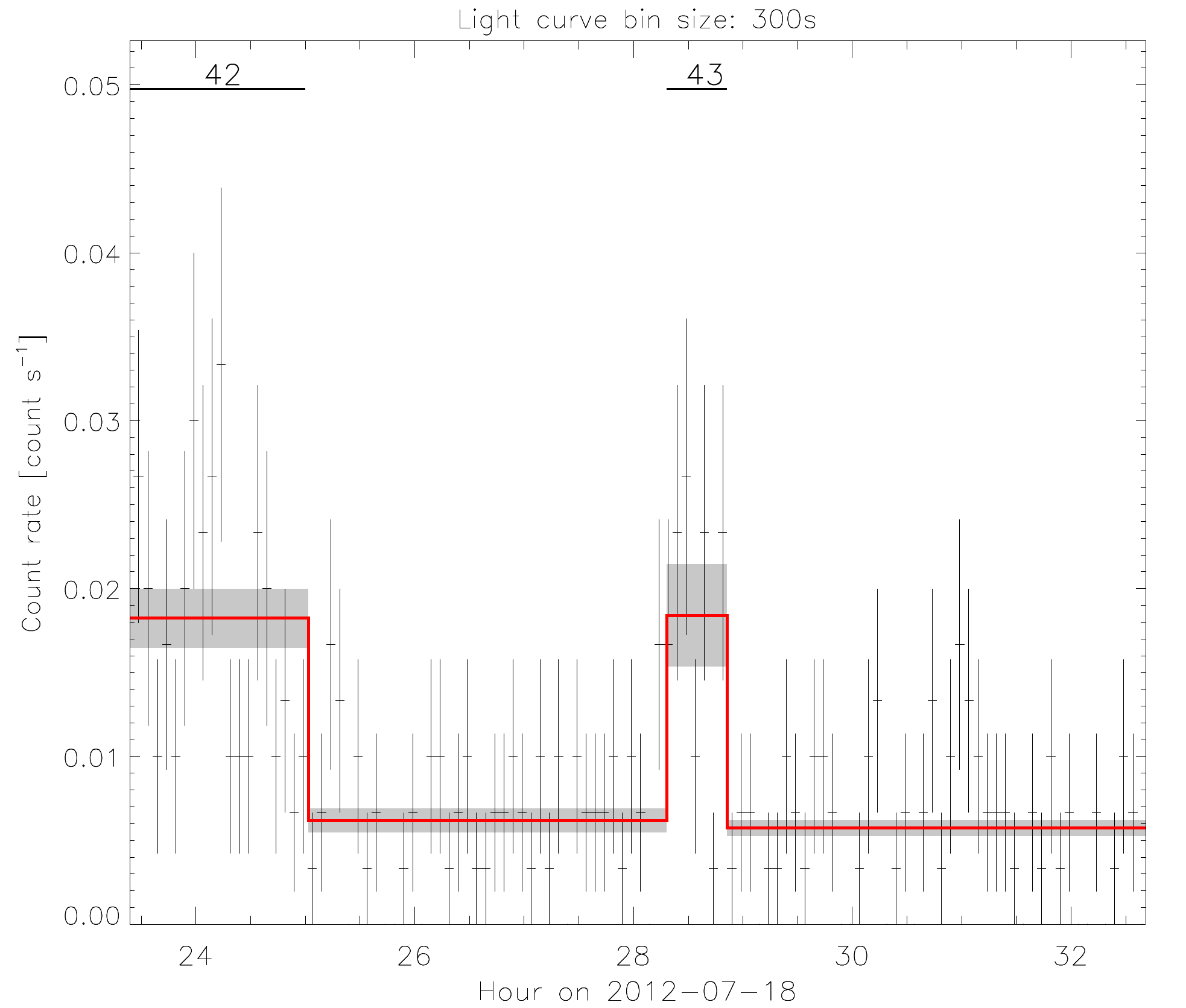}&
\includegraphics[width=0.20\textwidth,angle=90]{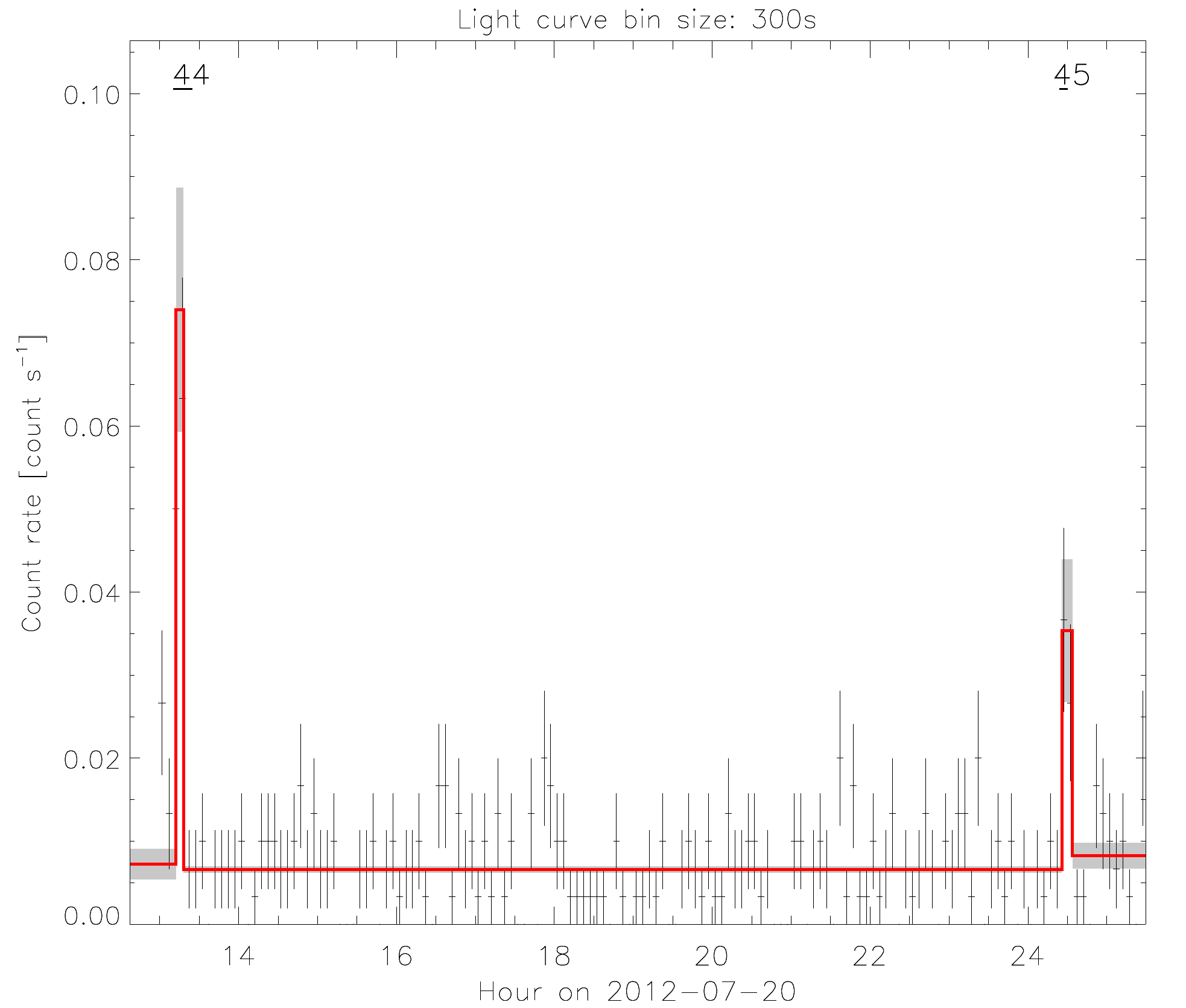}&
\includegraphics[width=0.20\textwidth,angle=90]{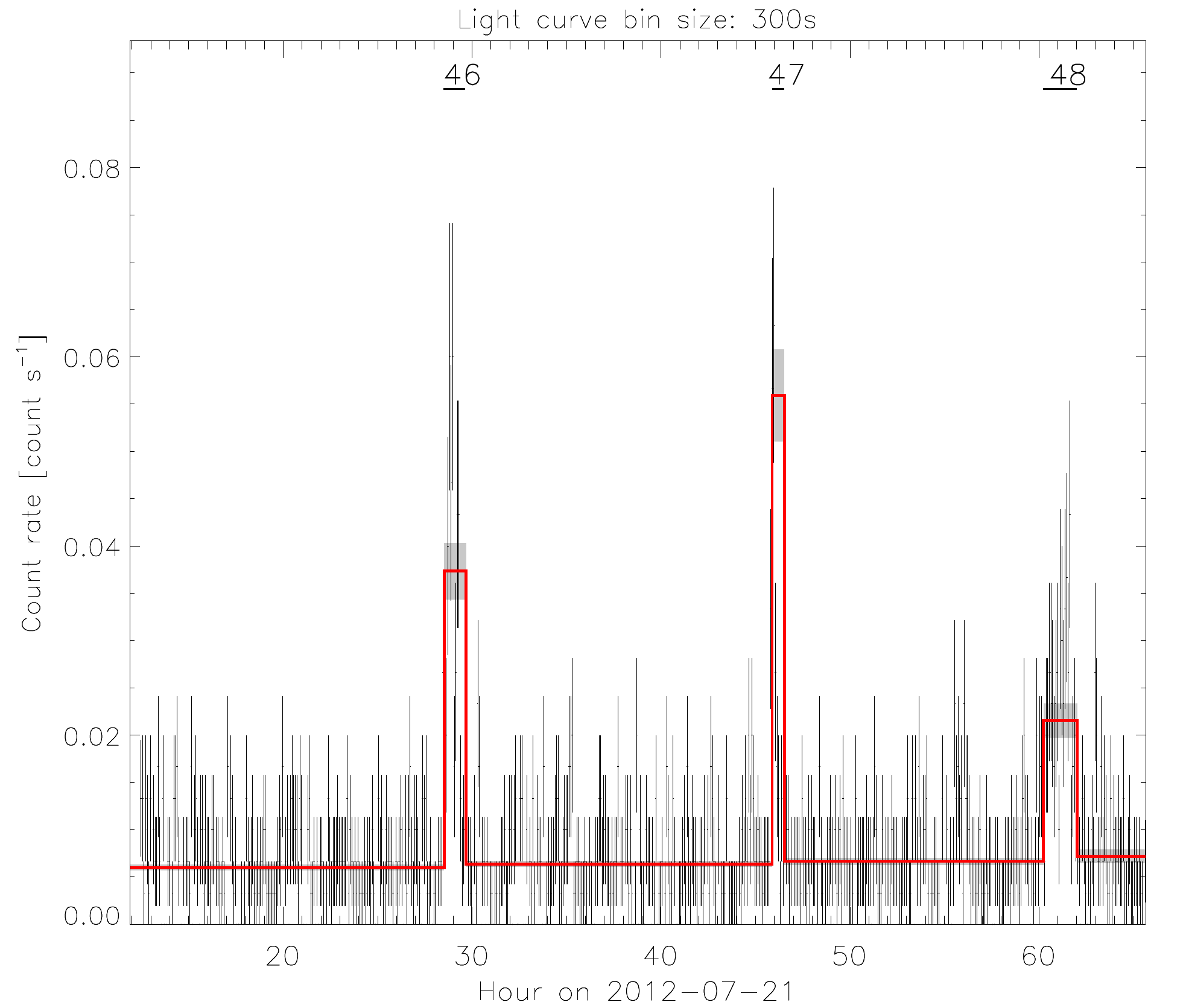}&
\includegraphics[width=0.20\textwidth,angle=90]{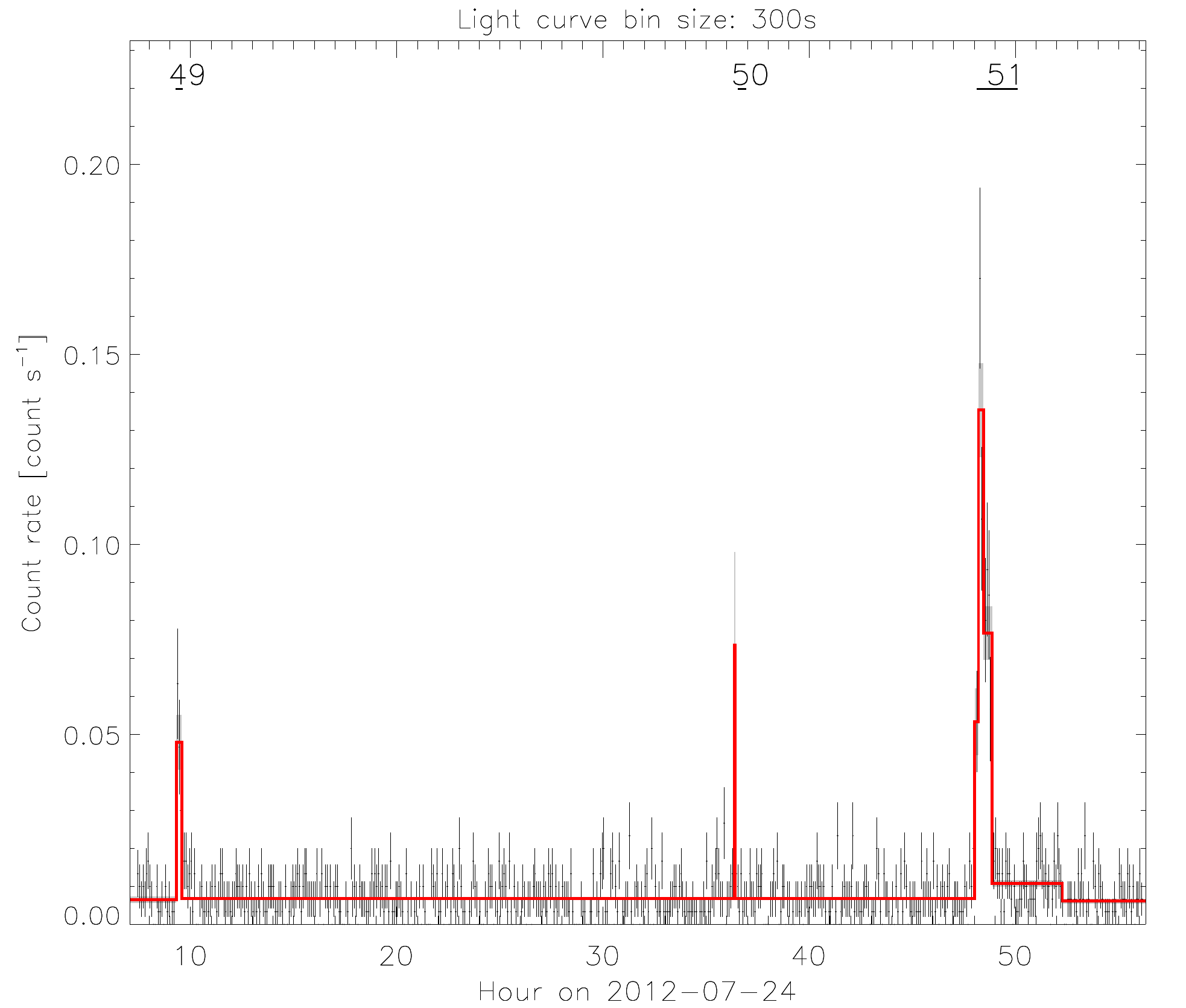}\\
\includegraphics[width=0.20\textwidth,angle=90]{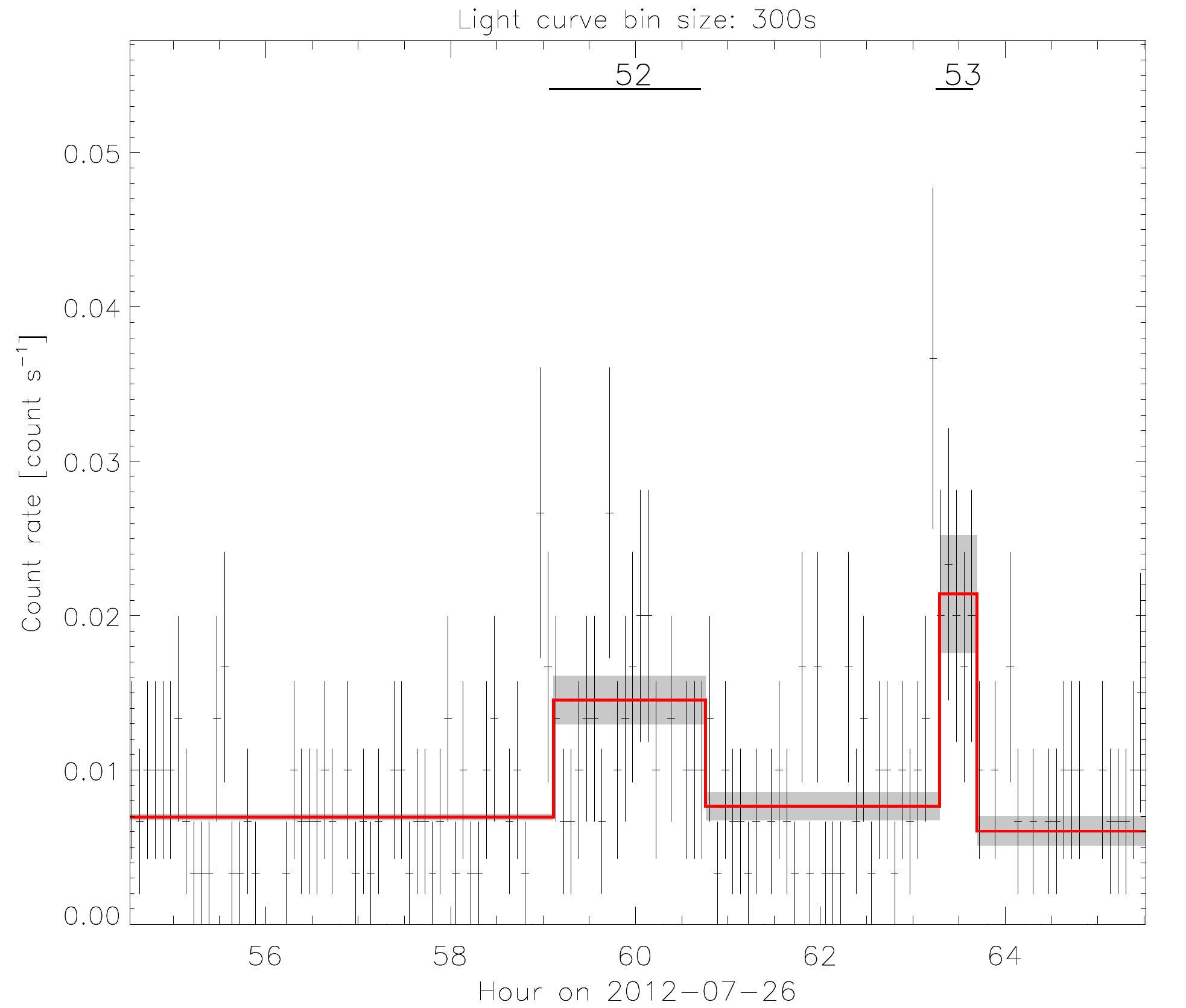}&
\includegraphics[width=0.20\textwidth,angle=90]{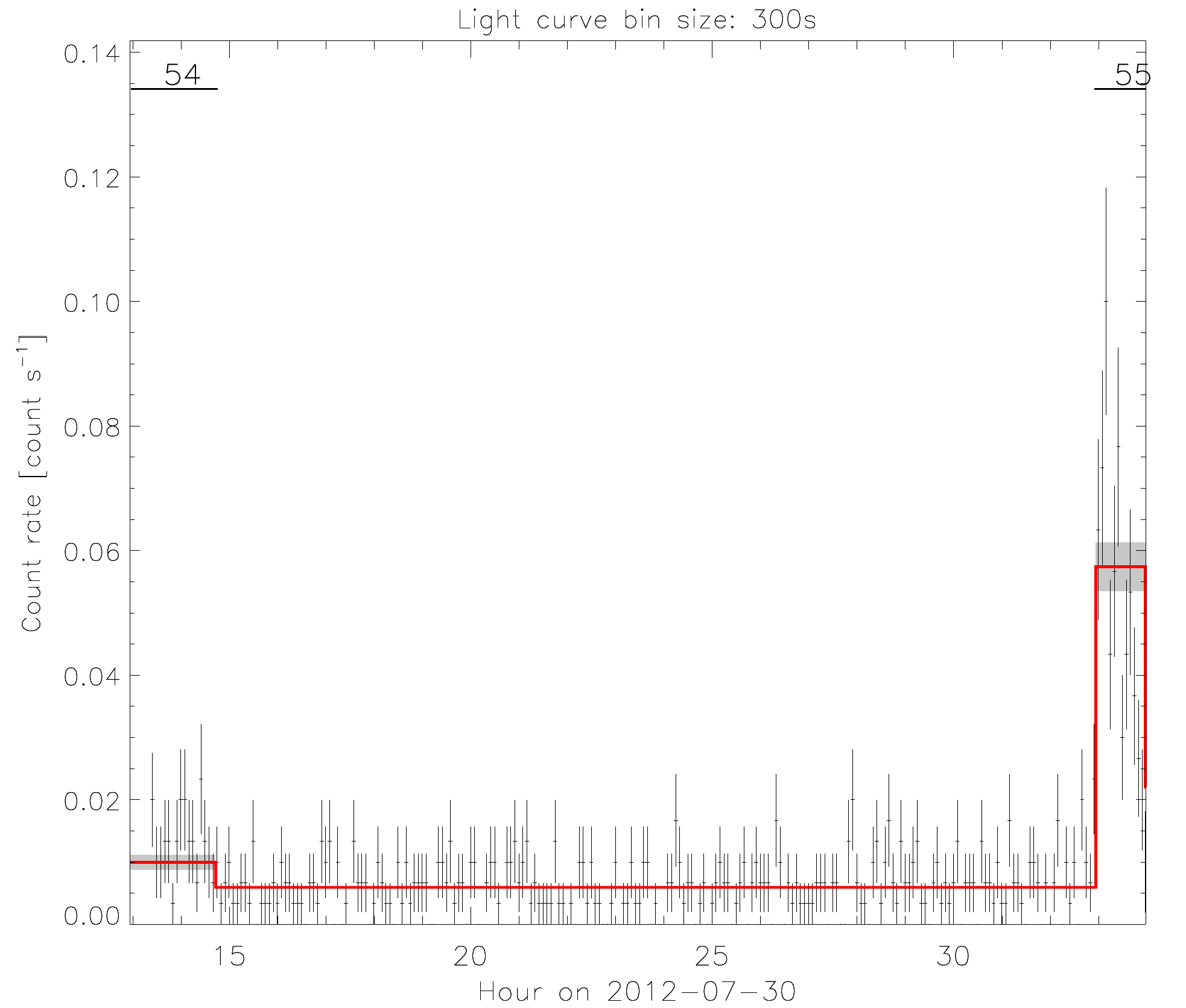}&
\includegraphics[width=0.20\textwidth,angle=90]{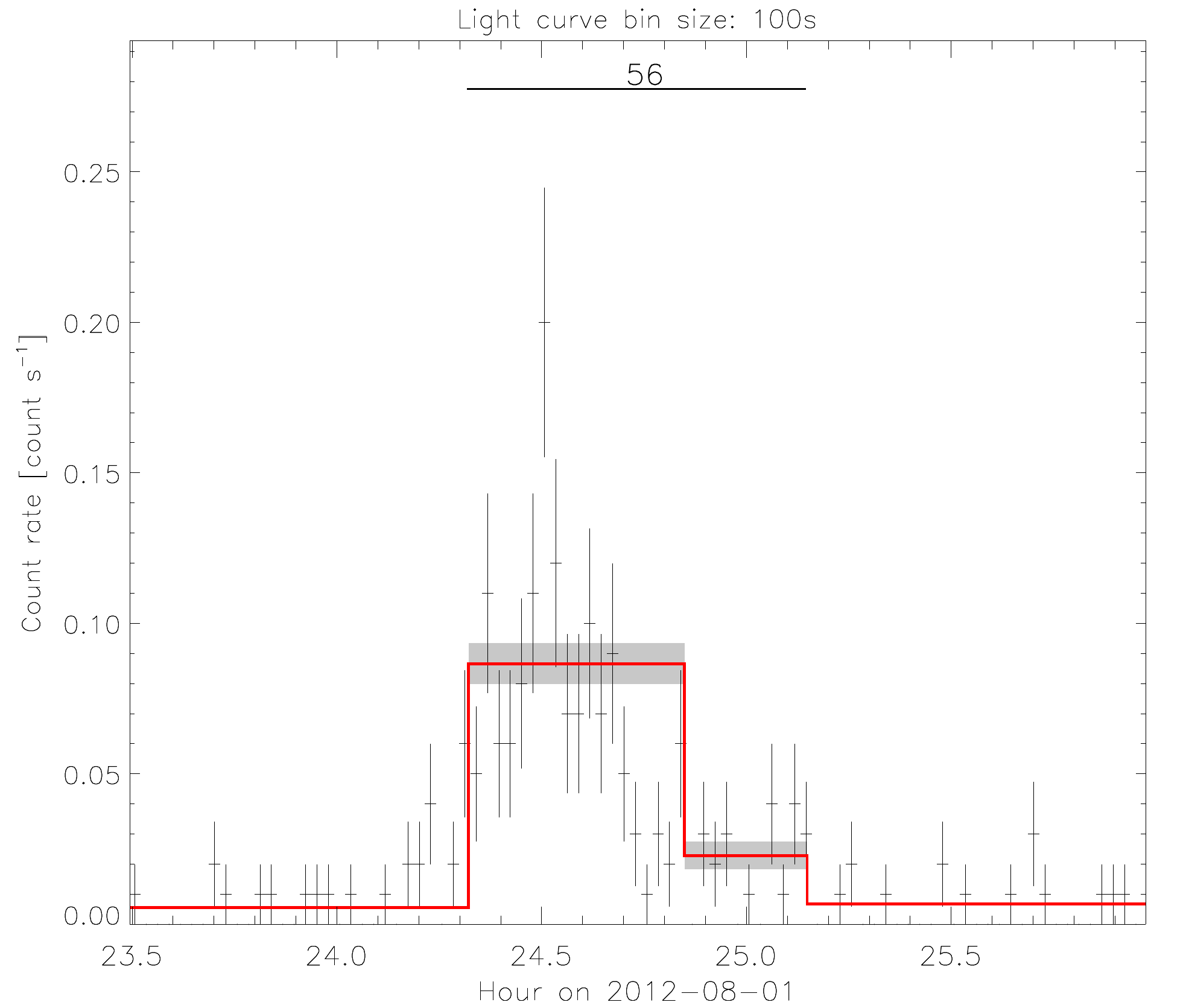}&
\includegraphics[width=0.20\textwidth,angle=90]{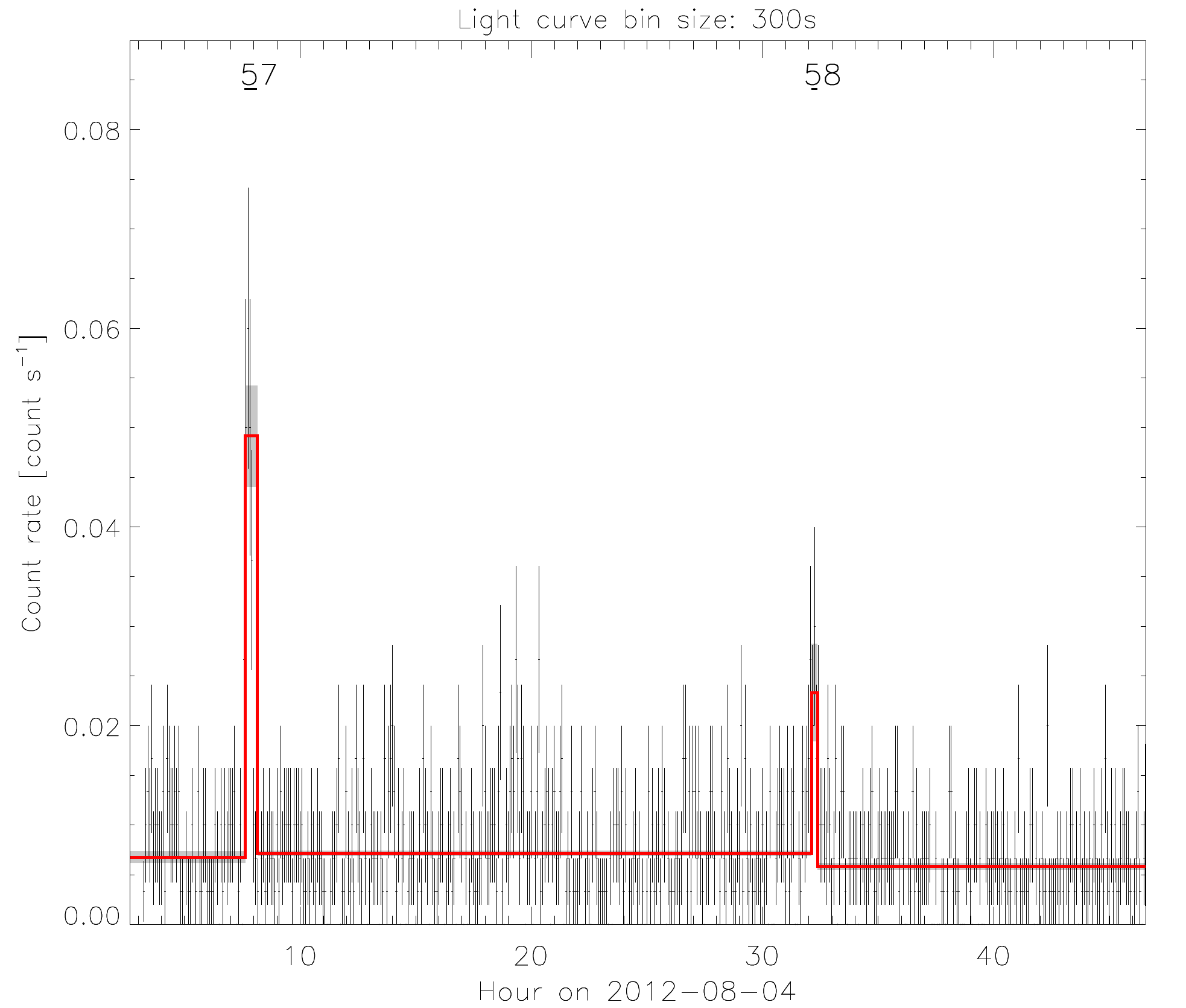}\\
\includegraphics[width=0.20\textwidth,angle=90]{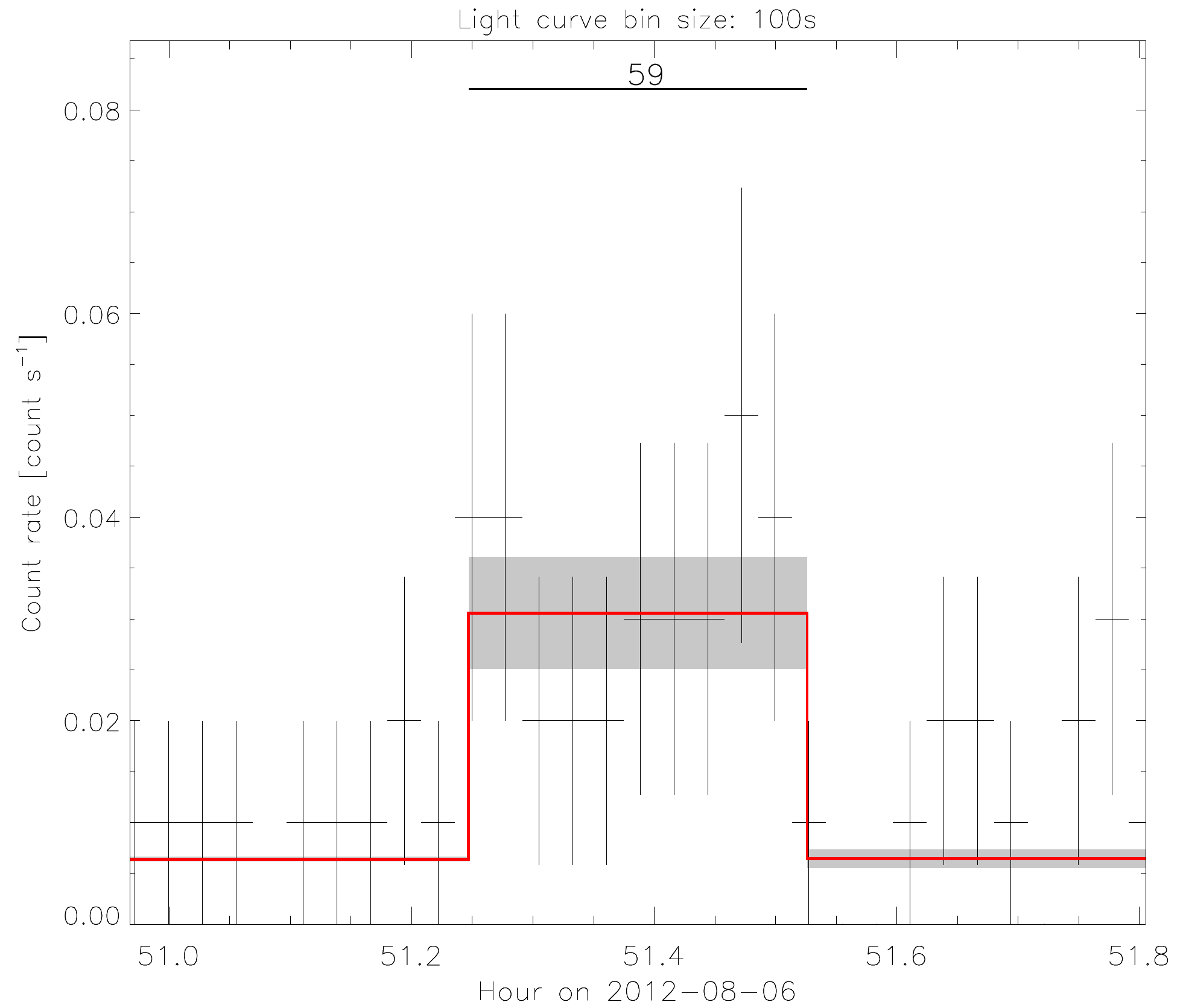}&
\includegraphics[width=0.20\textwidth,angle=90]{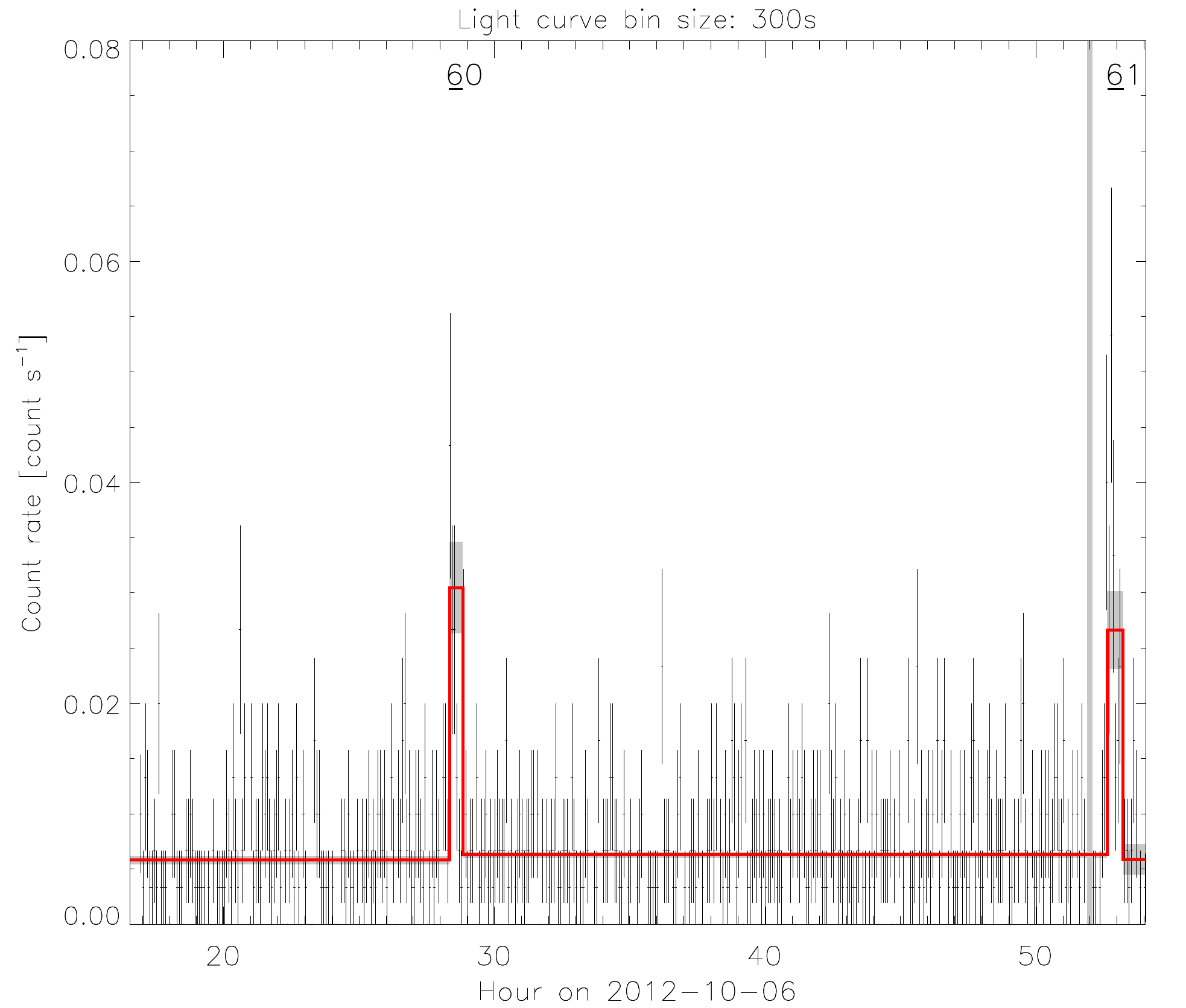}&
\includegraphics[width=0.20\textwidth,angle=90]{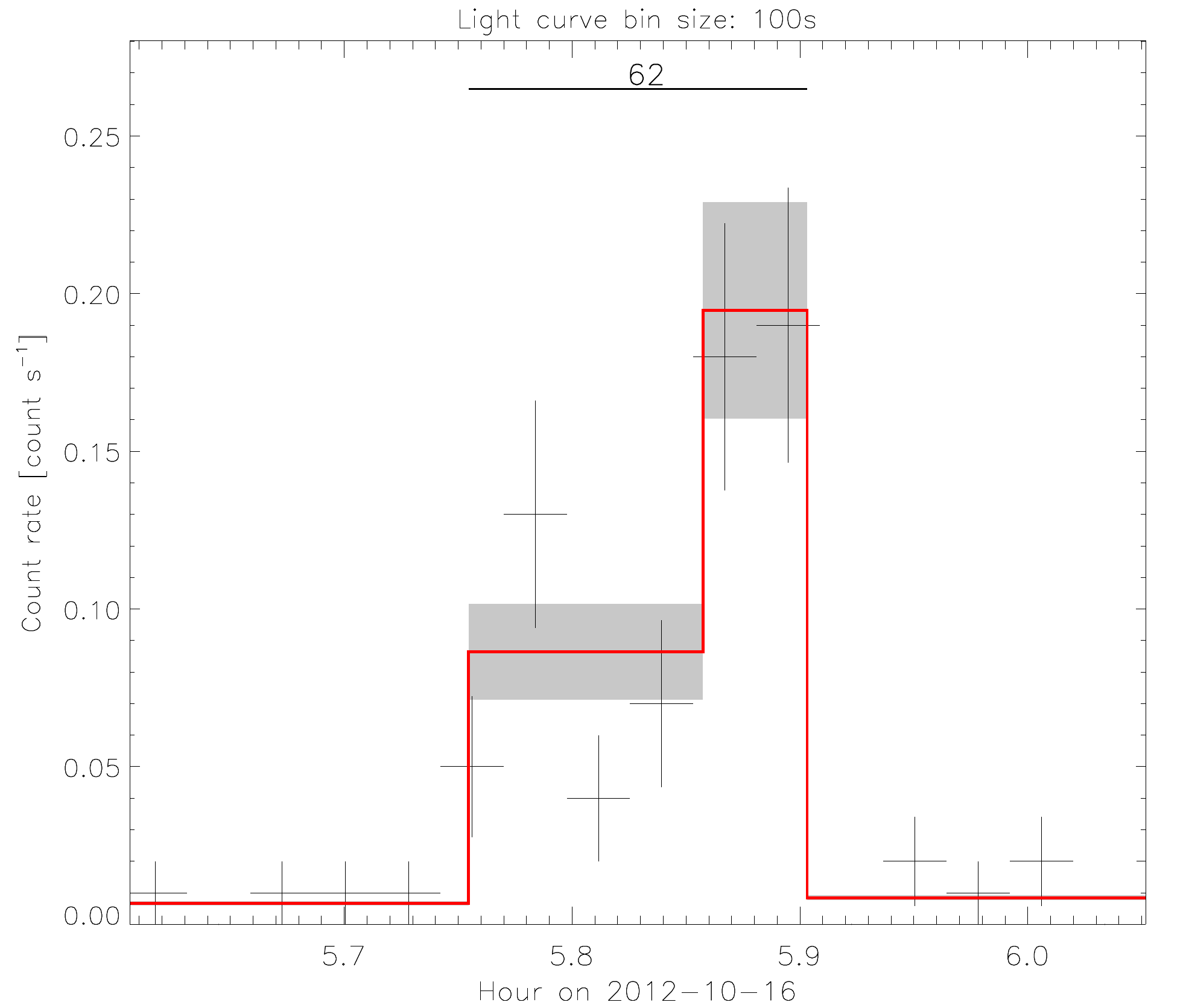}&
\includegraphics[width=0.20\textwidth,angle=90]{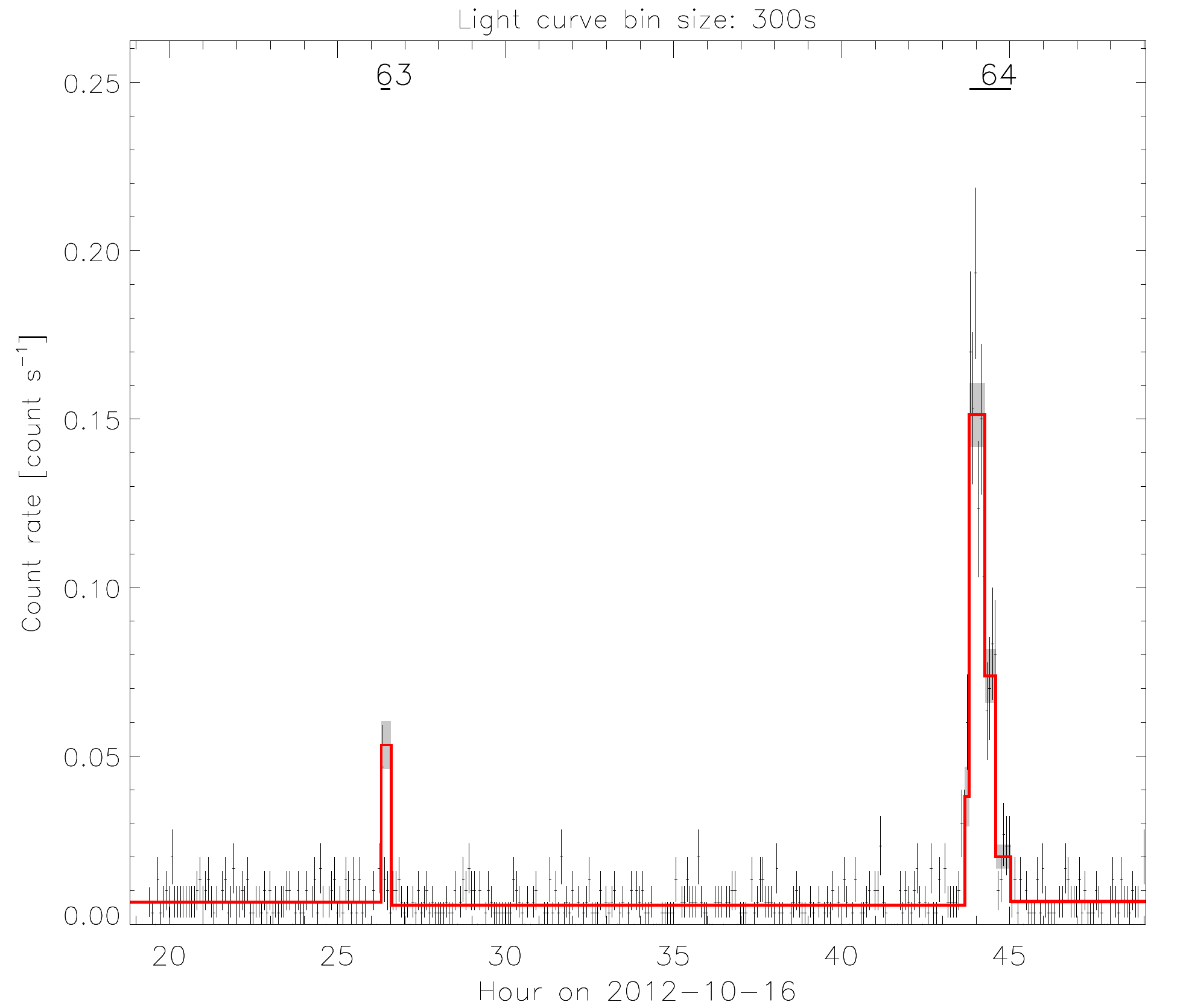}\\
\includegraphics[width=0.20\textwidth,angle=90]{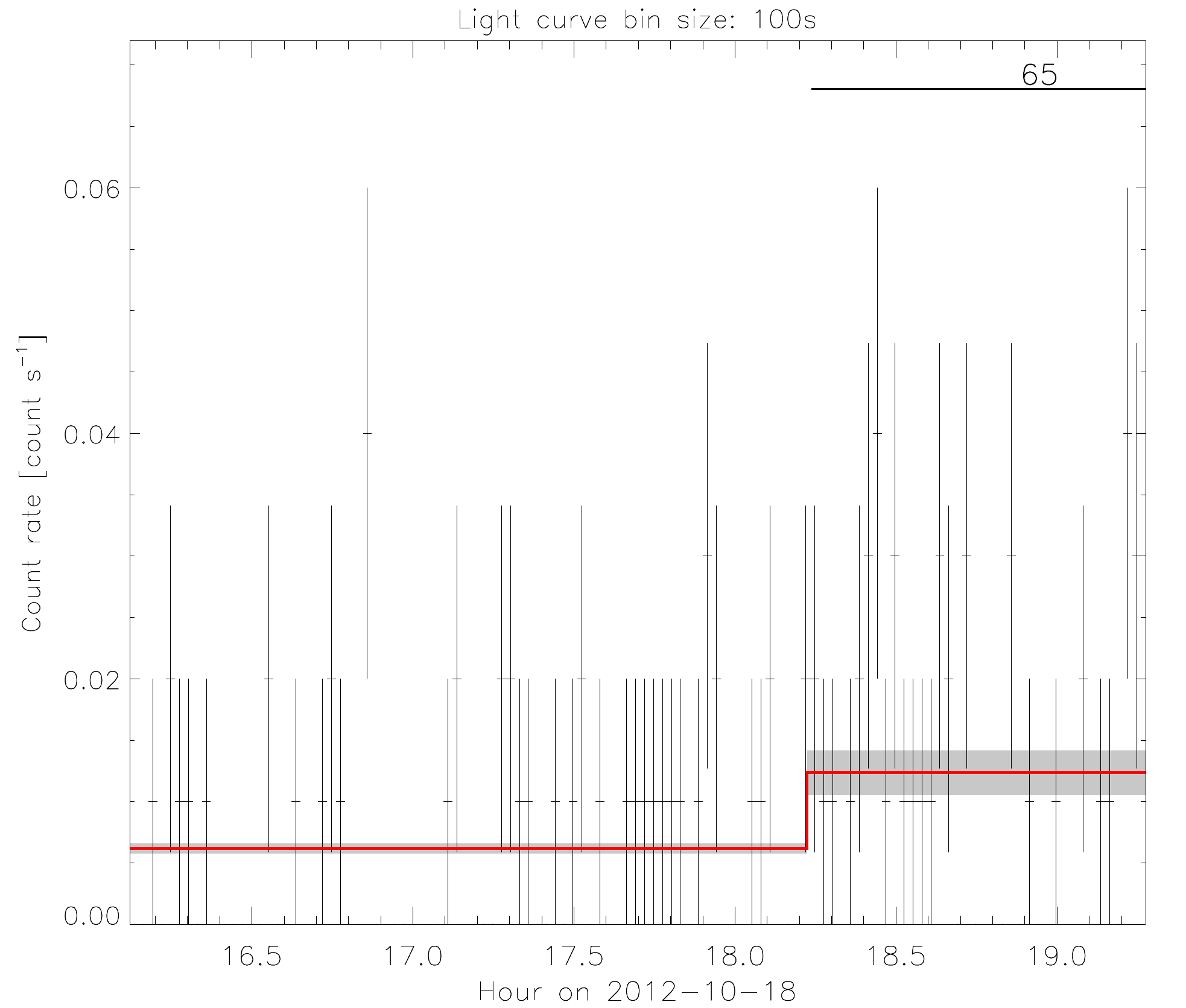}&
\includegraphics[width=0.20\textwidth,angle=90]{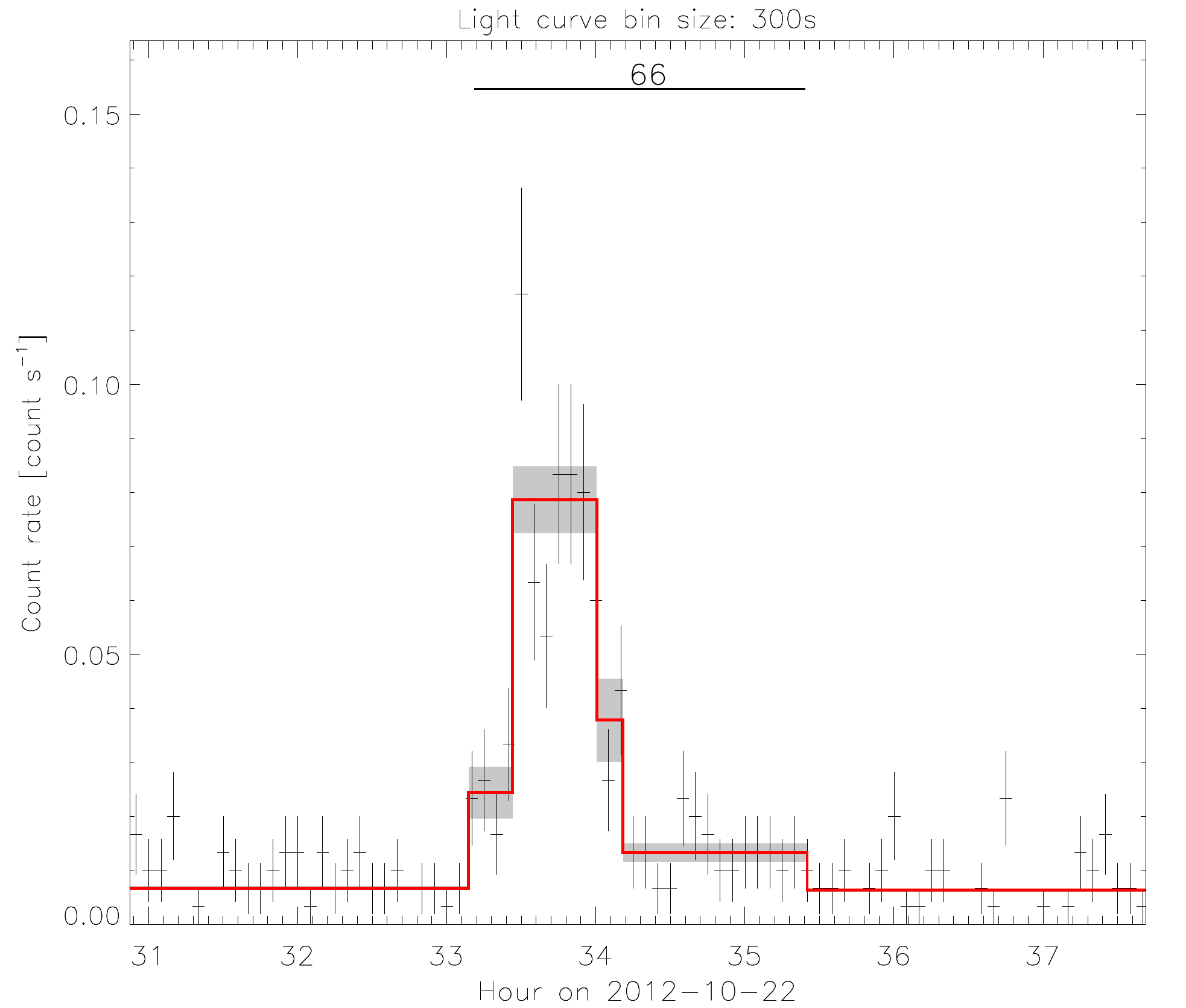}&
\includegraphics[width=0.20\textwidth,angle=90]{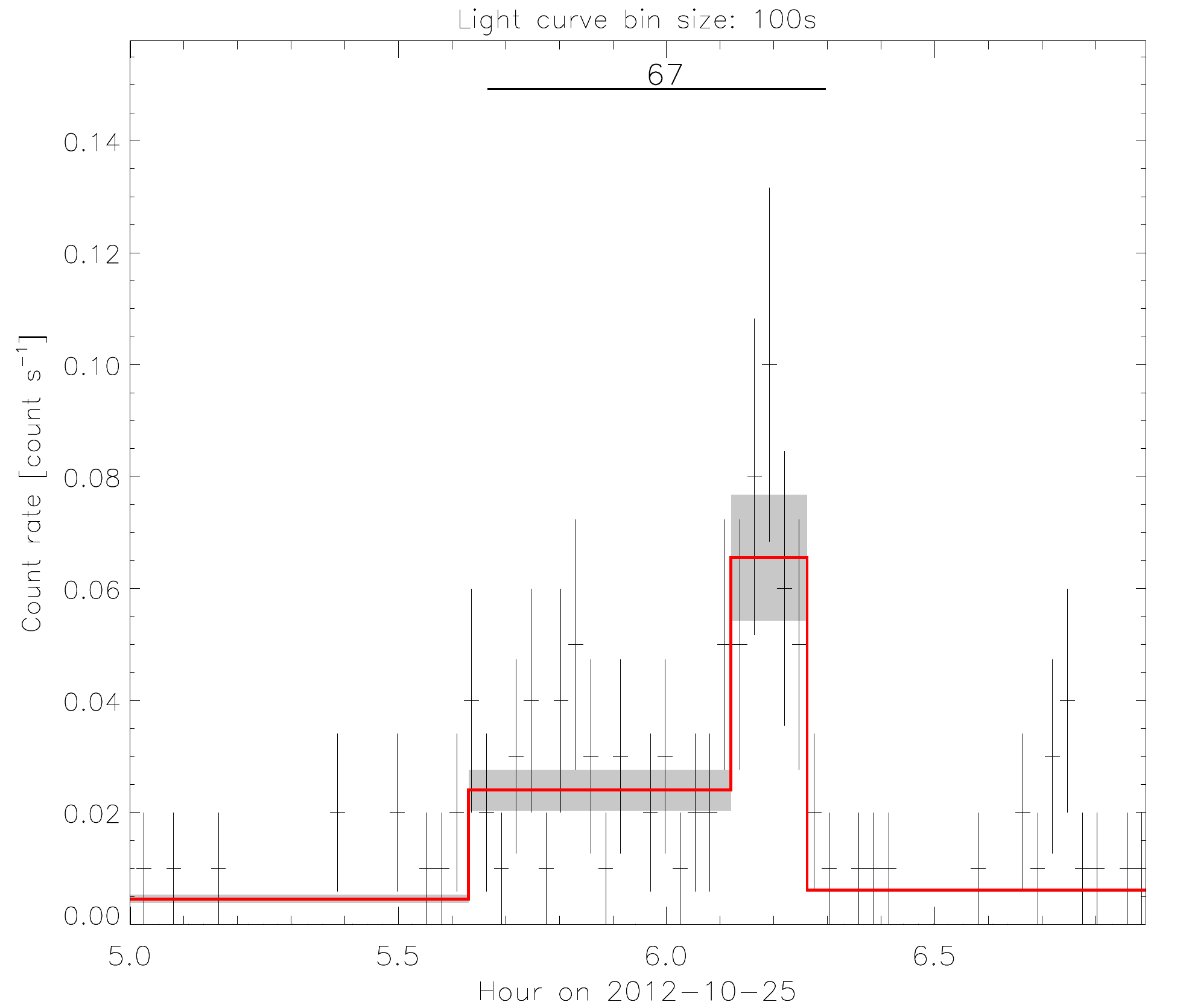}&
\includegraphics[width=0.20\textwidth,angle=90]{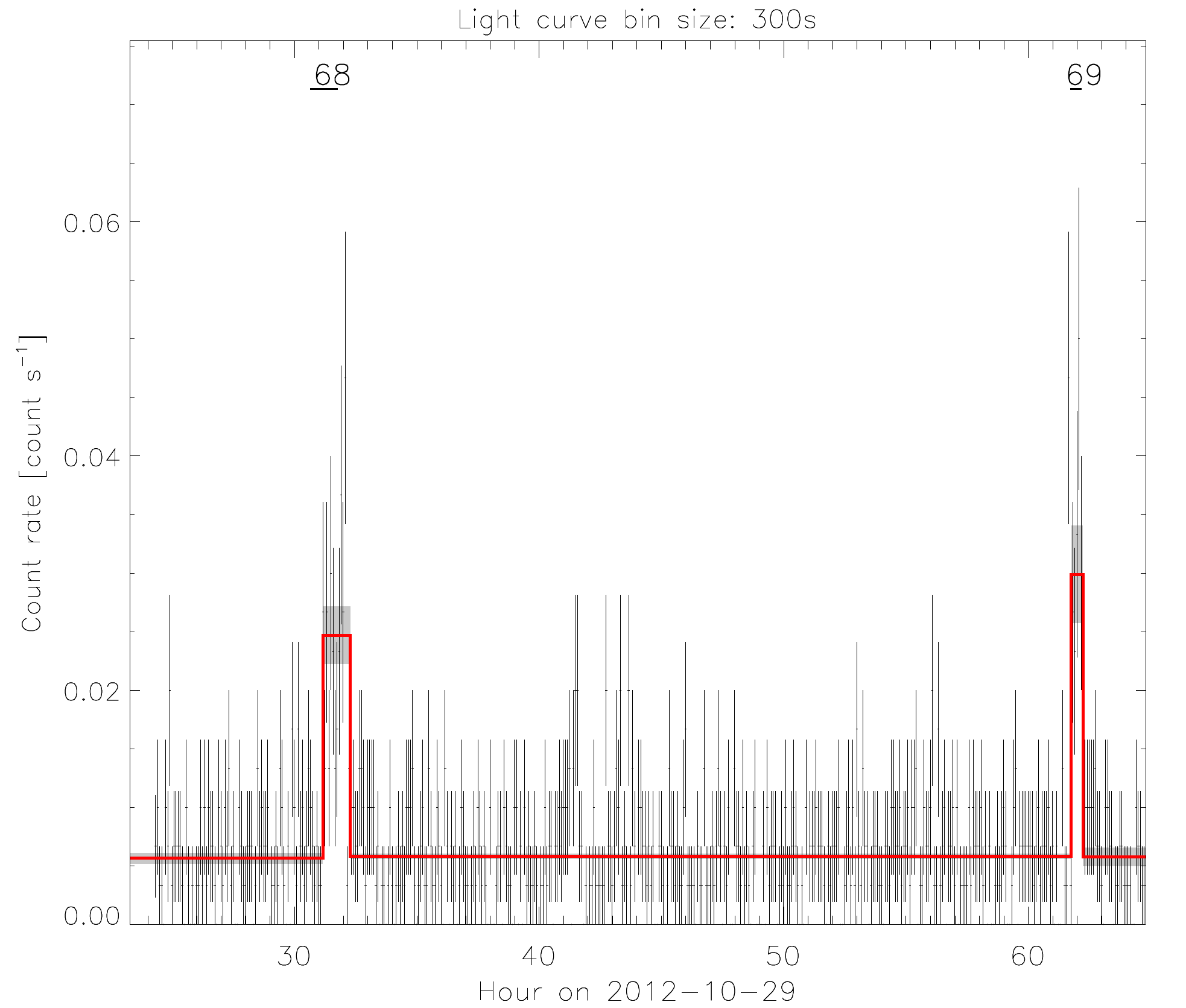}\\
\includegraphics[width=0.20\textwidth,angle=90]{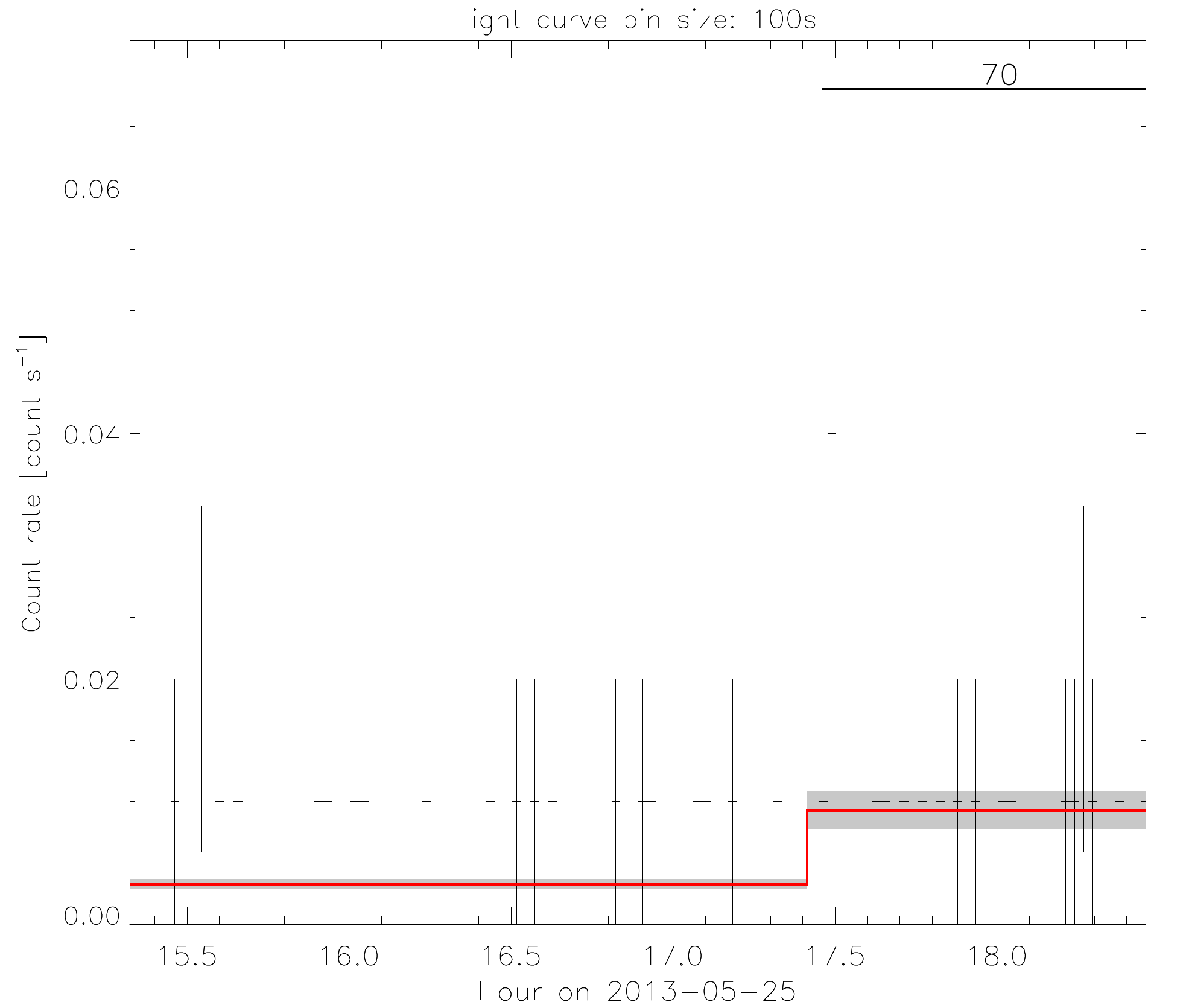}&
\includegraphics[width=0.20\textwidth,angle=90]{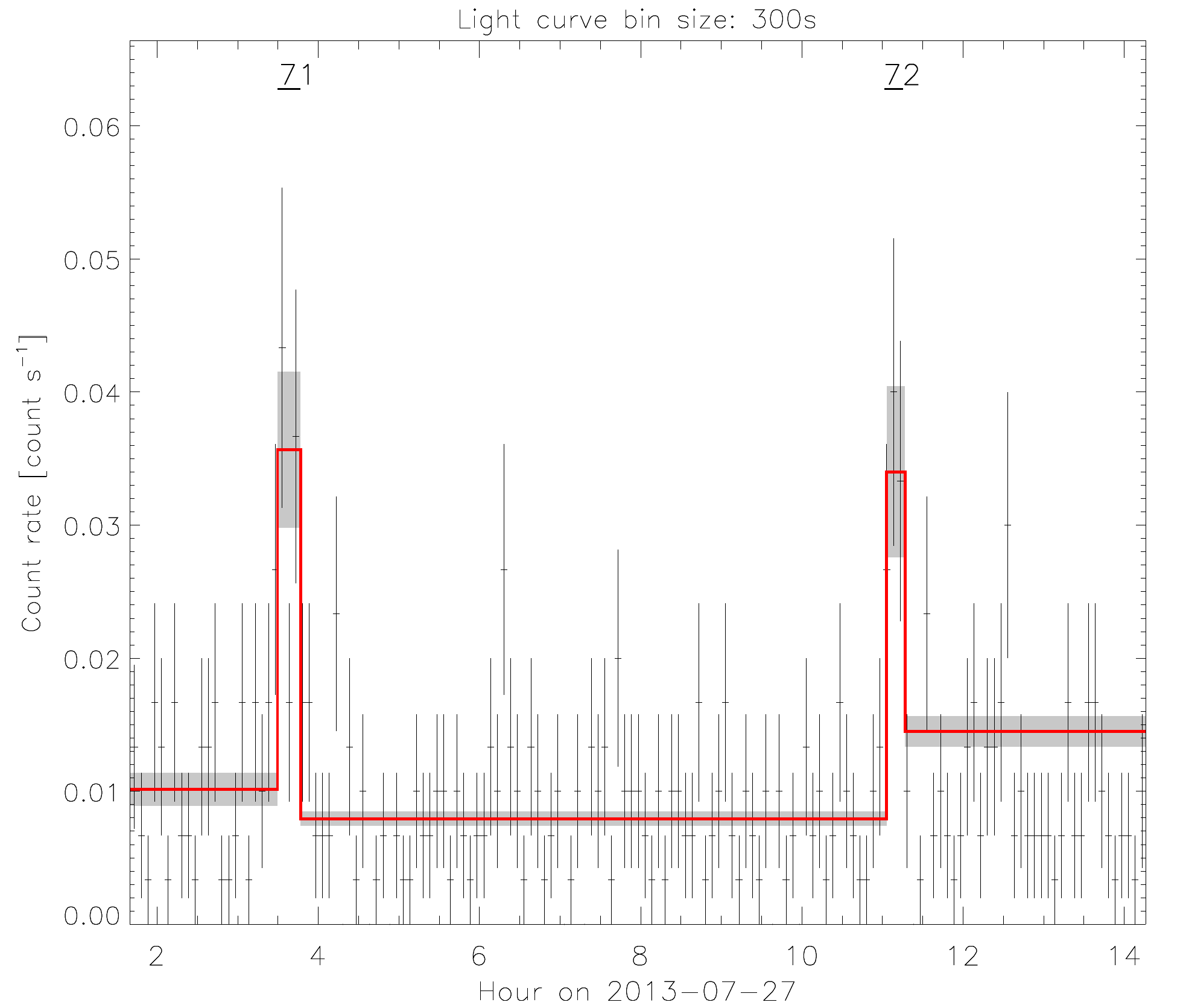}&
\includegraphics[width=0.20\textwidth,angle=90]{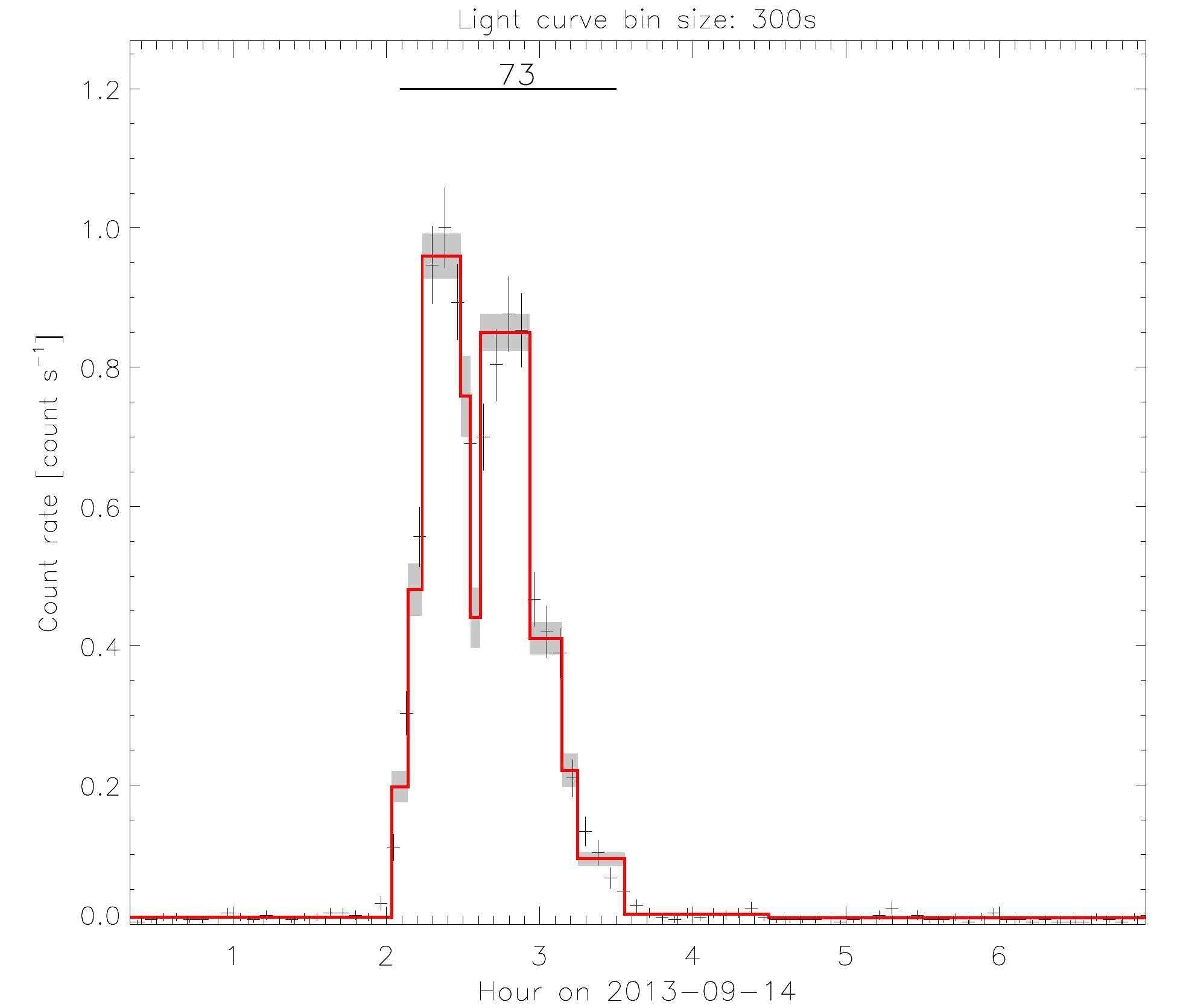}&
\includegraphics[width=0.20\textwidth,angle=90]{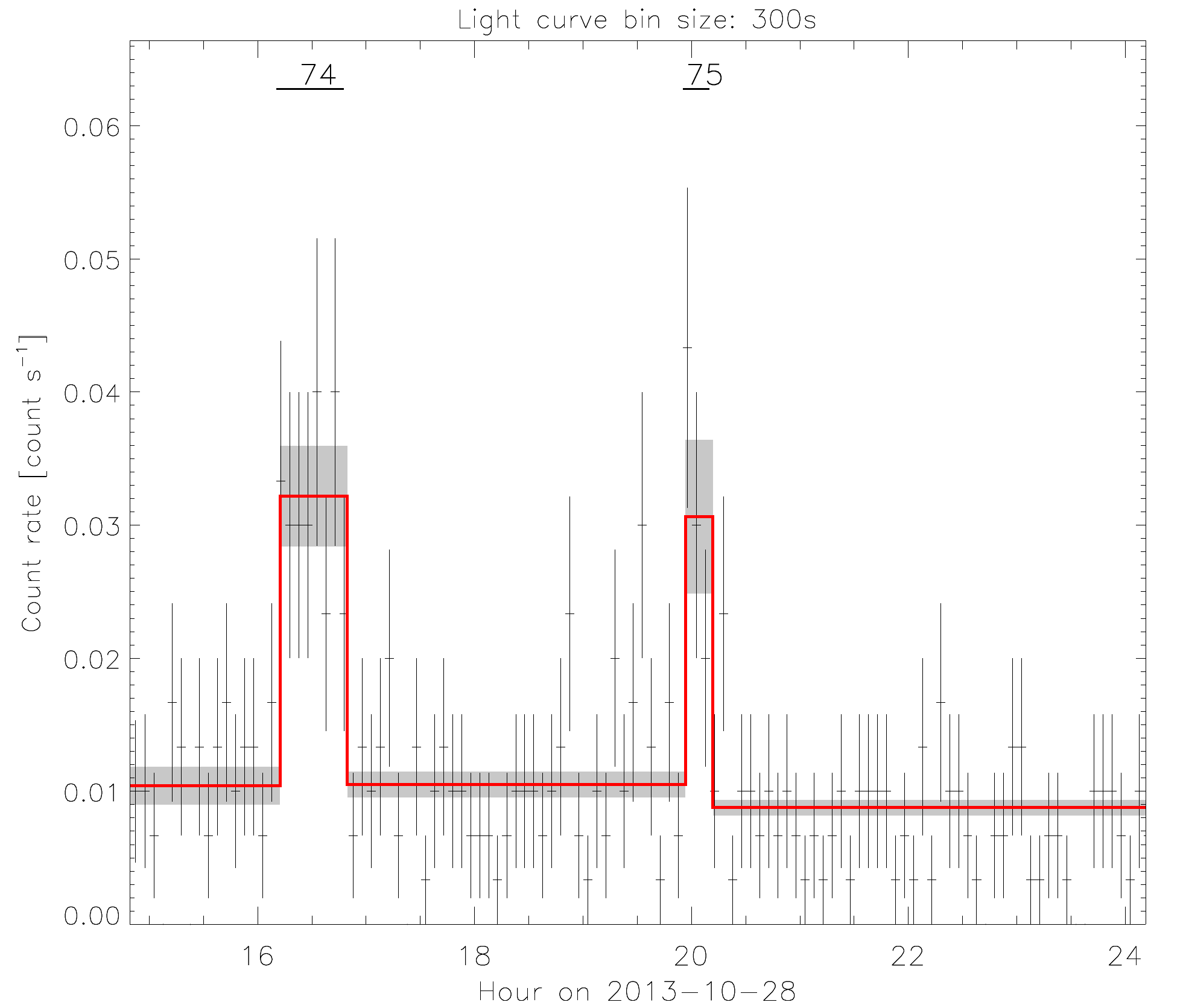}\\
\includegraphics[width=0.20\textwidth,angle=90]{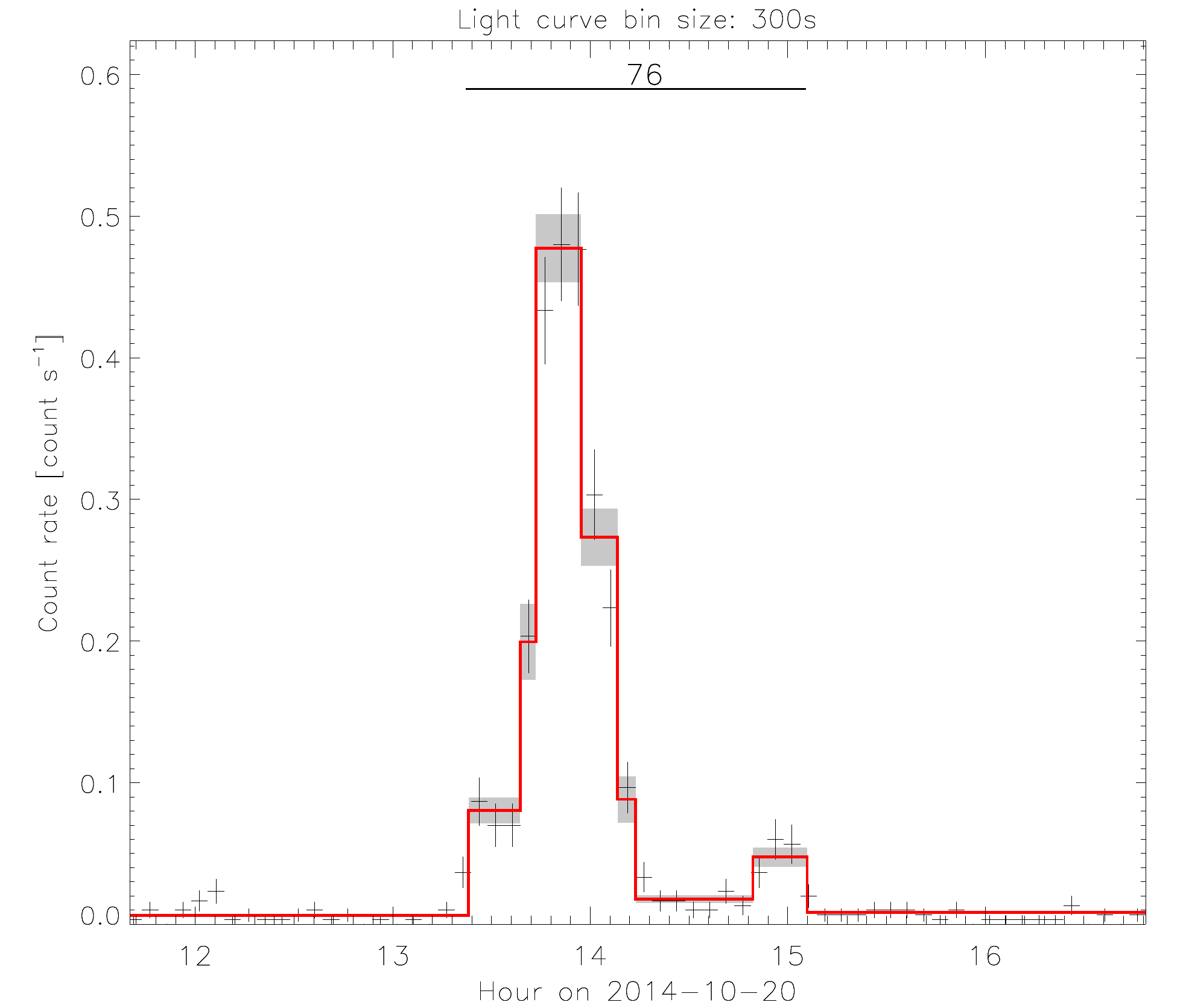}&
\includegraphics[width=0.20\textwidth,angle=90]{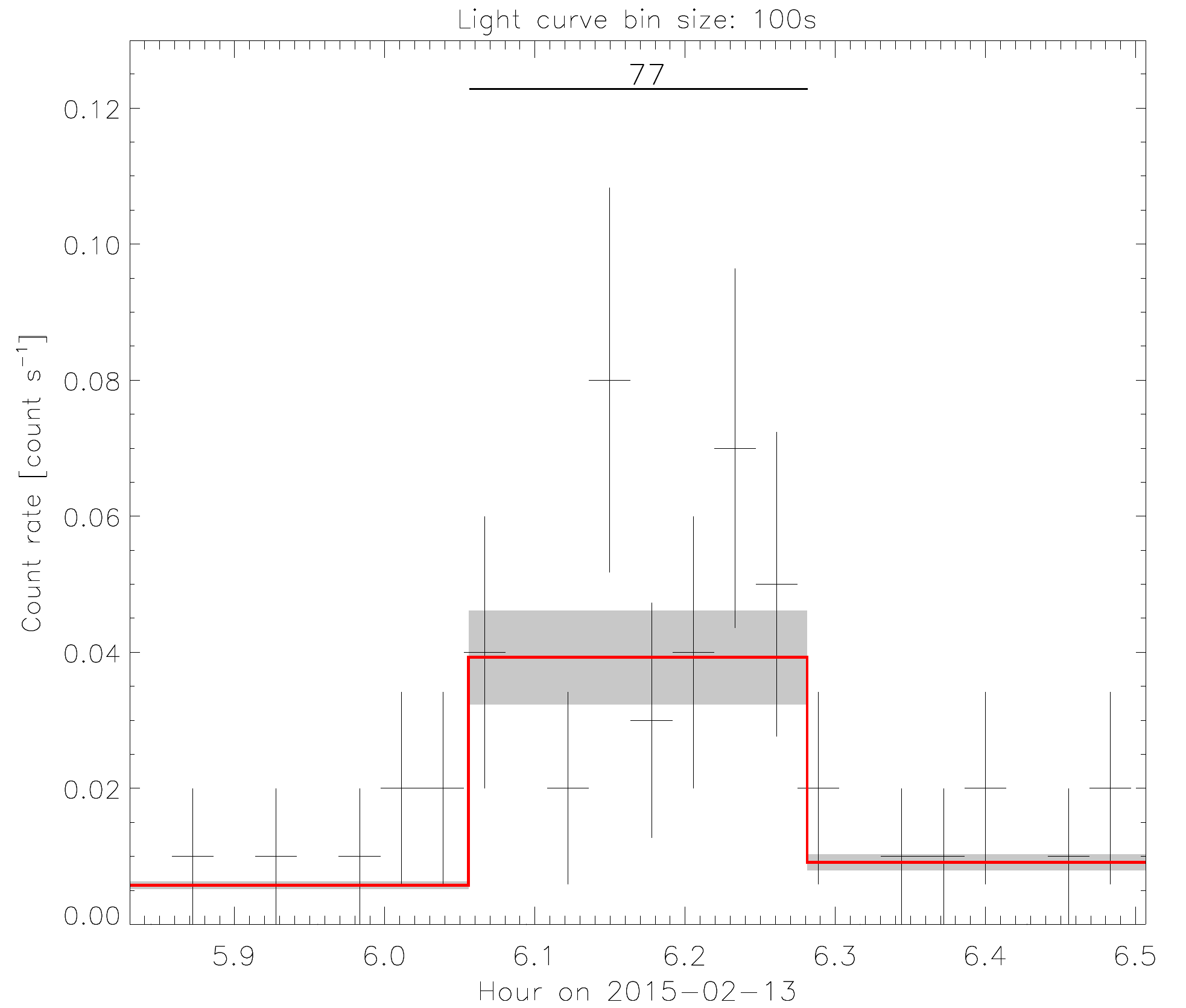}&
\includegraphics[width=0.20\textwidth,angle=90]{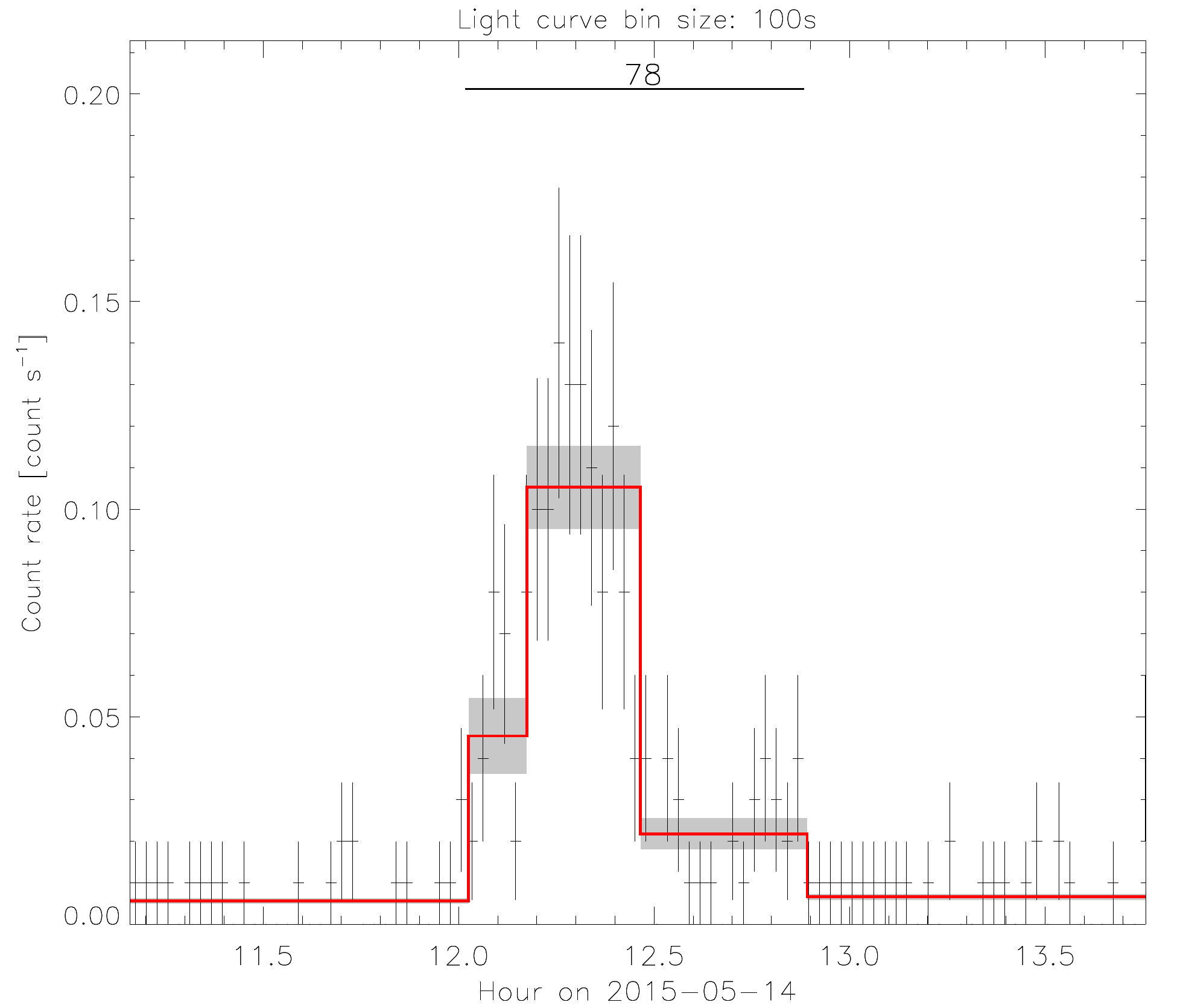}&
\includegraphics[width=0.20\textwidth,angle=90]{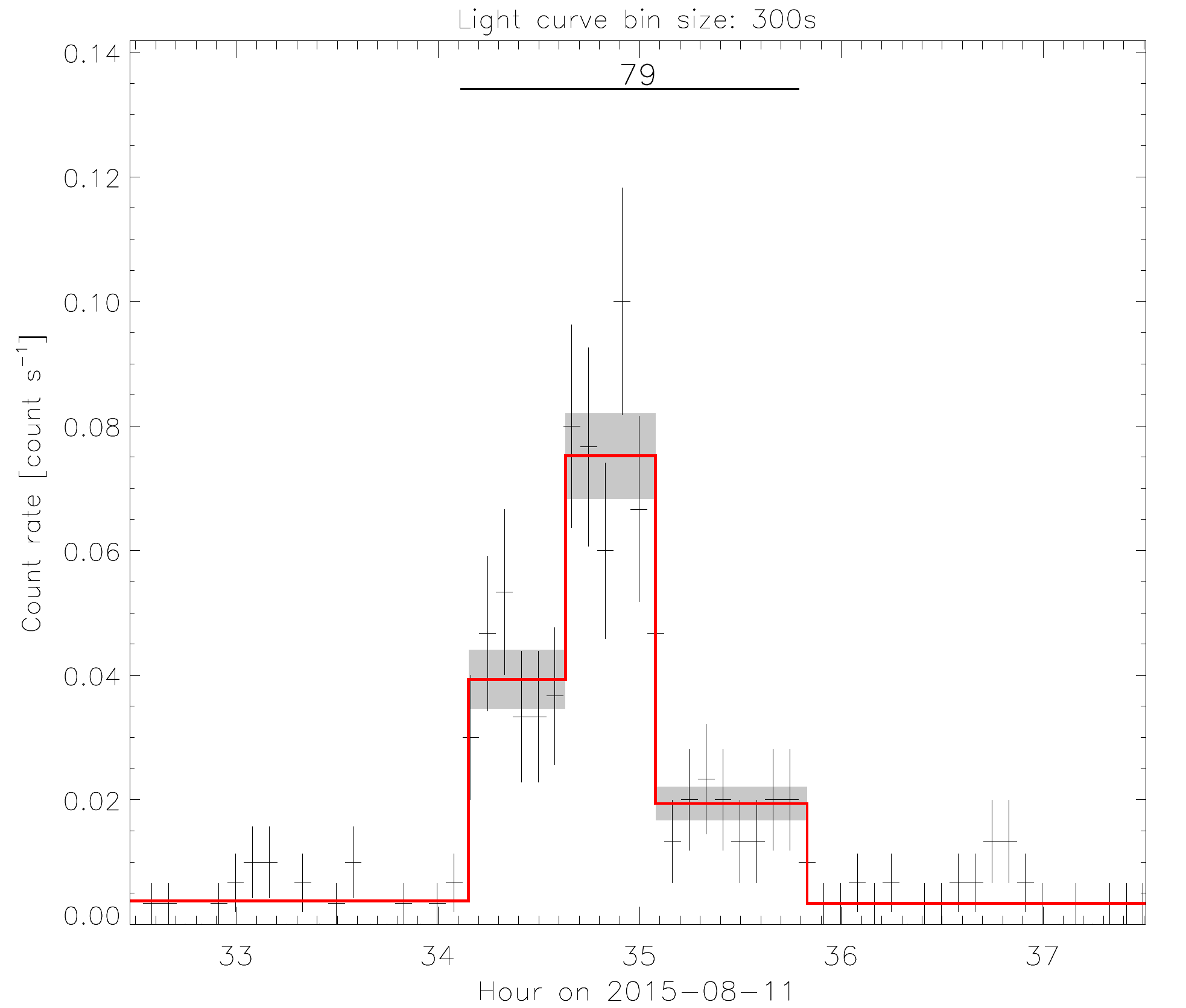}
\end{tabular}
\caption{Continued.}
\end{figure*}

\setcounter{figure}{1}
\begin{figure}[!ht]
\begin{tabular}{@{}ccccc@{}}
\includegraphics[width=0.20\textwidth,angle=90]{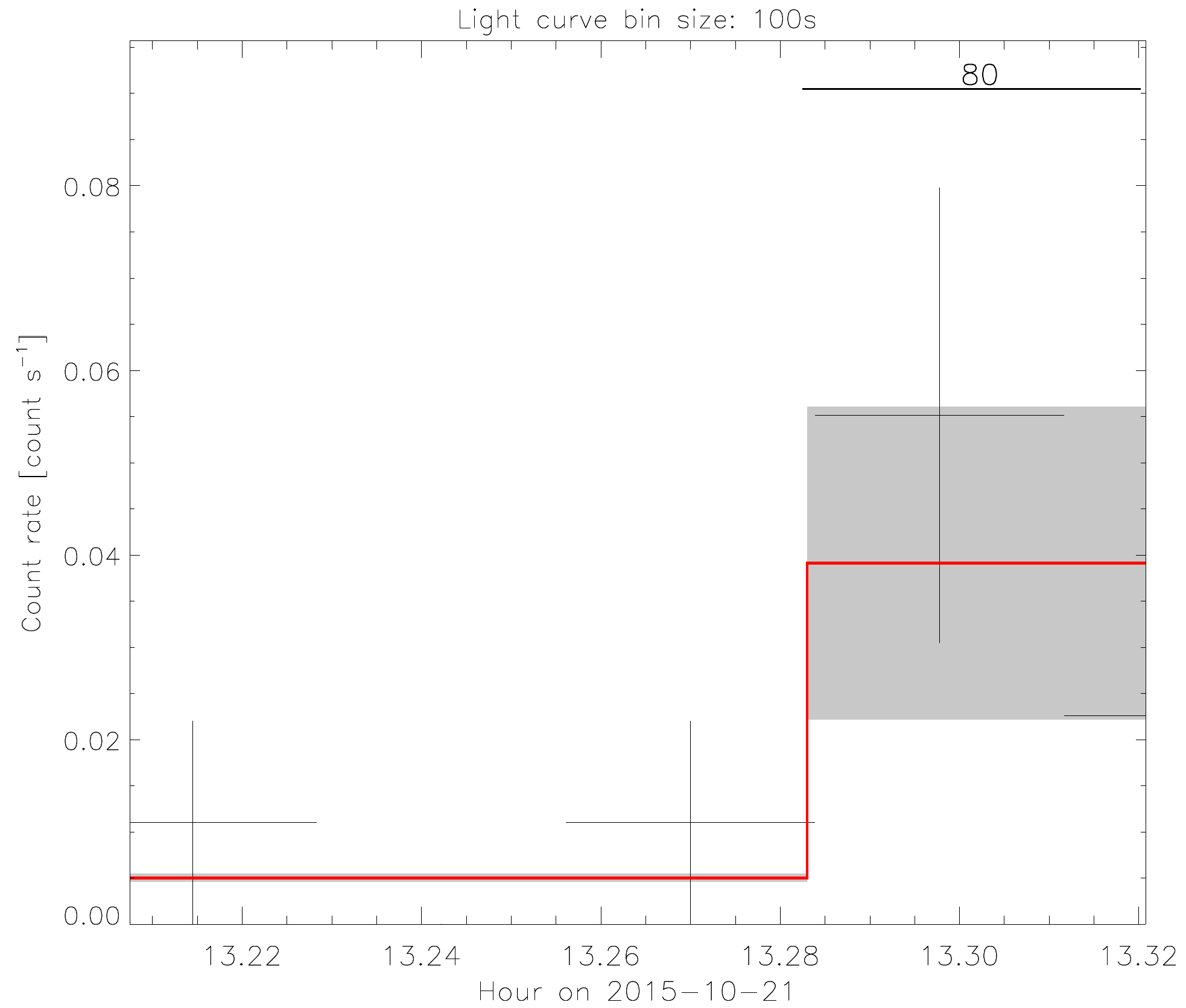}
\end{tabular}
\caption{Continued.}
\end{figure}

\newpage
\textbf{}
\newpage
\textbf{}
\newpage
\textbf{}
\section{Comparison with previous works}
\label{app_comp}
\begin{figure}
\centering
\includegraphics[width=5.1cm,angle=90]{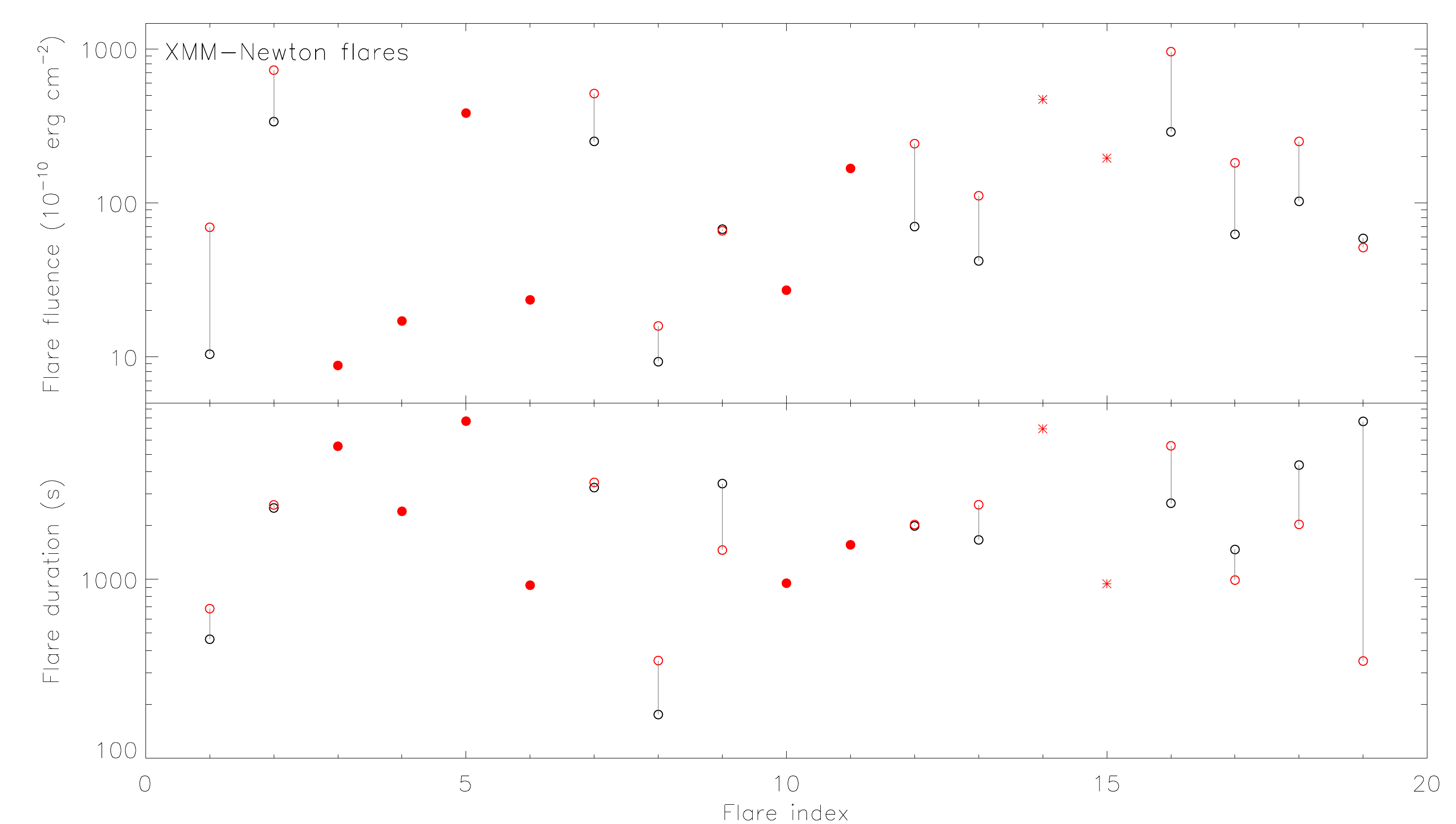}
\caption[]{X-ray flare fluences (top panel) and durations (bottom panel) observed with XMM-Newton from 1999 to 2014.
The red circles indicate the values computed in this work.
The black circles indicate the values computed by \citet{ponti15}.
The red filled circles indicate the flares detected in our work that \citet{ponti15} missed.
The gray lines connect the same flares.
The x-axis reports the flare index corresponding to its numbering in Table~\ref{table:xmm} and Fig.~\ref{fig:xmm_flare}.
The asterisks denote the flares detected in the observations recently released.}
\label{fig:comp_xmm}
\centering
\includegraphics[width=5.1cm,angle=90]{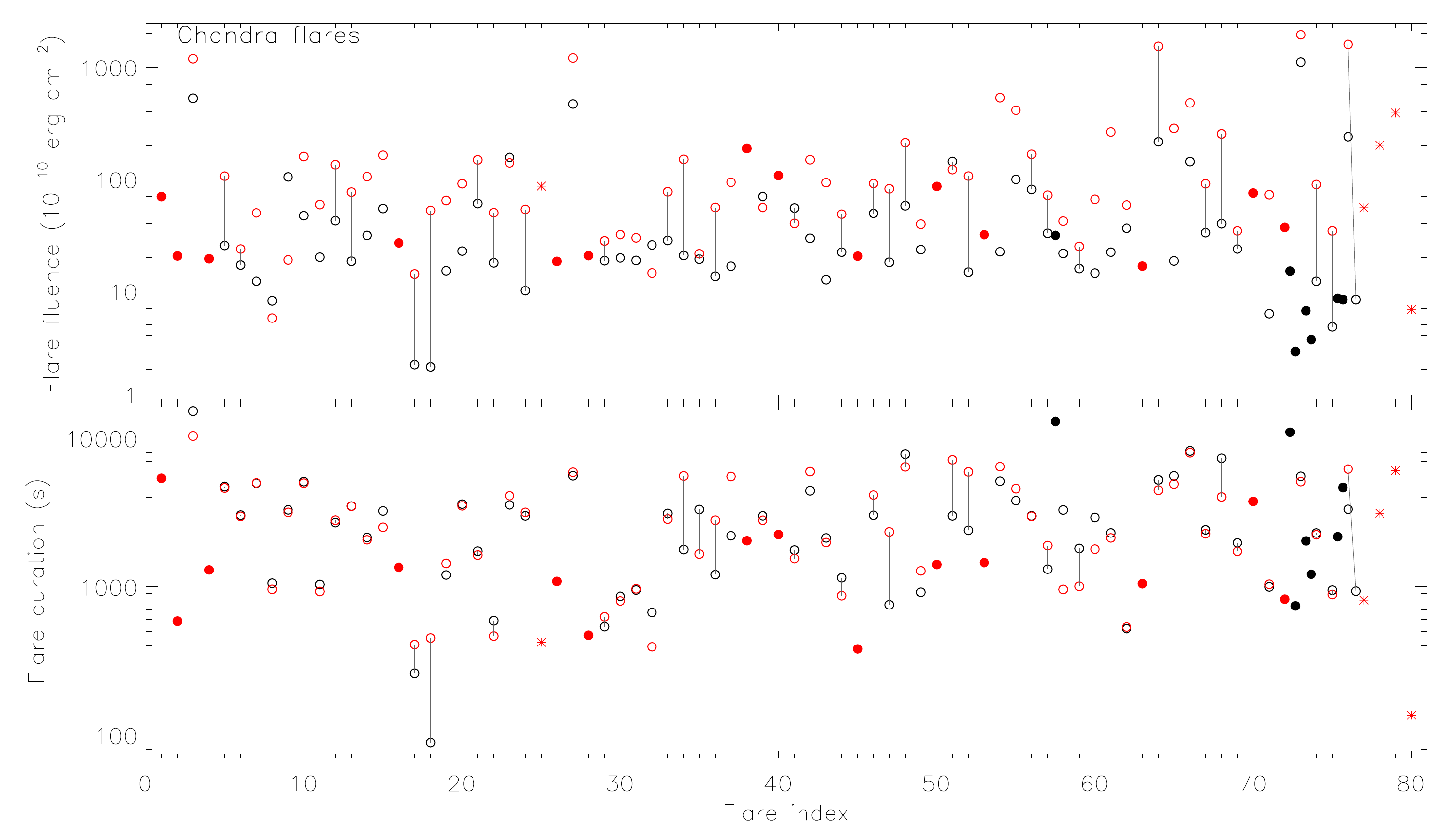}
\caption[]{X-ray flare fluences (top panel) and durations (bottom panel) observed with Chandra from 1999 to 2014.
The red circles indicate the values computed in this work.
The black circles indicate the values computed by \citet{ponti15}.
The black filled circles indicate their putative flares.
The red filled circles indicate the flares detected in our work that \citet{ponti15} missed.
The gray lines connect the same flares.
The x-axis reports the flare index corresponding to its numbering in Table~\ref{table:chandra1} and Fig.~\ref{fig:chandra_flare1}.
The asterisks denote the flares detected in the observations recently released or where \sgra{} has an off-axis angle larger than 2$\arcmin$.}
\label{fig:comp_chandra}
\end{figure}

As explained in the introduction, \citet{ponti15} used the Python implementation of the Bayesian block algorithm to detect the X-ray flares observed by Chandra and XMM-Newton from 1999 to 2014.
We thus compare the duration and fluence of the flares detected here using the two-step Bayesian blocks algorithm (the red circles in Fig.~\ref{fig:comp_chandra}) and those that \citet{ponti15} detected (the black circles in Fig.~\ref{fig:comp_chandra}).
The flares that were detected in both works are connected with a gray line.

During the 37 XMM-Newton observations from 2000 to 2014 that we have in common with \citet{ponti15}, we detected 19 flares with a false positive rate for the flare detection of 0.1\% whereas they detected only 11 flares with a false positive rate for the flare detection of 0.25\%.
\citet{ponti15} missed six of our flares (the red filled circles in Fig.~\ref{fig:comp_xmm})\footnote{Their missed XMM-Newton flares are on 2004 March 31 (flare \# 3 and \#4 in Table~\ref{table:xmm} and Fig.~\ref{fig:xmm_flare}); August\ 31 (\#5); September\ 1 (\#6); 2007 April\ 4 (\#10) and 2009 April\ 3 (\#11).}.
Finally, two additional flares were observed during the eight XMM-Newton observations of 2014 and 2015 recently released (the two asterisks in Fig.~\ref{fig:comp_xmm}).
We missed the flare labeled \#1 in Fig.~3 of \citet{porquet08}. 
However, this flare was detected by those authors when combining the three instruments of EPIC (MOS1, MOS1, and pn) and was confirmed by the simultaneous observations of the near-infrared counterpart with the Hubble Space Telescope.

During the 112 Chandra observations from 1999 to 2014 that we have in common with them, we detected 75 flares whereas they detected only 69 flares.
\citet{ponti15} missed 13 of our flares (the red filled circles in Fig.~\ref{fig:comp_chandra})\footnote{Their missed Chandra flares are on 1999 September\ 21 (flare \#1 in Table~\ref{table:chandra1} and Fig.~\ref{fig:chandra_flare1}); 2000 October\ 27 (\#2); 2006 September\ 25 (\#16); 2012 February\ 9 (\#26);
March\ 17 (\#28); May 12 (\#38); May 19 (\#40); July 21 (\#45); July 25 (\#50) and 28 (\#52); October\ 17 (\#63); 2013 May 25 (\#70); and July 27 (\#72).}.
We considered the 2014 October\ 20 Chandra flare as a single flare (\#15) since the Bayesian block between 14.2 and 14.8 hour is significantly well above the non-flaring level whereas \citet{ponti15} considered it as two flares.
Moreover, since \citet{ponti15} only considered observations where \sgra{} was at less than $2\arcmin$ off-axis angle, they did not study the 2011 July 21 Chandra observation where we detect one flare (\#25, the asterisk at the corresponding flare index in Fig.~\ref{fig:comp_chandra}).
Finally, four additional flares (\#77 to \#80) were observed during the last Chandra observations of 2015 recently released (the three last asterisks in Fig.~\ref{fig:comp_chandra}).

The majority of the flares missed by \citet{ponti15} have already been reported in previous works.
The XMM-Newton flares on 2004 March 31 (\#3 and \#4) and August\ 31 (\#5) were detected by \citet{belanger05} and \citet{porquet05}.
The XMM-Newton flare on 2007 April\ 4  (\#10) was detected and labeled \#5 by \citet{porquet08}.
The Chandra flares on 2012 February\ 9 (\#26), May 12 (\#38), July 25 (\#50), and October\ 17 (\#63) were detected by \citet{neilsen13} and \citet{yuan16}.
The Chandra flares on 2000 October\ 27 (\#2), 2006 September\ 25 (\#16), 2012 May 19 (\#40), July 21 (\#45), and July 28 (\#52) were also detected by \citet{yuan16}.

\citet{ponti15} reported seven putative Chandra flares\footnote{Their putative Chandra flares are on 2012 August\ 4 19:32:37, 2013 August\ 11 10:04:15, 2013 August\ 31 16:07:43, 2013 September\ 20 11:21:00, 2013 October\ 17 16:12:36, 2014 February\ 21 00:51:18, 2014 August\ 30 12:26:19} that we do not confirm with our more robust method (the black filled circles in Fig.~\ref{fig:comp_chandra}).
Their inconsistency in flare detection with respect to previous studies and our work may be explained by their blind use of the geometric prior.
This effect is clearly visible with their putative Chandra flares: the black filled circles in Fig.~\ref{fig:comp_chandra} are clustered when the contribution of the Galactic center magnetar \magn{} in the \sgra{} event lists is high (i.e., between 2013 and 20114).
Owing to the higher noise level of these observations, the absence of calibration of the prior may lead to spurious detection of blocks with a very small increase of the count rate compared to the non-flaring level and thus very low fluence.
Conversely, due to the low signal-to-noise ratio of the Chandra data from \sgra{} before 2013, the calibration of the prior is highly sensitive to the number of events in each observation.
Therefore, \citet{ponti15} missed several Chandra flares owing to the inconsistency between their false positive rate, the number of events in the Chandra observations, and the prior.
This effect has a smaller impact on the XMM-Newton observations due to their higher signal-to-noise ratio.

For the flares in common, the flare durations are roughly consistent but the improved fluences computed in this work are typically larger than those computed in \citet{ponti15} because of their utilization of WebPIMMS for the computation of the flare flux.
Indeed, WebPIMMS considers the effective area and the redistribution matrix computed for an on-axis source and for the entire field of view.
However, the flare events were extracted from a circular region of 1.25 and 10$\arcsec$ radius centered on the source (with a maximum off-axis angle of 2$\arcmin$).
Since the PSF extraction fraction is not corrected by WebPIMMS, the inferred unabsorbed flux is systematically underestimated by \citet{ponti15}.

\section{Simulation of Poisson flux to determine the flare detection efficiency}
\label{app:poisson}
We recall that for homogeneous Poisson flux, i.e., a constant mean count rate $CR$, the average number of recorded events during an exposure $T$ is $N=CR\times T$ with a standard deviation of $\sqrt{N}$.
Therefore, we simulate a constant Poisson flux by first drawing the total number $M$ of events in the simulated event list following a Poisson probability distribution, i.e.,
\begin{equation}
 P(M)=\frac{N^M}{M!}\,e^{-N}\,,
\end{equation}
and then by drawing $M$ values uniformly distributed between 0 and 1 and sorted by ascending order and multiplying them by $T$. 

This two-step method is equivalent to the iterative method of \citet{klein84}, which determines the waiting time before the next event considering their decreasing exponential distribution until the simulated arrival time of the event exceeds the exposure time.
Their resulting total number of events thus follows a Poisson distribution.

To determine the flare detection efficiency, we consider a Gaussian-shaped flare superimposed on a constant level leading to a non-homogeneous Poisson process.
The constant level is characterized by a constant Poisson flux of mean count rate $CR$ during a total observing time $T$ leading to an average number of events $N_\mathrm{c}=CR\times T,$ whereas the flare light curve peaks at $t_\mathrm{peak}$ with a count rate amplitude $A_\mathrm{peak}$ leading to an average number of events,
\begin{equation}
 N_\mathrm{g} = A_\mathrm{peak}  \int_{0}^{T} \! e^{-\frac{(t-t_\mathrm{peak})^2}{2  \sigma^2}} \, dt\,.
\end{equation}
The total number of events $M$ in each simulation thus follows a Poisson distribution of mean $N=N_\mathrm{g}+N_\mathrm{c}$.

We use the inverse method (see \citealt{klein84}, Chapter~7 of \citealt{numerical_recipes} and Fig.~2 of \citealt{harrod13}) based on the reciprocal of the cumulative distribution function (CDF) to simulated the arrival times of these $M$ events.
The CDFs for the non-flaring level and for the flare are, respectively,
\begin{equation}
 CDF_\mathrm{c}(t) = t/T\, ,
\end{equation}
and
\begin{equation}
 CDF_\mathrm{g}(t) = \frac{A_\mathrm{peak}\,\sigma}{N_\mathrm{g}}  \sqrt{\frac{\pi}{2}} \left(\erf\left(\frac{t_\mathrm{peak}}{\sqrt{2}\,\sigma}\right)+\erf\left(\frac{t-t_\mathrm{peak}}{\sqrt{2}\,\sigma}\right)\right)
.\end{equation}
\begin{figure}
\centering
\includegraphics*[angle=90,width=9cm]{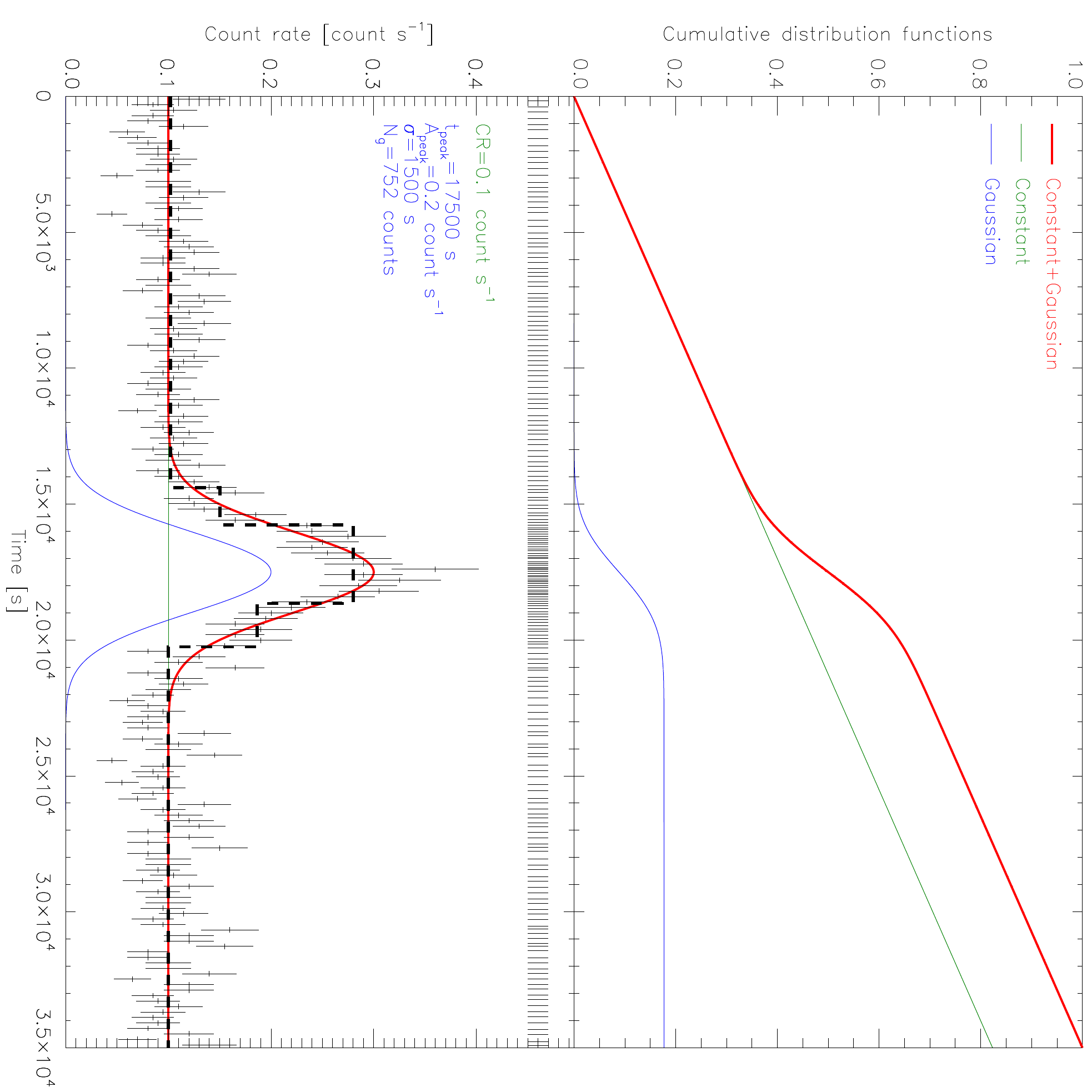}
\caption[The flare light curve simulation]{Simulation of an X-ray light curve with a flare.
\textit{Top panel:} The cumulative distribution function (CDF) of a constant function (green line) representing the non-flaring emission and a Gaussian function (blue line) representing the flaring emission.
The non-flaring emission has a count rate $CR=0.1\,\mathrm{count\,s^{-1}}$.
The flare is defined with a peak amplitude $A_\mathrm{peak}=0.2\,\mathrm{count\,s^{-1}}$ at $t_\mathrm{peak}=17500\,$s and $\sigma=1500\,$s (corresponding to a full width at half maximum of $3532\,$s) leading to a number of counts in the flare of $N_\mathrm{g}=752$.
\textit{Bottom panel:} The constant and Gaussian light curve models are represented in green and blue, respectively.
The red line indicates the model of the light curve with flare.
The ticks at the top of this panel represent 5\% of the simulated arrival times.
The resulting light curve and its error bars are computed for a bin time of 100$\,$s.
The results of the Bayesian block algorithm with a false positive rate for the flare detection of $0.1$\% are shown with dashed lines.}
\label{fig:CDF}
\end{figure}
We combine the constant and Gaussian CDFs as
\begin{equation}
CDF_\mathrm{c+g}(t)=CDF_\mathrm{c}(t)\, \frac{N_\mathrm{c}}{N_\mathrm{g}+N_\mathrm{c}} + CDF_\mathrm{g}(t)\, \frac{N_\mathrm{g}}{N_\mathrm{g}+N_\mathrm{c}}\,.
\end{equation}
We then draw $M$ values of $y$ uniformly distributed between 0 and 1 and sort these values in ascending order.
The corresponding arrival times of the events are finally obtained from $CDF^{-1}_\mathrm{c+g}(y)$.

The top panel of Fig.~\ref{fig:CDF} shows these CDFs for typical exposure of $35\,$ks with a non-flaring level of $CR=0.1\,\mathrm{count\,s^{-1}}$, which corresponds to those observed by XMM-Newton EPIC/pn, and a flare peaking at the exposure center with an amplitude of $A_\mathrm{peak}=0.2\,\mathrm{count\,s^{-1}}$, which corresponds to the mean amplitude measured in the X-ray flares, thus leading to $N_\mathrm{g}=752\,$counts.
The corresponding constant and Gaussian light curve models are shown with the corresponding color in the bottom panel.
The simulated arrival times are the black ticks at the top of the bottom panel of Fig.~\ref{fig:CDF} (only 1 arrival time in 20 are shown here for clarity purpose).
The resulting simulated light curve binned on 100$\,$s is shown in the bottom panel of this figure.
For illustration purpose, the Bayesian blocks computed for a false positive rate for the flare detection of $0.1$\% are also represented in this figure.

\end{document}